%% file: phd_master.tex
\documentclass[12pt]{report}
\include{jens}

\allowdisplaybreaks

\input{phd_newco}

\usepackage{mcite}
\begin{document}

\setlength{\baselineskip}{0.65cm}
\setlength{\parskip}{1ex}
\renewcommand{\arraystretch}{1.3}  

\include{phd_title}

\clearemptydoublepage
\include{phd_motto}

\clearemptydoublepage
\pagenumbering{roman} 
\tableofcontents
\clearemptydoublepage


\include{phd_preface}
\clearemptydoublepage
\pagenumbering{arabic} 
\part{Heavy Quark Production in CC and NC DIS}
\clearemptydoublepage
\include{phd_hqintro}
\clearemptydoublepage
\include{phd_hqcc}
\clearemptydoublepage
\include{phd_hqini}

\clearemptydoublepage
\include{phd_hqfrag}
\clearemptydoublepage
\include{phd_hqsummary}

\clearemptydoublepage
\part{The Structure of Real and Virtual Photons}
\clearemptydoublepage
\include{phd_gam_intro}

\clearemptydoublepage
\include{phd_gam_kinematics}

\clearemptydoublepage
\include{phd_gam_lobox}
\clearemptydoublepage
\include{phd_gam_parton}
\clearemptydoublepage
\include{phd_gam_pion}
\clearemptydoublepage
\include{phd_gam_photon}

\clearemptydoublepage
\include{phd_gam_vgam}

\clearemptydoublepage
\include{phd_gam_summary}

\clearemptydoublepage
\begin{appendix}
\include{phd_app_hqini2}
\clearemptydoublepage
%
\include{phd_app_hqfrag}
\clearemptydoublepage
\include{phd_app_box}

\clearemptydoublepage
\include{phd_app_para}

\end{appendix}
\clearemptydoublepage
%
\bibliographystyle{/home/schien/PhD/Bibliography/test}
\bibliography{/home/schien/PhD/Bibliography/phd_hq,/home/schien/PhD/Bibliography/phd}
\clearemptydoublepage


\include{phd_acknowledgements}

\clearemptydoublepage
\end{document}

%% file: jens.tex
\usepackage{a4}
\usepackage{fancyheadings}
\newcommand{\clearemptydoublepage}{%
    \newpage{\pagestyle{empty}\cleardoublepage}}
\usepackage{amssymb}
\usepackage{amsfonts}
\usepackage{varioref}
\usepackage{citesort}
\usepackage{mcite}
\usepackage{amsmath}
\newlength{\diracchlen}

\newlength{\vecchlen}
\newlength{\vecchhgt}

\usepackage{subfigure}
\usepackage{array}
\usepackage{epic}
\usepackage{eepic}
\usepackage{epsfig}
\usepackage{afterpage}
\usepackage{axodraw}
\makeatletter
\renewcommand{\@makecaption}[2]{%
  \vskip\abovecaptionskip
  \sbox\@tempboxa{#1: \emph{#2}}%
  \ifdim \wd\@tempboxa >\hsize
    #1: \emph{#2}\par
  \else
    \global \@minipagefalse
    \hb@xt@\hsize{\hfil\box\@tempboxa\hfil}%
  \fi
  \vskip\belowcaptionskip}
\makeatother

%% file: phd_newco.tex
\renewcommand{\clearemptydoublepage}{%
    \newpage{\pagestyle{empty}\cleardoublepage}}

\def\Mathematica{{\tt Mathematica}}
\def\Tracer{{\tt Tracer}}

\def\spp{\Sigma_{\scriptscriptstyle ++}}
\def\spm{\Sigma_{\scriptscriptstyle +-}}
\def\smp{\Sigma_{\scriptscriptstyle -+}}

\def\Sp{S_{\scriptscriptstyle +}}
\def\Sm{S_{\scriptscriptstyle -}}
\def\Rp{R_{\scriptscriptstyle +}}
\def\Rm{R_{\scriptscriptstyle -}}
\def\zp{z^\prime}

\def\as{\frac{\alpha_s}{4 \pi}}
\def\xip{\xi^\prime}
\def\zip{z^\prime}
\def\hsi{\hat{s}_1}
\def\hti{\hat{t}_1}
\def\hats{\hat{s}}

\def\delp{{\Delta^\prime}}
\def\LN{\ln\left(\frac{\spp+\hsi-\delp}{\spp+\hsi+\delp}\right)}

\def\msbar{\ensuremath{{\rm{\overline{MS}}}}} 
\def\disg{\ensuremath{{\rm{DIS_\gamma}}}} 
\def\alpsi{\ensuremath{{\cal{O}}(\alpha_s^1)}}

\def\alpsq{\ensuremath{{\cal{O}}(\alpha_s^2)}}
\def\Qt{{\tilde{Q}}}

\def\Oalp{\ensuremath{{\cal{O}}(\alpha)}}
\newcommand{\epsp}[3][ ]{\ensuremath{\epsilon^{#2}_{#3}(#1)}}

\newcommand{\Norm}{\ensuremath{N_c e_q^4 \frac{4 \pi \alpha^2}{Q^2} x\ }}
\newcommand{\Normh}{\ensuremath{N_c e_h^4 \frac{4 \pi \alpha^2}{Q^2} x\ }}
\newcommand{\Normb}{\ensuremath{N_c (\Sigma e_q^4) \frac{4 \pi \alpha^2}{Q^2} x\ }}
\newcommand{\Normc}{\ensuremath{\frac{Q^2}{4 \pi^2 \alpha}}}
\newcommand{\vgvg   }{\ensuremath{\gamma^\star(Q^2)\gamma^\star(P^2)\rightarrow q \bar{q}}}
\newcommand{\vgg   }{\ensuremath{\gamma^\star(Q^2)\gamma \rightarrow q \bar{q}}}
\newcommand{\mev    }{\ensuremath{\mathrm{MeV}}}

\newcommand{\gev    }{\ensuremath{\mathrm{GeV}}}
\newcommand{\gevsq  }{\ensuremath{\mathrm{GeV^2}}}
\newcommand{\me     }{\ensuremath{m_{\mathrm{e}}}}
\newcommand{\vone     }{\ensuremath{\beta}}
\newcommand{\vtwo     }{\ensuremath{\bar{\beta}}}
\newcommand{\xp       }{\ensuremath{\delta}}

\newcommand{\der}{\ensuremath{{\operatorname{d}}}}
\newcommand{\vep}{\ensuremath{{\varepsilon}}}
\newcommand{\Ord}{\ensuremath{{\cal O}}}
\newcommand{\imag}{\ensuremath{{\operatorname{Im}}}}
\newcommand{\mi}{\ensuremath{\mathrm{i}}}
\newcommand{\DISeg}{\ensuremath{\mathrm{DIS}_{e\gamma}}}

\newcommand{\fivg}[1][f]{\ensuremath{{#1}^{\gam(P^2)}}}
\newcommand{\fvg}{\fivg}
\newcommand{\qvg}{\fivg[q]}
\newcommand{\gamvg}{\fivg[\Gamma]}
\newcommand{\gvg}{\fivg[g]}

\newcommand{\fv}{\ensuremath{f^{\gamma}}}
\newcommand{\aemb}{\ensuremath{\frac{\alpha}{2 \pi}}}
\newcommand{\asb}[1][]{\ensuremath{\frac{\alpha_s #1}{2 \pi}}}
\newcommand{\asbq}{\asb[(Q^2)]}
\newcommand{\asbqz}{\asb[(Q_0^2)]}
\newcommand{\pdfb}[3][q]{\ensuremath{{#1}_{\mathrm{#2}}^{#3}}}
\newcommand{\pdf}[2][q]{\ensuremath{{#1}_{\mathrm{#2}}^{\,\gam(P^2)}}}
\newcommand{\qns}{\pdf{NS,\ell}}
\newcommand{\qs}{\pdf[{\vec{q}}]{S}}
\newcommand{\pdfn}[2][q]{\ensuremath{{#1}_{\mathrm{#2}}^{\,\gam(P^2),n}}}
\newcommand{\qnsn}{\pdfn{NS,\ell}}
\newcommand{\qsn}{\pdfn[{\vec{q}}]{S}}
\newcommand{\qsnhad}{\pdfn[{\vec{q}}]{S,had}}
\newcommand{\qsnpl}{\pdfn[{\vec{q}}]{S,pl}}
\newcommand{\qpl}{\ensuremath{q_{\mathrm{pl}}}}
\newcommand{\qhad}{\ensuremath{q_{\mathrm{had}}}}

\newcommand{\pl}{\ensuremath{\mathrm{pl}}}
\newcommand{\had}{\ensuremath{\mathrm{had}}}
\newcommand{\Op}[1][]{\ensuremath{\mathrm{O}_{#1}}}
\newcommand{\mGam}{\ensuremath{\pdfb[\Gamma]{}{n=2}}}
\newcommand{\mg}{\ensuremath{\pdfb[g]{}{n=2}}}
\newcommand{\mS}{\ensuremath{\pdfb[\Sigma]{}{n=2}}}
\newcommand{\mkS}[1][n=2]{\ensuremath{\mathrm{k}_{\Sigma}^{#1}}}
\newcommand{\mkg}[1][n=2]{\ensuremath{\mathrm{k}_{g}^{#1}}}
\newcommand{\mPgamgam}[1][n=2]{\ensuremath{\mathrm{P}_{\gamma\gamma}^{#1}}}
\newcommand{\dn   }{\ensuremath{\hat{d}_n}}
\newcommand{\Pbar      }{\ensuremath{\bar{\mathrm{P}}}}
\newcommand{\Pqigambar   }{\ensuremath{\Pbar_{q_i \gamma}}}
\newcommand{\Pqiqkbar    }{\ensuremath{\Pbar_{q_i q_k}}}
\newcommand{\Pqigbar     }{\ensuremath{\Pbar_{q_i g}}}
\newcommand{\Pggambar    }{\ensuremath{\Pbar_{g \gamma}}}
\newcommand{\Pgqkbar     }{\ensuremath{\Pbar_{g q_k}}}
\newcommand{\Pggbar      }{\ensuremath{\Pbar_{g g}}}
\newcommand{\Pgamgambar  }{\ensuremath{\Pbar_{\gamma \gamma}}}
\newcommand{\Pgamqkbar  }{\ensuremath{\Pbar_{\gamma q_k}}}
\newcommand{\Pgamgbar  }{\ensuremath{\Pbar_{\gamma g}}}

\newcommand{\Pqigam   }{\ensuremath{{\mathrm{P}}_{q_i \gamma}}}
\newcommand{\Pqiqk    }{\ensuremath{{\mathrm{P}}_{q_i q_k}}}
\newcommand{\Pqig     }{\ensuremath{{\mathrm{P}}_{q_i g}}}
\newcommand{\Pggam    }{\ensuremath{{\mathrm{P}}_{g \gamma}}}
\newcommand{\Pgqk     }{\ensuremath{{\mathrm{P}}_{g q_k}}}
\newcommand{\Pgq     }{\ensuremath{{\mathrm{P}}_{g q}}}
\newcommand{\Pgg      }{\ensuremath{{\mathrm{P}}_{g g}}}
\newcommand{\Pqq      }{\ensuremath{{\mathrm{P}}_{q q}}}
\newcommand{\Pgamgam  }{\ensuremath{{\mathrm{P}}_{\gamma \gamma}}}
\newcommand{\Pgamqk  }{\ensuremath{{\mathrm{P}}_{\gamma q_k}}}
\newcommand{\Pgamg  }{\ensuremath{{\mathrm{P}}_{\gamma g}}}

\newcommand{\Kqi  }{\ensuremath{{\mathrm{k}}_{q_i}}}
\newcommand{\Kq  }{\ensuremath{{\mathrm{k}}_{q}}}
\newcommand{\Kg  }{\ensuremath{{\mathrm{k}}_{g}}}
\newcommand{\Ksig  }{\ensuremath{{\mathrm{k}}_{\Sigma}}}
\newcommand{\XiL}{\ensuremath{\alpha_s(Q^2)/\alpha_s(Q_0^2)}}

\newcommand{\X    }{\ensuremath{\mathrm{X}}}

\newcommand{\PWTT}{\ensuremath{\mathrm{P_{\WTT}}^{\mu^\prime\nu^\prime,\mu\nu}}}
\newcommand{\PWTS}{\ensuremath{\mathrm{P_{\WTS}}^{\mu^\prime\nu^\prime,\mu\nu}}}
\newcommand{\PWST}{\ensuremath{\mathrm{P_{\WST}}^{\mu^\prime\nu^\prime,\mu\nu}}}
\newcommand{\PWSS}{\ensuremath{\mathrm{P_{\WSS}}^{\mu^\prime\nu^\prime,\mu\nu}}}
\newcommand{\PtTT}{\ensuremath{\mathrm{P_{\tTT}}^{\mu^\prime\nu^\prime,\mu\nu}}}
\newcommand{\PtTS}{\ensuremath{\mathrm{P_{\tTS}}^{\mu^\prime\nu^\prime,\mu\nu}}}
\newcommand{\PaTT}{\ensuremath{\mathrm{P_{\aTT}}^{\mu^\prime\nu^\prime,\mu\nu}}}
\newcommand{\PaTS}{\ensuremath{\mathrm{P_{\aTS}}^{\mu^\prime\nu^\prime,\mu\nu}}}

\newcommand{\ronepp}{\ensuremath{\rho_1^{++}}}
\newcommand{\ronemm}{\ensuremath{\rho_1^{--}}}
\newcommand{\ronepm}{\ensuremath{\rho_1^{+-}}}
\newcommand{\ronezz}{\ensuremath{\rho_1^{00}}}
\newcommand{\ronepz}{\ensuremath{\rho_1^{+0}}}
\newcommand{\rtwopp}{\ensuremath{\rho_2^{++}}}
\newcommand{\rtwomm}{\ensuremath{\rho_2^{--}}}
\newcommand{\rtwopm}{\ensuremath{\rho_2^{+-}}}
\newcommand{\rtwozz}{\ensuremath{\rho_2^{00}}}
\newcommand{\rtwopz}{\ensuremath{\rho_2^{+0}}}

\newcommand{\ri}{\ensuremath{\rho_{\mathrm{i}}}}
\newcommand{\ripp}{\ensuremath{\rho_{\mathrm{i}}^{++}}}
\newcommand{\ripm}{\ensuremath{\rho_{\mathrm{i}}^{+-}}}
\newcommand{\ripz}{\ensuremath{\rho_{\mathrm{i}}^{+0}}}
\newcommand{\rizz}{\ensuremath{\rho_{\mathrm{i}}^{00}}}

\newcommand{\ricontra}{\ensuremath{\rho_{\mathrm{i}}^{\alpha \beta}}}

\newcommand{\sigiil}{\ensuremath{\sigma_{\mathrm{2L}}}}
\newcommand{\sigiit}{\ensuremath{\sigma_{\mathrm{2T}}}}

\newcommand{\sigtt}{\ensuremath{\sigma_{\mathrm{TT}}}}
\newcommand{\sigtl}{\ensuremath{\sigma_{\mathrm{TL}}}}
\newcommand{\siglt}{\ensuremath{\sigma_{\mathrm{LT}}}}
\newcommand{\sigll}{\ensuremath{\sigma_{\mathrm{LL}}}}
\newcommand{\tautt}{\ensuremath{\tau_{\mathrm{TT}}}}
\newcommand{\tautl}{\ensuremath{\tau_{\mathrm{TL}}}}
\newcommand{\tauatt}{\ensuremath{\tau^{\mathrm{a}}_{\mathrm{TT}}}}
\newcommand{\tauatl}{\ensuremath{\tau^{\mathrm{a}}_{\mathrm{TL}}}}

\newcommand{\Feff}{\ensuremath{F_{\mathrm{eff}}}}
\newcommand{\WIIT}{\ensuremath{W_{\mathrm{2T}}}}
\newcommand{\WIIS}{\ensuremath{W_{\mathrm{2L}}}}
\newcommand{\WTT}{\ensuremath{W_{\mathrm{TT}}}}
\newcommand{\WTS}{\ensuremath{W_{\mathrm{TL}}}}
\newcommand{\WST}{\ensuremath{W_{\mathrm{LT}}}}
\newcommand{\WSS}{\ensuremath{W_{\mathrm{LL}}}}
\newcommand{\tTT}{\ensuremath{W^\tau_{\mathrm{TT}}}}
\newcommand{\tTS}{\ensuremath{W^\tau_{\mathrm{TL}}}}
\newcommand{\aTT}{\ensuremath{W^{\mathrm{a}}_{\mathrm{TT}}}}
\newcommand{\aTS}{\ensuremath{W^{\mathrm{a}}_{\mathrm{TL}}}}

\newcommand{\Wcontra}{\ensuremath{W^{\mu^\prime\nu^\prime,\mu\nu}}}
\newcommand{\Wco}{\ensuremath{W_{\mu^\prime\nu^\prime,\mu\nu}}}
\newcommand{\Tcontra}{\ensuremath{T^{\mu^\prime\nu^\prime,\mu\nu}}}

\newcommand{\ksl}[1][]{\ensuremath{{k\!\!\!\!\;/}_{#1}}}

\newcommand{\ponesl}{\ensuremath{{p\!\!\!\!\;/}_1}}

\newcommand{\ptwosl}{\ensuremath{{p\!\!\!\!\;/}_2}}

\newcommand{\pisl}{\ensuremath{{p\!\!\!\!\;/}_{i}}}
\newcommand{\pipsl}{\ensuremath{{p\!\!\!\!\;/}_{i}^\prime}}

\newcommand{\e}{\ensuremath{e}}
\newcommand{\ee}{\ensuremath{\e^+ \e^-}}
\newcommand{\eetoeeX}{\ensuremath{\e\e \to \e \e X}}
\newcommand{\Fee}{\ensuremath{F_{\e\e}}}
\newcommand{\ft}{\ensuremath{f_{\gamma_T/ \e}}}
\newcommand{\fl}{\ensuremath{f_{\gamma_L/ \e}}}
\newcommand{\qp}{\ensuremath{\xi_q^+}}
\newcommand{\qm}{\ensuremath{\xi_q^-}}
\newcommand{\ppl}{\ensuremath{\xi_p^+}}
\newcommand{\pmi}{\ensuremath{\xi_p^-}}

\newcommand{\qt}{\ensuremath{q_\perp}}
\newcommand{\pt}{\ensuremath{p_\perp}}
\newcommand{\piit}{\ensuremath{{p_2}_\perp}}
\newcommand{\pit}{\ensuremath{{p_1}_\perp}}

\newcommand{\AWTT}{\ensuremath{A_{W_{\mathrm{TT}}}}}
\newcommand{\CWTT}{\ensuremath{C_{W_{\mathrm{TT}}}}}
\newcommand{\EWTT}{\ensuremath{E_{W_{\mathrm{TT}}}}}

\newcommand{\AWTS}{\ensuremath{A_{W_{\mathrm{TL}}}}}
\newcommand{\CWTS}{\ensuremath{C_{W_{\mathrm{TL}}}}}
\newcommand{\EWTS}{\ensuremath{E_{W_{\mathrm{TL}}}}}

\newcommand{\AWST}{\ensuremath{A_{W_{\mathrm{LT}}}}}
\newcommand{\CWST}{\ensuremath{C_{W_{\mathrm{LT}}}}}
\newcommand{\EWST}{\ensuremath{E_{W_{\mathrm{LT}}}}}

\newcommand{\AWSS}{\ensuremath{A_{W_{\mathrm{LL}}}}}
\newcommand{\CWSS}{\ensuremath{C_{W_{\mathrm{LL}}}}}
\newcommand{\EWSS}{\ensuremath{E_{W_{\mathrm{LL}}}}}

\newcommand{\AtTT}{\ensuremath{A_{W^\tau_{\mathrm{TT}}}}}
\newcommand{\CtTT}{\ensuremath{C_{W^\tau_{\mathrm{TT}}}}}
\newcommand{\EtTT}{\ensuremath{E_{W^\tau_{\mathrm{TT}}}}}
\newcommand{\AtTS}{\ensuremath{A_{W^\tau_{\mathrm{TL}}}}}
\newcommand{\CtTS}{\ensuremath{C_{W^\tau_{\mathrm{TL}}}}}
\newcommand{\EtTS}{\ensuremath{E_{W^\tau_{\mathrm{TL}}}}}

\newcommand{\AaTT}{\ensuremath{A_{W^{\mathrm{a}}_{\mathrm{TT}}}}}
\newcommand{\CaTT}{\ensuremath{C_{W^{\mathrm{a}}_{\mathrm{TT}}}}}
\newcommand{\EaTT}{\ensuremath{E_{W^{\mathrm{a}}_{\mathrm{TT}}}}}
\newcommand{\AaTS}{\ensuremath{A_{W^{\mathrm{a}}_{\mathrm{TL}}}}}
\newcommand{\CaTS}{\ensuremath{C_{W^{\mathrm{a}}_{\mathrm{TL}}}}}
\newcommand{\EaTS}{\ensuremath{E_{W^{\mathrm{a}}_{\mathrm{TL}}}}}

\newcommand{\NT}{\ensuremath{\frac{(2 \pi \mu)^{4-n}}{i \pi^2}}}
\newcommand{\eulergam}{\ensuremath{\gamma_{\mathrm{E}}}}

\newcommand{\DUV}{\ensuremath{\Delta_{\mathrm{UV}}}}

\newcommand{\CF}{\ensuremath{\mathrm{C_F}}}
\newcommand{\dps}[2][]{\ensuremath{{\operatorname{d}}^{#1} #2}}

\newcommand{\XX}{\ensuremath{{\cal X}}}

\newcommand{\sfs}[3][F]{\ensuremath{#1_{\text{#2}}^{#3}}}
\newcommand{\gam}[1][]{\ensuremath{\gamma_{\mathrm{#1}}}}

\newcommand{\cq}[1][]{\ensuremath{C_{q,\mathrm{#1}}}}
\newcommand{\cg}[1][]{\ensuremath{C_{g,\mathrm{#1}}}}
\newcommand{\cgam}[1][]{\ensuremath{C_{\gamma,\mathrm{#1}}}}

\newcommand{\bp}[1][]{\ensuremath{\bar{p}_{#1}}}

%% file: phd_title.tex
\begin{titlepage}
\begin{flushright}
\vspace*{-2.cm}
DO-TH 2001/17  \\
hep-ph/0110292\\
October 2001 \\
\end{flushright}
\begin{center}
\vspace*{3cm}
{\LARGE 
{\bf Heavy Quark Production in CC and NC DIS\\
and
\\
The Structure of Real and Virtual Photons
\\
in NLO QCD}\\
}

\vspace*{3.0cm} 
{\Large 
{\bf Dissertation} \\
zur Erlangung des Grades eines\\
Doktors der Naturwissenschaften\\
der Abteilung Physik\\
der Universit\"{a}t Dortmund\\
}

\vspace*{3.cm}
vorgelegt von
\\
\vspace*{0.2cm}
{\Large{\bf Ingo Jan Schienbein}}

\vspace{1.cm}
{\large Juli 2001}
\end{center}
\end{titlepage}

%% file: phd_motto.tex
\thispagestyle{empty}
\vspace*{3cm}
\begin{center}
{\Large
{\bf{ F{\"u}r meinen Vater}}}
\end{center}

\vspace*{3.0cm} 
{\noindent\sf Granger stood looking back with Montag. ``Everyone must leave
something behind when he dies, my grandfather said. A child or a book
or a painting or a house or a wall built or a pair of shoes made.
Or a garden planted. Something your hand touched some way so your soul
has somewhere to go when you die, and when people look at that tree
or that flower you planted, you're there.
It doesn't matter what you do, he said, so long as you change something
from the way it was before you touched it into something that's like you
after you take your hands away. The difference between the man who just
cuts lawns and a real gardener is in the touching, he said.
The lawn--cutter might just as well not have been there at all;
the gardener will be there a life--time.''}

\vspace*{1cm}
\begin{flushright}
Ray Bradbury, Fahrenheit 451
\end{flushright}


%% file: phd_preface.tex
\chapter*{Preface}
\label{preface}
This thesis has been divided into two parts
both being applications of perturbative Quantum Chromo Dynamics (QCD).

The first part 'Heavy Quark Production in CC and NC DIS' is a continuation
of work presented in \cite{Schienbein:dipl} and has been done
in collaboration with S.\ Kretzer.
Part I is based on the following publications 
\cite{Kretzer:1997pd,Kretzer:1998ju,Kretzer:1998nt}:
\begin{itemize}
\item  S.\ Kretzer and I.\ Schienbein, 
    {\it Charged--current leptoproduction of D mesons in the variable flavor scheme}, 
    Phys. Rev. {\bf{D56}}, 1804 (1997) [Chapter~\ref{chap_hqcc}].
\item  S.\ Kretzer and I.\ Schienbein, 
    {\it Heavy quark initiated contributions to deep inelastic structure  functions},
    Phys. Rev. {\bf{D58}}, 094035 (1998) [Chapter~\ref{hqcontrib}].
\item S.\ Kretzer and I.\ Schienbein, 
    {\it Heavy quark fragmentation in deep inelastic scattering},
    Phys. Rev. {\bf{D59}}, 054004 (1999) [Chapter~\ref{hqffdis}].
\end{itemize}

The second part 'The Structure of Real and Virtual Photons' has emerged
from a collaboration with Prof.\ M.\ Gl{\"u}ck and Prof.\ E.\ Reya.
Part II is based on the following publications 
\cite{cit:GRSc99pi,cit:GRSc99,cit:smallx,Gluck:2000sn}:
\begin{itemize}
\item M.\ Gl{\"u}ck, E.\ Reya, and I.\ Schienbein, 
    {\it Pionic parton distributions revisited},
     Eur.\ Phys.\ J. {\bf{C10}}, 313 (1999) [Chapter~\ref{pipdf}].
\item M.\ Gl{\"u}ck, E.\ Reya, and I.\ Schienbein, 
    {\it Radiatively generated parton distributions for real and virtual photons},
    Phys. Rev. {\bf{D60}}, 054019 (1999); (E) {\bf D62} 019902 (2000) 
    [Chapter~\ref{chap:bc}].
\item  M.\ Gl{\"u}ck, E.\ Reya, and I.\ Schienbein, 
    {\it The Photon Structure Function at small--$x$},
    Phys. Rev. {\bf{D64}}, 017501 (2001) [Chapter~\ref{chap:bc}]. 
\item M.\ Gl{\"u}ck, E.\ Reya, and I.\ Schienbein, 
    {\it Has the QCD RG--improved parton content of virtual photons been  observed?},
    Phys. Rev. {\bf{D63}}, 074008 (2001) [Chapter~\ref{chap:vgam}].
\end{itemize}
The results of Refs.~\cite{cit:GRSc99pi,cit:GRSc99} have also been summarized
in \cite{Schienbein:1999af,Schienbein:2000fd}.
The work in Section \ref{sec:fact1} is completely new and has been done 
in collaboration with C.~Sieg.

%% file: phd_hqintro.tex
\chapter{Introduction and Survey}
\label{chap_hqintro}

Part I of this thesis is devoted to heavy quark production
in charged current (CC) and neutral current (NC) deep inelastic
scattering (DIS) where a quark $h$ with mass $m_h$ is viewed as heavy if  
$m_h \gg \Lambda_{\rm QCD}$
contrary to the light $u,d,s$ quarks with $m_{u,d,s} \ll \Lambda_{\rm QCD}$.
Thus, the heavy quark mass provides a hard scale allowing for a perturbative
analysis.
In deep inelastic heavy quark production a second hard scale is given by
the virtuality $Q^2$ of the probing boson (photon, Z--boson, W--boson) such 
that we have to deal with the theoretically interesting problem of 
describing a two--hard--scale process within perturbative QCD (pQCD).
In the following, we will prominently deal with  
charm quarks being the lightest of the heavy quarks ($h=c,b,t$) in 
the standard model.

In addition to these general theoretical arguments,
there are several important phenomenological reasons for studying 
heavy quark production in DIS:
\begin{itemize}
\item Charm contributes up to $30 \%$  
to the total structure function $F_2^p$ at small Bjorken--$x$ 
as measurements at the $ep$ collider HERA at DESY have 
shown \cite{Adloff:1996xq,Breitweg:1997mj,Breitweg:1999ad}. 
For this reason a proper treatment of charm contributions in DIS is 
essential 
for a global analysis of structure function data and
a precise extraction of the parton densities in the proton.
\item NC charm production offers the possibility to extract the gluon 
distribution in the proton from a measurement of $F_2^c$ 
\cite{Vogt:1996wr,Adloff:1998vb}.
%
\item CC charm production is sensitive to the nucleon's strange sea.
The momentum ($z$) distributions of D--mesons from the fragmentation of 
charm quarks produced in neutrino deep inelastic scattering have 
been used recently to determine 
the strange quark distribution of the nucleon $s(x,Q^2)$ at leading order
(LO) \cite{Rabinowitz:1993xx} and next--to--leading order (NLO) 
\cite{Bazarko:1995tt,*Bazarko:1994hm}.
\end{itemize}

In the past few years considerable effort has been devoted to including
heavy quark effects in DIS. 
If one could sum the whole perturbation series (keeping the full mass 
dependence) one would arrive at unique perturbative QCD predictions.
However, at any finite order in perturbation theory
differences arise due to 
distinct
prescriptions (schemes) 
\cite{cit:GRS94,Aivazis:1994kh,Aivazis:1994pi,Collins:1998rz,Thorne:1998ga,*Thorne:1998uu,Buza:1998wv,Martin:1996ev,Kramer:2000hn}
of how to order terms in the perturbative expansion. 
%
These schemes, in turn, enter global analyses of parton distributions
\cite{cit:GRV98,Martin:1998sq,Lai:1999wy} and hence feed back on the resulting 
light parton distributions \cite{Lai:1999wy}.
It is therefore necessary to work out the various schemes as much as 
possible and to compare them  
since a most complete possible understanding 
or --even better-- reduction of the theoretical uncertainties is required
to make concise tests of pQCD predictions against heavy flavor tagged 
deep inelastic data.

In this thesis we concentrate on the ACOT variable flavor number scheme
(VFNS) \cite{Aivazis:1994kh,Aivazis:1994pi} which we work out to full
order $\alpsi$.
Recently, this scheme has been proven by Collins 
\cite{Collins:1998rz} to work at all orders of QCD factorization theory. 
We will perform all required calculations with general couplings and masses
in order to be able to describe both CC and NC processes in one framework.

The outline of Part I will be as follows:
\begin{itemize}
\item In Chapter~\ref{chap_hqcc} 
we present formulae for the momentum ($z$) distributions of D--mesons
produced in neutrino deep inelastic scattering off strange partons. The 
expressions are derived within the variable flavor scheme of Aivazis
et al.\ (ACOT scheme) \cite{Aivazis:1994pi}, which is extended from its fully inclusive
formulation to one--hadron inclusive leptoproduction.   
The dependence of the results on the assumed strange quark
mass $m_s$ is investigated and the $m_s \to 0$ limit is compared to
the corresponding $\overline{\rm{{MS}}}$ results. The importance of 
${\cal{O}}(\alpha_s)$ quark--initiated corrections is demonstrated for the
$m_s=0$ case. 

\item In Chapter~\ref{hqcontrib} 
we perform an explicit calculation of the before missing 
${\cal{O}}(\alpha_s^1 )$ corrections to deep inelastic
scattering amplitudes on massive quarks 
within the ACOT scheme using general masses and couplings
thereby completing
the ACOT formalism up to full ${\cal{O}}(\alpha_s)$.
After identifying the correct subtraction term
the convergence of these contributions towards
the analogous coefficient functions for massless quarks, obtained within
the modified minimal subtraction scheme (${\overline{\rm{MS}}}$), is 
demonstrated. 
Furthermore, the importance of these contributions 
to neutral current and charged current structure functions is 
investigated for several choices of the mass factorization scale $\mu$ 
as well as the relevance of mass corrections.  

\item In Chapter~\ref{hqffdis} we turn to an analysis of
semi--inclusive production of charm 
(momentum ($z$) distributions of D--mesons) in 
neutral current  and charged current deep inelastic scattering 
at full $\alpsi$.
For this purpose we generalize the results of Chapter~\ref{hqcontrib} 
from its fully inclusive formulation to one--hadron inclusive 
leptoproduction.
We review the relevant massive formulae and subtraction terms and discuss 
their massless limits. 
We show how the charm fragmentation function can be measured in CC DIS
and we 
investigate whether the charm production dynamics
may be tested in NC DIS. 
Furthermore, we also discuss finite initial state quark mass effects
in CC and NC DIS.

\item Finally, we summarize our main results in Chapter~\ref{chap_hqsummary}.
Some details of the calculation and lengthy formulae are relegated to the
Appendices \ref{hqini} and \ref{rgediff}.
\end{itemize}


%% file: phd_hqcc.tex
\chapter{Charged Current Leptoproduction of D--Mesons 
in the Variable Flavor Scheme}
\label{chap_hqcc}

\section{Introduction}
The momentum ($z$) distributions of D--mesons from the fragmentation of 
charm quarks produced in neutrino deep inelastic scattering (DIS) have 
been used recently to determine 
the strange quark distribution of the nucleon $s(x,Q^2)$ at leading order
(LO) \cite{Rabinowitz:1993xx} and next--to--leading order (NLO) 
\cite{Bazarko:1995tt,*Bazarko:1994hm}.
A proper QCD calculation of this quantity requires the 
convolution of a perturbative hard scattering charm production cross 
section with a nonperturbative $c \rightarrow D$ fragmentation function
$D_c(z)$ leading at ${\cal{O}}(\alpha_s)$ to the breaking of factorization
in Bjorken--$x$ and $z$ as is well known for light quarks 
\cite{Altarelli:1979kv,cit:FP-8201}. 
So far experimental analyses have assumed a factorized cross section
even at NLO \cite{Bazarko:1995tt,*Bazarko:1994hm}. 
This shortcoming has been pointed out in \cite{Gluck:1997sj} and
the hard scattering convolution kernels needed for a correct NLO analysis
have been calculated there in the  $\overline{\rm{{MS}}}$ scheme with three 
massless flavors ($u,d,s$) using dimensional regularization.  
In the experimental NLO analysis in \cite{Bazarko:1995tt,*Bazarko:1994hm} 
the variable flavor scheme (VFS) 
of Aivazis, Collins, Olness and Tung (ACOT) \cite{Aivazis:1994pi} 
for heavy flavor 
leptoproduction has been utilized. In this formalism one considers, in 
addition to the quark scattering (QS) process, 
e.g.\  $W^+ s \rightarrow c$,
the contribution from the gluon fusion (GF) process 
$W^+ g \rightarrow c \bar{s}$ with its full $m_s$--dependence.
The collinear logarithm  which is already contained in the renormalized
$s(x,Q^2)$ is subtracted off numerically. The quark--initiated 
contributions from the subprocess $W^+ s \rightarrow c g$ 
(together with virtual corrections) which were included in the complete 
NLO ($\overline{\rm{{MS}}}$) analysis in \cite{Gluck:1997sj} are usually 
neglected in the ACOT formalism.
The ACOT formalism has been formulated explicitly only for 
fully inclusive 
leptoproduction \cite{Aivazis:1994pi}. 
It is the main purpose here to fill the gap and
provide the expressions needed for a correct calculation of one--hadron 
(D--meson) inclusive leptoproduction also in this formalism.

\section{Gluon--Fusion}
In the following we will stick closely to the ACOT formalism as formulated
in \cite{Aivazis:1994pi} except that we are not working in the helicity basis 
but prefer the standard tensor basis implying the usual structure functions
$F_{i=1,2,3}$. We are not considering kinematical effects arising from an 
initial state quark mass in 
the $W^+ s \rightarrow c$ quark scattering contribution,
i.e., $s(x,Q^2)$ represents massless initial state strange quarks.
This latter 
choice must be consistently kept in the 
subtraction term \cite{Aivazis:1994pi} to be identified below from the 
$m_s \to 0$
limit of the $W^+ g \rightarrow c \bar{s}$ gluon fusion contribution.
The fully massive partonic matrix elements have been calculated for the 
general boson--gluon fusion process 
$B g \rightarrow {\bar{Q}}_1 Q_2$ in \cite{Leveille:1979px} where 
$B=\gamma^*,\ W^{\pm},\ Z$.
When they are 
convoluted with a nonperturbative gluon distribution $g(x,\mu^2)$ and a 
fragmentation function $D_{Q_2}(z)$, one obtains the GF part of the 
hadronic structure
function $F_i(x,z,Q^2)$ describing the momentum ($z$) distribution of a 
hadron $H$ containing the heavy quark $Q_2$:
\begin{eqnarray}  \nonumber
F_{1,3}^{GF}(x,z,Q^2) &=& \int_{ax}^{1} \frac{dx'}{x'} 
              \int_{max[z,\zeta_{min}(x/x')]}^{\zeta_{max}(x/x')}
      \ \frac{d\zeta}{\zeta}\ g(x',\mu^2)\ f_{1,3}(\frac{x}{x'},\zeta,Q^2)
      \ D_{Q_2}(\frac{z}{\zeta})  \\ \nonumber \\
F_{2}^{GF}(x,z,Q^2) &=& \int_{ax}^{1} \frac{dx'}{x'} 
              \int_{max[z,\zeta_{min}(x/x')]}^{\zeta_{max}(x/x')}
      \ \frac{d\zeta}{\zeta}\ x'g(x',\mu^2)\ f_{2}(\frac{x}{x'},\zeta,Q^2)
      \ D_{Q_2}(\frac{z}{\zeta})
\label{FiGF}
\end{eqnarray}
\noindent
with the fractional momentum variables  $z=p_H\cdot p_N / q\cdot p_N$
and $\zeta =p_{Q_1}\cdot p_N / q\cdot p_N$,
$p_N$ and $q$ being the momentum of the nucleon and the virtual boson,
respectively. 
The structure functions $F_i(x,z,Q^2)$ generalize the usual 
fully inclusive structure functions $F_i(x,Q^2)$, 
if one considers one--hadron (H) inclusive leptoproduction.  
The partonic structure functions $f_i(x',\zeta,Q^2)$ are given by
\begin{equation}
f_{i=1,2,3}\left(x',\zeta,Q^2\right)=
\frac{\alpha_s(\mu^2)}{\pi}\ \left[\ \frac{A_i}
{(1-\zeta)^2}+\frac{B_i}{\zeta^2}+\frac{C_i}{1-\zeta}
+\frac{D_i}{\zeta}+E_i\ \right] 
\label{fiGFz}
\end{equation}
\noindent
with
\begin{eqnarray*}
A_1\left(x',Q^2\right)&=&q_{+}\ \frac{{x'}^2}{4}\ \frac{m_1^2}{Q^2}
\ \left(\ 1+
\frac{\Delta m^2}{Q^2}-\frac{q_{-}}{q_{+}}\frac{2 m_1 m_2}{Q^2}\right)\\
C_1\left(x',Q^2\right)&=&\frac{q_{+}}{4}\ \Bigg[\ \frac{1}{2}-x'(1-x')
-\frac{\Delta m^2\,x'}
{Q^2}(1-2x')+\left(\frac{\Delta m^2\,x'}{Q^2}\ \right)^2  \\
&+& \frac{q_{-}}{q_{+}}\ \frac{m_1 m_2}{Q^2}
\ 2x'\ (1-x'-x'\ \frac{m_1^2+m_2^2}{Q^2})\ \Bigg] \\
E_1\left(x',Q^2\right)&=&\frac{q_{+}}{4}\ (\ -1+2x'-2{x'}^2\ )\\
A_2\left(x',Q^2\right)&=&q_{+}\ x'\ \left[{x'}^2\ \frac{m_1^2}{Q^2}
\left(\frac{1}{2}
\left(\frac{\Delta m^2}{Q^2}\right)^2+\frac{\Delta m^2\,-m_1^2}{Q^2}
+\frac{1}{2}\right)\right]\\
C_2\left(x',Q^2\right) &=& q_{+}\ \frac{x'}{4}\ \Bigg[\ 1-2x'(1-x')
+\frac{m_1^2}{Q^2}
\left(1+8x'-18{x'}^2\right) \\
&+& \frac{m_2^2}{Q^2}\left(1-4x'+6{x'}^2\right) 
- \frac{m_1^4+m_2^4}{Q^4}\ 2x'(1-3x')+\frac{m_1^2 m_2^2}{Q^4}\ 4x'(1-5x')\\
&+&\frac{\Delta m^4\,\Delta m^2}{Q^6}2{x'}^2 
- \frac{q_{-}}{q_{+}}\ \frac{2 m_1 m_2}{Q^2}\ \Bigg] \\
E_2\left(x',Q^2\right)&=&q_{+}\ x'\ \left[-\frac{1}{2}+3x'(1-x')\right]\\
A_3\left(x',Q^2\right)&=&R_q\ m_1^2\ {x'}^2\ \frac{\Delta m^2 +Q^2}{Q^4}\\
C_3\left(x',Q^2\right)&=&R_q\left[\ \frac{1}{2}-x'(1-x')
-\frac{\Delta m^2}{Q^2}
\ x'(1-2x')+\frac{\Delta m^4}{Q^4}\ {x'}^2\right]\\
E_3\left(x',Q^2\right)&=&0 \\
B_{i={1,2 \atop 3}}\left(x',Q^2\right)&=&\pm A_i\left(x',Q^2\right)
[m_1\leftrightarrow m_2]\\
D_{i={1,2 \atop 3}}\left(x',Q^2\right)&=&\pm C_3\left(x',Q^2\right)
[m_1\leftrightarrow m_2]
\end{eqnarray*}
\noindent
where $\Delta m^n \equiv m_2^n-m_1^n\ ,\ m_{1,2}$ being the mass of
the heavy quark $Q_{1,2}$. The kinematical boundaries of phase space 
in the convolutions in Eq.~(\ref{FiGF}) are
\begin{equation}
a x = \left[1+\frac{(m_1+m_2)^2}{Q^2}\right]\ x\ \ \ ,\ \  
\ \zeta_{min,max}(x')=\frac{1}{2} \left[ 1+
\frac{\Delta m^2}{Q^2} \frac{x'}{1-x'} \pm v {\bar{v}} \right]
\label{kseq3}
\end{equation}
\noindent
with $ \displaystyle \quad
v^2=1-\frac{(m_1+m_2)^2}{Q^2} \frac{x'}{1-x'} \ \ ,    
\ \ {\bar{v}}^2=1-\frac{(m_1-m_2)^2}{Q^2} \frac{x'}{1-x'} \quad $.
The vector ($V$) and axialvector ($A$) couplings of the 
$\gamma_{\mu}(V-A \gamma_5)$ quark current enter via 
$q_{\pm}=V^2\pm A^2$, $R_q=V A$. 
\footnote{For the $\gamma^* Z$ interference term the couplings read: 
$q_{\pm}=V^{\gamma}V^{Z}\pm A^{\gamma}A^{Z}$, $R_q=1/2(V^{\gamma}
A^{Z}+V^{Z}A^{\gamma})$.}
If the partonic structure functions in Eq.~(\ref{fiGFz}) are integrated over
$\zeta$ the well known inclusive structure functions 
\cite{Gluck:1988uk,Schuler:1988wj,Baur:1988ai} 
for heavy flavor production are recovered:
\begin{equation}
\int_{\zeta_{min}(x')}^{\zeta_{max}(x')} d\zeta
\ f_{i={1,2 \atop 3}}(x',\zeta,Q^2) = \pm f_i(x',Q^2)
\ \ \ ,
\label{fiGF}
\end{equation}
where the   $f_i(x',Q^2)$ can be found in \cite{Gluck:1988uk}.

\section{CC Leptoproduction within ACOT}
In the following we will consider the special case of charged current
charm production, i.e., $m_1=m_s$, $m_2=m_c$ ($q_{\pm}=2,0$; $R_q=1$
assuming a vanishing Cabibbo angle). Of course, all formulae below can be 
trivially adjusted to the general case of Eqs.\ (\ref{FiGF}) and (\ref{fiGFz}).
The $m_s \to 0$ limit of the partonic structure functions in 
Eq.~(\ref{fiGFz}) is obtained by keeping terms up to ${\cal{O}}(m_s^2)$  in the 
$A_i$,
$C_i$ and in $\zeta_{max}$ due to the singularity of the phase space
integration stemming from  $\zeta \to 1$. One obtains
\begin{eqnarray} \nonumber
\lim_{m_s \to 0}\ 
\frac{\pi}{\alpha_s}\ f_{i}(x',\zeta,Q^2) &=&
c_i\ H_i^g(\frac{x'}{\lambda},\zeta,m_s^2,\lambda) \\
&=& c_i\ \delta (1-\zeta )\ P_{qg}^{(0)}(\frac{x'}{\lambda}) 
\ \ln \frac{Q^2+m_c^2}{m_s^2} + {\cal{O}}(m_s^0)
\label{mlimit}
\end{eqnarray}
\noindent
where $\displaystyle P_{qg}^{(0)}(x') = \frac{1}{2} [{x'}^2+(1-{x'}^2)]$, 
$\lambda=Q^2/(Q^2+m_c^2)$, $c_1=1/2$, $c_2=x'/\lambda$, $c_3=1$
and the $H_i^g$ are the dimensionally regularized 
$\overline{\rm{{MS}}}$ ($m_s=0$) gluonic 
coefficient functions obtained in \cite{Gluck:1997sj}. The $c_i$ arise from 
different normalizations of the $f_i$ and the $H_i^g$ and are such that
the infrared--safe subtracted [see Sec.~\ref{hqcc_sec_numres}] 
convolutions in Eq.~(\ref{FiGF})
converge towards the corresponding ones in \cite{Gluck:1997sj} as
$m_s \to 0$ if one realizes
that  $x'/\lambda=\xi'$, $x/\lambda=\xi\equiv x(1+m_c^2/Q^2)$.
Taking also the limit $m_c \to 0$ in Eq.~(\ref{mlimit}) gives --besides the 
collinear logarithm already present in Eq.~(\ref{mlimit})-- finite expressions
which agree 
with the massless results of 
\cite{Altarelli:1979kv,cit:FP-8201}. 

In the ACOT formalism the GF convolutions in Eq.~(\ref{FiGF}) coexist with the
Born level quark scattering contributions 
$F_i^{QS}(x,z,Q^2) = k_i\ s(\xi,\mu^2)\ D_c(z)$, 
$k_{i=1,2,3}=1,\ 2\xi,\ 2$. The overlap between the QS and the GF 
contributions is removed by introducing a subtraction term (SUB) 
\cite{Aivazis:1994pi} which is obtained from the massless limit in 
Eq.~(\ref{mlimit}) 
\begin{equation}
F_i^{SUB}=k_i\  \frac{\alpha_s(\mu^2)}{2\pi}\ \ln\frac{\mu^2}{m_s^2}\
\left[
\int_{\xi}^1 \frac{dx'}{x'}\ g(x',\mu^2)
\ P_{qg}^{(0)}\left(\frac{\xi}{x'}\right)\right]\ D_c(z)\ \ \ .
\label{FiSUB}
\end{equation}
The complete ${\cal{O}}(\alpha_s)$ structure functions for the $z$
distribution of charmed hadrons (i.e., dominantly D--mesons) produced
in charged current DIS are then given in the ACOT formalism 
\cite{Aivazis:1994pi} by
\begin{equation}
F_i^{ACOT}=F_i^{QS}-F_i^{SUB}+F_i^{GF}\ \ \ .
\label{FiACOT}
\end{equation}

It is worthwhile noting that 
general results for polarized partonic structure functions 
analogous to the unpolarized ones in Eq.~(\ref{fiGFz})
have been
obtained in \cite{Schienbein:1997sb} allowing for a formulation of 
polarized one--hadron inclusive heavy flavor leptoproduction within the
ACOT scheme along the same lines.

\section{Numerical Results}
\label{hqcc_sec_numres}
\begin{figure}[ht]
\vspace*{-1.cm}
\hspace*{2.cm}
\epsfig{figure=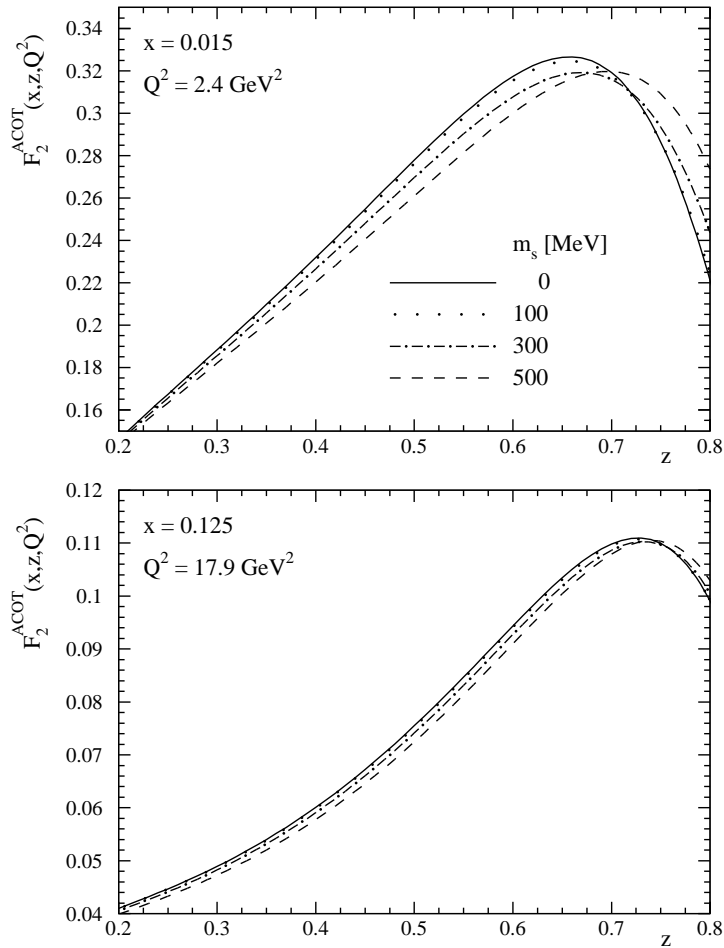,width=11.cm}
\vspace*{-2.cm}
\caption{\label{ksfig1}\sf
The structure function $F_2^{ACOT}(x,z,Q^2)$ as defined
in Eq.~(\ref{FiACOT}) 
using the GRV94(HO) parton densities \protect\cite{cit:GRV94} with 
$m_c=1.5\ {\rm{GeV}}$ and a Peterson fragmentation function 
\protect\cite{Peterson:1983ak}
with $\varepsilon_c = 0.06$. Several finite choices for $m_s$ are shown as
well as the asymptotic $m_s \to 0$ limit. }
\end{figure}

In Fig.~\ref{ksfig1} 
we show the structure function 
$F_2^{ACOT}$ at experimentally 
relevant \cite{Rabinowitz:1993xx} values of $x$ and $Q^2$  
for several finite choices of $m_s$ 
together with the asymptotic $m_s \to 0$ limit. For $D_c$  we use
a Peterson fragmentation function \cite{Peterson:1983ak} 
\begin{equation}
D_c(z) = N \left\{ z \left[ 1-z^{-1}-\varepsilon_c/(1-z)
\right]^2\right\}^{-1}
\label{Peterson-Frag}
\end{equation}
with $\varepsilon_c=0.06$ \cite{Chrin:1987yd,Bethke:1985sg,Hernandez:1990yc} 
normalized to 
$\int_0^1 dz D_c(z) = 1$ 
and we employ the GRV94(HO) parton distributions \cite{cit:GRV94} with
$m_c=1.5\ {\rm{GeV}}$. Our choice of 
the factorization scale is $\mu^2=Q^2+m_c^2$ which ensures that there is
no large $\ln (Q^2+m_c^2)/\mu^2$ present in the difference GF-SUB.    
As can be seen from Fig.~\ref{ksfig1}  the effects of a finite
strange mass are small and converge rapidly towards the massless
$\overline{\rm{{MS}}}$ limit provided $m_s \lesssim 200\ \mev$ as
is usually assumed \cite{Bazarko:1995tt,*Bazarko:1994hm}.

\begin{figure}[t]
\vspace*{-1.cm}
\hspace*{2.cm}
\epsfig{figure=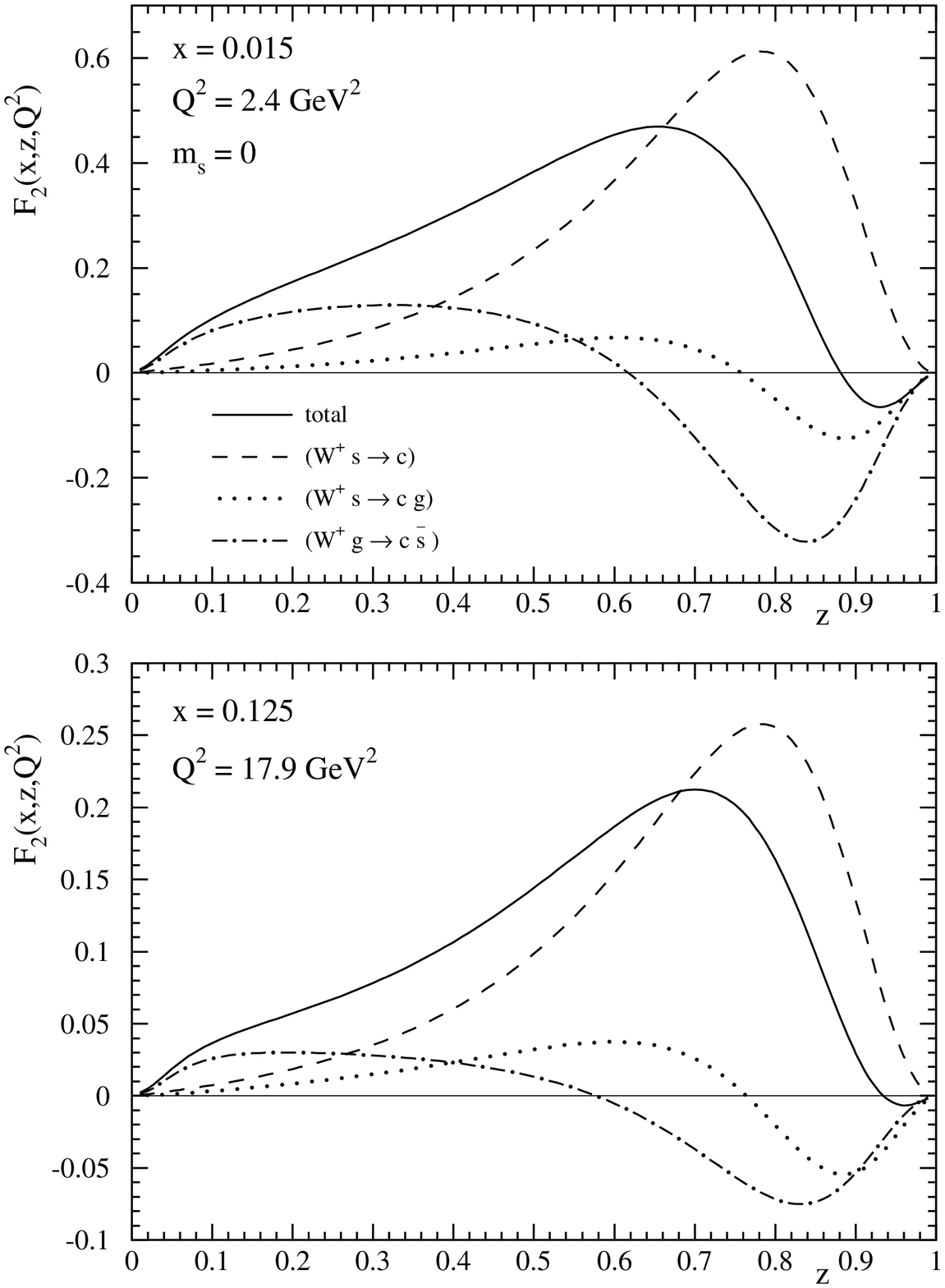,width=11.cm}
\vspace*{-2.cm}
\caption{\label{ksfig2}\sf
The structure function $F_2(x,z,Q^2)$ for charged current
leptoproduction of D--mesons at ${\cal{O}}(\alpha_s)$ for $m_s=0$ using
the CTEQ4($\overline{\rm{{MS}}}$) parton distributions 
\protect\cite{Lai:1997mg} and
a Peterson fragmentation function \protect\cite{Peterson:1983ak} 
with $\varepsilon_c = 0.06$.
The full
${\cal{O}}(\alpha_s)$ result is shown as well as the individual 
contributions from the distinct quark-- and gluon--initiated processes.}
\end{figure}
In Fig.~\ref{ksfig2} 
we show the effects of adding the quark--initiated 
${\cal{O}}(\alpha_s)$ correction 
from the process  $W^+ s \rightarrow c g$ (together with virtual 
corrections) to the asymptotic ($m_s\to 0$) $F_2^{ACOT}$,
employing the CTEQ4($\overline{\rm{{MS}}}$)
densities \cite{Lai:1997mg} with $m_c=1.6\ {\rm{GeV}}$.
The  ${\cal{O}}(\alpha_s)$ quark contribution is usually neglected in the 
ACOT formalism since 
it is assumed to be effectively suppressed by one order of 
$\alpha_s$ with respect to the gluon fusion contribution due to
$s(x,\mu^2)/g(x,\mu^2)\simeq {\cal{O}}(\alpha_s)$. To check this assumption
for the quantity $F_2(x,z,Q^2)$ we show, besides the full result, the 
contributions from the distinct processes (using again $\mu^2=Q^2+m_c^2$).
The $W^+ g \rightarrow c {\bar{s}}$ contribution corresponds to 
GF-SUB in Eq.~(\ref{FiACOT}).
The quark--initiated ${\cal{O}}(\alpha_s)$ contribution has been calculated
in the $\overline{\rm{{MS}}}$ scheme according to \cite{Gluck:1997sj} 
which is consistent with the 
asymptotic gluon--initiated correction in the ACOT scheme due to 
Eq.~(\ref{mlimit}).
It can be seen that the quark--initiated correction is comparable in size
to the gluon--initiated correction around the maximum of $F_2$. Since 
most of the experimentally measured D--mesons originate from this region
the ${\cal{O}}(\alpha_s)$ quark contributions should not be neglected in a 
complete NLO calculation. 

\section{Summary}
To summarize we have given formulae which extend the ACOT scheme 
\cite{Aivazis:1994pi} for the
leptoproduction of heavy quarks from its fully inclusive formulation 
to one--hadron inclusive leptoproduction. We have applied 
this formulation to 
D--meson production in charged current DIS and studied finite $m_s$
corrections to the asymptotic $m_s \to 0$ limit. The corrections 
turned out to be small for reasonable choices of 
$m_s \lesssim 200\ {\rm{MeV}}$ and we have shown that the  
$m_s \to 0$ limit  
reproduces the dimensionally regularized $\overline{\rm{{MS}}}$
($m_s=0$) gluonic coefficient functions \cite{Gluck:1997sj}.   
Furthermore we have investigated the
quark--initiated ${\cal{O}}(\alpha_s)$ corrections for $m_s=0$ using
the relevant $\overline{\rm{{MS}}}$ fermionic coefficient functions
\cite{Gluck:1997sj}. 
The latter corrections turned out to be numerically important 
at experimentally relevant values of $x$ and $Q^2$ \cite{Rabinowitz:1993xx} 
and should
be included in a complete NLO calculation of charged current 
leptoproduction of D--mesons.
In the next chapter, the quark--initiated diagrams will be calculated 
up to $\Ord(\alpha_s)$ in the ACOT scheme
allowing 
to study finite $m_s$ effects stemming from these contributions as
has been done in this chapter for the ${\cal{O}}(\alpha_s)$ gluon 
contributions.

%% file: phd_hqini.tex
\chapter{Quark Masses in Deep Inelastic Structure Functions}
\label{hqcontrib}

In this chapter we consider heavy quark contributions to inclusive
deep inelastic structure functions within the ACOT variable flavor 
number scheme to which we contribute the calculation of the 
before missing Bremsstrahlung corrections (incl.\ virtual graphs)
off initial state massive quarks. The calculation exemplifies 
factorization with massive quark--partons as has recently been proven 
to all orders in perturbation theory by Collins \cite{Collins:1998rz}.   
After identifying the correct subtraction term
the convergence of these contributions towards
the analogous coefficient functions for massless quarks, obtained within
the modified minimal subtraction scheme (\msbar), is 
demonstrated. Furthermore, the phenomenological relevance of the 
contributions to neutral current (NC) and charged current (CC) 
structure functions is investigated for several choices of the 
factorization scale.  
The results presented in this chapter are taken from 
Ref.~\cite{Kretzer:1998ju}.
\section{Introduction}
\label{intro2}

Leptoproduction of heavy quarks has become a subject of major interest
in QCD 
phenomenology both for experimental and theoretical reasons.
Heavy quark contributions are an important component of measured
neutral current (NC) 
\cite{Adloff:1996xq,Breitweg:1997mj,Breitweg:1999ad} 
and charged current (CC) \cite{Seligman:1997mc,*Seligman:1997fe} 
deep inelastic (DI) structure 
functions at lower values of Bjorken--$x$, accessible to present experiments.  
Charm tagging in NC and CC deep inelastic scattering (DIS) offers the 
possibility
to pin down the nucleon's gluon \cite{Blumlein:1996ef} and strange sea 
\cite{Abramowicz:1982zr,Rabinowitz:1993xx,Bazarko:1995tt,*Bazarko:1994hm,Vilain:1998uw}
density, respectively, 
both of which
are nearly unconstrained by global fits to inclusive DI data. 
Theoretically it is challenging to understand the production mechanism
of heavy quarks within perturbative QCD. 
The cleanest and most predictive method \cite{cit:GRS94}
of calculating heavy quark contributions to structure
functions seems to be fixed order perturbation theory (FOPT) where
heavy quarks are produced exclusively by operators built from light
quarks (u,d,s) and gluons (g) and no initial state heavy quark lines show 
up in any Feynman diagram. Heavy quarks produced via FOPT are 
therefore also
called `extrinsic' since no contractions of heavy quark operators
with the nucleon wavefunction are considered (which in turn would
be characteristic for `intrinsic' heavy quarks). 
Besides FOPT much effort has been spent on formulating
variable flavor number  
schemes (VFNS) 
\cite{Aivazis:1994kh,Aivazis:1994pi,Thorne:1998ga,*Thorne:1998uu,Martin:1996ev,Buza:1998wv} 
which aim at resumming  the quasi--collinear logs [$\ln (Q^2/m^2)$; $Q$ and
$m$ being the virtuality of the mediated gauge boson and the 
heavy quark mass, respectively] arising at
any order in FOPT. All these schemes have in common that 
extrinsic FOPT induces
the boundary condition \cite{Collins:1986mp,*Qian:1984kf,Buza:1998wv}
\begin{equation}
\label{boundary}
q(x,Q_0^2=m^2)=0+{\cal{O}}(\alpha_s^2) 
\end{equation}
for an intrinsic heavy quark
density, which then undergoes massless renormalization group (RG)
evolution. Apart from their theoretical formulation VFNS have to be
well understood phenomenologically for a comparison with FOPT and with
heavy quark tagged DI data. 
We will concentrate here on the scheme developed
by Aivazis, Collins, Olness and Tung (ACOT) 
\cite{Aivazis:1994kh,Aivazis:1994pi,Collins:1998rz}. 
In the ACOT scheme full
dependence on the heavy quark mass is kept in graphs containing heavy
quark lines. This gives rise to the above mentioned quasi--collinear logs
as well as to power suppressed terms of ${\cal{O}} [(m^2/Q^2)^k]$. 
While the latter give mass corrections to the massless, dimensionally
regularized, standard coefficient functions (e.g. in the 
${\overline{{\rm{MS}}}}$ scheme), the former are removed by numerical 
subtraction since the collinear region of phase space is already contained in 
the RG evolution of the heavy quark density. 
Up to now explicit expressions
in this scheme exist for DIS on a heavy quark at ${\cal{O}}(\alpha_s^0 )$ 
\cite{Aivazis:1994kh} as well as for the production of heavy quarks
via virtual boson gluon fusion (GF) at ${\cal{O}}(\alpha_s^1 )$
\ \cite{Aivazis:1994pi}.
In Section \ref{heavy}
we will give expressions which complete the scheme
up to ${\cal{O}}(\alpha_s^1 )$ and calculate DIS on a heavy quark at
first order in the strong coupling, i.e. $B^\ast Q_1 \rightarrow Q_2 g$
(incl.\ virtual corrections to $B^\ast Q_1 \rightarrow Q_2$) with general 
couplings of the virtual boson $B^\ast$ to the heavy quarks, keeping
all dependence on the masses $m_{1,2}$ of the quarks $Q_{1,2}$. 
It is unclear whether (heavy) quark scattering (QS) and GF at
${\cal{O}}(\alpha_s^1 )$ should be considered on the same level in the
perturbation series. Due to its extrinsic prehistory 
${\rm{QS}}^{(1)} $ (bracketed upper indices count 
powers\footnote{
For the reasons given here we refrain in this chapter
in most cases from using the standard terminology of 
`leading' and `next--to--leading' contributions and count explicit powers of $\alpha_s$.}
of $\alpha_s$)
includes a collinear subgraph of ${\rm{GF}}^{(2)}$,
e.g.\ $\gamma^\ast c\rightarrow c g$ contains the part of 
$\gamma^\ast g \rightarrow c {\bar{c}} g$, where the gluon splits into an
almost on--shell $c {\bar{c}}$ pair.
Therefore QS at ${\cal{O}}(\alpha_s^1 )$ can be considered on the level
of GF at ${\cal{O}}(\alpha_s^{2} )$. On the other hand the standard
counting for light quarks is in powers of $\alpha_s$ 
and heavy quarks should fit in. We therefore suggest that 
the contributions obtained in Section \ref{heavy}
should be included in complete
experimental and theoretical NLO--analyses 
which make use of the ACOT scheme. Theoretically the inclusion is required 
for a complete renormalization of the heavy quark density at 
${\cal{O}}(\alpha_s^{1} )$.
However, we leave an ultimate
decision on that point to numerical relevance and present
numerical results in Section \ref{results}. 
Not surprisingly they will depend
crucially on the exact process considered (e.g. NC or CC) and the
choice of the factorization scale.
Finally, in Section \ref{conclusions} 
we draw our conclusions. Some technical aspects and lengthy formulae 
are relegated to the Appendix \ref{hqini}.

\section{Heavy Quark Contributions to ACOT Structure Functions}
\label{heavy}

In this section we will present all contributions to heavy
quark structure functions up to ${\cal{O}}(\alpha_s^1)$. They
are presented analytically  in their fully massive form 
together with the relevant numerical 
subtraction terms which are needed to remove the collinear
divergences in the high $Q^2$ limit. Section \ref{dis} and \ref{gluon} 
contain no new results and are only included for completeness.
In Section \ref{dis2} we present our results for the massive analogue
of the massless ${\rm{{\overline{MS}}}}$ coefficient functions
$C_i^{q,{\rm{\overline{MS}}}}$. 
 
\subsection{DIS on a Massive Quark at ${\cal{O}}(\alpha_s^0)$}
\label{dis}

The ${\cal{O}}(\alpha_s^0)$ results for $B^\ast Q_1 \rightarrow Q_2$, 
including mass effects,
have been obtained in \cite{Aivazis:1994kh} within a helicity basis 
for the hadronic and partonic
structure functions. For completeness and in order to define our normalization
we repeat these results here within 
the standard tensor basis implying the usual structure functions $F_{i=1,2,3}$.
The helicity basis seems to be advantageous since 
in the tensor basis partonic structure functions mix 
to give hadronic structure functions in the presence of masses \cite{Aivazis:1994kh}. 
However, the mixing matrix is 
diagonal \cite{Aivazis:1994kh} for the experimental relevant structure functions $F_{i=1,2,3}$ and 
only mixes $F_4$ with $F_5$ which are both suppressed by 
two powers of the lepton
mass. We neglect target (nucleon) mass corrections
which are important at larger values of Bjorken--$x$ \cite{Aivazis:1994kh} 
where heavy quark
contributions are of minor importance.

\begin{figure}[t]
\vspace*{1cm}
\begin{center}
\begin{picture}(250,100)(0,-50)
\ArrowLine(50,20)(100,40)
\Vertex(100,40){2.0}
\ArrowLine(100,40)(150,20)
\Photon(100,80)(100,40){3}{6}
\Text(50,10)[]{$p_1$, $m_1$}
\Text(150,10)[]{$p_2$, $m_2$}
\Text(110,60)[]{$q$}
\Text(100,-20)[]{(a)}
\end{picture}
\end{center}
\begin{center}
\begin{picture}(450,90)(0,-50)
\ArrowLine(60,0)(80,20)
\Line(80,20)(100,40)
\Vertex(100,40){2.0}
\ArrowLine(100,40)(170,40)
\Photon(80,80)(100,40){3}{6}
\Gluon(80,20)(125,20){-2}{6}
\Vertex(80,20){2.0}
\Text(60,-10)[]{$p_1$, $m_1$}
\Text(170,30)[]{$p_2$, $m_2$}
\Text(110,60)[]{$q$}
\Text(200,-20)[]{(b)}
\Text(125,10)[]{$k$}
\ArrowLine(260,0)(300,40)
\Vertex(300,40){2.0}
\Line(300,40)(335,40)
\ArrowLine(335,40)(370,40)
\Photon(280,80)(300,40){3}{6}
\Gluon(335,40)(350,10){-2}{6}
\Vertex(335,40){2.0}
\Text(260,-10)[]{$p_1$, $m_1$}
\Text(370,30)[]{$p_2$, $m_2$}
\Text(310,60)[]{$q$}
\Text(350,0)[]{$k$}
\end{picture}
\end{center}
\begin{center}
\begin{picture}(550,90)(0,-50)
\ArrowLine(50,20)(78,31)
\Line(78,31)(100,40)
\Vertex(100,40){2.0}
\Line(100,40)(122,31)
\ArrowLine(122,31)(150,20)
\Photon(100,80)(100,40){3}{6}
\GlueArc(100,40)(25,-157,-23){-3}{6}
\Vertex(78,31){2.0}
\Vertex(122,31){2.0}
\Text(50,5)[]{$p_1$, $m_1$}
\Text(150,5)[]{$p_2$, $m_2$}
\Text(110,60)[]{$q$}
\Text(100,-20)[]{(c1)}
\Line(200,20)(211,25)
\ArrowLine(211,25)(238,35)
\Line(238,35)(250,40)
\Vertex(250,40){2.0}
\Photon(250,80)(250,40){3}{6}
\ArrowLine(250,40)(300,20)
\GlueArc(225,30)(15,-156,24){-2}{10}
\Vertex(211,25){2.0}
\Vertex(238,35){2.0}
\Text(200,5)[]{$p_1$, $m_1$}
\Text(300,5)[]{$p_2$, $m_2$}
\Text(260,60)[]{$q$}
\Text(250,-20)[]{(c2)}
\ArrowLine(350,20)(400,40)
\Vertex(400,40){2.0}
\Photon(400,80)(400,40){3}{6}
\Line(400,40)(411,35)
\ArrowLine(411,35)(438,25)
\Line(438,25)(450,20)
\GlueArc(425,30)(15,-204,-24){-2}{10}
\Vertex(411,35){2.0}
\Vertex(438,25){2.0}
\Text(350,5)[]{$p_1$, $m_1$}
\Text(450,5)[]{$p_2$, $m_2$}
\Text(410,60)[]{$q$}
\Text(400,-20)[]{(c3)}
\end{picture}
\end{center}
\vspace*{-1cm}
\caption{\label{qsfeyn}\sf
Feynman diagrams for the $QS^{(0)}$ (a) and $QS^{(1)}$ 
[(b), (c)]
contributions to ACOT structure functions in Eqs.\ (\ref{LO}) and (\ref{QS1}),
respectively.}
\end{figure}
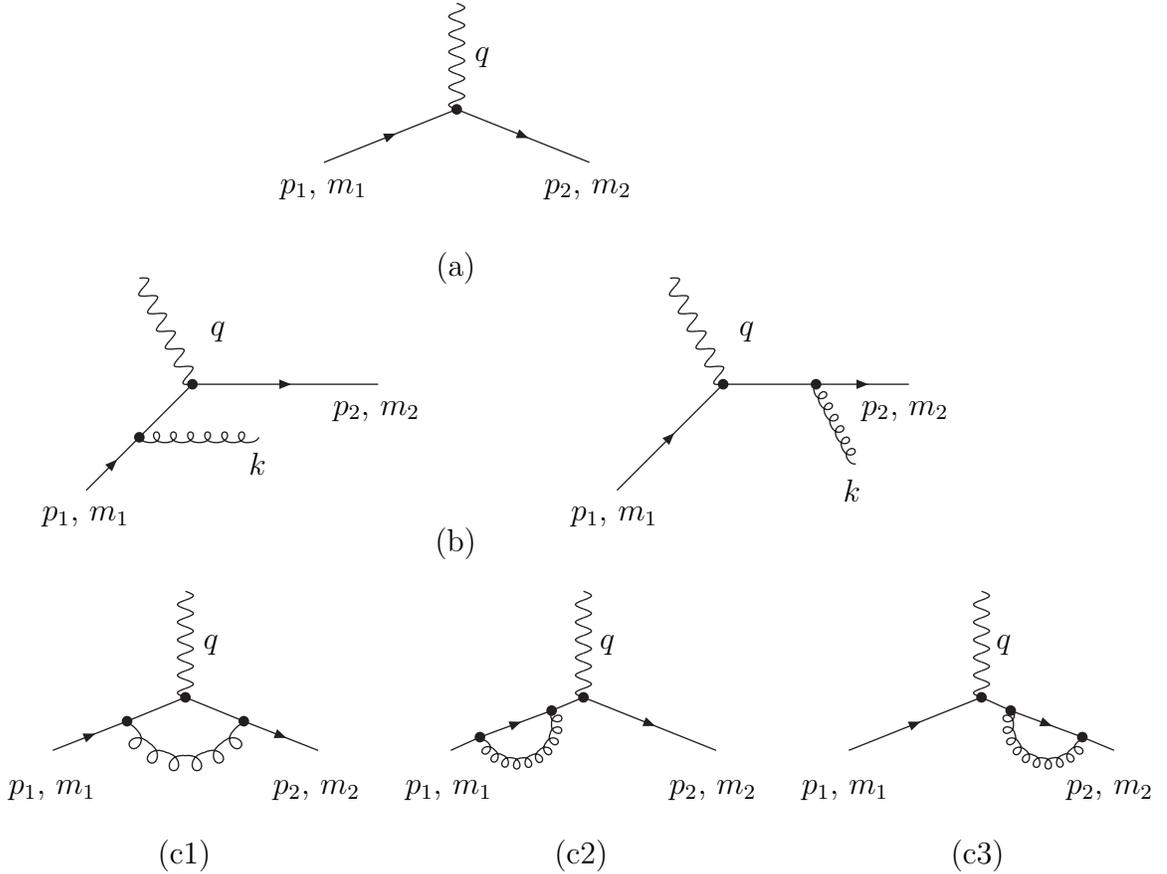

We consider DIS of the virtual Boson $B^\ast$ on the quark $Q_1$ with mass $m_1$
producing the quark $Q_2$ with mass $m_2$. At order ${\cal{O}}(\alpha_s^0)$ 
this proceeds through the parton model diagram in Fig.\ \ref{qsfeyn} (a). \\

Finite mass corrections to the massless parton model expressions are
taken into account by adopting the Ansatz given in Eq.\ (17) of \cite{Aivazis:1994kh}
\begin{equation}
\label{ansatz}
W^{\mu\nu} = \int\ \frac{d\xi}{\xi}\ Q_1(\xi,\mu^2)\ 
{\hat{\omega}}^{\mu\nu}|_{p_1^+=\xi P^+}\ \ \ .
\end{equation}
$W^{\mu\nu}$ is the usual hadronic tensor and ${\hat{\omega}}^{\mu\nu}$ 
is its partonic analogue.
Here as in the following a hat on partonic quantities refers to 
unsubtracted amplitudes, i.e. expressions which still contain mass
singularities in the massless limit.  
$p_1^+$ and $P^+$ are the light--cone momentum components of the incident
quark $Q_1$ and the nucleon, respectively. Generally the `$+$' light--cone component
of a vector $v$ is given by $v^+\equiv (v^0 +v^3)/\sqrt{2}$. 

Contracting the convolution in Eq.\ (\ref{ansatz}) with the projectors in 
Appendix \ref{rge} 
gives the individual hadronic structure functions $F_{i=1,2,3}$. 
In leading order (LO) the latter are given by \cite{Aivazis:1994kh}
\begin{eqnarray}
F_1^{QS^{(0)}}(x,Q^2)&=&
\frac{\Sp \spp- 2 m_1 m_2 \Sm}{2 \Delta}\ Q_1(\chi,Q^2) 
\nonumber \\
F_2^{QS^{(0)}}(x,Q^2)&=&\frac{\Sp \Delta}{2 Q^2}\ 2 x\ Q_1(\chi,Q^2) 
\nonumber \\
F_3^{QS^{(0)}}(x,Q^2)&=&2 \Rp\ Q_1(\chi,Q^2) 
\label{LO}
\end{eqnarray}
with 
\begin{equation}
\label{pmpm}
\Sigma_{\pm \pm}=Q^2\pm m_2^2\pm m_1^2\ \ \ .
\end{equation}
In Eq.\ (\ref{LO}) we use the shorthand 
$\Delta \equiv \Delta[m_1^2,m_2^2,-Q^2]$ 
, where the usual
triangle function is defined by 
\begin{equation}
\Delta[a,b,c]=\sqrt{a^2+b^2+c^2-2(a b+ b c + c a)}\ \ \ .
\end{equation}
The vector ($V$) and axial vector ($A$) couplings of the 
${\overline{Q}}_2 \gamma_\mu 
(V-A \gamma_5) Q_1$ quark current enter via the following combinations:
\begin{eqnarray}
\label{couplings}
S_{\pm}&=&V{V}^{\prime} \pm A {A}^{\prime}
\nonumber\\
R_{\pm}&=&(V A^{\prime}\pm V^{\prime} A)/2\
\end{eqnarray}
where $V,A \equiv V^{\prime},A^{\prime}$ in the case of pure $B$ scattering
and $V,A \neq V^{\prime},A^{\prime}$ in the case of $B, B^{\prime}$ 
interference (e.g. $\gamma, Z^0$ interference in the standard model). 
The scaling variable $\chi$ generalizes the usual Bjorken--$x$ in the presence
of parton masses and is given by \cite{Aivazis:1994kh}:
\begin{equation}
\label{chi}
\chi = \frac{x}{2 Q^2}\ \left(\ \spm\ +\ \Delta\ \right)\ \ \ . 
\end{equation}
The mass dependent structure functions in Eq.\ (\ref{LO}) 
motivate the following definitions
\begin{equation}
\label{def}
\left.
\begin{array}{ccc}
{\cal{F}}_1&=&\frac{2 \Delta}{\Sp \spp- 2 m_1 m_2 \Sm}\ F_1 \\
{\cal F}_2&=&\frac{2 Q^2}{\Sp \Delta}\ \frac{1}{2 x}\ F_2 \\
{\cal F}_3&=&\frac{1}{2 \Rp}\ F_3 
\end{array}
\right\}\ =\ Q_1(\chi,Q^2)\ +\ {\cal{O}}(\alpha_s^1)
\end{equation}
such that ${\cal F}_i-{\cal F}_j, \ i,j=1,2,3$, will be finite
of ${\cal O}(\alpha_s)$ in the limit $m_{1,2} \rightarrow 0$.
 
\subsection{DIS on a Massive Quark at ${\cal{O}}(\alpha_s^1)$}
\label{dis2}

At ${\cal{O}}(\alpha_s^1)$ contributions from real gluon emission
[Fig.\ \ref{qsfeyn} (b)] and virtual corrections [Fig.\ \ref{qsfeyn} (c)] have to be 
added
to the ${\cal{O}}(\alpha_s^0)$ results of Section \ref{dis}.  
The vertex correction with general masses and couplings [Fig.\ \ref{qsfeyn} (c)]
does to our knowledge not exist in the literature 
and is presented in some detail
in Appendix \ref{vcorr}. The final result (virtual+real) can 
be cast into the following form:
\begin{eqnarray} 
\label{QS1}
{\hat{{\cal F}}}_{i=1,2,3}^{QS^{(0+1)}} (x,Q^2,\mu^2) &\equiv& 
{\cal F}_i^{QS^{(0)}} (x,\mu^2)+
{\hat{{\cal F}}}_i^{QS^{(1)}} (x,Q^2,\mu^2) \\ \nonumber
&= & Q_1(\chi,\mu^2) 
+\frac{\alpha_s(\mu^2)}{2 \pi}\ \int_{\chi}^1\frac{d\xip}{\xip}
\
\left[Q_1\left(\frac{\chi}{\xip},\mu^2\right) {\hat{H}}_i^q(\xip,m_1,m_2)
\right]
\end{eqnarray}
with $\xip\equiv\frac{\chi}{\xi}$ and
\begin{equation}
\label{coefficient}
{\hat{H}}_i^q(\xip,m_1,m_2)=C_F\ \left[(S_i+V_i)\ \delta(1-\xip)\ + 
\frac{1-\xip}{(1-\xip)_{+}}
\  \frac{{\hat{s}}-m_2^2}{8 {\hat{s}}}
\ N_i^{-1}\ \hat{f}_i^Q(\xip)\right]
\end{equation}
where $\hats=(p_1+q)^2$ and the $S_i$, $V_i$, $N_i$ and $\hat{f}_i^Q$ 
are given in Appendix \ref{rvstruct}.
The factorization scale $\mu^2$ will be taken equal to the
renormalization scale throughout.
The `$+$' distribution in Eq.\ (\ref{coefficient}) is a 
remnant of the cancellation of the soft
divergences from the real and virtual contributions. It is defined as usual:
\begin{equation}
\int_0^1\ d\xip\ f(\xip)\ \left[ g(\xip)\right]_+ \equiv \int_0^1\ 
d\xip\ \left[f(\xip)-f(1)\right]\ g(\xip) \ \ \ .
\end{equation}      
As indicated by the hat on ${\hat{H}}_i^q$, the full massive 
convolution in  Eq.\ (\ref{QS1})
still contains the mass singularity arising from quasi--collinear 
gluon emission from
the initial state quark leg. The latter has to be removed by 
subtraction in
such a way that in the asymptotic limit $Q^2\rightarrow\infty$ 
the well known massless 
${\overline{{\rm{MS}}}}$ expressions are recovered. 
The ${\overline{{\rm{MS}}}}$ limit is
mandatory since all modern parton distributions 
--and therefore all available heavy
quark densities-- are defined in this particular scheme (or in the DIS scheme 
\cite{cit:AEM-7901}, which can be straightforwardly derived from \msbar). 
The correct subtraction
term can be obtained from the following limit
\begin{eqnarray} \nonumber
\lim_{m_1\rightarrow 0} \int_{\chi}^1 \frac{d \xip}{\xip}
Q_1\left(\frac{\chi}{\xip},\mu^2\right) {\hat{H}}_i^q(\xip,m_1,m_2) &=&
\int_{x/\lambda}^1 \frac{d \xip}{\xip}
Q_1\left(\frac{x}{\lambda\xip},\mu^2\right)  
\Bigg\{ H_i^{q,{\overline{{\rm{MS}}}}}(\xip,\mu^2,\lambda)
\\* \nonumber 
&+& C_F 
\ \left[\frac{1+{\xip}^2}{1-\xip}\left(\ln \frac{\mu^2}{m_1^2}-1
-2 \ln(1-\xip)\right)\right]_{+} \Bigg\} 
\\* 
\label{limit1}
&+& {\cal{O}}\left(\frac{m_1^2}{Q^2}\right)
\end{eqnarray}
where $\lambda = Q^2/(Q^2+m_2^2)$,   
$x/\lambda=\chi|_{m_1=0}$ and 
the $H_i^{q,{\overline{{\rm{MS}}}}}$ 
can be found in \cite{Gluck:1996ve,Kretzer:PhD}. 
Obviously the \msbar\ subtraction term for a `heavy quark inside a heavy
quark'  is given not only by   
the splitting function $P_{qq}^{(0)} = C_F [(1+{\xi^{\prime}}^2)/(1-\xip)]_+$
times the collinear log $\ln (\mu^2/m_1^2)$
but also comprises a constant term. Herein we agree with Eq.\ (3.15) in
\cite{Mele:1991cw}\footnote{
We also agree
with the quark initiated coefficient functions in \cite{vanderBij:1991ju} where
quark masses have been used as regulators.}, where this was first
pointed out in the framework of perturbative fragmentation functions
for heavy quarks. 
We therefore define 
\begin{equation}
{\cal F}_i^{SUB_q} (x,Q^2,\mu^2)=\frac{\alpha_s(\mu^2)}{2 \pi}C_F 
\int_{\chi}^1\frac{d\xip}{\xip}
\left[\frac{1+{\xip}^2}{1-\xip}\left(\ln \frac{\mu^2}{m_1^2}-1
-2 \ln(1-\xip)\right)\right]_{+} Q_1\left(\frac{\chi}{\xip},\mu^2\right)
\label{SUBq}
\end{equation}
such that
\begin{equation}
\label{limit2}
\lim_{m_1\to 0} \left[{\hat{{\cal F}}}_i^{QS^{(1)}} (x,Q^2,\mu^2)
-{\cal F}_i^{SUB_q} (x,Q^2,\mu^2)\right]=
{\cal F}_i^{Q_1^{(1)},\overline{\rm{{MS}}}} (x,Q^2,\mu^2)\ \ \ ,
\end{equation} 
where the superscript $Q_1$ on 
${\cal F}_i^{Q_1^{(1)},\overline{\rm{{MS}}}}$ refers to that
part of the inclusive structure function 
${\cal F}_i$ which is initiated by the 
heavy quark $Q_1$, i.e.\ which is obtained from a convolution with the heavy
quark parton density. 
Note that the limit in Eq.\ (\ref{limit1}) guarantees 
that Eq.\ (\ref{limit2}) is also fulfilled when
$m_1=m_2 \rightarrow 0$ (e.g.\ NC leptoproduction of charm) since 
\begin{equation}
\lim_{m_2 \rightarrow 0} H_i^{q,{\overline{{\rm{MS}}}}}(\xip,\mu^2,\lambda) 
= C_i^{q,{\overline{{\rm{MS}}}}}(\xip,\mu^2)
+ {\cal{O}}\left(\frac{m_2^2}{Q^2}\right) 
\end{equation}
where $C_i^{q,{\overline{{\rm{MS}}}}}$ are the 
standard massless coefficient functions 
in the ${\overline{\rm{{MS}}}}$ scheme, e.g.\ in 
\cite{cit:AEM-7901,cit:FP-8201}.

\subsubsection{Comparison to Existing NC and CC Results}
\label{comparison}

We have performed several cross checks of our results against well known
calculations that exist in the literature \cite{Gluck:1996ve,vanderBij:1991ju,cit:AEM-7901,cit:FP-8201,Hoffmann:1983ah,*Hoffmann:PhD}. 
The checks can be partly inferred from the above paragraph. 
Nevertheless we present here a 
systematic list for the reader's convenience and to point out 
discrepancies of our calculation with \cite{Hoffmann:1983ah,*Hoffmann:PhD}.

In the charged current case $V=A=1$ our results in Eq.\ (\ref{QS1}) reduce 
in the limit $m_1\rightarrow 0$ to the corresponding
expressions in \cite{vanderBij:1991ju}, or in 
\cite{Gluck:1996ve,Kretzer:PhD}
if the scheme dependent term represented by Eq.\ (\ref{SUBq}) is
taken into account. The latter
agrees with Eq.\ (3.15) in \cite{Mele:1991cw}. For $m_{1,2}\rightarrow 0$ 
we reproduce the well 
known \msbar\ coefficient functions, e.g., in 
\cite{cit:BBDM-7801,cit:AEM-7901,cit:FP-8201}. 
The vertex correction in Appendix \ref{vcorr} is implicitly tested because it 
contributes to any of the final results. However, as an independent
cross check the well known QED textbook result can be reproduced 
for $m_1=m_2$, $A=0$.

Initial state parton mass effects in NC DIS at $\alpsi$ have been first 
considered in \cite{Hoffmann:1983ah,*Hoffmann:PhD} within the scenario 
\cite{Brodsky:1980pb,*Brodsky:1981se} of intrinsic
nonperturbative $c\bar{c}$ pairs stemming from fluctuations of the nucleon 
Fock space wavefunction. Although we do not consider such a scenario here we 
note that our results could be easily transferred to corresponding applications
\cite{Harris:1996jx,*Gunion:1997mx}. The main difference would be an inclusion of kinematical
target mass effects which are important at larger $x$ \cite{Aivazis:1994kh}
where a possible nonperturbative charm component is expected \cite{Brodsky:1980pb,*Brodsky:1981se} to 
reside. We list a detailed comparison of our calculation to the one in 
\cite{Hoffmann:1983ah,*Hoffmann:PhD} in Appendix \ref{homocomp}. Since 
our calculation does not fully agree with \cite{Hoffmann:1983ah,*Hoffmann:PhD} for reasons
which we are completely able to trace back and given the amount of 
successful independent 
tests of our results we regard the disagreement with 
\cite{Hoffmann:1983ah,*Hoffmann:PhD} as
a clear evidence that the results in \cite{Hoffmann:1983ah,*Hoffmann:PhD} 
should be updated by our calculation.    

\subsection{Gluon Fusion Contributions at ${\cal{O}}(\alpha_s^1)$}
\label{gluon}

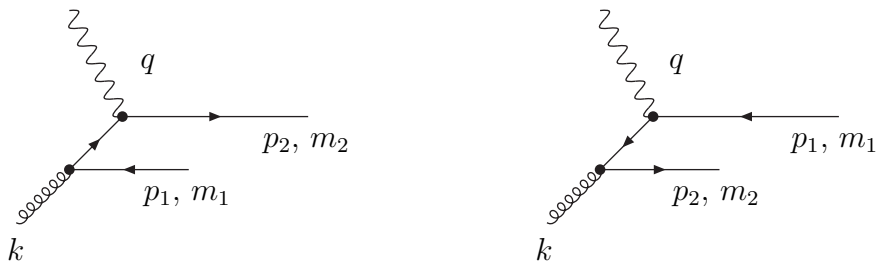
\begin{figure}[h]
\begin{center}
\begin{picture}(450,110)(0,-30)
\Gluon(60,-10)(80,10){-2}{6}
\ArrowLine(80,10)(100,30)
\Vertex(100,30){2.0}
\ArrowLine(100,30)(170,30)
\Photon(80,70)(100,30){3}{6}
\ArrowLine(125,10)(80,10)
\Vertex(80,10){2.0}
\Text(60,-20)[]{$k$}
\Text(170,20)[]{$p_2$, $m_2$}
\Text(110,50)[]{$q$}
\Text(125,0)[]{$p_1$, $m_1$}
\Gluon(260,-10)(280,10){-2}{6}
\ArrowLine(300,30)(280,10)
\Vertex(300,30){2.0}
\ArrowLine(370,30)(300,30)
\Photon(280,70)(300,30){3}{6}
\ArrowLine(280,10)(325,10)
\Vertex(280,10){2.0}
\Text(260,-20)[]{$k$}
\Text(370,20)[]{$p_1$, $m_1$}
\Text(310,50)[]{$q$}
\Text(325,0)[]{$p_2$, $m_2$}
\end{picture}
\end{center}
\caption{\label{gffeyn}\sf
Feynman diagrams for the production of a massive quark--antiquark pair via
boson--gluon fusion.}
\end{figure}

The gluon fusion contributions to heavy quark structure functions, depicted in 
Fig.\ \ref{gffeyn},
($B^\ast g \rightarrow {\bar{Q}}_1 Q_2$)
are known for a long time \cite{Gluck:1988uk,Schuler:1988wj,Baur:1988ai}
and have been reinterpreted in \cite{Aivazis:1994pi}
within the helicity basis for structure functions.
Here we only briefly recall the corresponding formulae in the tensor basis 
for completeness. The GF
component of DI structure functions is given by
\begin{eqnarray} \nonumber
F_{1,3}^{GF}(x,Q^2) &=& \int_{ax}^1\ \frac{d \xip}{\xip}\ g(\xip,\mu^2)\ 
f_{1,3}\left(\frac{x}{\xip},Q^2\right) \\
\label{GF}
F_{2}^{GF}(x,Q^2) &=& \int_{ax}^1\ \frac{d \xip}{\xip}\ \xip g(\xip,\mu^2)\ 
f_{2}\left(\frac{x}{\xip},Q^2\right) 
\end{eqnarray}
where $ax=[1+(m_1+m_2)^2/Q^2]x$ and the $f_i$ can be found for general
masses and couplings in \cite{Gluck:1988uk}. 
The corresponding ${\cal{F}}_i^{GF}$ are obtained from the
$F_i^{GF}$ by using the same normalization factors as
in Eq.\ (\ref{def}).  
Along the lines of \cite{Aivazis:1994pi} the GF contributions
coexist with the QS contributions which are calculated from the heavy
quark density, which is evolved via the massless RG equations in the
${\rm{{\overline{MS}}}}$ scheme. As already pointed out in Section \ref{dis2} the
quasi--collinear log of the fully massive GF term has to be subtracted 
since the corresponding  mass singularities are
resummed to all orders in the massless RG evolution. The subtraction
term for the GF contribution is given by \cite{Aivazis:1994pi}
\begin{equation}
\label{SUBg}
{\cal F}_i^{SUB_g} (x,Q^2,\mu^2)=
\sum_{k}
\ \frac{\alpha_s(\mu^2)}{2 \pi}
\ \ln \frac{\mu^2}{m_k^2} 
\ \int_{\chi}^1\frac{d\xip}{\xip}\ P_{qg}^{(0)}(\xip)   
\ g\left(\frac{\chi}{\xip},\mu^2\right)\ \ \ ,
\end{equation}  
where $P_{qg}^{(0)}(\xip) = 1/2\ [{\xip}^2+(1-\xip)^2]$. 
Note that Eq.\ (\ref{SUBg}) as well as Eq.\ (\ref{SUBq}) are 
defined relative to the 
${\cal{F}}_i$
in Eq.\ (\ref{def}) and not with respect to the experimental structure
functions $F_i$. The sum in Eq.\ (\ref{SUBg}) runs over the indices of 
the quarks $Q_k$ for which the quasi--collinear logs
are resummed by massless evolution of a heavy quark density, i.e.\
$k=1$, $k=2$ or $k=1,2$.

\subsection{ACOT Structure Functions at ${\cal{O}}(\alpha_s^1)$}
\label{acot}

As already mentioned in the introduction to this chapter
it is not quite clear how
the perturbation series should be arranged for massive quarks, i.e.\
whether the counting is simply in powers of $\alpha_s$ as for light
quarks or whether an intrinsic heavy quark density carries an extra
power of $\alpha_s$ due to its prehistory as an extrinsic particle 
produced by pure GF. We are here interested in the $QS^{(1)}$
component of heavy quark structure functions. Usually the latter is
neglected in the ACOT formalism since it is assumed to be suppressed
by one order of $\alpha_s$ with respect to the GF contribution as
just explained above. 
We have, however, demonstrated in Chapter~\ref{chap_hqcc} 
within \msbar\ that
this naive expectation is quantitatively not supported  in the special 
case of semi--inclusive production of charm (dimuon events) in CC DIS. 
We therefore want to investigate the numerical relevance of the
$QS^{(1)}$ contribution to general heavy quark structure functions. In
this chapter we present results for the fully inclusive case, relevant
for inclusive analyses and fits to inclusive data. We postpone
experimentally more relevant semi--inclusive ($z$--dependent) results
to Chapter \ref{hqffdis}. 
Our results at full ${\cal{O}}(\alpha_s^1)$
will be given by
\begin{equation}
\label{complete}
F_i^{(1)} = F_i^{QS^{(0+1)}} + F_i^{GF} - F_i^{SUB_q} - F_i^{SUB_g}
\end{equation}
with  $F_i^{QS^{(0+1)}}$, $F_i^{GF}$, $F_i^{SUB_q}$ and $F_i^{SUB_g}$
given in Eqs.\ (\ref{QS1}), (\ref{GF}), (\ref{SUBq}), and (\ref{SUBg}), 
respectively. 
Furthermore, we will also consider a perturbative expression for
$F_i$ which is constructed along the expectations of the original
formulation of the ACOT scheme, i.e.\ $QS^{(1)}$ is neglected and
therefore $F_i^{SUB_q}$ need not be introduced
\begin{equation}
\label{incomplete}
F_i^{(0)+GF-{SUB_g}} = F_i^{QS^{(0)}} + F_i^{GF} - F_i^{SUB_g}\ \ \ .
\end{equation}

\section{Results for NC and CC Structure Functions}
\label{results}

In this section we present results which clarify the numerical
relevance of $QS^{(1)}$ contributions to inclusive heavy quark structure
functions in the ACOT scheme. We will restrict ourselves to NC and CC
production of charm since bottom contributions are insignificant to
present DI data. 
Our canonical parton distributions for the NC case
will be CTEQ4M \cite{Lai:1997mg}
(Figs.\ \ref{ncq2} and \ref{ncmu2} below), which include `massless heavy partons'
$Q_k$ above the scale $Q^2=m_k^2$. 
Figures \ref{qsx} and \ref{qsq2}, however, 
have been obtained from the older GRV92 \cite{Gluck:1992ng} 
distributions. 
The newer GRV94 \cite{cit:GRV94} parametrizations do not include a 
resummed charm density
since they are constructed exclusively along FOPT.
GRV94 is employed in the CC case, Section \ref{ccsec}. 
The radiative strange sea of GRV94 seems to be closest
to presently available CC charm production data \cite{Gluck:1997sj}.
Furthermore, the low
input scale of GRV94 allows for a wide range of variation of the
factorization scale around the presently relevant experimental scales,
which are lower for CC DIS than for NC DIS.
Qualitatively all our results do not depend on the specific
set of parton distributions chosen.   

\subsection{NC Structure Functions}
\label{nc}

For our qualitative analysis we are only considering photon exchange
and we neglect the $Z^0$.
The relevant formulae are all given in Section \ref{heavy} with the following 
identifications:
\begin{eqnarray} \nonumber
Q_{1,2} &\rightarrow& c \\ \nonumber
m_{1,2} \rightarrow m_c &=& 1.6\ (1.5)\ {\rm{GeV\ \ for\ CTEQ4\ (GRV92)}}
\\ \nonumber
V=V^\prime,A=A^\prime &\rightarrow& \frac{2}{3},0
\end{eqnarray}
and we use $\mu^2=Q^2$ if not otherwise noted.
We consider contributions from 
charmed quarks and anti--quarks which are inseparably mixed by
the GF contribution. This means that in Eq.\ (\ref{SUBg}) the sum runs
over $k=1,2$ and the relevant expressions of Section \ref{dis} and \ref{dis2} have
to be doubled [since $c(x,\mu^2)={\bar{c}}(x,\mu^2)$].

First we investigate the importance of finite mass
corrections to the limit in Eq.\ (\ref{limit2}).
\begin{figure}[t]
\vspace*{-0.5cm}
\hspace*{-1.25cm}
\epsfig{figure=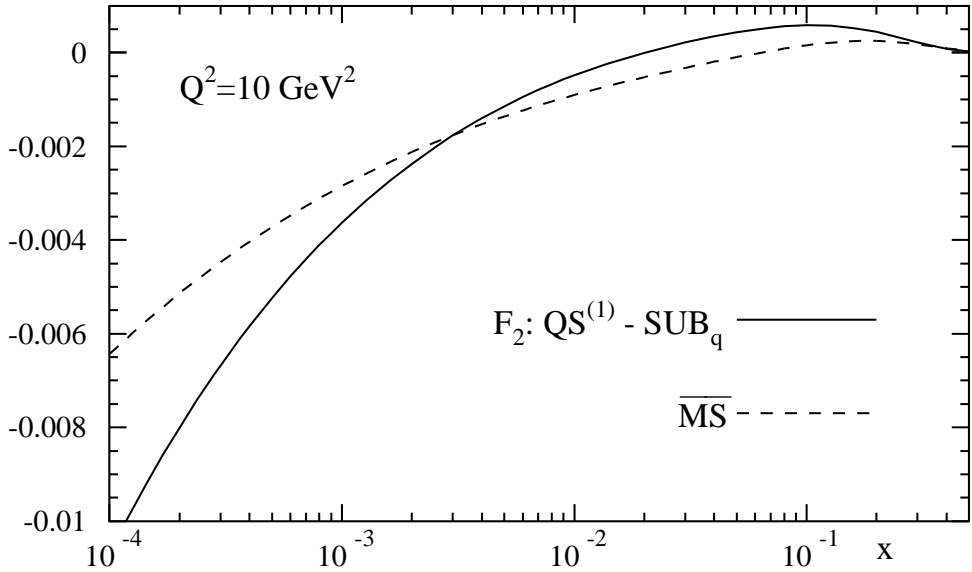,width=16cm}
\vspace*{-2.5cm}
\caption{\label{qsx}\sf
$x$--dependence of the subtracted $QS^{(1)}$ contribution to the NC charm 
structure function $F_2^c$ \ (solid line). 
$Q^2=\mu^2=10\ {\rm{GeV^2}}$ is fixed.
For comparison the \msbar\ analogue in 
Eq.\ (\ref{cqmsbar}) is shown (dashed line). 
The GRV92 parton distributions have been used.}
\end{figure} 
In Fig.\ \ref{qsx} the difference
${\hat{F}}_2^{QS^{(1)}}-F_2^{SUB_q}$ can be compared to its 
${\rm{{\overline{MS}}}}$ analogue which is 
\begin{equation}
\label{cqmsbar}
F_2^{(c+{\bar{c}})^{(1)},{\rm{\overline{MS}}}} = \frac{4}{9}\ x\  
\frac{\alpha_s(\mu^2)}{2\pi} \left[ (c+{\bar{c}})(\mu^2) \otimes
C_2^{q,{\rm{\overline{MS}}}}\left(\frac{Q^2}{\mu^2}\right)\right] (x,Q^2)
\end{equation}
where $\otimes$ denotes the usual (massless) convolution.
From Fig.\ \ref{qsx} it is obvious that the 
relative difference between ACOT and ${\rm{{\overline{MS}}}}$ depends crucially
on $x$. It can be large and only slowly convergent to the
asymptotic ${\rm{{\overline{MS}}}}$ limit as can be inferred from Fig.\ \ref{qsq2}.
\begin{figure}[t]
\hspace*{-1.25cm}
\epsfig{figure=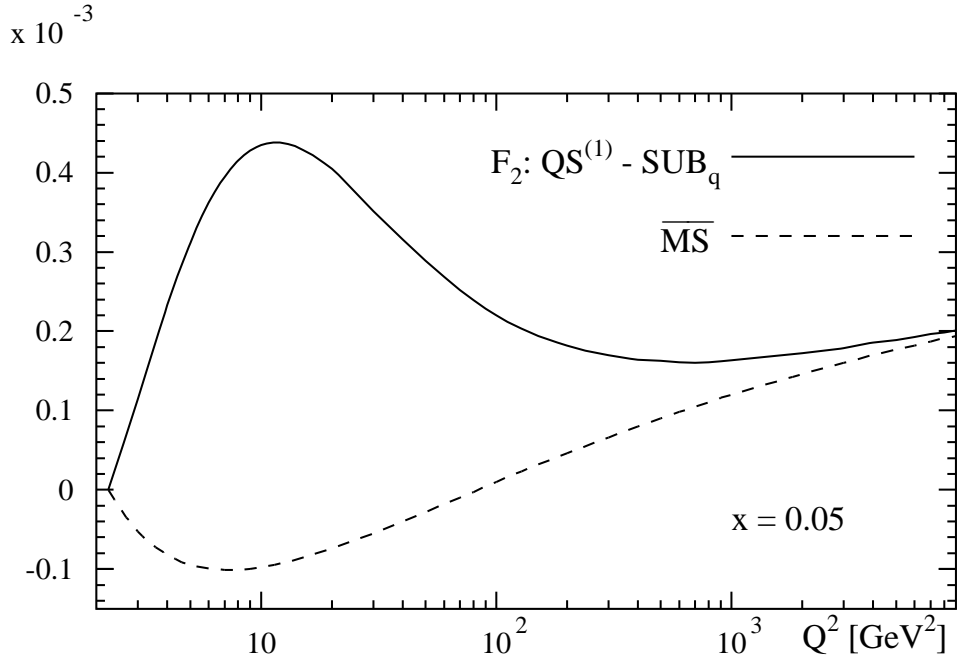,width=16cm}
\vspace*{-2.5cm}
\caption{\label{qsq2}\sf
The same as Fig.\ \ref{qsx} but varying $Q^2$($=\mu^2$) for fixed $x$.}
\end{figure}
Note that the solid curves in Figs.\ 
\ref{qsx}, \ref{qsq2} are extremely sensitive to the 
precise definition of the subtraction term in Eq.\ (\ref{SUBq}), 
e.g.\ changing $\chi \rightarrow x$ --which also removes the collinear 
singularity in the high $Q^2$ limit-- can change the ACOT result by about
a factor of $5$ around $Q^2 \sim  5\ {\rm{GeV}}^2$.\footnote{
The  subtracted gluon contribution GF changes by about a factor 
of $2$ under the same replacement.}  
This is an example of the ambiguities in defining a
variable flavor number scheme which have been formulated in a
systematic manner in \cite{Thorne:1998ga,*Thorne:1998uu}.    
  
\begin{figure}[t]
\vspace*{-0.5cm}
\hspace*{-1.25cm}
\epsfig{figure=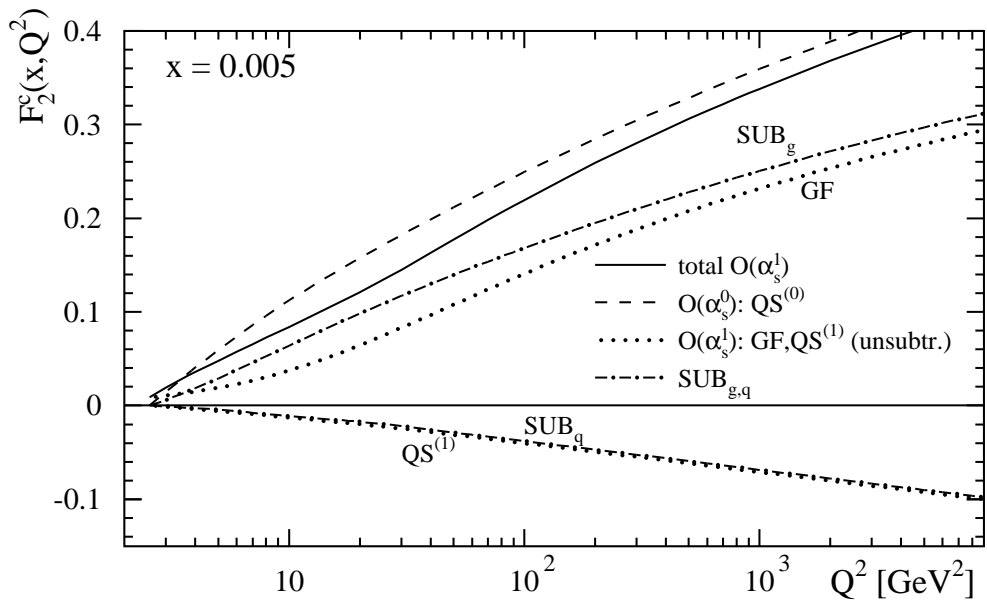,width=16cm}
\vspace*{-2.5cm}
\caption{\label{ncq2}\sf
The complete $\alpsi$ neutral current structure function $F_2^c$ and all
individual 
contributions over
a wide range of $Q^2$, calculated from the CTEQ4M distributions. 
Details of the distinct 
contributions are given in the text.}
\end{figure}

The relative difference  between the subtracted $QS^{(1)}$ contribution
calculated along ACOT and the corresponding ${\rm{{\overline{MS}}}}$
contribution in Eq.\ (\ref{cqmsbar})
appears, however, phenomenologically irrelevant   
if one considers the significance of these contributions 
to the total charm structure function in Fig.\ \ref{ncq2}. 
The complete
${\cal{O}}(\alpha_s^1)$ result (solid line) 
is shown over a wide range of $Q^2$ together with its individual 
contributions from Eq.\ (\ref{complete}). It can be clearly seen that 
the full massive $QS^{(1)}$ contribution is almost completely
cancelled by the subtraction term ${SUB_q}$ (Indeed the curves for
$QS^{(1)}$ and ${SUB_q}$ are hardly distinguishable on the scale of
Fig.\ \ref{ncq2}). 
The subtracted quark correction is numerically negligible and 
turns out to be indeed suppressed compared to the gluon
initiated contribution, which is also shown in Fig.\ \ref{ncq2}. 
Note, however,
that the quark initiated corrections are not unimportant because they
are intrinsically small. Rather the large massive contribution
$QS^{(1)}$ is
perfectly cancelled by the subtraction term ${SUB_q}$ provided that 
$\mu^2=Q^2$ 
is chosen.
This is not necessarily the case for different choices of 
$\mu^2$ as we will now 
demonstrate.  

\begin{figure}[t]
\vspace*{-0.5cm}
\hspace*{-1.25cm}
\epsfig{figure=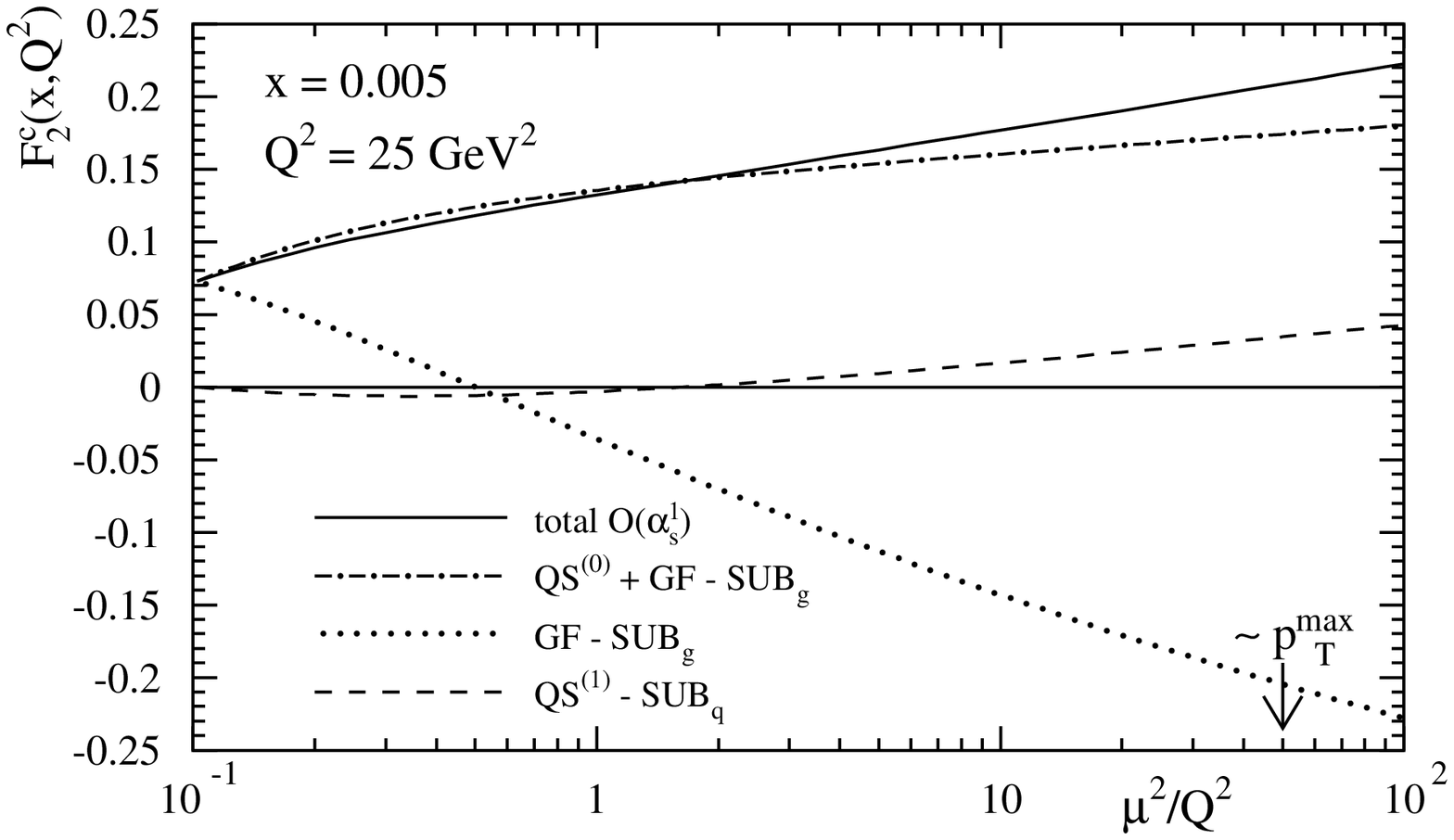,width=16cm}
\vspace*{-2.5cm}
\caption{\label{ncmu2}\sf
$\mu^2$ dependence of the complete $\alpsi$ NC structure function (CTEQ4M)
in Eq.\ (\ref{complete})
(solid line) and of the structure function in Eq.\ (\ref{incomplete}) 
(dot--dashed line)
where the subtracted $QS^{(1)}$ contribution is neglected. 
Also shown are the different
subtracted $\alpsi$ contributions $GF$ and $QS^{(1)}$.}
\end{figure} 
In Fig.\ \ref{ncmu2} we show the dependence of the complete structure function
and its components on the arbitrary factorization scale
$\mu^2$. 
Apart from the canonical choice $\mu^2=Q^2$ (which was used for all
preceding figures) also different scales have been proposed
\cite{Aivazis:1990pe,Collins:1990bu}  
like the maximum transverse momentum of the outgoing heavy quark which
is approximately given by $(p_T^{max})^2 \simeq (1/x-1)\ Q^2/4$. For
low values of $x$, where heavy quark structure functions are most
important, the scale $(p_T^{max})^2 \gg  Q^2$. 
The effect of choosing a $\mu^2$ which differs
much from $Q^2$ can be easiest understood for the massless coefficient
functions $C_i^{q,g,{\rm{\overline{MS}}}}$ which contain an
unresummed $\ln (Q^2/\mu^2)$. The latter is of course absent for
$\mu^2=Q^2$ but becomes numerically increasingly important, the more 
$\mu^2$ deviates from $Q^2$. This logarithmic contribution cannot
be neglected since it is the unresummed part of the collinear
divergence which is necessary to define the scale dependence of the
charm density. This expectation is confirmed by Fig.\ \ref{ncmu2}.
For larger values of $\mu^2$ the subtracted $QS^{(1)}$
contribution is indeed still suppressed relative to the subtracted $GF$
contribution. Nevertheless, its contribution to the total structure function
becomes numerically significant and reaches the $\sim$ 20 \% level around
$(p_T^{max})^2$.   
Note that in this regime the involved formulae of Section \ref{dis2} may be
safely approximated by the much simpler convolution in Eq.\
(\ref{cqmsbar}) because they are completely dominated by the universal
collinear logarithm and the finite differences 
${\rm{ACOT}} - {\rm{\overline{MS}}}$
from Figs.\ \ref{qsx} and \ref{qsq2} become immaterial.
In practice it is therefore always legitimate to approximate the ACOT
results of Section \ref{dis2} by their ${\rm{\overline{MS}}}$ analogues
because both are either numerically insubstantial or logarithmically dominated.  
Finally we confirm that
the scale dependence of the
full ${\cal{O}}(\alpha_s^1)$ structure function $F_2^{(1)}$ in
Eq.\ (\ref{complete}) is
larger than the scale dependence of $F_i^{(0)+GF-SUB_g}$ 
in Eq.\ (\ref{incomplete})
which was already pointed out in \cite{Schmidt:1997yr}. 
Nevertheless, the subtracted
$QS^{(1)}$ contribution should be respected for theoretical reasons
whenever $\alpha_s \ln (Q^2/\mu^2)\not\ll 1$.  
 
\subsection{CC Structure Functions}
\label{ccsec}

Charm production in CC DIS is induced by an $s \rightarrow c$
transition at the $W$--Boson vertex. The strange quark is not really a
heavy quark in the sense of the ACOT formalism, i.e., the production of
strange quarks cannot be calculated reliably at any scale using FOPT because
the strange quark mass is too small. It is nevertheless reasonable to
take into account possible finite $m_s$ effects into perturbative
calculations using ACOT since the subtraction terms remove all long
distance physics from the coefficient functions. Indeed the ACOT
formalism has been used for an experimental analysis of CC charm
production in order to extract the strange sea density of the
nucleon \cite{Bazarko:1995tt,*Bazarko:1994hm}. 
Along the assumptions of ACOT $QS^{(1)}$ contributions have
not been taken into account. This procedure is obviously questionable and has
been shown not to be justified within the ${\rm{\overline{MS}}}$
scheme in Chapter~\ref{chap_hqcc}. 
With our results in Section 
\ref{dis2} we can investigate the
importance of quark initiated ${\cal{O}}(\alpha_s^1)$ corrections
within the ACOT scheme for inclusive CC DIS. As already mentioned above,
results for the
experimentally more important case of semi--inclusive ($z$--dependent)
DIS will be presented in Chapter \ref{hqffdis} of this thesis. 
In the following we
only introduce subtraction terms for collinear divergencies correlated
with the strange mass and treat all logarithms of the charm mass along
FOPT. We do so for two reasons, one theoretical and one experimental:
First, at present experimental scales of CC charm production $\ln
(Q^2/m_c^2)$ terms can be safely treated along FOPT and no
introduction of an a priori unknown charm density is necessary. Second,
the introduction of a subtraction term for the mass singularity of the
charm quark would simultaneously require the inclusion of the 
$c \rightarrow s$ $QS$--transition at the $W$--vertex with no spectator--like 
${\bar{c}}$--quark as in $GF$. This contribution must, however,
be absent when experiments tag on charm in the final state.
CC DIS on massive charm quarks without final--state
charm tagging has been studied
in \cite{dalesio:PhD}.    

The numerics of this section can be obtained by the formulae of
Section \ref{heavy} with the following identifications:
\begin{eqnarray} \nonumber
Q_1 \rightarrow s\ \ \ &,& Q_2 \rightarrow c \\ \nonumber
m_{2} \rightarrow m_c &=& 1.5\ {\rm{GeV}}\ {\rm{(GRV94)}}
\\ \nonumber
V=V^\prime,A=A^\prime &\rightarrow& 1,1\ \ \ 
\end{eqnarray}
and the strange mass $m_1=m_s$ will be varied in order to show its
effect on the structure function $F_2^{c}$.

\begin{figure}[p]
\vspace*{-1.5cm}
\hspace*{-0.25cm}
\epsfig{figure=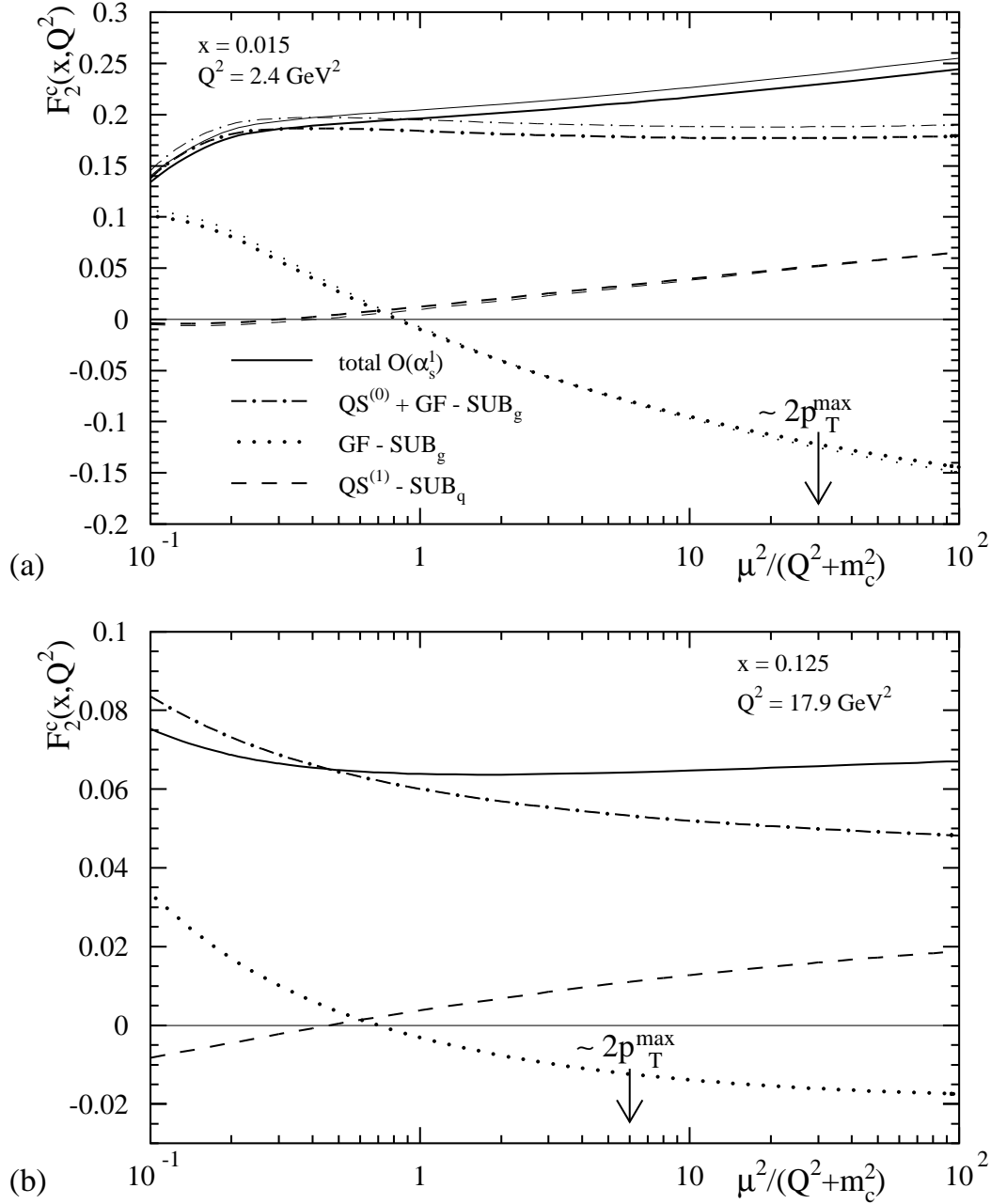,width=16cm}
\vspace*{-2.5cm}
\caption{\label{cc}\sf
The charm production contribution to the charged current structure 
function $F_2$
for a wide range of the factorization scale $\mu^2$ using GRV94. 
The curves are as for the
neutral current case in Fig.\ \ref{ncmu2}. 
In Fig.\ \ref{cc} (a) the thicker curves have been obtained with a 
(purely regularizing) 
strange mass of $10\ {\rm{MeV}}$ which according to Eq.\ (\ref{limit2}) 
(and to the analogous limit for the subtracted $GF$ term 
\protect\cite{Aivazis:1994pi}) numerically reproduces \msbar.
For the thinner curves a strange mass of 
$500\ {\rm{MeV}}$ has been assumed. In Fig.\ \ref{cc} (b) all curves 
correspond to
$m_s=10\ {\rm{MeV}}$ ($\equiv$ \msbar).}
\end{figure}
In Fig.\ \ref{cc} we show the structure function $F_2^{c}$ and its individual
contributions for two experimental values of $x$ and $Q^2$ 
\cite{Rabinowitz:1993xx}
under variation of the factorization scale $\mu^2$. Like in
the NC case we show the complete ${\cal{O}}(\alpha_s^1)$ result as
well as $F_2^{(0)+GF-SUB_g}$ where $QS^{(1)}$ has been neglected. 
The thick curves in Fig.\ \ref{cc} (a) have been obtained with a regularizing
strange mass of 10 MeV.
They are numerically indistinguishable from the $m_s=0$ \msbar\ results
along the lines of \cite{Gluck:1996ve}. 
For the thin curves a larger strange mass of
500 MeV has been assumed as an upper limit. Finite mass effects can
therefore be inferred from the difference 
between the thin and the thick curves.
Obviously they are very small for all contributions and can be safely
neglected. For the higher $Q^2$ value of Fig.\ \ref{cc} (b) they would be
completely invisible, so we only show the $m_s=$10 MeV 
results ($\equiv$\msbar). 
Since the finite mass corrections within the ACOT scheme turn out to
be negligible as compared to massless \msbar
\ it is not surprising that we confirm the findings of 
Chapter~\ref{chap_hqcc}
concerning the importance of quark initiated corrections.
They are --in the case of CC production of charm--
{\it{not}} suppressed with respect to gluon
initiated corrections for all reasonable values of the factorization
scale. Only for small choices of $\mu^2 \sim Q^2+m_c^2$ 
can the quark initiated correction be neglected. In this region of $\mu^2$ also
gluon initiated corrections are moderate and Born approximation 
holds within $\sim 10$\%. 
For reasons explained in Section \ref{nc}
the absolute value of both corrections --gluon and quark initiated--
become very significant when large factorization scales like
$p_T^{max}$ are chosen. This can be inferred by looking at the region
indicated by the arrow in Fig.\ \ref{cc} which marks the scale $\mu=2p_T^{max}$
which was used in \cite{Bazarko:1995tt,*Bazarko:1994hm}. 
Analyses which use ACOT with a
high factorization scale and neglect quark initiated corrections
therefore undershoot the complete $\alpsi$ result by the difference
between the solid and the dot--dashed curve, which can be easily as large as 
\mbox{$\sim 20$ \%}. 
For reasons explained in
the introduction to this section we have used the radiative strange
sea of GRV94. When larger strange seas like CTEQ4 are used the
inclusion of the quark initiated contributions is even more important.

%
%
%

\section{Conclusions}
\label{conclusions}

In this chapter we have calculated and analyzed DIS on massive quarks
at $\alpsi$ within the ACOT scheme for heavy quarks. 
For NC DIS this contribution differs significantly from its massless
\msbar\ analogue for $\mu^2=Q^2$. Both give, however, a very small
contribution to the total charm structure function such that the large
relative difference is phenomenologically immaterial. At higher values of
the factorization scale $\mu^2 \sim (p_T^{max})^2$ the contributions
become significant and their relative difference vanishes. The
$QS^{(1)}$ contribution of Section \ref{dis2} can therefore be safely
approximated by its much simpler \msbar\ analogue at any scale. For CC
DIS quark initiated corrections should always be taken into account on
the same level as gluon initiated corrections. Due to the smallness of
the strange quark mass ACOT gives results which are almost identical 
to \msbar.

%% file: phd_hqfrag.tex
\chapter{Charm Fragmentation in Deep Inelastic Scattering}
\label{hqffdis}

This chapter generalizes the theoretical considerations
of Chapter \ref{hqcontrib} to the semi--inclusive deep inelastic 
production of heavy flavored hadrons. We apply these results 
to the CC and NC production of charm where we investigate
the universality of charm fragmentation functions
and the charm production dynamics, respectively. 
The work in this chapter is based on \cite{Kretzer:1998nt}.
\section{Introduction}
\label{introdiff}
In the preceeding chapter we have analyzed heavy quark initiated 
contributions to fully inclusive deep inelastic (DI) structure functions.
Towards lower values of the Bjorken variable $x$,
heavy (charm) quarks are produced in about $20 \%$ of the neutral current (NC)
\cite{Adloff:1996xq,Breitweg:1997mj,Breitweg:1999ad}
and charged current (CC) 
\cite{Seligman:1997mc,*Seligman:1997fe}
deep inelastic events in lepton--nucleon collisions. Therefore in this 
kinematical range, heavy quark events contribute an important component to 
the fully inclusive DI structure functions of the nucleon.  
However, due to acceptance losses
this component can usually not be measured directly by inclusively tagging on 
charm events and more differential observables have to be considered
like $E_D$ 
\cite{Abramowicz:1982zr,Rabinowitz:1993xx,Bazarko:1995tt,*Bazarko:1994hm}, 
$p_T$ or $\eta$ 
\cite{Adloff:1996xq,Breitweg:1997mj,Breitweg:1999ad,Adloff:1998vb}
distributions, where $E_D$, $p_T$ and $\eta$ are
the energy, transverse momentum and pseudorapidity of the charmed hadron
produced, i.e.\ mainly of a $D^{(\ast)}$ meson. 
In this chapter we consider $E_D$ spectra represented by the usual 
scaling variable $z$ defined below. Within DIS 
the charmed hadron energy spectrum is the distribution which is most sensitive
to the charm fragmentation process and may give complementary information to
one hadron inclusive $e^+e^-$ annihilation which is usually chosen to define
fragmentation functions (FFs) \cite{Altarelli:1979kv,Nason:1994xx}. 
A well understanding of charm fragmentation is
essential for any charm observable, e.g.\ the normalization of $p_T$ and $\eta$
distributions in photoproduction is substantially influenced by the hardness
of the FF \cite{Cacciari:1997du,Breitweg:1998yt}.
The $z$ distribution of charm 
fragments is directly measured in CC neutrinoproduction 
\cite{Abramowicz:1982zr,Rabinowitz:1993xx,Bazarko:1995tt,*Bazarko:1994hm}
and may give insight into details of the charm production dynamics in NC 
electroproduction. It has, e.g., been shown in 
\cite{Adloff:1996xq,Breitweg:1997mj} that the energy 
spectrum of $D^{(\ast)}$--mesons produced at the $ep$ collider 
HERA may be able to discriminate between
intrinsic and extrinsic production of charm quarks. 

Intrinsic heavy quark densities may arise due to a nonperturbative
component of the nucleon wave function \cite{Brodsky:1980pb,*Brodsky:1981se}
or due to a perturbative resummation 
\cite{Aivazis:1994pi,Thorne:1998ga,*Thorne:1998uu,Martin:1996ev,Buza:1998wv}
of large quasi--collinear logs [$\ln (Q^2/m^2)$; $Q$ and
$m$ being the virtuality of the mediated gauge boson and the 
heavy quark mass, respectively] arising at any order in fixed order extrinsic
production (or fixed order perturbation theory: FOPT). 
Here we will only consider the latter possibility for inducing
an intrinsic charm density $c(x,Q^2)$ which is concentrated at small $x$ and 
we will ignore nonperturbative components which are expected to be located at
large $x$ \cite{Brodsky:1980pb,*Brodsky:1981se}. 
Technically the resummation of large perturbative logs proceeds
through calculating the boundary conditions for a transformation of the
factorization scheme 
\cite{Collins:1986mp,*Qian:1984kf,Buza:1998wv,Mele:1991cw}, 
which is switched from $n_f$ to $n_f+1$ active,
massless flavors, canonically at $Q^2=m^2$. For fully inclusive DIS 
the Kinoshita--Lee--Nauenberg theorem confines all quasi--collinear logs 
to the initial 
state such that they may be absorbed into $c(x,Q^2)$.
For semi--inclusive DIS (SI DIS) also final state collinearities 
arise which are resummed in 
perturbative fragmentation functions $D_i^c(z,Q^2)$ 
(parton $i$ decaying into charm quark $c$)
along the lines of \cite{Mele:1991cw,Cacciari:1997du}. The scale dependence of 
$c(x,Q^2)$ and  $D_i^c(z,Q^2)$ is governed by massless renormalization
group (RG) evolution.

Besides this {\it{zero mass variable flavor number scheme}}, where
mass effects are only taken care of by the boundary conditions for $c(x,Q^2)$
and $D_i^c(z,Q^2)$, variable flavor number schemes have been formulated 
\cite{Aivazis:1994pi,Thorne:1998ga,*Thorne:1998uu,Martin:1996ev,Buza:1998wv}, 
which aim at resumming the quasi--collinear logs as outlined above
while also keeping power suppressed terms of
${\cal{O}} [(m^2/Q^2)^k]$ in the perturbative coefficient functions. 
Our reference
scheme for this type of schemes will be the one developed by  
Aivazis, Collins, Olness and Tung (ACOT) 
\cite{Aivazis:1994kh,Aivazis:1994pi,Collins:1998rz}. 
In the ACOT scheme full dependence on the heavy quark mass is kept in graphs 
containing heavy quark lines. This gives rise to the above mentioned 
quasi--collinear 
logs and to the power suppressed terms. While the latter are regarded as mass 
corrections to the massless, dimensionally regularized, standard coefficient 
functions (e.g.\ in the 
${\overline{{\rm{MS}}}}$ scheme), the former are removed numerically by  
subtraction terms, which are obtained from the small mass limit of the
massive coefficient functions.

The outline of this chapter will be the following: In Section \ref{sec2} 
we will shortly
overview the relevant formulae for SI DIS for general masses
and couplings including quark scattering (QS) and boson gluon fusion (GF)
contributions up to $\alpsi$. 
We will thereby
present our ACOT based calculation for the $QS^{(1)}$\footnote{
Bracketed upper indices count powers of $\alpha_s$.
}
component of SI structure functions. 
In Section \ref{chffsi} 
we will analyze the charm fragmentation function in CC and
NC DIS. In Section \ref{siconc} 
we draw our conclusions and
some uncomfortably long formulae are relegated to Appendix \ref{rgediff}.

\section{Semi--Inclusive Heavy Quark Structure Functions}
\label{sec2}

This section presents the relevant formulae for one (heavy flavored) hadron
inclusive DIS structure functions. The contributions from scattering events
on massive quarks are given up to $\alpsi$ in Section \ref{scattq} and GF 
contributions are briefly recalled in Section \ref{glufu}. 
Section \ref{subterms}
presents all subtraction terms which render the structure functions infrared
safe and includes a discussion of these terms.

\subsection{Scattering on Massive Quarks} \label{scattq}

We consider DIS of the virtual Boson $B^\ast$ with momentum $q$
on the quark $Q_1$ with mass $m_1$
and momentum $p_1$
producing the quark $Q_2$ with mass $m_2$ and momentum $p_2$.
The latter fragments into a heavy quark
flavored hadron $H_{Q_2}$, 
e.g.\ a $\left|Q_2 \bar{q}_l\right>$ meson, $q_l$ being any light
quark. Phenomenologically most prominent are of course charm quarks fragmenting
into $D^{(\ast)}$--mesons which are the lightest heavy flavored hadrons.

We will strictly take over the formulae and notations of 
our inclusive analysis in Chapter \ref{hqcontrib}
whenever possible and merely extend them for SI DIS considered here.
In particular we take over the definition of the structure functions 
${\cal{F}}_i$
given in terms of the usual experimental structure functions $F_i$ in 
Eq.\ (\ref{def}).

Since we want to investigate the energy spectrum of charm fragments,
we introduce the Lorentz--invariant $z\equiv p_{H_{Q_2}}\cdot p_N/q\cdot p_N$ 
which reduces to the energy $E_{H_{Q_2}}$ scaled to its maximal value 
$\nu=q_0$ in the target rest
frame. Therefore in contrast to Eq. (\ref{hadronic}) we do not integrate 
the tensor ${\hat{\omega}}^{\mu\nu}$ over the 
full partonic phase space but keep it differential in the corresponding
partonic variable $z^\prime \equiv p_2\cdot p_1/q\cdot p_1$
or the mass corrected variable ${\hat{z}}$ which is defined below. 
In order to obtain hadronic observables we have to extend the Ansatz
of Eq.\ (17) in \cite{Aivazis:1994kh} such that it includes a 
nonperturbative hadronization function $D_{Q_2}$. 
In the limit of vanishing masses the massless parton model expressions
have to be recovered. Our Ansatz will be
\begin{equation}
\label{ansatzxz}
W^{\mu\nu} = \int\ \frac{d\xi}{\xi}\ \frac{d\zeta}{\zeta} 
\ Q_1(\xi,\mu^2)
\ D_{Q_2}(\zeta,[\mu^2])
\ {\hat{\omega}}^{\mu\nu}|_{\{p_1^+=\xi P^+; z=\zeta \hat{z}\}}\ \ \ ,
\end{equation}
where $\mu$ is the factorization scale,  
${\hat{z}}=z^\prime / z^\prime_{LO}$ with
$z^\prime_{LO}=\spp / \spm$
and $v^+\equiv (v^0+v^3)/\sqrt{2}$ for a general vector $v$.
$W^{\mu\nu}$ is the usual hadronic tensor and ${\hat{\omega}}^{\mu\nu}$ 
is its partonic analogue. 
Eq.\ (\ref{ansatzxz}) defines the fragmentation function $D_{Q_2}$ to be a 
multiplicative factor multiplying inclusive structure functions at LO/Born
accuracy, i.e.
\begin{equation}
{\cal F}_{i=1,2,3}^{QS^{(0)}} (x,z,Q^2) = 
{\cal F}_{i=1,2,3}^{QS^{(0)}} (x,Q^2)\ D_{Q_2}(z,[Q^2]) 
\ \ \ ,
\end{equation}
where the ${\cal F}_{i=1,2,3}^{QS^{(0)}}(x,Q^2)$ are defined and given in
(\ref{def}). The scale dependence of $D_{Q_2}$ is bracketed here and in the 
following because it is optional; a more detailed   
discussion on this point will be given at the end of Section \ref{subterms}.
We do not construct our Ansatz in Eq.\ (\ref{ansatzxz})
for the convolution of the fragmentation 
function along light front components for the outgoing particles which would
only be Lorentz--invariant for boosts along a specified axis. Since the final
state of DIS is spread over the entire solid angle it has no preferred axis as 
defined 
for the initial state by the beam direction of collider experiments.
Note that 
Eq.\ (\ref{ansatzxz}) is in agreement with usual factorized 
expressions for massless initial state quanta, as considered, e.g., in 
Chapter~\ref{chap_hqcc},
since there $m_1=0$ such that $z^\prime_{LO}=1$. 

Up to $\alpsi$ the hadronic structure functions for scattering on a 
heavy quark read 
\begin{eqnarray} \label{QS1xz}
{\hat{{\cal{F}}}}_{i=1,2,3}^{QS^{(0+1)}} (x,z,Q^2,\mu^2)
&=& Q_1(\chi,\mu^2)\ D_{Q_2}(z,[\mu^2])
+\frac{\alpha_s(\mu^2)}{2 \pi}\int_{\chi}^1\frac{d\xip}{\xip}
\int_z^1\frac{d{\hat{z}}}{{\hat{z}}} \\ \nonumber
&\times&
\left[Q_1\left(\frac{\chi}{\xip},\mu^2\right)\ {\hat{H}}_i^q(\xip,\zp,\mu^2)
\ \Theta_q\  \right]
\ D_{Q_2}\left(\frac{z}{{\hat{z}}},[\mu^2]\right)\ \ ,
\end{eqnarray}
with \cite{Aivazis:1994kh}
\begin{equation}
\chi = \frac{x}{2 Q^2}\ \left(\ \spm\ +\ \Delta\ \right)\ \ \ . 
\end{equation}
As throughout this thesis we set the renormalization scale
equal to the factorization scale In Eq.\ (\ref{QS1xz}).
The kinematical boundaries of the phase space in Eq.\ (\ref{QS1xz}) 
are introduced by
the theta function cut $\Theta_{q}$. In the massless limit 
$\Theta_{q}\rightarrow 1$. The precise arguments of $\Theta_{q}$
are  set by the kinematical requirement
\begin{equation}
\label{bounds} 
\zp_{\min} < \zp < \zp_{\max}\ \ \ ,
\end{equation}
with
\begin{eqnarray} \nonumber
\zp_{\max\atop \min} = \frac{
\pm \Delta[{\hat{s}},m_1^2,-Q^2]({\hat{s}}-m_2^2)+(Q^2+m_1^2+{\hat{s}})
({\hat{s}}+m_2^2)
}
{
2 {\hat{s}}({\hat{s}}-m_1^2+Q^2)
}\ \ \ ,
\end{eqnarray}
where ${\hat{s}}=(p_1+q)^2$.
Note that Eq.\ (\ref{bounds}) also poses an implicit constraint on $\xip$
via $\zp_{\min} < \zp_{\max}$.

The ${\hat{H}}_i^q$ for nonzero masses are obtained in exactly the same way 
as outlined for fully inclusive structure functions in Appendix \ref{hqini}
if the partonic phase space is not fully integrated over. 
They are given by
\begin{eqnarray} \nonumber
{\hat{H}}_i^q(\xip,\zp)&=&C_F\ \Big[
\ (S_i+V_i)\ \delta(1-\xip)\ \delta(1-{\hat{z}})  \\ \label{Hiq} 
&+& 
\frac{1-\xip}{(1-\xip)_{+}}
\ \frac{z^\prime_{LO}}{8}\ \frac{\hsi+\spm}{\Delta^\prime}
\ N_i^{-1}\ \hat{f}_i^Q(\hsi,\hti)\Big]\ ,
\end{eqnarray}
where the normalization factors $N_i$
are given in Appendix \ref{rvstruct}
and the Mandelstam variables read
\begin{eqnarray*}
\hsi(\xip)&\equiv& (p_1+q)^2-m_2^2 \\
&=&\frac{1-\xip}{2 \xip}[(\Delta-\spm)\xip+\Delta+\spm]\ \\
\hti(\xip,\zp)&\equiv& (p_1-p_3)^2-m_1^2 \\
&=&[\hsi(\xip)+\spm](\zp-z_0^\prime)\ , 
\end{eqnarray*}
where $\displaystyle z_0^\prime=\frac{\hsi+\spp}{\hsi+\spm}$ 
is the would--be pole of the ${\hat{t}}$--channel propagator
and we use $\Delta^\prime\equiv\Delta[{\hat{s}},m_1^2,-Q^2]$.
$S_i$ and $V_i$ are the soft real and virtual contribution to 
${\hat{H}}_i^q$, respectively.
They can be found in Appendix \ref{rvstruct} 
whereas the 
$\hat{f}_i^Q(\hsi,\hti)$ are listed in Appendix \ref{rgediff}.
In the massless limit the ${\hat{H}}_i^q$
reduce to the \msbar\ coefficient functions in
\cite{Altarelli:1979kv,cit:FP-8201} up to some divergent subtraction 
terms which we will specify in Section \ref{subterms}.

\subsection{Gluon Fusion Contributions at $\alpsi$} \label{glufu}

We have already obtained the semi--inclusive coefficient functions for GF 
production of massive quarks for general masses and couplings in 
Chapter \ref{chap_hqcc}. They are given by
\begin{eqnarray}  \nonumber
{\hat{F}}_{1,3}^{GF}(x,z,Q^2,\mu^2) &=& \int_{ax}^{1} \frac{dx'}{x'} 
              \int_z^1
      \ \frac{d\zeta}{\zeta}\ g(x',\mu^2)\ f_{1,3}(\frac{x}{x'},\zeta,Q^2)
      \ D_{Q_2}(\frac{z}{\zeta},[\mu^2])\ \Theta_g  \\ \nonumber \\
{\hat{F}}_{2}^{GF}(x,z,Q^2,\mu^2) &=& \int_{ax}^{1} \frac{dx'}{x'} 
              \int_z^1
      \ \frac{d\zeta}{\zeta}\ x'g(x',\mu^2)\ f_{2}(\frac{x}{x'},\zeta,Q^2)
      \ D_{Q_2}(\frac{z}{\zeta},[\mu^2])\ \Theta_g
\label{GFxz}
\end{eqnarray}
\noindent
with the $f_i$ from Eq.~(\ref{fiGFz}) and $ax=[1+(m_1+m_2)^2/Q^2]x$.
The $\Theta_g$ cut guarantees that $\zeta_{\min} < \zeta < \zeta_{\max}$ where 
$\zeta_{\min,\max}$ are given in Eq.\ (\ref{kseq3}). 
Similarly to 
Eq.\ (\ref{bounds}) $\zeta_{\min} < \zeta_{\max}$ may also constrain the phase
space available for the $x^\prime$ integration.

\subsection{Subtraction Terms}
\label{subterms}

It requires three \msbar\ subtraction terms to render the double convolutions
in Eqs.\ (\ref{QS1xz}), (\ref{GFxz}) infrared safe: 
\begin{eqnarray} \nonumber
{\cal F}_i^{SUB_q} (x,z,Q^2,\mu^2)&=&\frac{\alpha_s(\mu^2)}{2 \pi}C_F 
\int_{\chi}^1\frac{d\xip}{\xip}
\left[\frac{1+{\xip}^2}{1-\xip}\left(\ln \frac{\mu^2}{m_1^2}-1
-2 \ln(1-\xip)\right)\right]_{+} \\ 
& & \times \
Q_1\left(\frac{\chi}{\xip},\mu^2\right)\ D_{Q_2}(z,[\mu^2])
\end{eqnarray}
\begin{equation}
\label{SUBgxz}
{\cal F}_i^{SUB_g} (x,z,Q^2,\mu^2)= D_{Q_2}(z,[\mu^2]) 
\ \frac{\alpha_s(\mu^2)}{2 \pi}
\ \ln \frac{\mu^2}{m_1^2} 
\ \int_{\chi}^1\frac{d\xip}{\xip}\ P_{qg}^{(0)}(\xip)   
\ g\left(\frac{\chi}{\xip},\mu^2\right)\ \ \ 
\end{equation}  
\begin{eqnarray} \nonumber
{\cal F}_i^{SUB_D} (x,z,Q^2,\mu^2)&=&\frac{\alpha_s(\mu^2)}{2 \pi}C_F 
\int_{z}^1\frac{d\zip}{\zip}
\left[\frac{1+{\zip}^2}{1-\zip}\left(\ln \frac{\mu^2}{m_2^2}-1
-2 \ln(1-\zip)\right)\right]_{+} \\ 
& &\times \
D_{Q_2}\left(\frac{z}{\zip},\mu^2\right)\ Q_1\left(\chi,\mu^2\right)\ \ \ ,
\label{SUBD}
\end{eqnarray}
where $P_{qg}^{(0)}(\xip) = 1/2\ [{\xip}^2+(1-\xip)^2]$. 
Note that $SUB_g$ in Eq.\ (\ref{SUBgxz}) differs slightly
from Eq.~(\ref{FiSUB}) in Chapter~\ref{chap_hqcc} 
because we are allowing for
a nonzero initial state parton mass
$m_1$ here which we did not in 
Chapter~\ref{chap_hqcc}. 

The subtraction terms define the running of the initial state quark density
($SUB_q$, $SUB_g$) and the final state fragmentation function ($SUB_D$) in
the massless limit. They remove collinear logarithms and scheme defining
finite terms from the convolutions in  Eqs.\ (\ref{QS1xz}), (\ref{GFxz}) and  
they are constructed such that the massless 
\msbar\ results of \cite{Altarelli:1979kv,cit:FP-8201} are recovered 
in the limit 
\begin{equation}
\lim_{m_{1,2}\to 0} \left[{\hat{{\cal F}}}_i^{QS^{(0+1)}+GF} (x,z,Q^2,\mu^2)
-{\cal F}_i^{SUB_q+SUB_g+SUB_D} (x,z,Q^2,\mu^2)\right]=
{\cal F}_i^{(1),\overline{\rm{{MS}}}} (x,z,Q^2)\ \ \ ,
\end{equation}
where $SUB_q$ and $SUB_D$ regularize ${\hat{{\cal F}}}_i^{QS^{(0+1)}}$
whereas $SUB_g$ regularizes ${\hat{{\cal F}}}_i^{GF}$.
Contrary to the fully inclusive $SUB_g$ term in Eq.\ (\ref{SUBg}) 
there is no need to include an additional $\sim \ln (\mu^2/m_2^2)$ subtraction
in Eq.\ (\ref{SUBgxz}) because the ${\zeta}^{-1}$ ${\hat{u}}$--channel 
singularity of the massless limit of $GF^{(1)}$ is located at 
$\zeta=0$ and is outside the integration volume of Eq.\ (\ref{GFxz}).  
If only initial state subtractions, i.e.\ $SUB_q$ and $SUB_g$, are considered 
and final state subtractions, i.e.\ $SUB_D$, are not performed one
reproduces, in the limit $m_1\rightarrow 0$ the results in 
\cite{Gluck:1997sj,Kretzer:PhD} 
for
producing a heavy quark from a light quark. Note that in this case the 
fragmentation function $D_{Q_2}$ should be taken scale--{\em{in}}dependent,
say of the Peterson form \cite{Peterson:1983ak}.
In the limit where also the final state quark mass $m_2$ approaches zero 
and the final state subtraction term
$SUB_D$ is subtracted from the results in 
\cite{Gluck:1997sj,Kretzer:PhD}\footnote{
See Equation (B.11) in \protect\cite{Kretzer:PhD}.} 
the massless quark results in \cite{Altarelli:1979kv,cit:FP-8201} 
are obtained and a running of 
$D_{Q_2}$ is induced via a RG resummation of final state collinear logs 
as formulated for one hadron inclusive $e^+e^-$ annihilation 
in \cite{Mele:1991cw,Cacciari:1997du}.

Apart from removing the long distance physics from the coefficient functions 
the subtraction terms set the boundary conditions for the intrinsic
heavy quark density  $Q_1$ \cite{Collins:1986mp,*Qian:1984kf}
and the perturbative part of the heavy quark
fragmentation function $D_{Q_2}$ \cite{Mele:1991cw}: 
\begin{eqnarray} \label{boundq}
Q_1(x,Q_0^2)&=&
\frac{\alpha_s(Q_0^2)}{2 \pi}
\ \ln \frac{Q_0^2}{m_1^2} 
\ \int_{x}^1\frac{d\xi}{\xi}\ P_{qg}^{(0)}(\xi)   
\ g\left(\frac{x}{\xi},Q_0^2\right) \\ \nonumber
D_{Q_2}(z,\Qt_0^2)&=&
\frac{\alpha_s(\Qt_0^2)}{2 \pi}C_F 
\int_{z}^1\frac{d\zip}{\zip}
\left[\frac{1+{\zip}^2}{1-\zip}\left(\ln \frac{\Qt_0^2}{m_2^2}-1
-2 \ln(1-\zip)\right)\right]_{+}  
D_{Q_2}\left(\frac{z}{\zip}\right) \\ \label{boundd}
\end{eqnarray}
where $Q_0$, $\Qt_0$ are the transition scales at which the factorization 
scheme is switched from $n_f$ to $n_f+1$, $n_f+2$ active flavors, 
respectively (assuming here for simplicity that $m_1<m_2$; 
a generalization to $m_1\ge m_2$ is straightforward). 
For general $Q_0$ also the gluon density and $\alpha_s$ undergo a scheme
transformation. 
Canonically $Q_0$ is set equal to the heavy quark mass $m_1$ which guarantees 
\cite{Collins:1986mp,*Qian:1984kf} up to two loops a continuous 
evolution of $\alpha_s$ and the light
parton densities across $Q_0$. All available heavy quark densities 
are generated using $Q_0=m_1$ and we will therefore follow this choice here 
although a variation of $Q_0$ might substantially influence the heavy quark
results even far above the threshold \cite{cit:GRS94} as was found in 
\cite{Kretzer:1999zd}. 
Note that at three loops
a continuous evolution across $Q_0$ can no longer be achieved, neither for 
the parton distributions \cite{Buza:1998wv} nor for $\alpha_s$ 
\cite{Chetyrkin:1997sg,*Wetzel:1982qg,*Bernreuther:1982sg,*Bernreuther:1983zp,*Bernreuther:1983ds,*Larin:1995va}.
Analogously to $Q_0=m_1$ we use $\Qt_0=m_2$ throughout. 
In Eq.\ (\ref{boundd}) we have made the distinction between the scale
dependent FF $D_{Q_2}(z,\Qt_0^2)$ and the scale independent FF $D_{Q_2}(z)$
explicit. Following the terminology in \cite{Mele:1991cw,Cacciari:1997du} 
the latter
corresponds to the nonperturbative part of the former and describes
the hadronization process at the end of the parton shower which is
described perturbatively by the massless RG evolution. Alternatively, 
$D_{Q_2}(z)$ corresponds to a scale--independent FF within FOPT where no
collinear resummations are performed. These two points of view may
induce a scheme dependence if $D_{Q_2}(z)$ is fitted to data.
In principle, the massless evolution 
equations generate nonzero FFs also for light partons to decay into heavy
flavored hadrons. These light$\rightarrow$heavy contributions  
are important at LEP energies \cite{Cacciari:1997du} 
but can be safely neglected at the scales considered here\footnote{
We could therefore, in principle, restrict the evolution of the charm FF 
to the nonsinglet sector.}
and we will assume $D_{i\neq Q_2}(z,\mu^2)=0$ throughout. 

\subsection{SI Structure Functions at $\alpsi$}
\label{sisf1}

In the next section we will consider three types of $\alpsi$ VFNS structure
functions. The first two are constructed at full $\alpsi$
\begin{equation}
\label{full}
 F_i^{QS^{(0+1)}+GF} - F_i^{SUB_q+SUB_g+[SUB_D]}\ \ \ ,
\end{equation}
where the inclusion or omission of the bracketed $SUB_D$ term corresponds
to a running or scale--independent fragmentation function, respectively,  
as discussed in
the previous section. It is somewhat unclear whether $QS^{(1)}$ 
contributions (and the corresponding subtractions) 
should be considered on the same perturbative level as $GF^{(1)}$, see
the introduction to Chapter \ref{hqcontrib} 
for a more detailed discussion on that point. In the original
formulation of the ACOT scheme \cite{Aivazis:1994pi} $QS^{(1)}$ 
contributions are neglected 
at the level we are considering here and we therefore do also consider this
option via the partial $\alpsi$ structure function
\begin{equation}
\label{partial}
F_i^{QS^{(0)}+GF} - F_i^{SUB_g}\ \ \ .
\end{equation}   
For obtaining the numerical results of the next section the general 
formulae of this section have to be adjusted by choosing masses and 
couplings according 
to the relevant NC and CC values as in Section \ref{nc} and \ref{ccsec}.
For the CC case where $Q_1$ should be identified with strange quarks
the boundary condition in Eq.\ (\ref{boundq}) is inadequate
since $m_s^2\sim \Lambda_{QCD}^2$ is below the perturbative regime of
QCD. We will have recourse to standard strange seas from the literature 
\cite{cit:GRV94,Lai:1997mg} instead.

\section{The Charm Fragmentation Function in SI DIS} 
\label{chffsi}

We will investigate the charm fragmentation function in CC and in NC SI DIS.
In CC DIS the charm production mechanism is undebated since charm is
dominantly produced by scattering on light strange quanta.
Our reasoning will therefore be that $D_c$ is directly accessible in 
CC DIS at relatively low spacelike momentum transfer. An extracted
$D_c$ can then be applied to NC DIS,
where it might give insight into the details of the 
production dynamics \cite{Adloff:1996xq}. 
Also a test of the universality \cite{Kleinknecht:1983eb}
of the charm FF measured in CC DIS and $e^+e^-$ annihilation 
\cite{Cacciari:1997du} 
would be an important issue directly related to the factorization
theorems \cite{Collins:1989gx}
of perturbative QCD (pQCD). 

All $\varepsilon_c$ parameters 
discussed
below refer to a Peterson type \cite{Peterson:1983ak} functional form 
given in Eq.~(\ref{Peterson-Frag}) 

\subsection{CC DIS}
\label{ccsi} 

\begin{figure}[t]
\vspace*{-1.cm}
\hspace*{2.cm}
\epsfig{figure=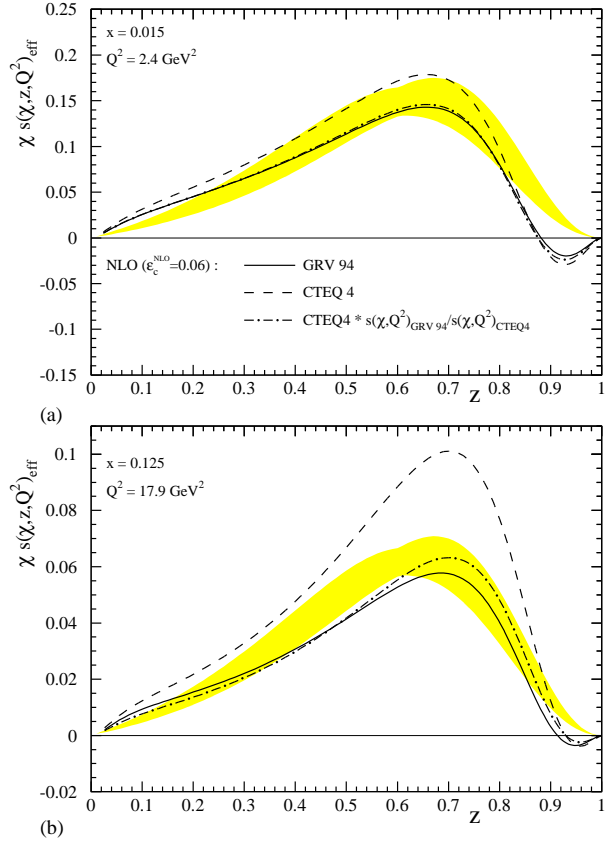,width=9.5cm}
\vspace*{-1.5cm}
\caption{\label{ks2fig1}\sf
The charm production cross section obtained in \protect\cite{Gluck:1997sj}
for two $x,Q^2$ points in the CCFR \protect\cite{Rabinowitz:1993xx} 
kinematical regime. 
Up to a constant of normalization which was
conveniently chosen in Eq.\ (4) of \protect\cite{Gluck:1997sj}
 $s_{eff}$ represents the triple--differential 
cross section $d^3 \sigma / dx dy dz$ where $x$ and $y$ are
standard and $z\equiv p_D\cdot p_N/q\cdot p_N$. Shown are the predictions
using GRV94 \protect\cite{cit:GRV94} (solid) and CTEQ4 
\protect\cite{Lai:1997mg} (dashed) partons and 
a curve (dot--dashed) where the normalization of the CTEQ4 prediction 
is changed by multiplying with
the ratio of the strange seas $s_{GRV}(\chi)/s_{CTEQ}(\chi)$. For all curves
a scale--independent Peterson FF with $\varepsilon_c=0.06$ has been used.}
\end{figure}

In CC DIS at fixed target energies
one does not expect to gain much insight into the charm
production process since 
charm is dominantly produced in scattering events on 
strange quarks\footnote{
We assume a vanishing Cabibbo angle. Our results remain, however, 
unchanged if the $CKM$ suppressed $d\rightarrow c$ background is included.
} 
in the nucleon, thereby permitting an experimental
determination of the strange quark content of the nucleon
\cite{Abramowicz:1982zr,Rabinowitz:1993xx,Bazarko:1995tt,*Bazarko:1994hm,Vilain:1998uw}.
On the other hand the well understood production mechanism makes
a direct determination of the charm fragmentation function feasible
by measuring the energy spectrum of final state charm fragments. 
This is obvious in leading order accuracy where\footnote{
We will suppress some obvious scale--dependences in the following 
formulae and in their discussion.
} 
\begin{equation}
\label{simple}
d \sigma_{LO} \propto s(\chi) D_c(z) 
\end{equation}
is directly proportional to $D_c$.
More precisely, it is not $z$ but the closely related energy of the 
$c\rightarrow \mu\nu$ decay muon which can be observed in iron detectors
\cite{Abramowicz:1982zr,Rabinowitz:1993xx,Bazarko:1995tt,*Bazarko:1994hm}. 
The smearing effects of the decay complicates the
determination of the precise shape of $D_c$ but only weakly influences an
extraction of $\left<z\right>$ \cite{Abramowicz:1982zr} which is valuable 
information if physically motivated one--parametric Ans\"{a}tze 
\cite{Peterson:1983ak,Collins:1985ms} for $D_c$ are {\it{assumed}}.  
At NLO the production cross section is no longer of the simple factorized
form of Eq.\ (\ref{simple}) and double convolutions 
(symbol $\otimes$ below) of the
form of Eqs.\ (\ref{QS1xz}), (\ref{GFxz}) have to be considered.
However, to a reasonable approximation
\begin{eqnarray} \nonumber
d \sigma_{NLO} &=& \left( \left[ s \otimes d{\hat{\sigma}}_s 
+ g \otimes d{\hat{\sigma}}_g \right ] \otimes D_c \right) 
(x,z,Q^2) \\ \nonumber
&\equiv& d \sigma_{LO}\ K(x,z,Q^2) \\ \nonumber
&\propto& s(\chi)\ D_c(z)\ K(x,z,Q^2) \\
&\simeq& s(\chi)\ {\cal{D}}_{x,Q^2}[D_c](z) 
\label{kfactor}
\end{eqnarray}
holds also at NLO accuracy within experimental errors
and for the limited kinematical range of present data
on neutrinoproduction of charm. In Eq.\ (\ref{kfactor}) the approximate 
multiplicative factor ${\cal{D}}$ absorbs the precise $K$--factor $K(x,z,Q^2)$
obtained from a full NLO QCD calculation. ${\cal{D}}$
is {\it{not}} a simple universal fragmentation
function but a nontrivial process--dependent functional 
which is, however, mainly sensitive on $D_c$ and shows little sensitivity on
the exact parton distributions considered. 
The occurrence of $x,Q^2$ and $z$ in Eq.\ (\ref{kfactor}) as indices and 
as a functional argument, respectively, reflects the fact that the dependence
on $x$ and $Q^2$ is much weaker than is on $z$. 
Eq.\ (\ref{kfactor}) tells us that $s(\chi)$ fixes the normalization of
$d \sigma$ once $K$ is known. On the other hand $K$ (or ${\cal{D}}$) can be
computed from $D_c$ with little sensitivity 
on $s(\chi)$, 
such that $s(\chi)$ and $D_c(z)$ decouple in the production
dynamics and can be simultaneously extracted. 
This point can be clearly inferred from Fig.\ \ref{ks2fig1} 
where it is shown
that the wide spread of CC charm production predictions which were obtained
in \cite{Gluck:1997sj} using GRV94 \cite{cit:GRV94} and 
CTEQ4 \cite{Lai:1997mg} strange seas
can be brought into good agreement by a mere
change of the normalization given by the ratio 
$s_{GRV}(\chi)/s_{CTEQ4}(\chi)$.
The remaining difference is not within present experimental
accuracy which can be inferred from the shaded band representing
a parametrization \cite{Gluck:1997sj} of CCFR data \cite{Rabinowitz:1993xx}.  
High statistics neutrino data therefore offer an ideal
scenario to measure $D_c$ complementary to an extraction
from LEP data on $e^+e^- \rightarrow D X$ 
\cite{Cacciari:1997du,Binnewies:1997gz,Binnewies:1998xq}.
This has been first noted in \cite{Kleinknecht:1983eb} where also a successful
test of the universality of the charm FF has been performed. 
With new data 
\cite{Abramowicz:1982zr,Rabinowitz:1993xx,Bazarko:1995tt,*Bazarko:1994hm}\footnote{Data 
from NuTeV \cite{ref:adams,Yu:1998su,*Yu:1998sw} are to be 
expected in the near future;
$\mu^\pm$ events observed at NOMAD await further analysis \cite{zuber}.} at hand and with 
a sounder theoretical understanding of neutrinoproduction of
charm it would be desirable to update the analysis in 
\cite{Kleinknecht:1983eb}.
Nowadays one can in principle 
examine the possibility of a uniform renormalization group
transformation from spacelike momenta near above the charm mass 
($\nu N\rightarrow D X$) to
timelike momenta at the $Z^0$ peak ($e^+e^-\rightarrow D X$). 
In \cite{Cacciari:1997du,Binnewies:1997gz,Binnewies:1998xq} charm 
fragmentation functions extracted from
LEP data have been tested against $p_T$ and $\eta$ distributions
measured in photoproduction at HERA. 
However, we believe that a comparison to $z$ differential neutrinoproduction 
data is worthwhile beyond, since the latter measure directly 
the fragmentation spectrum
whereas $p_T$ and $\eta$ shapes are rather indirectly influenced by
the precise hardness of the FF via their 
normalization \cite{Cacciari:1997du,Breitweg:1998yt}.
Unfortunately up to now no real production data are available
but only strange sea extractions 
\cite{Abramowicz:1982zr,Rabinowitz:1993xx,Bazarko:1995tt,*Bazarko:1994hm}
resulting from an analysis of the
former. We therefore strongly recommend that experiments produce
real production data such that the above outlined program can be executed
with rigor.    
Here we can only find an $\varepsilon_c$ parameter which
lies in the correct ball park and examine a few points which will
become relevant for an extraction of $D_c$ once data will become
available.

\begin{figure}[t]
\vspace*{-1.cm}
\hspace*{2.5cm}
\epsfig{figure=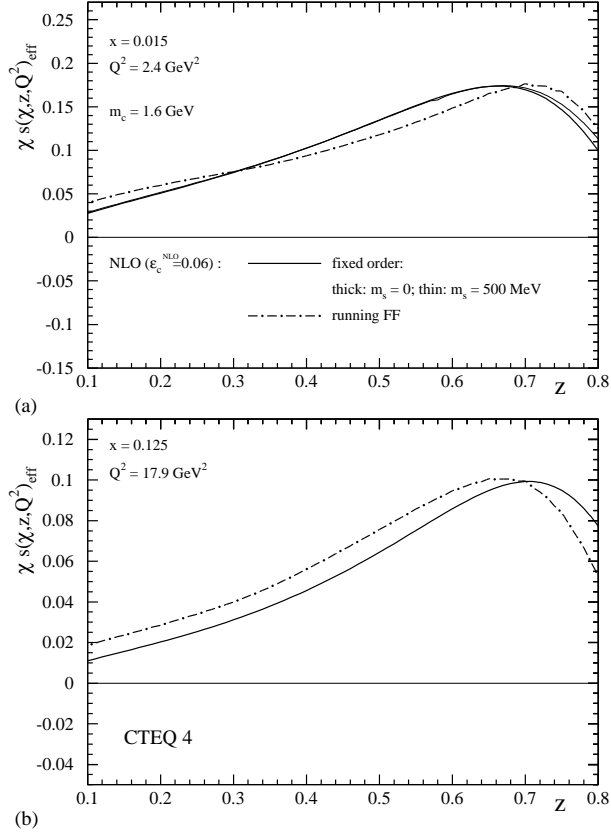,width=9.5cm}
\vspace*{-1.5cm}
\caption{\label{ks2fig2}\sf
The same quantity as in Fig.\ \ref{ks2fig1}. All predictions have been 
obtained for the CTEQ4 parton distributions. The solid curves result from
a scale--independent Peterson FF with $\varepsilon_c=0.06$. In (a) 
the thin solid curve has
been obtained using the formulae of Section \ref{sec2} and choosing 
a finite strange quark mass of $m_s=500\ {\rm{MeV}}$ whereas
the thicker curve corresponds to the asymptotic $m_s \rightarrow 0$ limit
($\equiv$\msbar). In (b) these two options would be completely
indistinguishable and we only show the \msbar\ result. 
For the dot--dashed curves
the final state quasi--collinear logarithm has been absorbed into a running of
the charm FF. A Peterson FF with $\varepsilon_c=0.06$ has been used as the
nonperturbative part of the input as given in Eq.\ (\ref{boundd}).}
\end{figure}

An outstanding question is the possible effect of a finite strange mass
on the full semi--inclusive charm production cross section 
including $\alpsi$ quark scattering contributions.
By comparing the thick and the thin solid curve in 
Fig.\ \ref{ks2fig2} (a) it is
clear that the effect of a finite $m_s$ can be neglected even at low
scales and for a maximally reasonable value of $m_s=500$ MeV.
For the larger scale of 
Fig.\ \ref{ks2fig2} (b) the effect of choosing a finite
$m_s$ would be completely invisible.
A further question which might influence the extraction of a universal
FF from neutrinoproduction is the one of the scheme dependence in handling
final state quasi--collinear logarithms $\ln (Q^2/m_c^2)$. If these are
subtracted from the coefficient functions as discussed in Section 
\ref{subterms}, the subtraction defines a running of the charm FF which 
becomes scale dependent\footnote{
Evolving the charm FF we adopt for consistency the evolution parameters 
$m_{c,b}$ and $\Lambda^{QCD}_{4,5}$ of the CTEQ4(M) \cite{Lai:1997mg}
parton distribution functions and we use $\Qt_0=m_c$.    
}
according to Eq.\ (\ref{boundd}).
In Fig.\ \ref{ks2fig2} we examine such resummation effects for 
CCFR kinematics \cite{Rabinowitz:1993xx}. We use the same Peterson FF with 
$\varepsilon_c=0.06$  once for a fixed order calculation
(solid lines) and once as the
nonperturbative part on the right hand side 
of the entire $c\rightarrow D$ FF on the left hand side of 
Eq.\ (\ref{boundd}) (dashed curves).
We note that towards intermediate scales around 
$Q^2\sim 20\ {\rm{GeV}}^2$
one begins to see the softening effects of the resummation
which are enhanced as compared to FOPT. However,
as one  would expect at these scales, the resummation effects are 
moderate  and could be compensated by only a {\it{slight}} shift of the
$\varepsilon_c$ parameter which is therefore, within experimental accuracy, 
insensitive to scheme transformations. 
We note that --as was already shown in \cite{Gluck:1997sj}--  
according to Fig.\ \ref{ks2fig1} an  $\varepsilon_c$ of around $0.06$ 
which we took from an older analysis in \cite{Chrin:1987yd} seems to 
reproduce the measured spectrum quite well. 
For $\left<E_\nu\right>=80\ {\rm{GeV}}$, $\left< Q^2\right>=20\ {\rm{GeV}}^2$
a value $\varepsilon_c \simeq 0.06$ gives an average $\left<z\right>\simeq0.6$
consistent with $\left<z\right>=0.68\pm 0.08$ measured by CDHSW 
\cite{Abramowicz:1982zr}.
In \cite{Cacciari:1997du}\footnote{
A similar $\varepsilon_c$ has been obtained in \cite{Binnewies:1997gz} 
in a related scheme.
The $\varepsilon_c$ value in \cite{Binnewies:1998xq} has no connection 
to a massive
calculation and cannot be compared to the values discussed here.}
a distinctly harder
value of $\varepsilon_c \simeq 0.02$ was extracted from LEP data
on $e^+e^-\rightarrow D^\ast X$. 
If the latter fit is
evolved down to fixed target scales it is --even within the
limited experimental accuracy-- incompatible with the  CCFR neutrino 
data represented
in Fig.\ \ref{ks2fig1}. From Fig.\ \ref{ks2fig2} 
it is clear that the difference cannot be attributed to
a scheme dependence of the $\varepsilon_c$ parameter which is too
small to explain the discrepancy. It would of course be interesting to know
how much the above mentioned smearing effect of the $c\rightarrow \mu\nu$ 
decay might dilute the discrepancy.   
In any case, charm fragmentation at LEP
has been measured by tagging on $D^\ast$s whereas neutrinoproduction
experiments observe mainly $D$s through their semileptonic decay--channel
(dimuon events). ARGUS \cite{Albrecht:1991ss} and CLEO 
\cite{Bortoletto:1988kw} data at 
$\sqrt{s}\simeq 10 {\rm{GeV}}$ indeed show \cite{Barnett:1996hr} 
a harder energy distribution of $D^\ast$s compared
to $D$s. It seems therefore to be possible within experimental accuracy 
to observe 
a nondegeneracy of the charm fragmentation functions into the lowest
charmed pseudoscalar and vector mesons.  
We note that an $\varepsilon_c$ value around $0.06$ which is in agreement
with neutrino data on $D$--production
is also compatible with the $D$ energy spectrum measured
at ARGUS where the evolution may be performed either via FOPT using 
expressions in \cite{Nason:1994xx} or via a RG transformation along the lines of
\cite{Mele:1991cw,Cacciari:1997du}. 
If forthcoming experimental analyses should confirm our 
findings the lower decade $m_c(\sim 1\ {\rm{GeV}})\rightarrow$ 
ARGUS$(10\ {\rm{GeV}})$ may be added to 
the evolution path ARGUS$(10\ {\rm{GeV}})\rightarrow$ LEP$(M_Z)$ paved in 
\cite{Cacciari:1997du} for the charm FF.

\subsection{NC DIS}
\label{ncsi}

The fragmentation function of charm quarks observed in NC DIS is of special 
interest since it allows for directly investigating 
\cite{Adloff:1996xq,Breitweg:1997mj}
the charm production
mechanism which is a vividly discussed issue in pQCD phenomenology 
\cite{cit:GRS94,Buza:1998wv,Olness:1995zn,*Lai:1997vu,*Daum:1996ec}. 
Whereas for intrinsic heavy quarks one expects to observe a Peterson--like
hard spectrum attributed 
to the dominance of the leading order quark scattering
contribution, one expects a much softer spectrum for extrinsic GF since
the gluon radiates the $c\bar{c}$ pair towards lower energies ($z$) during
the hard production process before the nonperturbative hadronization takes 
place. 
Experimental analyses have been performed in 
\cite{Adloff:1996xq,Breitweg:1997mj} and the steep 
spectrum\footnote{
The variable $x_D$ considered in \cite{Adloff:1996xq,Breitweg:1997mj} 
differs slightly from $z$ in definition. 
The variables are, however, identical at the 2\% level \cite{daum}.
} (best visible in Fig.\ 6 in \cite{Adloff:1996xq})
of observed $D$--mesons together with the missing of a hard 
component at larger $z$ seem to give clear evidence for the dominance of 
extrinsic $GF$ over $QS$ which was excluded at the 5\% 
level \cite{Adloff:1996xq}.
In a complete VFNS the $QS^{(0)}$ component makes, however, just one part 
of the
$\alpsi$ structure functions in Eqs.\ (\ref{full}), (\ref{partial}). 
Especially, there also exists a $GF^{(1)}$ component, albeit with 
the leading log part of it subtracted. 
Since the subtraction term in Eq.\ (\ref{SUBgxz}) is proportional to  
$D_c$ and therefore only removes a hard component from $GF^{(1)}$ one expects 
the rise towards lower $z$ to survive the subtraction. Furthermore a 
perturbative evolution of the
charm fragmentation function $D_c(z,Q^2)$ might soften somewhat the
hard QS term.

These expectations can be quantitatively confirmed in Fig.\ \ref{ks2fig3} where
we show for HERA kinematics \cite{Adloff:1996xq}
the total (solid line) normalized $\alpsi$ production cross section for 
transverse virtual 
photons ($F_L=0$)
on protons. We also show the individual components contributing to it:
\begin{figure}[t]
\vspace*{-0.5cm}
\hspace*{-1.25cm}
\epsfig{figure=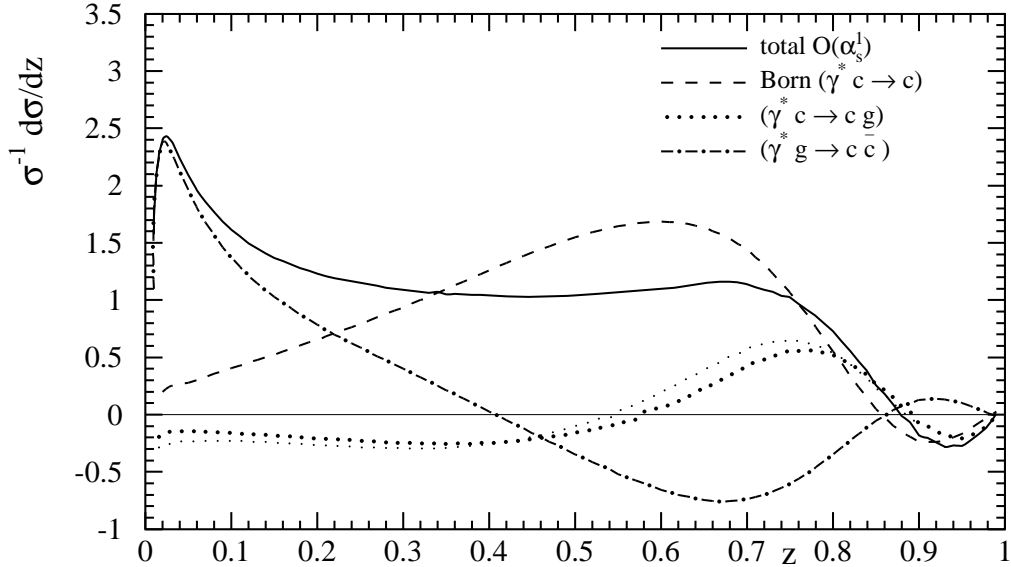,width=16cm}
\vspace*{-2.5cm}
\caption{\label{ks2fig3}\sf
The normalized charm production cross section 
$d \sigma / dz$ ($z\equiv p_D\cdot p_N/q\cdot p_N$) for HERA kinematics
($\sqrt{s}=314\ {\rm{GeV}}$). 
For a comparison with H1 data \protect\cite{Adloff:1996xq} the cross 
section has been integrated over $10\ {\rm{GeV}}^2< Q^2 < 100\ {\rm{GeV}}^2$
and $0.01<y<0.7$. Following the experimental analysis 
\protect\cite{Adloff:1996xq} only
contributions from transverse photons ($F_L=0$) are considered.   
Shown is the total $\alpsi$ result (solid line) and the individual 
contributions.
Details to the calculation of the total result and the individual contributions
are given in the text. The charm mass has been kept finite at the
$CTEQ4$ value of $m_c=1.6\ {\rm{GeV}}$ everywhere except for the thin dotted
curve where the $m_c\rightarrow 0$ limit has been taken.}
\end{figure}
The processes $\gamma^\ast g\rightarrow c {\bar{c}}$ 
(dot--dashed) and
$\gamma^\ast c \rightarrow c g$ 
(dotted; incl.\ virtual corrections) correspond
to the $GF^{(1)}-SUB_g$ and $QS^{(1)}-SUB_q-SUB_D$ terms, respectively,  
subtracted at $\mu=Q$. They
are {\it{not}} physically observable and only sensible if they are
added to the $QS^{(0)}$ Born term (dashed) as in Eqs.\ (\ref{full}), 
(\ref{partial}). 
We have perturbatively resummed all logarithms of the charm mass via massless 
evolution equations starting at the charm mass [$Q_0=\Qt_0=m_c$]
and using the standard boundary 
conditions in Eqs.\ (\ref{boundq}), (\ref{boundd}) for $\varepsilon_c=0.06$. 
Finite charm mass effects on the subtracted $QS^{(1)}$ contribution can be
inferred by comparing the thick and the thin dotted curves, where the 
$m_c \rightarrow 0$ limit has been taken for the latter. 
As has been
theoretically anticipated in \cite{Collins:1998rz} the charm mass 
can be safely set
to zero in $QS^{(1)}$ and the involved convolutions in Eq.\ (\ref{ansatzxz})
may be replaced by the massless expressions in 
\cite{Altarelli:1979kv,cit:FP-8201} which simplifies 
the numerics essentially and which we will therefore do for 
the $\mu=2 m_c$ curve in Fig.\ \ref{ks2fig4} below.
As also stressed in \cite{Collins:1998rz} it is, however, essential to keep
the charm mass finite in the $GF^{(1)}$ contribution since $m_c$ tempers
the strength of the ${z^\prime}^{-1}$ ${\hat{u}}$--channel 
propagator singularity.
Obviously the $\alpsi$ result of an ACOT based calculation deviates essentially
from the naive Born term expectation and it seems by no means legitimate
to treat $GF^{(1)}$ as a higher order correction here. Contrary to the
corresponding inclusive results in Chapter \ref{hqcontrib} and to 
the expectations in 
\cite{Aivazis:1994pi} also the subtracted $QS^{(1)}$ term is numerically 
significant
in the semi--inclusive case considered here. In the light of the huge 
$\alpsi$ corrections it seems, however, undecidable as to whether include
or omit $QS^{(1)}$ at $\alpsi$ without knowing \alpsq\ corrections within ACOT.

In Fig.\ \ref{ks2fig4} the resummed result (VFNS: solid lines)
can be compared to 
unsubtracted $\alpsi$\footnote{
An ${\cal{O}}(\alpha_s^2)$ NLO calculation \cite{Harris:1995tu} 
within FOPT gives very similar results \cite{daum}. 
} 
$GF^{(1)}$ (fixed order, dashed line). 
We show the total $\alpsi$ VFNS result for the choices
$\mu=Q,2 m_c$. 
\begin{figure}[t]
\vspace*{-0.5cm}
\hspace*{-1.25cm}
\epsfig{figure=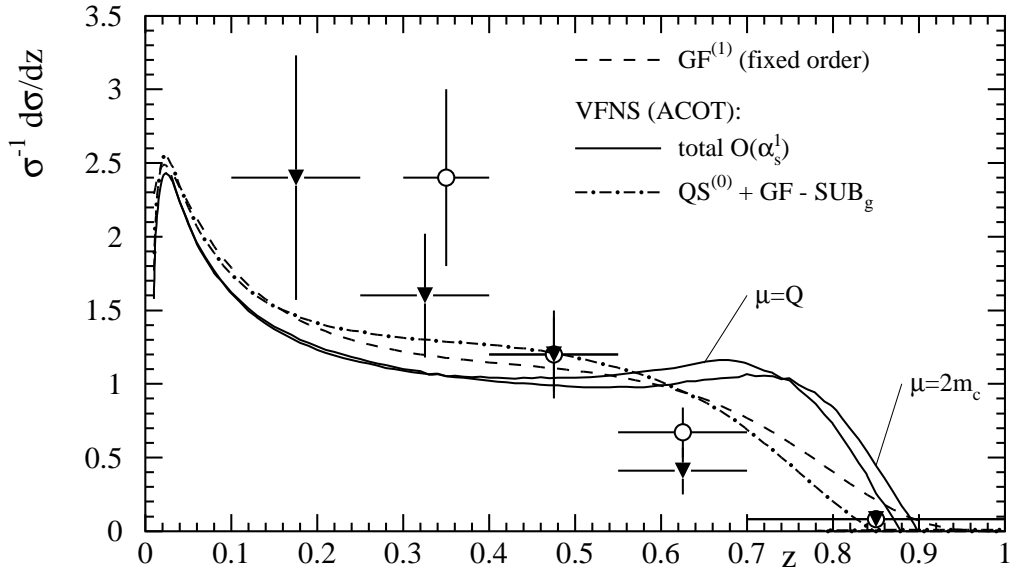,width=16cm}
\vspace*{-2.5cm}
\caption{\label{ks2fig4}\sf
A Comparison of the total $\alpsi$ result (solid lines)
to H1 data \protect\cite{Adloff:1996xq} on $D^0$ (circles) and ${D^+}^\ast$ 
(triangles; both including
charge conjugation) production. Shown are results for two choices of the
factorization/renormalization scale $\mu$. 
Also shown is the outcome of a fixed order $\alpsi$
$GF$ calculation for comparison (dashed line). The dot--dashed line follows
the suggestion in \protect\cite{Aivazis:1994pi} and neglects quark initiated 
contributions
at $\alpsi$, i.e. the difference between the solid ($\mu=Q$)
and the dot--dashed line is given
by the (thick) dotted line in Fig.\ \ref{ks2fig3}. 
For the dot--dashed curve as well as for the 
fixed order calculation (dashed)
only $\mu=Q$ is shown since the scale dependence is
completely insignificant.}
\end{figure}
The modest  scale dependence arises exclusively through the $QS^{(1)}$ term.
For any of the other curves a variation of $\mu$ is completely 
insignificant and we therefore only show $\mu=Q$.
A full VFNS calculation (solid lines)
seems hardly distinguishable from
fixed order perturbation theory (dashed line) within experimental accuracy. 
The two approaches are even closer if one follows the suggestion in 
\cite{Aivazis:1994pi} and does not include 
(dot--dashed line) the subtracted $QS^{(1)}$ term at the level of 
$QS^{(0)}+GF^{(1)}$. 
The data points in the figure correspond to H1 measurements \cite{Adloff:1996xq}
of $D^{0}$ (circles) and ${D^{\ast}}^+$ 
(triangles; both including charge conjugation) spectra.
The measurement is restricted to $\eta_D<1.5$. Extrapolating to the full
phase space gives rise to large acceptance corrections which are, however,
quite uniform \cite{daum} over the kinematical range considered and therefore
have a minor effect on the {\it{normalized}} spectrum.   
Since FOPT and ACOT based calculations are very close it seems improbable
that an experimental discrimination between the two approaches
will be possible. The Born term in Fig.\ \ref{ks2fig3} 
is far from being the dominant 
contribution and an intrinsic $c(x,Q^2)$ stemming from the resummation of
perturbative logs can therefore not be excluded.
The tendency of the data appears somewhat softer than any of the 
calculations and the tendency seems to be confirmed by preliminary
data in a lower $z$ bin \cite{Tzamariudaki:1998na}. 
The resummed calculation appears to be 
too hard at larger $z$ around $0.6$ if the $QS^{(1)}$ component is included. 
As already mentioned,
at the present stage of the calculations it cannot be decided whether this
hints at an intrinsic problem within VFNS calculations or whether this may
be cured by \alpsq\ corrections.

\section{Conclusions}
\label{siconc}
In this chapter we have performed an ACOT \cite{Aivazis:1994pi} based 
analysis of heavy quark 
fragmentation in DIS including a calculation of semi--inclusive scattering on 
massive quarks at $\alpsi$ for general masses and couplings. 
As in the inclusive case of Chapter \ref{hqcontrib}, effects from 
finite initial state
quark masses can be neglected for practical applications to charm
production in CC and NC DIS. The involved convolutions in Section 
\ref{scattq} can therefore safely be replaced by their analogues in
\cite{Altarelli:1979kv,cit:FP-8201} and in Appendix B of \cite{Kretzer:PhD}.  
Neutrinoproduction is an ideal environment to extract the charm FF
within DIS and a Peterson \cite{Peterson:1983ak} type FF with 
$\varepsilon_c\simeq 0.06$ seems to lie in the correct ball park, where the 
sensitivity on the choice of scheme is small and finite $m_s$ effects
are irrelevant. 
The $\varepsilon_c$ value above is compatible with $e^+e^-$ data 
if a nondegeneracy of charm quarks fragmenting into $D$ and $D^\ast$ mesons is
allowed for. 
For NC DIS it seems unlikely that a discrimination between
fixed order and resummed calculations will be possible at HERA. Both
approaches give similar results which show a spectrum that is somewhat harder
than the tendency of the data 
\cite{Adloff:1996xq,Breitweg:1997mj,Tzamariudaki:1998na}. 
The resummed calculation is made worse if the $\alpsi$ quark scattering 
contribution is included at the perturbative level considered here.     

%% file: phd_hqsummary.tex
\chapter{Summary}
\label{chap_hqsummary}
In the first part of this thesis we have studied deep inelastic 
production of heavy quarks in neutral current and charged current
processes.
For this purpose we have calculated the relevant partonic subprocesses
to order \alpsi\ (BGF, QS) for general masses and couplings taking into 
account massive initial state quark--partons as needed in the
variable flavor number scheme of ACOT \cite{Aivazis:1994pi}. 
Our calculation of the vertex correction
with general masses and couplings is, to our best knowledge, new.

The integrated partonic results could be used to formulate
and investigate heavy quark contributions to inclusive deep inelastic
structure functions within the ACOT variable flavor number scheme.
By the calculation of the before missing radiative corrections
to scattering amplitudes on massive quark partons 
(including virtual corrections)
the ACOT scheme could be completed to full order \alpsi.
Furthermore, we utilized the unintegrated partonic structure 
functions to extend the original ACOT scheme \cite{Aivazis:1994pi}
from its inclusive formulation to ($z$--differential)
one--hadron--inclusive leptoproduction. 

With help of these results we demonstrated in the charged current
sector that the effects of a finite strange mass are small and
rapidly converge towards \msbar\ in the limit $m_s \to 0$.
Furthermore, it turned out that the effects of finite initial
state masses can be neglected in all cases, such that the (massive) 
QS$^{(1)}$ contribution of Section~\ref{dis2} can safely be approximated
by its much simpler \msbar\ analogue at any scale.
Nevertheless, it should be stressed that radiative corrections to 
quark initiated processes are in general not negligible and should
be included in complete NLO analyses employing heavy quark
parton densities.

Finally, we have performed an ACOT based analysis of heavy quark
fragmentation in DIS led by the observation
that semi--inclusive CC DIS is well suited to extract
the charm fragmentation function. 
We reassured that theoretical uncertainties due to scheme choices
and finite $m_s$ effects are irrelevant and found out that 
a Peterson type fragmentation function with $\varepsilon_c \approx 0.06$
seems to lie in the correct ball park.
The $\varepsilon_c$ value above is compatible with $e^+e^-$ data 
if a nondegeneracy of charm quarks fragmenting into $D$ and $D^\ast$ mesons is
allowed for. 
In the case of semi--inclusive NC DIS 
it seems unlikely that a discrimination between
fixed order and resummed calculations will be possible at HERA. Both
approaches gave similar results exhibiting a spectrum that 
is somewhat harder than the tendency of the data 
\cite{Adloff:1996xq,Breitweg:1997mj,Tzamariudaki:1998na}. 
The resummed calculation is made worse if the $\alpsi$ quark scattering 
contribution is included at the perturbative level considered here.     


%% file: phd_gam_intro.tex
\chapter{Introduction and Survey}
\label{chap:phd_gam_intro}
Photon physics is an active field of research 
as is documented by and in a large number of reviews
\cite{cit:Bud-7501,Bauer:1978iq,Kolanoski:1984hu,cit:Berger-Rep,Abramowicz:1993xb,
Drees:1992eb,*Drees:1995kd,Brodsky:1995nf,Erdmann-9701,
Nisius:1999cv,Krawczyk:2000mf}.
Particularly in the past few years there has been much progress
due to the wealth of experimental results from the $e^+ e^-$ collider LEP  
and the $ep$ collider HERA \cite{Nisius:1999cv,Krawczyk:2000mf}.
%
Among the reactions initiated by high energy photons 
such processes providing a hard scale
are of particular interest 
since they can be described 
(at least partly) by means of perturbative QCD.
More precisely, in this thesis we are interested in processes
which can be described within the well known
framework of the parton model and in this case the
'hadronic nature of the photon' is quantitatively described 
by the partonic structure of the photon.
%
The classical way of measuring the photonic parton content is
deep inelastic electron--photon scattering (\DISeg)
which is the main source of information.
The $\DISeg$ data on the photon structure function $F_2^\gamma(x,Q^2)$
are mainly sensitive to the up--quark density $u^\gamma(x,Q^2)$ as can
be seen from the parton model expression for $F_2^\gamma$ (in LO),
$F_2^\gamma \propto 4 u^\gamma + d^\gamma + s^\gamma$,
whereas the gluon distribution $g^\gamma(x,Q^2)$ is only indirectly 
constrained at small values of $x$ due to the evolution.
Complementary information has become available in the last years due to 
'resolved photon processes' 
(e.g.\ production of (di--)jets or hadrons with large $p_T$ ($E_T$),
heavy quark production, isolated prompt photon production)
in $\gamma p$ and $\gamma \gamma$ collisions at HERA and LEP, respectively,
which are mainly sensitive to the gluon distribution 
in the photon. 
However, the experimental constraints on the gluon density are
still weak, especially at $x < 0.1$, 
and one has to resort to model assumptions about
the parton distributions of the photon.

It is the virtue of the phenomenologically successful 
radiative parton model (GRV model) 
\cite{cit:GRV90,Gluck:1992ng,Gluck:1993im,cit:GRV94,cit:GRV98}
to predict the small--$x$ behavior 
of parton distributions
by pure DGLAP--evolution
of valence--like input distributions at a low input scale
$\mu^2 \simeq 0.3\ \gevsq$.
Recently, the parton distributions of the proton have been
revised \cite{cit:GRV98}
due to precision measurements of the proton structure function
$F_2^p$ at HERA leading to slightly changed parameters 
(low input scale, $\alpha_s$) and 
a less 
steep small--$x$ increase
compared to the previous GRV94 \cite{cit:GRV94}.
In order to test quantitatively (within the GRV approach) if
protons, pions and photons show a similar small--$x$ (i.e.\ high energy) 
behavior it is necessary 
to update also
the parton distributions 
of pions and photons.

At present $e^+ e^-$ and $ep$ collider experiments the photon beams
consist of bremsstrahlung radiated off the incident lepton beam
resulting in a continuous spectrum of target photons $\gamma(P^2)$ where
$P^2$ is the photon's virtuality.
The bremsstrahlung spectrum is proportional to $1/P^2$ such that the
bulk of target photons is produced at $P^2 \simeq P^2_{\rm min} \simeq 0$.
The parton content of such (quasi--)real photons is 
well established both experimentally and theoretically
and it is quite natural to expect the parton distributions to decrease
smoothly with increasing $P^2$ and not to vanish immediately.
In this sense the real photon $\gamma \equiv \gamma(P^2 \simeq 0)$ is
just a
'primus inter pares' and
unified approaches to the parton content
of virtual photons $\gamma(P^2)$ which comprise the real photon
case in the limit $P^2 \to 0$ are highly desirable.
This is also reflected by the fact that measurements of
the real photon structure function $F_2^\gamma$ in single--tag events
integrate over the bremsstrahlung spectrum from $P^2_{\rm min}$ up to
a $P^2_{\rm max}$ which depends on the experimental details.
For instance, at LEP1(LEP2) $P^2_{\rm max}$ is as large as
$P^2_{\rm max} \simeq 1.5\ \gevsq$($4.5\ \gevsq$),
cf.\ Section 2.2 in \cite{Nisius:1999cv}.
Allthough the bulk of photons is produced at $P^2 \simeq P^2_{\rm min}$
the amount of ignorance of the $P^2$--dependence, mainly in the
range $P^2 \lesssim \Lambda^2$, feeds back on the determination
of the structure function $F_2^\gamma$ (parton distributions) 
of (quasi--)real photons.

%
It is the central goal of the second part of this thesis
to perform LO and NLO \mbox{analyses} of the parton content of pions and 
real and virtual photons within the 
latest fine--tuned/improved
setting of the GRV model \cite{cit:GRV98}.

%
The outline of Part II will be as follows:
\begin{itemize}
\item In Chapter \ref{chap:twogam}
we provide the basic kinematical background
for studying the structure of real and virtual photons
in two--photon scattering events.
Structure functions for virtual photons are defined in a
general (model independent) way
which will be studied in the following chapters either
in fixed order perturbation theory 
(Chapters \ref{chap:lobox}, \ref{chap:vgam}) 
or within the framework of the QCD--improved parton model 
(Chapters \ref{chap:partonmodel},\ldots, \ref{chap:vgam}).
Furthermore, we demonstrate the factorization of the
cross section for the process \eetoeeX\ into a flux of photons times
the cross section for deep inelastic scattering (DIS) on these ``target''
photons for arbitrary virtualities $P^2$ of the target photon
in the Bjorken limit $P^2 \ll Q^2$.
It should be noted that the factorization is essential for 
a theoretical description of two--photon processes in terms of 
structure functions of target photons $\gamma(P^2)$ which can be measured in
deep inelastic electron--photon scattering.
\item In Chapter \ref{chap:lobox}
we calculate the photon photon cross sections $\sigma_{ab}$ 
according to the doubly virtual box $\vgvg$ in lowest order
perturbation theory.
The general expressions 
are casted in a form which easily allows to read off various important
limits, e.g., the quark--parton model (QPM) results
for the structure functions of real and virtual photons and the
heavy quark contributions to the photon structure functions.
For deeply virtual target photons the 
perturbative results make reliable predictions
and we compare them ($\Feff$) with present $e^+e^-$ virtual photon data.
On the other hand,
for quasi--real target photons these results are plagued by
mass singularities which have to be subtracted and afford
the introduction of parton distributions which will be done
in the next chapter.
\item In Chapter \ref{chap:partonmodel}
the complete theoretical framework necessary for our 
phenomenological study of
the parton content of real and virtual photons in 
Chapter \ref{chap:bc} is provided where
special emphasis is laid on a unified treatment of
real and virtual photons also in NLO.
\item Chapter \ref{pipdf} is devoted to
an analysis of the parton content of the pion 
in LO and NLO QCD.
Since only the pionic valence density $v^{\pi}(x,Q^2)$ is experimentally rather
well known at present, we utilize a constituent quark model
\cite{Altarelli:1974ff,Hwa:1980pn}  
in order to unambiguously relate the pionic light sea and gluon
to the much better known (recently updated) parton distributions of the 
proton \cite{cit:GRV98} and $v^{\pi}(x,Q^2)$.
These results will serve via vector meson dominance (VMD)
as input for the hadronic component
of the photon in the next chapter.
\item In Chapter \ref{chap:bc} we turn to LO and NLO analyses
of the parton content of real and virtual photons
within the framework
of the radiative parton model.
Apart from utilizing the latest refined parameters of the radiative
parton model \cite{cit:GRV98} there are some 
novelties
as compared to the original $\text{GRV}_\gamma$ approach \cite{cit:GRV-9202}.
At first, the boundary conditions for the real photon are based
on a coherent superposition of vector mesons which maximally enhances
the up--quark contribution to the photon structure function
$F_2^\gamma(x,Q^2)$ as is favored by the experimental data.
As a result no extra normalization factor for the photonic boundary
conditions is needed.
Furthermore, in order to remove model ambiguities of the hadronic
light quark sea and gluon input distributions of the photon
(being related to the ones of the pion via VMD), inherent to the older
$\text{GRV}_\gamma$ \cite{cit:GRV-9202} and SaS \cite{cit:SaS-9501} 
parametrizations, 
we employ, as already mentioned, predictions for the pionic light quark sea
and gluon which follow from constituent 
quark model constraints.
%
The resulting predictions for the real photon structure functions
$F_2^\gamma(x,Q^2)$ will be compared with all presently (January 1999)
available relevant data.
Most recently the OPAL collaboration \cite{Abbiendi:2000cw} at the CERN--LEP  
collider has extended the measurements of the photon structure 
function $F_2^{\gamma}(x,Q^2)$ into the small--$x$ region down 
to $x\simeq 10^{-3}$, probing lower values of $x$ than ever 
before and we include a comparison with these data as well.
Finally, we construct LO and NLO boundary conditions 
for the virtual photon which allow for a smooth transition
to the real photon case and perform a careful study of the
resulting predictions for the partonic content and the
structure function $F_2^{\gamma(P^2)}(x,Q^2)$ of virtual photons.
%
\item In Chapter \ref{chap:vgam}
we test our model for the parton content of virtual photons
in more detail and compare these QCD--resummed results with
the fixed order box expressions of Chapter \ref{chap:lobox}.
It is demonstrated that present $e^+e^-$ and DIS $ep$ 
data on the structure of the virtual photon can be understood 
entirely in terms of the standard `naive' quark--parton model box 
approach.  Thus the QCD--resummed (renormalization group (RG) improved) 
parton distributions of virtual photons, in particular their gluonic 
component, have not yet been observed.  The appropriate 
kinematical regions for their future observation are pointed out 
as well as suitable measurements which may demonstrate their 
relevance. 
\item Finally, in Chapter \ref{chap:phd_gam_summary}
we summarize our main results and discuss open questions.
\item Various limits of the doubly virtual box calculation
and parametrizations of our pionic and photonic parton distributions
have been relegated to the Appendices \ref{app:limits} and 
\ref{app_para}, respectively.
\end{itemize}

%% file: phd_gam_kinematics.tex
\chapter{Photon--Photon Scattering}\label{chap:twogam}
In this chapter we provide the basic kinematical background
for studying the structure of real and virtual photons
in two--photon scattering events.

In Section \ref{sec:kinematics} we give a short introduction into
so--called two--photon processes, introduce the usual kinematical variables 
to describe them 
and derive the cross section for such events.
The notation follows mainly the report of Budnev et al.~\cite{cit:Bud-7501}.
Here we only present a selection of the topic needed 
throughout the thesis.
Additional information can be found in \cite{cit:Bud-7501} 
and, e.g., in the more recent
reviews \cite{cit:Berger-Rep,Nisius:1999cv}.

Next, in Section \ref{sec:photonsfs} we define
structure functions for virtual photons and relate them to the photon--photon
scattering cross sections. 
These structure functions will be studied in the following chapters either
in fixed order perturbation theory 
(Chapters \ref{chap:lobox}, \ref{chap:vgam}) 
or within the framework of the QCD--improved parton model 
(Chapters \ref{chap:partonmodel},\ldots, \ref{chap:vgam}).

Finally, in Section \ref{sec:fact1} we demonstrate the factorization of the
cross section for the process \eetoeeX\ into a flux of photons and
the cross section for deep inelastic scattering (DIS) on these ``target''
photons.
While the factorization is well known for (quasi--)real target photons
with virtuality $P^2\approx 0$ we deal with the general $P^2 \ne 0$ case
in the Bjorken limit.
It should be noted that the factorization is essential for a theoretical
description of two--photon processes in terms of structure functions
of the (real or virtual) photon.
%
\section{Kinematics}\label{sec:kinematics}
\begin{figure}[ht]
\begin{center}
\begin{picture}(200,200)(0,0)
\ArrowLine(20,180)(100,160)
\ArrowLine(100,160)(180,180)
\Vertex(100,160){1.5}
\Photon(100,160)(100,120){4}{5}
\ArrowLine(180,20)(100,40)
\ArrowLine(100,40)(20,20)
\Vertex(100,40){1.5}
\Photon(100,40)(100,80){4}{5}
\Line(120,100)(150,100)
\Line(117,110)(150,115)
\Line(117,90)(150,85)
\GCirc(100,100){20}{0.5}
%
\Text(95,60)[r]{\large $p \uparrow$}
\Text(95,140)[r]{\large $q \downarrow$}
\Text(160,100)[l]{$\Bigg\}\ $\large \X}
\Text(0,180)[l]{\large $p_1$}
\Text(0,20)[l]{\large $p_2$}
\Text(190,180)[l]{\large $p_1^\prime$}
\Text(190,20)[l]{\large $p_2^\prime$}
\Text(120,140)[l]{$-q^2 = Q^2 \equiv Q_1^2$}
\Text(-10,100)[l]{$W^2 = (q+p)^2$}
\Text(120,60)[l]{$-p^2 = P^2 \equiv Q_2^2$}
\end{picture}
\end{center}
\caption{\sf Two--photon particle production. The solid lines are the incoming and
outgoing leptons and the wavy lines are virtual photons which produce
a final state $\X$ consisting of hadrons (or leptons).}
\label{fig:kin}
\end{figure}
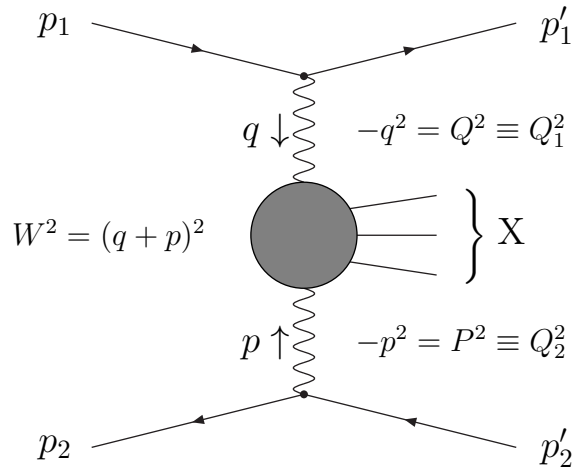

The kinematics of particle production via photon--photon scattering in
$\ee$ collisions 
\begin{equation}
\begin{aligned}
\e^{-}(p_1)\e^{+}(p_2)\rightarrow 
\e^{-}(p_1^{\prime})\e^{+}(p_2^{\prime}) \gamma^\star(q)\gamma^\star(p) 
\rightarrow 
\e^{-}(p_1^{\prime})\e^{+}(p_2^{\prime}) \X(p_{\X})
\end{aligned}
\label{eq:twogam}
\end{equation}
is depicted in Fig.~\ref{fig:kin}.
The momenta of the incoming and outgoing leptons are denoted by
$p_i \equiv (E_i,\vec{p_i})$ and 
$p_i^\prime \equiv (E_i^\prime,\vec{p_i}^\prime)$ ($i = 1,2$) respectively and
the momenta of the photons are given by
\begin{equation}\label{eq:kinematics1}
\begin{aligned}
q & \equiv p_1 - p_1^\prime\ ,\, \qquad \, Q^2 &= -q^2 \equiv Q_1^2&  \ ,
\\
p & \equiv p_2 - p_2^\prime\ ,\, \qquad \, P^2 &= -p^2  \equiv Q_2^2 &\ .
\end{aligned}
\end{equation} 
In general both photons have spacelike momenta 
and $P^2$ refers to the photon with smaller virtuality ($P^2 \le Q^2$).
\X\ denotes the final state produced in the 
$\gamma^\star(q)+\gamma^\star(p) \rightarrow \X$ subprocess.
For later use, we define the following variables:
\begin{gather}\label{eq:variables}
\nu = p \cdot q\ ,\, \qquad
x = \frac{Q^2}{2 \nu}\ ,\, \qquad \, \xp = \frac{P^2}{2 \nu}\ ,\, \qquad
y_1 =  \frac{p \cdot q}{p \cdot p_1}\ ,\, \qquad \, 
y_2 = \frac{p \cdot q}{q \cdot p_2}\ ,
\nonumber\\
W^2 \equiv (p+q)^2 = 2 \nu (1-x-\xp) 
= Q^2 \frac{1-x}{x} - P^2 \ .
\end{gather}

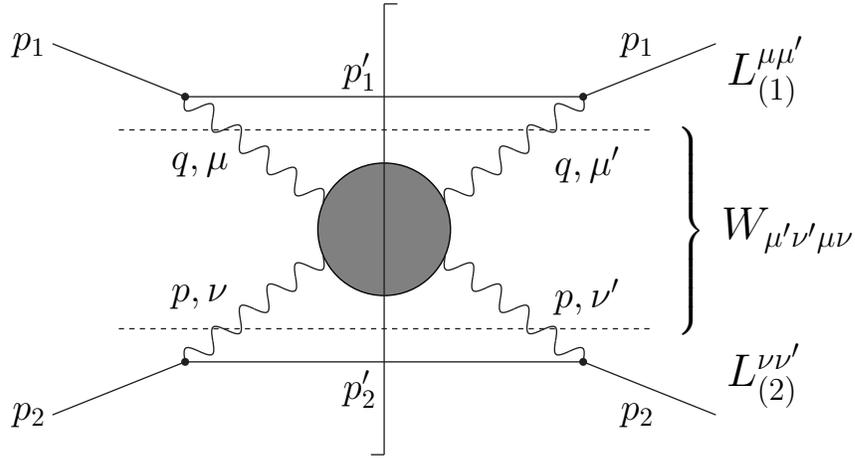
\begin{figure}[ht]
\begin{center}
\begin{picture}(300,200)(0,0)

\GCirc(150,100){25}{0.5}
\Photon(75,150)(127,110){-4}{5}
\Photon(75,50)(127,90){4}{5}
\Photon(173,110)(225,150){4}{5}
\Photon(173,90)(225,50){-4}{5}
%
\Line(75,150)(225,150)
\Line(25,170)(75,150)
\Line(225,150)(275,170)
\Vertex(75,150){1.5}
\Vertex(225,150){1.5}
\Line(75,50)(225,50)
\Line(25,30)(75,50)
\Line(225,50)(275,30)
\Vertex(75,50){1.5}
\Vertex(225,50){1.5}
%
\Line(150,185)(150,15)
\Line(150,185)(155,185)
\Line(150,15)(145,15)
\DashLine(50,137.5)(250,137.5){2}
\DashLine(50,62.5)(250,62.5){2}

\Text(280,160)[l]{\Large $L_{(1)}^{\mu\mu^\prime}$}
\Text(280,45)[l]{\Large $L_{(2)}^{\nu\nu^\prime}$}
\Text(260,100)[l]{\parbox{0.5cm}{$\left.\rule{0cm}{1.5cm} \right\}$}
{\Large $W_{\mu^{\prime}\nu^{\prime}\mu\nu}$}}

\Text(70,125)[l]{\large $q,\mu$}
\Text(70,75)[l]{\large $p,\nu$}
\Text(215,125)[l]{\large $q,\mu^\prime$}
\Text(215,75)[l]{\large $p,\nu^\prime$}
\Text(10,170)[l]{\large $p_1$}
\Text(10,30)[l]{\large $p_2$}
\Text(135,160)[l]{\large $p_1^{\prime}$}
\Text(135,40)[l]{\large $p_2^{\prime}$}
\Text(240,170)[l]{\large $p_1$}
\Text(240,30)[l]{\large $p_2$}

\end{picture}
\end{center}
\caption{\sf Squared matrix element of the process~(\ref{eq:twogam}).
Integration over the phase space of the system \X\ is implied
as indicated by the vertical cut.}
\label{fig:Xsec}
\end{figure}
The cross section for the process~(\ref{eq:twogam})
is given by
\begin{equation}
d\sigma = \frac{1}{F_{ee}} |M|^2 dQ^{(n+2)}
\end{equation}
with the invariant matrix element $M$, the M{\o}ller flux factor of the two incoming leptons
\begin{equation}
\Fee = 4 \sqrt{(p_1 \cdot p_2)^2 - m_e^2 m_e^2}\ ,
\end{equation}
and the Lorentz--invariant $(n+2)$--particle phase space
\begin{equation}
\begin{aligned}
d Q^{(n+2)}(p_1+p_2;p_1^{\prime},p_2^{\prime},k_1,\ldots,k_n)
& =
\frac{d^3 p_1^{\prime}}{(2 \pi)^3 2 E_1^{\prime}}
\frac{d^3 p_2^{\prime}}{(2 \pi)^3 2 E_2^{\prime}}
d Q^{(n)}(q+p;k_1,\ldots,k_n)
\\
& =
\frac{d^3 p_1^{\prime}}{(2 \pi)^3 2 E_1^{\prime}}
\frac{d^3 p_2^{\prime}}{(2 \pi)^3 2 E_2^{\prime}}
(2\pi)^4\delta^{(4)}\left(q+p-p_{\X} \right)d\Gamma\ .
\end{aligned}
\end{equation}
Here $p_{\X} = \sum_i k_i, i \in \X$ is the total momentum  
and $d\Gamma = \prod_i \frac{d^3 k_i}{2 k_i^0 (2 \pi)^3}, i \in \X$
the phase space volume of the produced system \X.

The cross section can be expressed in terms of the 
amplitudes $M^{\mu \nu}$ of the 
$\gamma^\star(q)+\gamma^\star(p) \rightarrow \X$ subprocess as follows
(see Fig.~\ref{fig:Xsec}):
\begin{equation}
d\sigma =    
\frac{d^3 p^{\prime}_1 d^3 p^{\prime}_2}{2 E^{\prime}_1 2 E^{\prime}_2 
(2 \pi)^6}
\frac{(4 \pi \alpha)^2}{Q^4 P^4} 
\frac{1} {F_{ee}} 
L_{(1)}^{\mu \mu^\prime} L_{(2)}^{\nu \nu^\prime} 
\Wco
\label{eqn:eeXsec1}
\end{equation}
with 
\begin{equation}
\Wcontra = \frac{1}{2} \int 
{M^\star}^{\mu^\prime \nu^\prime}M^{\mu\nu} (2 \pi)^4 \delta^{(4)}(q+p-p_{\X})
d\Gamma\ .
\label{eq:defw}
\end{equation}
For unpolarized leptons
the (leptonic) tensors $L_{(1)}$ and $L_{(2)}$ (see Fig.~\ref{fig:Xsec})
are given by
\begin{eqnarray}
L_{(i)}^{\alpha\beta} &=&
\frac{1}{2} Tr[(\pisl + \me) \gamma^\alpha (\pipsl + \me) \gamma^\beta]\ .
\label{eq:lalbe}
\end{eqnarray}
The factor $1/2$ is due to a spin average over the incoming leptons.
In addition, we introduce the dimensionless quantities \cite{cit:Bud-7501}
\begin{equation}
\ricontra = \frac{1}{Q_i^2} L_{(i)}^{\alpha\beta} 
= -\left(g^{\alpha \beta} - \frac{q_i^\alpha q_i^\beta}{q_i^2}\right) 
- \frac{(2 p_i - q_i)^\alpha (2 p_i - q_i)^\beta}{q_i^2}
\label{eq:rhoalbe}
\end{equation}
which have the interpretation of (unnormalized) density matrices for
the corresponding virtual photons.
\subsection{The Hadronic Tensor \Wcontra}
According to the optical theorem \Wcontra\ is the absorptive part of the
virtual $\gamma\gamma$ forward amplitude shown in Fig.~\ref{fig:walbemunu}.
\begin{figure}[ht]
\begin{center}
\begin{picture}(300,100)(0,0)
\GCirc(150,50){25}{0.5}
\Photon(75,100)(130,65){-4}{5}
\Photon(75,0)(130,35){4}{5}
\Photon(170,65)(225,100){4}{5}
\Photon(170,35)(225,0){-4}{5}
\Text(35,90)[l]{\large $q,\mu,m$}
\Text(35,10)[l]{\large $p,\nu,n$}
\Text(235,90)[l]{\large $q,\mu^\prime,m^\prime$}
\Text(235,10)[l]{\large $p,\nu^\prime,n^\prime$}
\end{picture}
\end{center}
\caption{\sf The photon--photon forward scattering amplitude \Tcontra;
$m$, $n$, $m^\prime$, and $n^\prime$ are helicity indices. 
The tensor 
$\Wcontra$ defined in (\protect\ref{eq:defw}) is the absorptive part of
the $\gamma\gamma$ forward amplitude: $\Wcontra = \frac{1}{\pi} \mathrm{Im} \Tcontra$.}
\label{fig:walbemunu}
\end{figure}

Taking into account P-- and  T--invariance
(symmetry $\mu^\prime\nu^\prime \leftrightarrow \mu\nu$) and
gauge invariance, i.e.
\begin{displaymath}
q_\mu \Wcontra =  q_{\mu^\prime} \Wcontra = p_\nu \Wcontra =
p_{\nu^\prime} \Wcontra = 0\ ,
\end{displaymath}
the Lorentz--tensor $\Wcontra$ can be expanded in terms of a basis of 8 independent 
tensors constructed from the vectors $q$, $p$ and the metric tensor $g$.
The choice of the tensor basis in which the expansion is carried out
is clearly arbitrary and different forms are discussed in the literature 
\cite{cit:Bud-7501,Brown:1971pk,Carlson:1971pk}.
In the following we will stick to the expansion
given in \cite{cit:Bud-7501}:
\begin{equation}
\begin{aligned}
\Wcontra = & R^{{\mu^\prime}\mu}R^{{\nu^\prime}\nu}\WTT
+R^{{\mu^\prime}\mu}Q_2^\nu Q_2^{\nu^\prime} \WTS + Q_1^{\mu^\prime} Q_1^\mu
R^{{\nu^\prime}\nu} \WST 
+ Q_1^{\mu^\prime} Q_1^\mu Q_2^{\nu^\prime} Q_2^\nu \WSS 
\\
& + \frac{1}{2}(R^{{\mu^\prime}{\nu^\prime}}R^{\mu\nu}
+R^{{\mu^\prime}\nu}R^{{\nu^\prime}\mu}
-R^{{\mu^\prime}\mu}R^{{\nu^\prime}\nu})\tTT\\
&-(R^{\mu\nu}Q_1^{\mu^\prime}
Q_2^{\nu^\prime}+R^{{\nu^\prime}\mu}Q_1^{\mu^\prime} Q_2^\nu
+R^{{\mu^\prime}{\nu^\prime}}Q_1^\mu
Q_2^\nu+R^{{\mu^\prime}\nu}Q_2^{\nu^\prime}
Q_1^\mu)\tTS \\ 
& +(R^{{\mu^\prime}{\nu^\prime}}R^{\mu\nu}-R^{{\nu^\prime}\mu}
R^{{\mu^\prime}\nu})\aTT
\\
& -(R^{\mu\nu}Q_1^{\mu^\prime}
Q_2^{\nu^\prime}-R^{{\nu^\prime}\mu}Q_1^{\mu^\prime} Q_2^\nu
+R^{{\mu^\prime}{\nu^\prime}}Q_1^\mu
Q_2^\nu-R^{{\mu^\prime}\nu}Q_2^{\nu^\prime} Q_1^\mu)\aTS\ .
\end{aligned}
\label{eq:budtensor}
\end{equation}
The tensor structures used in (\ref{eq:budtensor}) are connected to the
photon polarization vectors, see App.~B and C in \cite{cit:Bud-7501}:
\begin{gather}
\epsp[q]{\alpha}{0} = i Q_1^\alpha\ , \qquad \epsp[p]{\alpha}{0} = -i Q_2^\alpha\ , \qquad
\nonumber\\
\epsp[q]{\star \alpha}{\pm}\epsp[q]{\beta}{\pm} = 
\frac{1}{2} \left[R^{\alpha\beta} \pm i \frac{1}{\nu \vtwo} \varepsilon^{\alpha\beta\rho\sigma}
q_{\rho} p_{\sigma}\right]\ , \qquad
\epsp[p]{\star \alpha}{\pm}\epsp[p]{\beta}{\pm} = 
\epsp[q]{\star \alpha}{\mp}\epsp[q]{\beta}{\mp}   
\label{eq:photonpol}
\end{gather}
with $\vtwo^2 \equiv 1 - 4 x \xp$ and
\begin{gather}
R^{\alpha\beta}=  -g^{\alpha\beta}+\frac{\nu (p^\alpha q^\beta+q^\alpha p^\beta)
-p^2 q^\alpha q^\beta-q^2 p^\alpha p^\beta}{\nu^2 \vtwo^2}\, ,
\nonumber\\
Q_1^\alpha=
\frac{\sqrt{-q^2}}{\nu \vtwo}
\left(p^\alpha-\frac{\nu}{q^2}q^\alpha\right)
\ , \qquad
Q_2^\alpha=\frac{\sqrt{-p^2}}{\nu \vtwo}
\left(q^\alpha-\frac{\nu}{p^2}p^\alpha\right)\ .
\label{eq:rmunu}
\end{gather}
The photon momenta $q,p$, the unit vectors $Q_1,Q_2$, and the symmetric 
tensor $R^{\alpha \beta}$ satisfy the following (orthogonality) relations:
\begin{gather}
q \cdot Q_1 = p \cdot Q_2 = 0\ ,\qquad Q_{1,2}^2 = 1\ ,
\nonumber\\
q^{\alpha} R_{\alpha\beta} = p^{\alpha} R_{\alpha\beta} =
Q_{1,2}^{\alpha} R_{\alpha\beta} = 0
\ , \qquad
R_{\alpha\beta} R^{\alpha\beta} = 2\ ,\qquad 
R^{\alpha}_{\beta} R^{\beta\gamma} = -R^{\alpha\gamma}\ .
\label{eq:orthogonality}
\end{gather}
With help of these relations it is easy to see that the various
tensors in front of the invariant functions $W_{ab}$ in 
Eq.~(\ref{eq:budtensor}) are mutually orthogonal and therefore 
can be used to project out the invariant functions.
\pagebreak
With obvious notation ($\PWTT \Wco = \WTT$ etc.) the projectors read:
\begin{align}
\PWTT &= \frac{1}{4} R^{{\mu^\prime}\mu}R^{{\nu^\prime}\nu}
\ , 
\nonumber\\
\PWTS &= \frac{1}{2} R^{{\mu^\prime}\mu}Q_2^\nu Q_2^{\nu^\prime} \ , 
\nonumber\\
\PWST &=  \frac{1}{2} Q_1^{\mu^\prime} Q_1^\mu R^{{\nu^\prime}\nu} 
\ , 
\nonumber\\
\PWSS &= Q_1^{\mu^\prime} Q_1^\mu Q_2^{\nu^\prime} Q_2^\nu \ , 
\nonumber\\
\PtTT &=   \frac{1}{4}(R^{{\mu^\prime}{\nu^\prime}}R^{\mu\nu}
+R^{{\mu^\prime}\nu}R^{{\nu^\prime}\mu}
-R^{{\mu^\prime}\mu}R^{{\nu^\prime}\nu})\ , 
\nonumber\\
\PtTS &= 
\frac{-1}{8}(R^{\mu\nu}Q_1^{\mu^\prime}
Q_2^{\nu^\prime}+R^{{\nu^\prime}\mu}Q_1^{\mu^\prime} Q_2^\nu
+R^{{\mu^\prime}{\nu^\prime}}Q_1^\mu
Q_2^\nu+R^{{\mu^\prime}\nu}Q_2^{\nu^\prime}
Q_1^\mu)\ , 
\nonumber\\ 
\PaTT &= 
\frac{1}{4}(R^{{\mu^\prime}{\nu^\prime}}R^{\mu\nu}-R^{{\nu^\prime}\mu}
R^{{\mu^\prime}\nu})\ , 
\nonumber\\
\PaTS &= 
\frac{-1}{8}(R^{\mu\nu}Q_1^{\mu^\prime}
Q_2^{\nu^\prime}-R^{{\nu^\prime}\mu}Q_1^{\mu^\prime} Q_2^\nu
+R^{{\mu^\prime}{\nu^\prime}}Q_1^\mu
Q_2^\nu-R^{{\mu^\prime}\nu}Q_2^{\nu^\prime} Q_1^\mu)\ .
\label{eq:budprojectors}
\end{align}

The dimensionless invariant functions $W_{ab}$ depend only on the
invariants $W^2$, $Q^2$ and $P^2$ and are related to the 
$\gamma\gamma$--helicity amplitudes\footnote{The photon helicities can adopt the 
values $m^\prime,n^\prime,m,n = +1,-1,0$. Total helicity conservation 
for forward $\gamma\gamma$--scattering implies 
$m^\prime-n^\prime = m - n$ and due to P-- and T--invariance
($W_{m^\prime n^\prime,m n} 
\overset{P}{=} (-1)^{m^\prime-n^\prime+m-n} W_{-m^\prime -n^\prime,-m -n}
= W_{-m^\prime -n^\prime,-m -n} 
\overset{T}{=}W_{m n,m^\prime n^\prime}$) 
there exist 8 independent helicity amplitudes.
See App.~C in Ref.~\protect\cite{cit:Bud-7501} and Ref.~\protect\cite{Carlson:1971pk} 
for further details.}
$W_{m^\prime n^\prime,mn}$ in the $\gamma\gamma$--CMS 
via \cite{cit:Bud-7501}
\begin{equation}
\begin{aligned}
\WTT &= \frac{1}{2}(W_{++,++}+W_{+-,+-})\ ,\qquad &\WTS &= W_{+0,+0}\ ,
\\
\WST &= W_{0+,0+}\ , &\WSS &= W_{00,00}\ ,
\\
\tTT &= W_{++,--}\ , &\tTS &= \frac{1}{2}(W_{++,00}+W_{0+,-0})\ ,
\\
\aTT &= \frac{1}{2}(W_{++,++}-W_{+-,+-})\ , 
&\aTS &= \frac{1}{2}(W_{++,00}-W_{0+,-0})\ .
\end{aligned}
\label{eq:amptoheli}
\end{equation}
The amplitudes $\tTT$, $\tTS$, and $\aTS$ correspond to transitions with spin flip for
each of the photons (with total helicity conservation).
As we will see in the next section only 6 of these amplitudes 
($\WTT$, $\WTS$, $\WST$, $\WSS$, $\tTT$, $\tTS$)
enter the cross section for unpolarized lepton beams because the tensors in
(\ref{eq:rhoalbe}) are symmetric whereas the tensor structures multiplying
$\aTT$ and $\aTS$ in (\ref{eq:budtensor}) are anti--symmetric such that these 
terms do not contribute when
the leptonic and the hadronic tensors are contracted.
Only if the initial leptons are polarized, can the amplitudes $\aTT$ and $\aTS$ 
be measured as well \cite{cit:Bud-7501}.
\subsection{Derivation of the Cross Section}
Using Eqs.~(\ref{eq:lalbe}), (\ref{eq:rhoalbe}) and (\ref{eq:budtensor}) 
one obtains by a straightforward (but tedious) calculation 
\begin{equation}
\begin{aligned}
L_{(1)}^{\mu \mu^\prime} L_{(2)}^{\nu \nu^\prime} \Wco & = Q^2 P^2
\Big[4 \ronepp \rtwopp \WTT + 2 |\ronepm \rtwopm| \tTT \cos 2 \bar{\phi}
+ 2 \ronepp \rtwozz \WTS
\\
&\phantom{= Q^2 P^2\Big[} + 2\ronezz \rtwopp \WST + \ronezz \rtwozz \WSS 
-8 |\ronepz \rtwopz| \tTS \cos \bar{\phi}\Big]
\label{eq:contraction}
\end{aligned}
\end{equation}
where 
$\bar{\phi}$ is the angle between the scattering planes of
the $\e^-$ and the $\e^+$ in the center--of--mass system (CMS)
of the colliding photons
and the \ri's are elements of the photon density matrix:
\begin{align}\label{eqn:rhos}
 2\ronepp  &= 2\ronemm = \rho_1^{\alpha\beta} R_{\alpha\beta} =
\frac{\left(2 p_1\cdot p-p\cdot q\right)^2}
                {(p\cdot q)^2 - Q^2 P^2} + 1 - 4\frac{\me^2}{Q^2}\, ,
\nonumber\\
 2\rtwopp  &= 2\rtwomm = \rho_2^{\alpha\beta} R_{\alpha\beta} =
\frac{\left(2 p_2\cdot q-p\cdot q\right)^2}
                     {(p\cdot q)^2 -Q^2P^2} + 1 - 4\frac{\me^2}{P^2}\, ,
\nonumber\\
 \ronezz   &= \rho_1^{\alpha\beta} {Q_1}_{\alpha}{Q_1}_{\beta} =
2\ronepp - 2 + 4\frac{\me^2}{Q^2}\, , 
\nonumber\\
 \rtwozz   &= \rho_2^{\alpha\beta} {Q_2}_{\alpha}{Q_2}_{\beta} =
2\rtwopp - 2 + 4\frac{\me^2}{P^2}\, , 
\nonumber\\
2 |\ronepm \rtwopm| \cos 2 \bar{\phi} & = 
\frac{C^2}{Q^2 P^2} - 2 (\ronepp-1)(\rtwopp-1)\, ,
\nonumber\\
8 |\ronepz \rtwopz| \cos \bar{\phi} &= \frac{4C}{\sqrt{Q^2 P^2}} 
\frac{\left(2 p_1\cdot p-p\cdot q\right)
\left(2 p_2\cdot q-p\cdot q\right)}{(p\cdot q)^2 - Q^2 P^2}\, 
\nonumber\\
\text{with} \quad C & = (2 p_1 - q)^\alpha (2 p_2 - p)^\beta R_{\alpha\beta} =
-(2 p_1 - q) \cdot (2 p_2 - p)
\nonumber\\
&\quad + \frac{p \cdot q}{(p\cdot q)^2 - Q^2 P^2}
\left(2 p_1\cdot p-p\cdot q\right)\left(2 p_2\cdot q-p\cdot q\right)\, ,
\nonumber\\
 |\ripm|   &= \ripp - 1,
\nonumber\\
 |\ripz|   &= \sqrt{\left(\rizz+1\right)|\ripm|}.
\end{align}
Note that all these quantities are expressed in terms of the measurable 
momenta $p_1,p_2$ and $p_1^\prime,p_2^\prime$ (respectively $q,p$) only
and therefore are entirely known.

With help of Eqs.~(\ref{eqn:eeXsec1}) and (\ref{eq:contraction}) we easily find the 
fully general final result for the $ee \rightarrow ee X$ cross section 
\cite{cit:Bud-7501,cit:Berger-Rep,Nisius:1999cv}:
\begin{equation}
\begin{aligned}
d^6\sigma (ee \rightarrow ee X) &=
\frac{d^3 p^{\prime}_1 d^3 p^{\prime}_2}{E^{\prime}_1 E^{\prime}_2}
\frac{\alpha^2}{16 \pi^4 Q^2 P^2} 
\frac{F_{\gamma\gamma}}{F_{ee}} 
\Big[4 \ronepp \rtwopp \sigtt + 2\ronezz \rtwopp \siglt 
+ 2 \ronepp \rtwozz \sigtl 
\\
&\quad
+ \ronezz \rtwozz \sigll 
+ 2 |\ronepm \rtwopm| \tautt \cos 2 \bar{\phi}
-8 |\ronepz \rtwopz| \tautl \cos \bar{\phi}\Big]
\\
& \text{with}
\\
\frac{F_{\gamma\gamma}}{F_{ee}} &=
\left[\frac{(p \cdot q)^2 - Q^2 P^2} {(p_1 \cdot p_2)^2 - m_e^2 m_e^2} 
\right]^{1/2}\, . 
\label{eq:eeXsec}
\end{aligned}
\end{equation}
Here the cross sections $\sigma_{ab}$ (used as a shorthand for
\sigtt, \sigtl, \siglt, \sigll, \tautt, \tautl, \tauatt, \tauatl)
are identical to the corresponding structure functions $W_{ab}$ 
up to a division by the appropriate flux factor of the two 
incoming photons\footnote{Note 
that a factor $\frac{1}{2}$ has already been absorbed into
the definition of \Wcontra\ in Eq.~(\ref{eq:defw}).}, i.e.
\begin{equation}\label{eq:sigma_ab}
\sigma_{ab} = \frac{1}{2 \sqrt{(p \cdot q)^2 - Q^2 P^2}} W_{ab}
= \frac{1}{2 \nu \vtwo} W_{ab} \ .
\end{equation}

The cross section in (\ref{eq:eeXsec}) considerably simplifies in certain
kinematical regions \cite{cit:Bud-7501,cit:Berger-Rep,Nisius:1999cv}.
For instance, if both photons are highly virtual
Eq.~(\ref{eq:eeXsec}) can be evaluated in the limit $Q^2, P^2 \gg \me^2$ 
\cite{Nisius:1999cv} in which some relations between the elements of the
photon density matrix exist.
Of special interest in this thesis is the case where one of the 
lepton scattering angles becomes
small leading to a small virtuality $P^2\approx 0$ of the corresponding 
photon while the
other photon provides a hard scale $Q^2 \gtrsim 1\ \gevsq$.
In this limit the cross section factorizes into a product of a 
flux of quasi--real target photons
times the cross section for deep inelastic electron--photon scattering,
see for example \cite{Nisius:1999cv}.
This process is the classical way of measuring the structure of 
(quasi--real) photons.
The findings in the latter limit can be generalized to the case of 
photons with non--zero virtuality
$P^2 \ne 0$ as we will see in Section \ref{sec:fact1}.
This allows to study the structure of virtual photons in deep inelastic 
$\e \gamma(P^2)$ scattering processes in a continuous range of the scale $P^2$.
\section{Photon Structure Functions}\label{sec:photonsfs}
It is the aim of this section to relate the structure functions
of a (virtual) {\em target} photon to the  invariant functions
$W_{ab}$. The defining relations are generally valid for arbitrary
$P^2$. However, they only have a meaningful interpretation as structure
functions of a target photon probed by a deeply virtual photon $\gamma(Q^2)$
in the limit $P^2 \ll Q^2$.
The following expressions are simplified if one introduces 
the transverse components 
of a four--vector $x_\mu$ and of $g_{\mu\nu}$ 
\begin{equation}
x^T_\mu=x_\mu-\frac{q\cdot x}{q^2}q_\mu\, , \qquad
g^T_{\mu\nu}=g_{\mu\nu}-\frac{1}{q^2}q_\mu q_\nu \, ,
\end{equation}
where 'transverse' refers to $q$:
$q\cdot x^T =0$, $q^\mu g^T_{\mu\nu}=q^\nu g^T_{\mu\nu}=0$.
\subsection{Structure Functions for a Spin--Averaged Photon}\label{sec:spinav}
Usually one introduces structure functions for a {\em spin--averaged}
target photon. The corresponding structure tensor can be obtained
by contracting $\Wco$ given in Eq.~(\ref{eq:budtensor}) with the metric 
tensor $g^{\nu\nu^\prime}$.
With the help of Eqs.~(\ref{eq:rmunu}) and (\ref{eq:orthogonality}) one obtains\footnote{Of 
course the virtual photon has three ($+,-,0$) degrees of freedom.
The factor $1/2$ guarantees the {\em conventional} normalization in the 
real photon limit with only two ($+,-$) transverse degrees of freedom.}:
\begin{equation}\label{eq:wsymspinav1}
\begin{aligned}
W_{\mu^\prime \mu}^{<\gamma>} & \equiv \frac{-g^{\nu{\nu^\prime}}}{2} \Wco
\\
&=R_{\mu^\prime\mu} \left[\WTT-\frac{1}{2} \WTS\right] + Q_{1\mu^\prime}Q_{1\mu} 
\left[\WST - \frac{1}{2} \WSS \right]
\\
&=-g_{\mu^\prime \mu}^T \left[\WTT-\frac{1}{2}\WTS\right] 
+p_{\mu^\prime}^T p_{\mu}^T \frac{Q^2}{\nu^2 \vtwo^2} 
\left[\WIIT-\frac{1}{2}\WIIS \right]
\end{aligned}
\end{equation}
where $\WIIT \equiv \WTT+\WST$ and $\WIIS \equiv \WTS +\WSS$.

Alternatively the spin--averaged tensor can be expressed in standard form in terms of the
structure functions $F_1\equiv W_1$ and $F_2 \equiv \nu W_2$:
\begin{equation}\label{eq:wsymspinav2}
\frac{1}{8 \pi^2 \alpha} W^{<\gamma>}_{\mu^\prime\mu}=-g_{\mu^\prime\mu}^T F_1^{<\gamma>} 
+ p_{\mu^\prime}^T p_{\mu}^T \frac{1}{\nu} F_2^{<\gamma>}\ .
\end{equation}

Comparing Eqs.~(\ref{eq:wsymspinav1}) and (\ref{eq:wsymspinav2})
we find
\begin{equation}
\begin{aligned}
2 x F_1^{<\gamma>} &=\frac{1}{8 \pi^2 \alpha}\frac{Q^2}{\nu}
\left[\WTT-\frac{1}{2}\WTS\right]
\\
F_2^{<\gamma>} &=
\frac{1}{8 \pi^2 \alpha}\frac{Q^2}{\nu}\frac{1}{\vtwo^2} \left[\WIIT-\frac{1}{2} \WIIS\right]\ .
\end{aligned}
\end{equation}
These relations can be re--expressed in terms of the photon--photon cross sections 
$\sigma_{ab}= W_{ab}/(2 \nu \vtwo)\, (a,b = 2,\mathrm{L},\mathrm{T})$  
\cite{Nisius:1999cv,cit:Bud-7501,cit:Berger-Rep} 
\begin{equation}
\begin{aligned}\label{eq:sfs_av}
2 x F_1^{<\gamma>} &=\frac{Q^2}{4 \pi^2 \alpha}\vtwo
\left[\sigtt-\frac{1}{2}\sigtl\right]
\\
F_2^{<\gamma>} &=
\frac{Q^2}{4 \pi^2 \alpha}\frac{1}{\vtwo} \left[\sigiit-\frac{1}{2} \sigiil\right]\ .
\end{aligned}
\end{equation}

Finally, $F_{\mathrm{L}}$ satisfies the usual relation
\begin{equation}
F_\mathrm{L}^{<\gamma>} = \vtwo^2 F_2^{<\gamma>} - 2 x F_1^{<\gamma>}\ .
\end{equation}
This can be seen by contracting
$W_{\mu^\prime \mu}^{<\gamma>}$ with the polarization vectors of longitudinal 
{\em probe} photons given in Eq.~(\ref{eq:photonpol}) thereby employing again
the orthogonality relations in Eq.~(\ref{eq:orthogonality})
\begin{displaymath}
\begin{aligned}
W_{\mathrm{L}}^{<\gamma>} &\equiv 
\epsp[q]{\star \mu^\prime\!\!}{0}\epsp[q]{\mu}{0}
W_{\mu^\prime \mu}^{<\gamma>}
= Q_1^{\mu^\prime}Q_1^{\mu}W_{\mu^\prime \mu}^{<\gamma>}=\WST-\frac{1}{2}\WSS
\\
&= 8 \pi^2 \alpha \left[-F_1^{<\gamma>}+ \frac{\vtwo^2}{2 x}F_2^{<\gamma>}\right]
\end{aligned}
\end{displaymath}
followed by the appropriate normalization:
\begin{displaymath}
F_\mathrm{L}^{<\gamma>} = \frac{1}{8 \pi^2 \alpha} 2 x W_{\mathrm{L}}^{<\gamma>}\ .
\end{displaymath}
\subsection{Longitudinal and Transverse Target Photons}
Since the fluxes of transverse and longitudinal virtual photons 
will turn out to be {\em different} 
(see  Eq.~(\ref{eq:photonfluxes}) below)
it is most convenient to introduce structure
functions of transverse respectively longitudinal target photons
(instead of spin--averaged target photons).
The procedure is completely analogous to the one in the previous section 
using Eqs.~(\ref{eq:photonpol})--(\ref{eq:orthogonality})
and for this reason the description will be brief.

\noindent\underline{I. Transverse Photons}\\
With the help of Eq.~(\ref{eq:photonpol}) we can construct the structure tensor
for a transverse photon target which can
can be cast again into different forms
\begin{equation}
\begin{aligned}
W_{\mu^\prime \mu}^{\gam[T]} &\equiv 
\frac{1}{2}\big[\epsp[p]{\star\nu^\prime\!}{+} \epsp[p]{\nu}{+} 
+\epsp[p]{\star\nu^\prime\!}{-} \epsp[p]{\nu}{-}\big]\Wco 
=\frac{1}{2} R^{\nu^\prime\nu}\Wco
\\
& = R_{\mu^\prime\mu} \WTT + Q_{1\mu^\prime}Q_{1\mu} \WST
=-g_{\mu^\prime \mu}^T \WTT +p_{\mu^\prime}^T p_{\mu}^T \frac{Q^2}{\nu^2 \vtwo^2} \WIIT
\\
&\overset{!}{=} 8 \pi^2 \alpha \left[ -g_{\mu^\prime\mu}^T F_1^{\gam[T]} 
+ p_{\mu^\prime}^T p_{\mu}^T \frac{1}{\nu} F_2^{\gam[T]} \right]
\end{aligned}
\end{equation}
and we can directly read off the structure functions: 
\begin{equation}\label{eq:sfsgamt}
\begin{aligned}
2 x F_1^{\gam[T]} &=\frac{1}{8 \pi^2 \alpha}\frac{Q^2}{\nu}\WTT
=\frac{Q^2}{4 \pi^2 \alpha} \vtwo \sigtt
\\
F_2^{\gam[T]} &=
\frac{1}{8 \pi^2 \alpha}\frac{Q^2}{\nu}\frac{1}{\vtwo^2} \WIIT
=\frac{Q^2}{4 \pi^2 \alpha}\frac{1}{\vtwo} \sigiit\ .
\end{aligned}
\end{equation}
Repeating the steps in Sec.~\ref{sec:spinav} to determine $F_\mathrm{L}^{<\gamma>}$, 
we obtain $W_{\mathrm{L}}^{\gam[T]}=\WST$ implying
\begin{equation}\label{eq:flgamt}
F_\mathrm{L}^{\gam[T]} = \frac{1}{8 \pi^2 \alpha} 2 x W_{\mathrm{L}}^{\gam[T]}
= \vtwo^2 F_2^{\gam[T]} - 2 x F_1^{\gam[T]}\ .
\end{equation}
 
\noindent\underline{II. Longitudinal Photons}\\
The structure tensor for a longitudinal photon target 
is given by (using again Eq.~(\ref{eq:photonpol}))
\begin{equation}
\begin{aligned}
W_{\mu^\prime\mu}^{\gam[L]} &\equiv
\epsp[p]{\star \nu^\prime\!}{0}\epsp[p]{\nu}{0}\Wco
= Q_2^{\nu^\prime} Q_2^\nu \Wco
\\
& = R_{\mu^\prime\mu} \WTS + Q_{1\mu^\prime}Q_{1\mu} \WSS
=-g_{\mu^\prime \mu}^T \WTS +p_{\mu^\prime}^T p_{\mu}^T \frac{Q^2}{\nu^2 \vtwo^2} \WIIS
\\
&\overset{!}{=} 8 \pi^2 \alpha \left[ -g_{\mu^\prime\mu}^T F_1^{\gam[L]} 
+ p_{\mu^\prime}^T p_{\mu}^T \frac{1}{\nu} F_2^{\gam[L]} \right]
\end{aligned}
\end{equation}
and we find the following result for a longitudinal target photon:
\begin{equation}
\begin{aligned}\label{eq:sfsgaml}
2 x F_1^{\gam[L]} &=\frac{1}{8 \pi^2 \alpha}\frac{Q^2}{\nu}\WTS
=\frac{Q^2}{4 \pi^2 \alpha} \vtwo \sigtl
\\
F_2^{\gam[L]} &=
\frac{1}{8 \pi^2 \alpha}\frac{Q^2}{\nu}\frac{1}{\vtwo^2} \WIIS
=\frac{Q^2}{4 \pi^2 \alpha}\frac{1}{\vtwo} \sigiil\ .
\end{aligned}
\end{equation}
Finally, we have (as could be expected)
\begin{equation}\label{eq:flgaml}
W_{\mathrm{L}}^{\gam[L]}=\WSS \qquad \Rightarrow \qquad
F_\mathrm{L}^{\gam[L]} =\frac{1}{8 \pi^2 \alpha} 2 x W_{\mathrm{L}}^{\gam[L]}
= \vtwo^2 F_2^{\gam[L]} - 2 x F_1^{\gam[L]}\ .
\end{equation}

Further inspection of Eqs.~(\ref{eq:wsymspinav1})--(\ref{eq:flgaml})
reveals a relation 
[``$\displaystyle <\gamma> = \gam[T] - \tfrac{1}{2} \gam[L]$'']
between the spin--averaged, transverse
and longitudinal target photons
\begin{equation}
\begin{aligned}
W_{\mu^\prime \mu}^{<\gamma>}& = 
W_{\mu^\prime \mu}^{\gam[T]} - \frac{1}{2} W_{\mu^\prime \mu}^{\gam[L]} 
\\
F_{i}^{<\gamma>}& = F_{i}^{\gam[T]} - \frac{1}{2} F_{i}^{\gam[L]}\qquad (i = 1,2,\mathrm{L})
\end{aligned}
\end{equation}
which is a consequence of the completeness relation for {\em spacelike} photons
(cf.~\cite{cit:Bud-7501}, Eq.~(B.1)):
\begin{equation}
\begin{matrix}
\underbrace{\epsp[p]{\star\mu}{+} \epsp[p]{\nu}{+} +\epsp[p]{\star\mu}{-} \epsp[p]{\nu}{-}} 
&-& \underbrace{\epsp[p]{\star\mu}{0} \epsp[p]{\nu}{0}} & = & 
\underbrace{- g^{\mu \nu} + \frac{p^\mu p^\nu}{p^2}}&\ . \\
 2 \gam[T] &- & \gam[L] &=& 2 <\gamma>&
\end{matrix}
\end{equation}
\section{QED--Factorization}\label{sec:fact1}
\begin{figure}[htb]
\begin{center}
\setlength{\unitlength}{1pt}
\SetScale{1.8}
\vspace*{1.5cm}
\begin{picture}(80,70)
\SetColor{Black}
   \Line(0,50)(25,50)
   \Line(25,50)(70,65)
   \Photon(25,50)(50,40){-1.3}{6}
   \rText(45,75)[][]{\normalsize$q$}
\SetColor{NavyBlue}
   \Line(0,15)(25,15)
   \Line(25,15)(70,10)
   \Photon(25,15)(35,20){1.3}{2}
\SetColor{Black}
   \Photon(35,20)(50,27.5){1.3}{3}
   \rText(45,40)[][]{\normalsize$p$}
   \Line(55,40)(70,45)
   \Line(55,35)(70,35)
   \Line(55,30)(70,25)
%
   \GOval(55,35)(8.6,8.6)(0){0}

\SetColor{Black}
\SetWidth{1.0}
\Line(34,24)(38,17)
\Line(32,23)(36,16)
\SetWidth{1.0}
\end{picture}
\end{center}
\caption{\sf Factorization of the $ee\to ee \gamma\gamma \to ee X$ 
cross section into a flux of ``target'' photons radiated off 
the lower lepton line times the cross section
for deep inelastic electron--photon scattering (black part). 
The cut in the photon line
indicates a time order between the two subprocesses 
(photon emission {\em followed} by deep inelastic $\e \gamma$ scattering)  
and implies also that these two factors are independent of each other.}
\label{fig:fact1}
\end{figure}
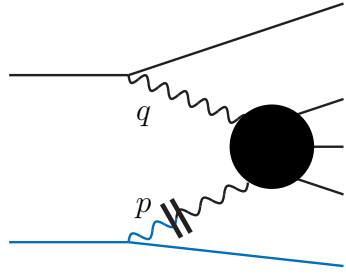
It is well known that for $P^2 \approx 0$ the general cross section
for the process $ee\to ee \gamma\gamma \to ee X$ factorizes
into a product of a flux of target photons (radiated off the electron)
with the deep inelastic electron--photon scattering cross section
\cite{cit:Berger-Rep,Nisius:1999cv}, 
see Fig.~\ref{fig:fact1} for a graphical representation: 
\begin{equation}\label{eq:fact1}
\frac{d \sigma(ee \to ee X)}{dx dQ^2 dz dP^2} = 
\ft(z,P^2) \frac{d\sigma(e \gamma \to e X)}{dx dQ^2}\ \qquad (P^2 \approx 0)
\end{equation}
where $z= E_\gamma/E \approx y_2$ is the fraction of the lepton energy carried by the photon
(in the $\ee$--CMS).
The cross section for deep inelastic electron--photon scattering in Eq.~(\ref{eq:fact1})
reads
\begin{equation}\label{eq:egamWQ}
\begin{aligned}
\frac{d\sigma(e \gamma \to e X)}{dx dQ^2} & = 
\frac{4 \pi^2 \alpha y}{x Q^2}
\frac{\alpha}{2 \pi}
\left[\frac{1+(1-y)^2}{y} \frac{1}{Q^2}\right] 
\left[2 x \sfs[F]{1}{\gamma} + \frac{2 (1-y)}{1+(1-y)^2}\sfs{L}{\gamma}
\right]
\\
&=\frac{2 \pi \alpha^2}{x Q^4}
\left[(1+(1-y)^2) 2 x \sfs[F]{1}{\gamma} + 2 (1-y)\sfs{L}{\gamma}
\right]
\end{aligned}
\end{equation}
with the usual variable $\displaystyle y = \frac{p \cdot q}{p \cdot p_1} = y_1$.
Furthermore, $\ft$ denotes the flux factor
of {\em transversely} (or circularly) polarized photons with virtuality $P^2$.
For use below we also provide the  
flux factor $\fl$ of longitudinally polarized photons:
\begin{equation}\label{eq:photonfluxes}
\begin{aligned}
\ft(z,P^2) &= 
\frac{\alpha}{2 \pi}
\left[\frac{1+(1-z)^2}{z} \frac{1}{P^2}- \frac{2 \me^2 z}{P^4}\right] 
\\
\fl(z,P^2) &= \frac{\alpha}{2 \pi}
\left[\frac{2(1-z)}{z} \frac{1}{P^2}\right]\ .
\end{aligned}
\end{equation}
%
%
%

The factorization in (\ref{eq:fact1}) is {\em essential}\ for 
relating the concept of the structure of a (real) photon 
to experimental measurements of two--photon processes.
For this reason we want to generalize Eq.~(\ref{eq:fact1}) for photons 
with virtuality $P^2 \ne 0$ and show that, in the Bjorken limit, factorization 
holds for virtual ``target'' photons as well.
For definiteness, as the Bjorken limit we consider
\begin{equation}
Q^2\equiv Q_1^2 \to \infty\ , \ \nu \to \infty\ , \ 
x = Q^2/2 \nu = \mathrm{fixed}\ .
\end{equation}
Practically, this means $P^2\equiv Q_2^2 \ll Q^2,\nu$ such that 
$\xp \equiv P^2/2 \nu = x P^2/Q^2$ is a small quantity which can be neglected.

The starting point is the general cross section in Eq.~(\ref{eq:eeXsec}).
Employing Eqs.~(\ref{eq:sfsgamt}), (\ref{eq:flgamt}), 
(\ref{eq:sfsgaml}), and (\ref{eq:flgaml})
and rearranging the terms inside the square brackets 
it can be written as
\begin{equation}\label{eq:eeXsec2}
\begin{aligned}
d^6\sigma 
&=  \frac{d^3 p^{\prime}_1 d^3 p^{\prime}_2}{E^{\prime}_1 E^{\prime}_2}
\frac{\alpha^2}{16 \pi^4 Q^2 P^2}\frac{F_{\gamma\gamma}}{F_{ee}} 
\Bigg[\frac{4 \pi^2 \alpha}{Q^2 \vtwo}\Big(2 \rtwopp 2 \ronepp 
[2 x \sfs{1}{\gam[T]} + \vep \sfs{L}{\gam[T]}]
\\
&\, 
 +  \rtwozz 2 \ronepp [2 x \sfs{1}{\gam[L]} + \vep \sfs{L}{\gam[L]}]\Big)
+2 |\ronepm \rtwopm| \tautt \cos 2 \bar{\phi}
  -8 |\ronepz \rtwopz| \tautl \cos \bar{\phi} \Bigg]
\\
& \text{with} \qquad \vep = \frac{\ronezz}{2 \ronepp}\ .
\end{aligned}
\end{equation}

The general strategy will be to demonstrate that
$\rtwopp$ and $\rtwozz$ are proportional to the flux factors of 
transverse and longitudinal photons, respectively, radiated off an electron
and that the interference terms disappear after having performed
an appropriate angular integration.

In the Bjorken limit it is useful to perform a ``light cone decomposition''
of the 4--momenta of the two photons \cite{cit:Bud-7501}:\\
\begin{equation}\label{eq:lcd}
\begin{gathered}
q  = \qp p_1 + \qm p_2 + \qt\,
\qquad \,
p = \ppl p_1 + \pmi p_2 + \pt
\\
\text{with} 
\\
p_i^2 = 0,\ p_i\cdot \qt = p_i\cdot \pt = 0,\ p_1 \cdot p_2 = S/2
\end{gathered}
\end{equation}
where $S$ is the square of the $\e\e$--CMS energy.
The momentum fractions $\qp$, $\qm$, $\ppl$ and $\pmi$ and
the transverse momenta can be easily calculated:
\begin{equation}\label{eq:lcd2}
\begin{gathered}
\begin{array}{lcl}
2 p_1 \cdot q = S \qm = -Q^2 &\qquad \Rightarrow &\qquad \qm = -Q^2/S
\\
2 p_2 \cdot q = S \qp & \qquad \Rightarrow & \qquad
\qp = 2 p_2\cdot q/S\, (\approx y_1)
\\
2 p_1 \cdot p  = S \pmi & \qquad \Rightarrow & \qquad
\pmi = 2 p_1\cdot p/S\, (\approx y_2)
\\
2 p_2 \cdot p  = S \ppl = -P^2 & \qquad \Rightarrow &\qquad \ppl = -P^2/S
\\
Q^2 =- S \qp \qm - \qt^2 & \qquad \Rightarrow &\qquad \qt^2 = -Q^2 (1-\qp)
\\
P^2 =- S \ppl \pmi - \pt^2 & \qquad \Rightarrow &\qquad \pt^2 = -P^2 (1-\pmi)
\end{array}
\\
\qt \cdot \pt  \equiv - \sqrt{\qt^2 \pt^2} \cos \phi 
                = - \sqrt{Q^2 P^2 (1-\qp)(1-\pmi)} \cos \phi \ .
\end{gathered}
\end{equation}
Here $\phi$ is
the angle between the scattering planes of
the $\e^-$ and $\e^+$ in the $\ee$--CMS.
Obviously $\ppl$ is negligibly small such that we can use\footnote{Of course, in the calculation 
of quantities which are themselves small of the order $P^2$ 
(e.g. $\pt^2$ in the $\ee$--CMS or $\piit^2$ in the $\gamma \gamma$--CMS) $\ppl$ must be 
taken into account.} 
\begin{equation}\label{eq:lcone}
p = \pmi p_2 + \pt \ .
\end{equation}
(On the other hand $\qm$ {\em cannot} be neglected in the Bjorken limit.)
In the $\ee$--CMS the 4--momenta of the incoming leptons can be written as
$p_1 = (E,0,0,E)$ and $p_2 = (E,0,0,-E)$ where $E = \tfrac{\sqrt{S}}{2}$
(neglecting terms of the order $\Ord(\tfrac{\me^2}{S})$).
Since the transverse 4--vector is given by $\pt = (0,{\pt}_x,{\pt}_y,0)$
we can infer from Eq.~(\ref{eq:lcone}) that $E_\gamma = \pmi E$, i.e., in the $\ee$--CMS 
$\pmi$ is the energy fraction of the lepton energy transferred to the photon. 
For a {\em real} ($P^2 = 0$) photon we recover the familiar relation
$p = \pmi p_2$ between the {\em 4-momenta} $p$ and $p_2$ 
of the collinearly radiated photon and its (massless) ``parent'' lepton, respectively.

Before turning to the photon density matrix elements in the Bjorken limit it
is helpful to relate the variable 
$\nu \equiv p \cdot q$ to $\qp$, $\pmi$ and the 
transverse momenta:
\begin{equation}\label{eq:nu}
2\nu = S \pmi \qp (1+\rho)\qquad  \text{with}\qquad  \rho \equiv 
\frac{2 \pt \cdot \qt}{S \pmi \qp}
\approx - 2 \sqrt{x \xp (1-\qp)(1-\pmi)} \cos \phi \ ,
\end{equation}
where $\rho \propto \sqrt{\xp}$ is small in the Bjorken limit.
Employing Eq.~(\ref{eq:nu}) we find in addition
\begin{equation}\label{eq:y1y2}
y_1 \equiv \frac{\nu}{p \cdot p_1} = \qp (1+\rho)\ , \qquad 
y_2 \equiv \frac{\nu}{q \cdot p_2} = \pmi (1+\rho)\ . 
\end{equation} 

Introducing the variables 
$\omega_1 \equiv q\cdot (p_1+p_2)/\sqrt{S} = \tfrac{\sqrt{S}}{2} (\qm+\qp)$
and 
$\omega_2 \equiv p\cdot (p_1+p_2)/\sqrt{S} = \tfrac{\sqrt{S}}{2} (\pmi+\ppl)$
the phase space can be written as \cite{cit:Bud-7501} [Eq.(5.15b)]
(up to terms of the order $\Ord(\tfrac{\me^2}{S})$) 
\begin{equation}\label{eq:PS}
\begin{aligned}
\frac{d^3 p_1^\prime}{E_1^\prime} \frac{d^3 p_2^\prime}{E_2^\prime} 
&=\frac{2\pi}{S} dQ^2 dP^2 d\omega_1 d\omega_2 d\phi
=\frac{\pi}{2} dQ^2 d\qp dP^2 d\pmi d\phi
\\
&=\frac{\pi}{2} dQ^2 dy_1 dP^2 d\pmi d\phi 
(1+\Ord(\rho))
\end{aligned}
\end{equation}
where the third equality can be understood with the help of Eq.~(\ref{eq:y1y2}).

In the Bjorken limit the photon density matrix elements 
in Eq.~(\ref{eqn:rhos}) can be cast in a very compact form
using the symmetric notation $Q_1^2 = Q^2$ and $Q_2^2 = P^2$:
\begin{equation}\label{eq:rhos2}
\begin{aligned}
2 \ripp & = 
\frac{2}{y_i} Q_i^2 \left[\frac{1+(1-y_i)^2}{y_i} \frac{1}{Q_i^2}
- \frac{2 \me^2 y_i}{Q_i^4}\right] + \Ord(\xp)
\\
\rizz & =
\frac{2}{y_i} Q_i^2 \left[\frac{2(1-y_i)}{y_i} \frac{1}{Q_i^2}\right]
+ \Ord(\xp)
\\
|\ripm| & = 
\frac{1}{y_i} Q_i^2 \left[\frac{2(1-y_i)}{y_i} \frac{1}{Q_i^2}
- \frac{2 \me^2 y_i}{Q_i^4}\right] + \Ord(\xp)\ .
\end{aligned}
\end{equation}
These results have to be expressed by the independent integration
variables $Q^2$, $y_1$, $P^2$, and $\pmi$. Furthermore, from here
on we identify $y_1 \equiv y$.

Obviously $\rtwopp$ and $\rtwozz$ are proportional to the flux factors
of transverse respectively longitudinal photons 
in Eq.~(\ref{eq:photonfluxes})
\begin{equation}\label{eq:photonfluxes2}
\begin{aligned}
2 \rtwopp &= \frac{2}{\pmi} P^2 \frac{2\pi}{\alpha} \ft(\pmi,P^2) 
+ \Ord(\rho)
\\
\rtwozz & = \frac{2}{\pmi} P^2 \frac{2\pi}{\alpha} \fl(\pmi,P^2) 
+ \Ord(\rho)
\end{aligned}
\end{equation}
and similarly we can write
\begin{equation}\label{eq:rho1}
\begin{aligned}
2 \ronepp & = 
\frac{2}{y} Q^2 \left[\frac{1+(1-y)^2}{y} \frac{1}{Q^2}\right] 
\, , \qquad
\ronezz =
\frac{2}{y} Q^2 \left[\frac{2(1-y)}{y} \frac{1}{Q^2}\right]
\\
\vep &= \frac{\ronezz}{2 \ronepp}= \frac{2 (1-y)}{1+(1-y)^2}
\end{aligned}
\end{equation}
where we have discarded the mass terms due to $Q^2 \gg \me^2$ .

Inserting Eqs.~(\ref{eq:photonfluxes2}), 
(\ref{eq:rhos2}), and (\ref{eq:PS}) into Eq.~(\ref{eq:eeXsec2})
one straightforwardly obtains 
(using $\frac{F_{\gamma\gamma}}{F_{ee}}=2 \nu \vtwo/S = \pmi y \vtwo$) 
\begin{equation}
\begin{aligned}
d\sigma 
&= dQ^2 dy dP^2 d\pmi \frac{d\phi}{2 \pi} 
\bigg\{\ft(\pmi,P^2) 
\frac{d\sigma(e \gam[T] \to e X)}{dy dQ^2}|_{\hat{S}=\pmi S}
+ [\gam[T] \rightarrow \gam[L]] 
\\
&\phantom{= dQ^2 dy dP^2 d\pmi \frac{d\phi}{2 \pi} \bigg\{} 
+ \Ord(\rho)
+ \text{interference-terms} 
\bigg\}
%
\end{aligned}
\end{equation}
with the cross sections for deep inelastic electron--photon scattering given
by [cf. Eq.~(\ref{eq:egamWQ})]
\begin{equation}\label{eq:egamWQ2}
\frac{d\sigma(e \gam[T,L] \to e X)}{dy dQ^2}|_{\hat{S}=\pmi S}=
\frac{4 \pi^2 \alpha}{Q^2} \frac{\alpha}{2 \pi} 
\left[ \frac{1+(1-y)^2}{y} \frac{1}{Q^2}\right]
\left[2 x \sfs{1}{\gam[T,L]}(x,Q^2) + \vep \sfs{L}{\gam[T,L]}(x,Q^2)
\right]
\end{equation}
where $\vep$ can be found in Eq.~(\ref{eq:rho1}).
Note that only two of the variables $x$, $y$, and $Q^2$ are independent since
they are related via $Q^2 = \hat{S} x y$.
The terms proportional to $\rho$ vanish after $\phi$--integration like terms
of the order $\Ord(\xp)$.

Unfortunately, the interference terms are proportional to
$\cos \bar{\phi}$ and $\cos 2\bar{\phi}$ 
where
$\bar{\phi}$ is the angle of the electron scattering 
planes in the $\gamma\gamma$--CMS while $\phi$ is  
the angle of the electron scattering planes in the $e^+e^-$--CMS
($\equiv$ laboratory system).
However, we show below that
\begin{equation}\label{eq:cosphi}
\cos \bar{\phi} = \cos \phi (1 + \Ord(\rho))\ .
\end{equation}
Therefore, we can also get rid of the interference terms 
(proportional to $\cos \bar{\phi} \approx \cos \phi$)
by integrating over $\phi$, leaving a remainder of the order $\Ord(\sqrt{\xp})$.
The latter becomes obvious if we remember that the variable $\rho$ defined in (\ref{eq:nu})
is proportional to $\sqrt{\xp} \cos \phi$ such that terms
$\rho \cos \phi$ occurring for example in Eq.~(\ref{eq:cosphi}) are proportional to
$\sqrt{\xp} \cos^2 \phi$ which do {\em not} vanish
by integrating over $\phi$. 

Before calculating $\cos \bar{\phi}$ let us state the final
factorization formula:
\begin{equation}\label{eq:fact2}
\begin{aligned}
\frac{d \sigma(ee \to ee X)}{dy dQ^2 d\pmi dP^2} &= \phantom{+{}}
\ft(\pmi,P^2) \frac{d\sigma(e \gam[T] \to e X)}{dy dQ^2}|_{\hat{S}=\pmi S} + \Ord(\xp)
\\
&\phantom{={}}+\fl(\pmi,P^2) \frac{d\sigma(e \gam[L] \to e X)}{dy dQ^2}|_{\hat{S}=\pmi S}
+ \Ord(\xp)
\\
&\phantom{={}} +\Ord(\sqrt{\xp})\ .
\end{aligned}
\end{equation}
A few comments are in order:
\begin{itemize}
\item In the limit $P^2 \to 0$ the contributions from longitudinal target photons
have to vanish since a real photon has only two transverse physical degrees of 
freedom and we recover Eq.~(\ref{eq:fact1}).\footnote{More precisely, assuming that 
the structure functions of a longitudinal target photon vanish like 
$P^2/\Lambda^2$ where $\Lambda$
is a typical hadronic scale, say $\Lambda = m_\rho$, we obtain a {\em non--zero} result 
because $\fl \propto 1/P^2$ which is, however, negligible compared to the contribution from
transverse target photons.}
\item The factorized result in Eq.~(\ref{eq:fact2}) is valid up to terms
$\Ord(\sqrt{\xp})$ which formally go to zero in the Bjorken limit.
However, for practical purposes it is not clear when $\xp= x P^2/Q^2$ is small
enough for Eq.~(\ref{eq:fact2}) to be a good approximation of the exact
cross section in Eq.~(\ref{eq:eeXsec}). Here, a numerical comparison 
of the factorization formula with the exact cross section would be interesting.
\end{itemize}
%
%


To complete our derivation we still have to show that 
$\cos \bar{\phi} = \cos \phi + \Ord(\rho)$
where $\cos \bar{\phi}$ is given by Eq.~(A.4) in
\cite{cit:Bud-7501}
\begin{equation}\label{eq:phibar}
\cos \bar{\phi} \equiv \frac{- \pit \cdot \piit}{\sqrt{\pit^2 \piit^2}}
\qquad \text{with} \qquad {p_i}_\perp^\mu = - {p_i}_\nu R^{\mu \nu}(q,p)
\end{equation}
and where $R^{\mu \nu}(q,p)$ has been defined in Eq.~(\ref{eq:rmunu}).
Using the decomposition in Eq.~(\ref{eq:lcd}) it is straightforward to obtain
\begin{equation}
\begin{aligned}
\pit \cdot  \piit &= \frac{\pt \cdot \qt}{\pmi \qp}
\\
\pit^2 &=-\frac{Q^2}{{\qp}^2}(1-\qp) + \Ord(\rho) 
\\
\piit^2 &=-\frac{P^2}{{\pmi}^2}(1-\pmi) + \Ord(\rho) \ .
\end{aligned}
\end{equation}
Inserting these relations into Eq.~(\ref{eq:phibar}) and comparing with
the definition of $\cos \phi$ in Eq.~(\ref{eq:lcd2}) we find the above stated result 
\begin{equation}
\cos \bar{\phi} = \frac{- \pt \cdot \qt}{\sqrt{\pt^2 \qt^2}} + \Ord(\rho)
= \cos \phi + \Ord(\rho)\ .
\end{equation}

%% file: phd_gam_lobox.tex
\chapter{The Doubly Virtual Box in LO}\label{chap:lobox}
\begin{figure}[h]
\begin{center}
\begin{picture}(140,100)(0,0)
\Photon(10,90)(40,70){-4}{5}
\Photon(10,10)(40,30){4}{5}
\Photon(130,90)(100,70){4}{5}  
\Photon(130,10)(100,30){-4}{5}
\SetWidth{1.0}
\Line(40,70)(100,70)
\Line(40,30)(100,30)
\Line(40,70)(40,30)
\Line(100,70)(100,30)
\Vertex(40,70){1.5}
\Vertex(40,30){1.5}
\Vertex(100,30){1.5}
\Vertex(100,70){1.5}
\SetWidth{0.5}
\Line(70,5)(70,95)
\Line(70,95)(75,95)
\Line(70,5)(65,5)
%
\end{picture}
\end{center}
\caption{\sf Box--diagram for 
$\gamma^\star(q,\mu) \gamma^\star(p,\nu)\ \rightarrow
\gamma^\star(q,\mu^\prime) \gamma^\star(p,\nu^\prime)$ 
.There are 4 possibilities to attach the photons to the vertices.
(2 for the initial state times 2 for the final state.)}
\label{fig:Box}
\end{figure}
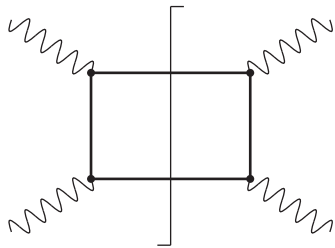

In this chapter we calculate the invariant amplitudes
$W_{ab}$ (or the cross sections $\sigma_{ab}$) 
in lowest order perturbation theory 
and compare them with present $e^+e^-$ virtual photon data.
These expressions are usually referred
to as box results due to the diagram representing the tensor
\Wcontra, where the photons are attached
to a fermion box, see Fig.~\ref{fig:Box}.

\begin{figure}[h]
\begin{center}
\begin{picture}(320,100)(-20,0)

\Photon(10,90)(40,70){-4}{5}
\Photon(10,10)(40,30){4}{5}
\SetWidth{1.0}
\ArrowLine(100,30)(40,30)
\ArrowLine(40,30)(40,70)
\ArrowLine(40,70)(100,70)
\SetWidth{0.5}
\Vertex(40,70){1.5}
\Vertex(40,30){1.5}

\Text(-20,90)[l]{\large $q,\mu$}
\Text(-20,10)[l]{\large $p,\nu$}
\Text(110,70)[l]{\large $k_1,m$}
\Text(110,30)[l]{\large $k_2,m$}
\Text(150,50)[l]{\Large $+$}
%
%
\Photon(200,90)(230,70){-4}{5}
\Photon(200,10)(230,30){4}{5}
\SetWidth{1.0}
\ArrowLine(290,30)(260,50)
\Line(260,50)(230,70)
\ArrowLine(230,70)(230,30)
\Line(230,30)(260,50)
\ArrowLine(260,50)(290,70)
\SetWidth{0.5}
\Vertex(230,70){1.5}
\Vertex(230,30){1.5}

\Text(170,90)[l]{\large $q,\mu$}
\Text(170,10)[l]{\large $p,\nu$}
\Text(300,70)[l]{\large $k_1,m$}
\Text(300,30)[l]{\large $k_2,m$}
\end{picture}
\end{center}
\caption{\sf Amplitude $M^{\mu\nu}$ for the process
$\gamma^\star(q,\mu) \gamma^\star(p,\nu)\ \rightarrow
\mathrm{f}(k_1)\mathrm{\bar{f}}(k_2)$ 
where $\mathrm{f}$ is a fermion of mass $m$.}
\label{fig:Mmunu}
\end{figure}
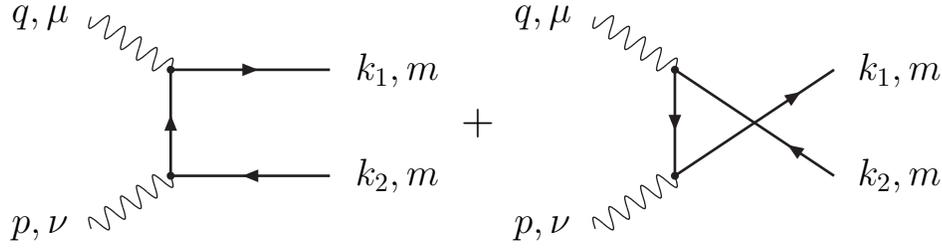

The calculation is a straightforward application of Feynman rules and 
will only be shortly outlined
here.
In order to obtain \Wcontra\ according to Eq.~(\ref{eq:defw})
one has to build up the amplitude $M^{\mu\nu}$ of the process
$\gamma^\star(q,\mu) \gamma^\star(p,\nu)\ \rightarrow \mathrm{f}(k_1)
\mathrm{\bar{f}}(k_2)$ 
shown in Fig.~\ref{fig:Mmunu} where $\mathrm{f}$ is either a lepton
or a quark of mass $m$.
The kinematics can be described 
with the help of the 4--momentum conservation relation $q + p = k_1 + k_2$
and the usual Mandelstam variables 
\begin{gather}
s = (q+p)^2 \ , \qquad
t = (q-k_1)^2 = (p-k_2)^2\ , \qquad u = (p-k_1)^2 = (q-k_2)^2\ ,
\nonumber\\
s_1 \equiv 2 q \cdot p=s + Q^2 + P^2 \ , \qquad
t_1 \equiv t - m^2\ , \qquad u_1 \equiv u - m^2 
\end{gather}
satisfying $s_1+t_1+u_1 = 0$.  

The external fermion lines are on--shell, i.e. $k_1^2 = k_2^2 = m^2$, whereas
the two photons are virtual (space-like): $q^2 = -Q^2 <0$, $p^2 = -P^2 <0$.
The Dirac traces occurring in \Wcontra\ (due to the closed fermion loop) have been
evaluated with the help of the \Mathematica\ \cite{cit:math} package
\Tracer\ \cite{Jamin:1993dp}.
Finally, the individual structure functions $W_{ab}$ have been projected
out using the projection operators given in Eq.~(\ref{eq:budprojectors}).
%
\section{Unintegrated Structure Functions}\label{sec:unintegrated}
The general structure of the boson--boson fusion cross section (in lowest order) 
is given by\footnote{The $t-$channel amplitude has a propagator $1/t_1$ 
and therefore has the form $(a+b t_1)/t_1$. The $u-$channel is obtained by
$t_1 \leftrightarrow u_1 = -(s_1+t_1)$
and thus reads $(a+b u_1)/u_1 = (c+d t_1)/(s_1+t_1)$.
Squaring the amplitudes generates the structure in Eq.~(\protect\ref{eq:bgf}).} 
\begin{equation}
\frac{dW_{ab}}{dz_1} = 16 \pi^2 \alpha^2 e_q^4 N_c
\left[\frac{A_{W_{ab}}}{(1-z_1)^2}+\frac{B_{W_{ab}}}{z_1^2}+
   \frac{C_{W_{ab}}}{1-z_1}+\frac{D_{W_{ab}}}{z_1}+E_{W_{ab}}\right] 
\label{eq:bgf}
\end{equation}
where $z_1 = 1+t_1/s_1$, $N_c=3$ is the number of colors
and $e_q$ is the quark charge.
The QED case can be obtained from Eq.~(\ref{eq:bgf}) by setting
$N_c e_q \to 1$.
Note that $z_1$ is very similar to the fractional
momentum variables used in Part I of this thesis to describe energy 
spectra of heavy quarks (mesons):
\begin{displaymath}
z_1 = \frac{k_1 \cdot p}{q \cdot p} - \frac{p^2}{2 q \cdot p}\ .
\end{displaymath}
On the other hand, $z_1=-u_1/s_1$ (or $1-z_1=-t_1/s_1$) can be viewed as a natural 
dimensionless combination of the Mandelstam variables of the problem.
%
Furthermore, it is convenient to use the dimensionless variables
\begin{equation}
x = \frac{Q^2}{s_1} =\frac{Q^2}{2 \nu}\, ,\,
\xp = \frac{P^2}{s_1} =\frac{P^2}{2 \nu}\, ,\,
\lambda = \frac{4 m^2}{s}
\end{equation}
in order to write the coefficients in a form which is
manifestly symmetric under $x \leftrightarrow \xp$ \footnote{Note however that 
neither $\WTS$ nor $\WST$ but only
the {\em sum} $\WTS+\WST$ is invariant under exchanging $x$ and $\xp$.}
and which
easily allows to read off the important ``massless limits'' $P^2 \to 0$
or $m^2 \to 0$ to be discussed in 
Appendix \ref{app:limits}.

With $\vtwo = \sqrt{1-4 x \xp}$ the coefficients read:
\begin{align}\label{eq:ace}
\AWTT & =
\frac{-1}{32 \pi \vtwo^5}
\Big[2 x \vtwo^2 - (1-x-\xp)(4x\xp+\lambda \vtwo^2)\Big]
\Big[2 \xp \vtwo^2 - (1-x-\xp)(4x\xp+\lambda \vtwo^2)\Big]
\nonumber\\
\CWTT & =
\frac{-1}{16 \pi \vtwo^5}
\Big[\lambda^2 \vtwo^4 (1-x-\xp)^2 -2\lambda \vtwo^2 (1-x-\xp)^2  
-2 \Big(1-2x(1-x)-2\xp(1-\xp)
\nonumber\\*  
&
-4 x \xp(x^2+\xp^2)+8 x^2 \xp^2 \left[1+(1-x-\xp)^2\right]\Big)\Big]
\nonumber\\
\EWTT & =
\frac{-1}{4 \pi \vtwo^5}\Big[1-2x(1-x)-2\xp(1-\xp)+4 x \xp(1-2x)(1-2\xp)\Big]
\nonumber\\
\nonumber\\
\AWTS & = 
\frac{\xp (1-x-\xp)}{4 \pi \vtwo^5}  
x \Big[2 x \vtwo^2 - (1-x-\xp)(4x\xp+\lambda \vtwo^2)\Big]
\nonumber\\
\CWTS & = \frac{\xp(1-x-\xp)}{-4 \pi \vtwo^5}
\Big[\lambda \vtwo^2 \Big(1+2x(-1+x-\xp)\Big)+4x \Big(-1+2x+2\xp-2x\xp(1+x+\xp)\Big)\Big] 
\nonumber\\
\EWTS & = \frac{\xp(1-x-\xp)}{\pi \vtwo^5}(1-2x)^2   
\nonumber\\
\nonumber\\
\AWSS & = -\frac{2 x^2 \xp^2 (1-x-\xp)^2}{\pi \vtwo^5} 
\nonumber\\
\CWSS & = \frac{2 x \xp (1-x-\xp)^2}{\pi \vtwo^5}(1+ 2 x \xp) 
\nonumber\\
\EWSS & = -\frac{8 x \xp (1-x-\xp)^2}{\pi \vtwo^5}
\nonumber\\
\nonumber\\
\AtTT & =
\frac{-1}{32 \pi \vtwo^5}(1-x-\xp)^2 [4 x \xp+\lambda \vtwo^2]^2
\nonumber\\
\CtTT & =
\frac{-1}{16 \pi \vtwo^5}
\Big[\lambda^2 \vtwo^4 (1-x-\xp)^2 -4 \lambda \vtwo^2 (1-x-\xp)\Big(-x-\xp+2 x \xp(1+x+\xp)\Big)
\nonumber\\*
&
-x \xp \Big(1-2x(1-x)-2\xp(1-\xp)+2x\xp(1-2x-2\xp-x^2-\xp^2+6 x \xp)\Big)\Big]
\nonumber\\
\EtTT & =\frac{-1}{2 \pi \vtwo^5}(x+\xp-4 x \xp)^2
\nonumber\\
\nonumber\\
\AtTS & =
\frac{\sqrt{x\xp}(1-x-\xp)}{16 \pi \vtwo^5}
(1-2x)(1-2\xp)(4x\xp+\lambda \vtwo^2)
\nonumber\\
\CtTS & =
\frac{\sqrt{x\xp}(1-x-\xp)}{4 \pi \vtwo^5}
\Big[4x\xp(-3+x+\xp)+2x+2\xp-\lambda \vtwo^2 (1-x-\xp)\Big]
\nonumber\\
\EtTS & =
\frac{\sqrt{x\xp}(1-x-\xp)}{2 \pi \vtwo^5}(1-4x-4\xp+12 x\xp)
\nonumber\\
\nonumber\\
\AaTT & =
\frac{1}{16 \pi \vtwo^3}
\Big[-2x\xp\vtwo^2+(1-x-\xp)(4x\xp+\lambda \vtwo^2)\Big]
\nonumber\\
\CaTT & =
-\frac{(1-2x)(1-2\xp)}{8 \pi \vtwo^3}
\nonumber\\
\EaTT & =
\frac{(1-2x)(1-2\xp)}{4 \pi \vtwo^3}
\nonumber\\
\nonumber\\
\AaTS & =
\frac{\sqrt{x\xp}(1-x-\xp)}{16 \pi \vtwo^3}
(-4x\xp+\lambda \vtwo^2)
\nonumber\\
\CaTS & =
\frac{\sqrt{x\xp}(1-x-\xp)}{\pi \vtwo^3}
x\xp
\nonumber\\
\EaTS & =
-\frac{\sqrt{x\xp}(1-x-\xp)}{2 \pi \vtwo^3}
\nonumber\\
\nonumber\\
B_{W_{ab}} &=  A_{W_{ab}}\, , \qquad  D_{W_{ab}} =  C_{W_{ab}}\ .
\end{align}
The coefficients of $\WST$ can be obtained from the corresponding 
ones of $\WTS$ by exchanging $x \leftrightarrow \xp$:
$\AWST = \AWTS[x \leftrightarrow \xp]$,
$\CWST = \CWTS[x \leftrightarrow \xp]$, and
$\EWST = \EWTS[x \leftrightarrow \xp]$.

It is noteworthy that 
our results in Eq.~(\ref{eq:bgf}) generalize
the $z$--differential expressions for the (QED) structure functions
$F_2^\gamma$, $F_\mathrm{L}^\gamma$, and $F_\mathrm{T}^\gamma$
of real photons given in \cite{Nisius:1998ue} 
to the $P^2 \ne 0$ case.
\clearpage
\section{Inclusive Structure Functions}
The desired inclusive structure functions are obtained
by integrating over the kinematically allowed range in $z_1$.
The boundaries for the $z_1$--integration are given by
\begin{equation}
z_{1,\pm} = (1 \pm \vone\vtwo)/2
\end{equation}
with
$\vone^2 = 1- 4 m^2/{s}= 1-\lambda$
and $\vtwo^2 = 1-4 x^2 P^2/Q^2= 1-4 x \xp$.
%

Noticing that 
\begin{gather}
z_{1,+} - z_{1,-}= \vone\vtwo\ , \qquad 
1-z_{1,+}=z_{1,-}\ , \qquad  1-z_{1,-}=z_{1,+}\ ,
\nonumber\\
z_{1,+}z_{1,-}=(1-z_{1,+})(1-z_{1,-})=\frac{4x\xp+\lambda \vtwo^2}{4} 
\end{gather}
the required integrals can be immediately obtained
\begin{equation}\label{eq:z1-integrals}
\begin{aligned}
\int_{z_{1,-}}^{z_{1,+}}\frac{dz_1}{(1-z_1)^2} &= 
\int_{z_{1,-}}^{z_{1,+}}\frac{dz_1}{z_1^2} 
= \frac{\vone\vtwo}{z_{1,+}z_{1,-}}
= \frac{4 \vone\vtwo}{4x\xp+\lambda \vtwo^2}
\\
\int_{z_{1,-}}^{z_{1,+}}\frac{dz_1}{(1-z_1)} &= 
\int_{z_{1,-}}^{z_{1,+}}\frac{dz_1}{z_1} = \ln \frac{z_{1,+}}{z_{1,-}} 
= \ln \frac{1+ \vone\vtwo}{1- \vone\vtwo}
\\
\int_{z_{1,-}}^{z_{1,+}} dz_1 &= \vone\vtwo \ .
\end{aligned}
\end{equation}
Now Eq.~(\ref{eq:bgf}) can be integrated using Eq.~(\ref{eq:z1-integrals})
and we arrive at the following result for
the inclusive structure functions expressed
by the coefficients given in the previous Section~\ref{sec:unintegrated}:
\begin{eqnarray}\label{eq:Wab}
W_{ab} &=& 
16 \pi^2 \alpha^2 e_q^4 N_c \Theta(\vone^2) \left\{ 2 C_{W_{ab}} L + \vone\vtwo 
\left(2 A_{W_{ab}} \frac{4}{4x\xp+\lambda \vtwo^2} + E_{W_{ab}}\right)\right\}
\end{eqnarray} 
with
\begin{equation}\label{eq:log}
L =  \ln \frac{1+ \vone\vtwo}{1- \vone\vtwo}\ .
\end{equation} 
The $\Theta$-function guarantees that the physical threshold
condition $s \ge 4 m^2$ is satisfied.\footnote{It is easy to see 
that $\vone^2 \ge 0$ also implies $\vtwo^2 \ge 0$.}

Recalling the relation $\sigma_{ab} = \frac{1}{2 \nu \vtwo} W_{ab}$
we can rewrite Eq.~(\ref{eq:Wab}) for the photon--photon cross sections
\begin{equation}\label{eq:sigab}
\sigma_{ab} = N \left\{ 2 C_{W_{ab}} L + \vone\vtwo 
\left(2 A_{W_{ab}} \frac{4}{4x\xp+\lambda \vtwo^2} + E_{W_{ab}}\right)\right\}
\ , \ \quad
N\equiv \frac{16 \pi^2 \alpha^2 N_c e_q^4}{2 \nu \vtwo}\Theta(\vone^2)\ .
\end{equation} 
Inserting the coefficients given in Eq.~(\ref{eq:ace}) 
into Eq.~(\ref{eq:sigab})
one obtains the \underline{final result} for the doubly virtual
box $\vgvg$ in leading order:
\begin{align}\label{eq:LO-box}
\sigtt &= \frac{N}{4 \pi}\frac{1}{\vtwo^5}
 \Biggl\{ \biggl[ 1-2x(1-x)-2\xp(1-\xp) - 4x\xp(x^2+\xp^2)  + 8 x^2\xp^2 \left[1+(1-x-\xp)^2\right]
\nonumber\\*
&  
+\lambda\vtwo^2(1-x-\xp)^2 -\frac{1}{2}\lambda^2\vtwo^4 (1-x-\xp)^2\biggr] \, L
+ \beta\vtwo\, \Biggl[ 4x(1-x)-1
\nonumber\\* 
& +4\xp(1-\xp)-8x\xp (1-x^2-\xp^2) 
-(4x\xp+\lambda\vtwo^2)(1-x-\xp)^2 - \frac{4x\xp\vtwo^4}{4x\xp + \lambda\,\vtwo^2}\Biggr] 
\Biggr\}
\nonumber\\
\nonumber\\
\siglt & = \frac{N}{4 \pi} \frac{4}{\vtwo^5} (1-x-\xp)\,
\Biggl\{ x\biggl[\, - \frac{1}{2}\lambda\,\vtwo^2 \Bigl( 1-2\xp(1+x-\xp)\Bigr) 
-2\xp \Bigl(-1+2x+2\xp 
\nonumber\\* 
&  
-2x\xp(1+x+\xp) \Bigr) \biggr]\,  L
+\beta\vtwo\left[ x(1-6\xp +6\xp^2 +2x\xp)  
+ \xp\vtwo^2\, \frac{4x\xp} {4x\xp +\lambda\, \vtwo^2} \right] \Biggr\} 
\nonumber\\
\nonumber\\
\sigtl & = \siglt [x\leftrightarrow \xp] 
\nonumber\\
%
\nonumber\\
\sigll & =\frac{N}{4 \pi} \frac{16}{\vtwo^5} x \xp (1-x-\xp)^2
\left\{(1+ 2 x \xp)L -
2 \vone\vtwo \frac{6x\xp+\lambda \vtwo^2}{4x\xp+\lambda \vtwo^2}\right\}
\nonumber\\
\nonumber\\
\tautt & = \frac{N}{4 \pi} \frac{1}{\vtwo^5}
\Bigg\{
\bigg[\frac{1}{2}x \xp \Big(1-2x(1-x)-2\xp(1-\xp)+2x\xp(1-2x-2\xp-x^2-\xp^2+6 x \xp)\Big)
\nonumber\\
&-2 \lambda \vtwo^2 (1-x-\xp)\Big(x+\xp-2 x \xp(1+x+\xp)\Big)
-\frac{1}{2}\lambda^2 \vtwo^4 (1-x-\xp)^2 
\bigg]L
\nonumber\\*
&- \vone\vtwo \bigg[(1-x-\xp)^2 (4x\xp+\lambda\vtwo^2)+2 (x+\xp-4x\xp)^2\bigg]
\Bigg\}
\nonumber\\
\nonumber\\
\tautl & = \frac{N}{4 \pi} \frac{2}{\vtwo^5} \sqrt{x\xp}(1-x-\xp)
\Bigg\{\bigg[2x+2\xp-4x\xp(3-x-\xp)-\lambda \vtwo^2 (1-x-\xp)\bigg]L
\nonumber\\*
&+ \vone\vtwo (1-3x-3\xp+8x\xp)\Bigg\}
\nonumber\\
\nonumber\\
\tauatt & = \frac{N}{4 \pi} \frac{1}{\vtwo^3}
\Bigg\{(2x-1)(1-2\xp)L
+ \vone\vtwo \bigg[\frac{-4x\xp \vtwo^2}{4x\xp+\lambda \vtwo^2}
+3-4x-4\xp+4x\xp\bigg]\Bigg\}
\nonumber\\
\nonumber\\
\tauatl & = \frac{N}{4 \pi} \frac{4}{\vtwo^3} \sqrt{x\xp}(1-x-\xp)
\left\{2x\xp L - \vone\vtwo \frac{4x\xp}{4x\xp+\lambda \vtwo^2}\right\}
\end{align}
This recalculation is in agreement with the results of Ref.\
\cite{cit:Bud-7501} (with $N_c e_q^4 \to 1$)
with exception of a relative sign between the part containing
the logarithm $L$ and the part proportional
to $\vone\vtwo$ in $\tauatl$\footnote{This relative sign
has also been noted in \protect\cite{cit:Rossi-PhD} where in addition the
overall sign in $\tautl$ is different. Concerning the latter,
we agree with the results of \protect\cite{cit:Bud-7501}.}.

A derivation of various important limits of the doubly virtual box expressions 
in (\ref{eq:LO-box}) can be found in App.~\ref{app:limits}.
%
\section{Comparison with Present $e^+e^-$ Virtual Photon Data}  
\label{sec:feff}
We will now compare our LO--box expressions with present $e^+e^-$ virtual photon
data.
The physically measured effective 
structure function in the Bjorken limit is 
\cite{cit:Bud-7501,cit:Berger-Rep,Nisius:1999cv} 
\begin{equation}\label{eq:vgam2.1}
\Feff(x;\, Q^2,y_1;\, P^2,\,y_2)  
= \frac{Q^2}{4 \pi^2 \alpha} \frac{1}{\vtwo}\,
\Big[\sigtt+\varepsilon(y_1)\siglt+\varepsilon(y_2)\sigtl+\varepsilon(y_1)\varepsilon(y_2)\sigll \Big]
\end{equation} 
where the kinematical variables have been given in Eqs.~(\ref{eq:kinematics1}) and 
(\ref{eq:variables}) and where $\varepsilon(y_i)$
are the ratios of longitudinal to transverse 
photon fluxes, 
\begin{equation}\label{eq:vgam2.2} 
\varepsilon(y_i) = 2(1-y_i)/[1+(1-y_i)^2]\, . 
\end{equation} 
Furthermore, the photon--photon cross sections $\sigma_{ab} = \sigma_{ab}(x,Q^2,P^2)$
with $a=(\rm{L,T}),\quad b=(\rm{L,T})$ have been defined in (\ref{eq:sigma_ab}).
In the following we shall consider the kinematical region $y_i\ll 1$ relevant 
for double--tag experiments \cite{Acciarri:2000rw,Berger:1984xu} performed thus far 
where  
Eq.\ (\ref{eq:vgam2.1}) reduces to 
%
\begin{equation}\label{eq:vgam2.3}
\Feff(x,Q^2,P^2)\simeq 
\frac{Q^2}{4 \pi^2 \alpha} \frac{1}{\vtwo}\, \Big[\sigtt+\siglt+\sigtl+\sigll \Big]\, .
\end{equation} 
Recalling the general definitions of the structure functions of spin--averaged, transverse 
and longitudinal target photons in Sec.~\ref{sec:photonsfs} we can finally write
\begin{align}\label{eq:feff}
\Feff(x,Q^2,P^2) &\simeq \sfs{2}{\gam[T](P^2)}(x,Q^2) + \sfs{2}{\gam[L](P^2)}(x,Q^2)
\nonumber\\*
&= \sfs{2}{<\gam(P^2)>}(x,Q^2) + \frac{3}{2} \sfs{2}{\gam[L](P^2)}(x,Q^2)\ .
\end{align}
So far, our results are entirely general.

We shall furthermore introduce the decomposition 
%
\begin{equation}\label{eq:vgam2.5}  
\sigma_{ab} = \sigma_{ab}^{\ell} + \sigma_{ab}^h 
\end{equation} 
with $\sigma_{ab}^{\ell(h)}$ denoting the light (heavy) quark $q = u,\, d,\, s$ 
($h = c,\, b,\, t$) contributions.
The light $u,\, d,\, s$ contributions to $\sigma_{ab}^{\ell}$ are obtained 
from Eq.~(\ref{eq:LO-box}) 
by setting $m\equiv m_q = 0\, (\lambda = 0)$ and summing over 
$q = u,\, d,\, s$. 
(Note that the box expressions involving a real photon,  
$\gamma^*(Q^2)\, \gamma\, (P^2=0)\to q\bar{q}$, require on the contrary a finite 
regulator mass $m\equiv m_q \neq 0$; here one usually chooses $m_q$ to be,  
somewhat inconsistently, a constant, i.e.\ $Q^2$--independent effective constituent 
mass, $m_q \simeq 0.3$ GeV.) 
For each heavy quark flavor $h = c,\, b,\, t$ the heavy contribution $\sigma_{ab}^h$ 
in (\ref{eq:vgam2.5}) is obtained from Eq.~(\ref{eq:LO-box}) 
with $e_q \equiv e_h$ and $m\equiv m_h$. 
Only charm gives a non--negligible contribution for which we choose $m_c = 1.4$ 
GeV throughout. 

Finally, it is instructive to recall the asymptotic results of our virtual  
($P^2\neq 0$) box expressions for the light $q = u,\, d,\, s$ quarks derived  
from (\ref{eq:LO-box}) in the Bjorken limit $P^2/Q^2\ll 1$, see App.~\ref{sec:bjorken2}:   
\begin{align}\label{eq:vgam2.6}  
\sigtt^{\ell} & \simeq  \Normb \left\{[x^2+(1-x)^2]\ln \frac{Q^2}{P^2 x^2} +4x(1-x)-2\right\}
\nonumber\\
\sigtl^{\ell} & \simeq \siglt^{\ell} \simeq \Normb [4 x (1-x)]
\nonumber\\
\sigll^{\ell} & \simeq  0\ .
\end{align} 

\begin{figure}
\centering
\vspace*{-0.5cm}
\epsfig{figure=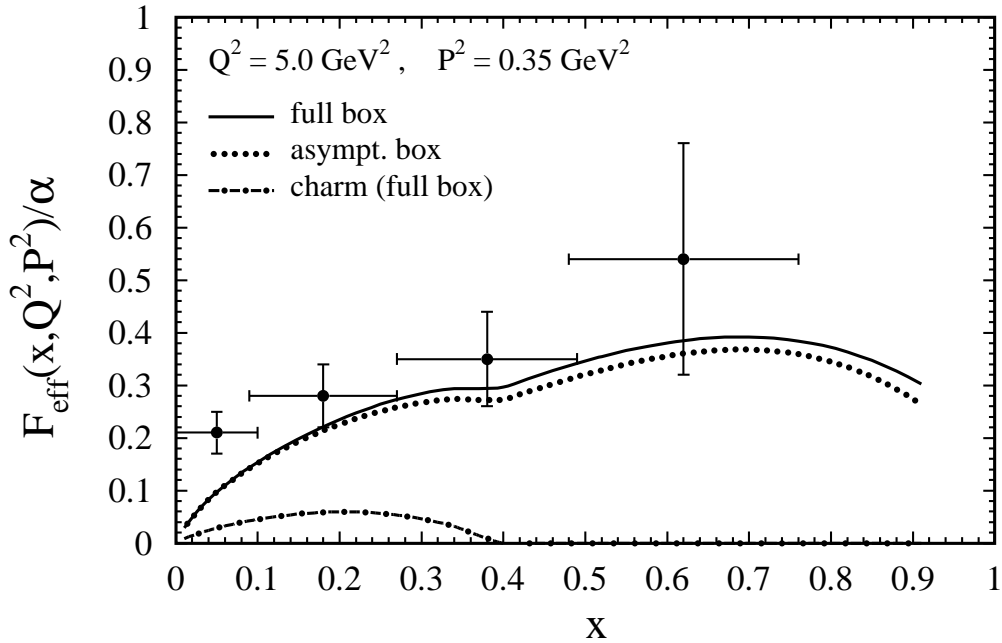,width=14.0cm}
\vspace*{-0.5cm}
\caption{\sf
      Predictions for $\Feff$ as defined in (\ref{eq:vgam2.3}).  The light 
      $(u,\, d,\, s)$ and heavy (charm) contributions in (\ref{eq:vgam2.5}) of the `full box'  
      expressions in (\ref{eq:LO-box}) are calculated 
      as explained in the text below Eq.\ (\ref{eq:vgam2.5}).  
      The `asymptotic box' results refer to the light quark contributions 
      being given by (\ref{eq:vgam2.6}).  
      The PLUTO data are taken from \protect\cite{Berger:1984xu}. 
}
\label{fig:feffbox_fig1}
\end{figure}

\begin{figure}
\centering
\vspace*{0.5cm}
\epsfig{figure=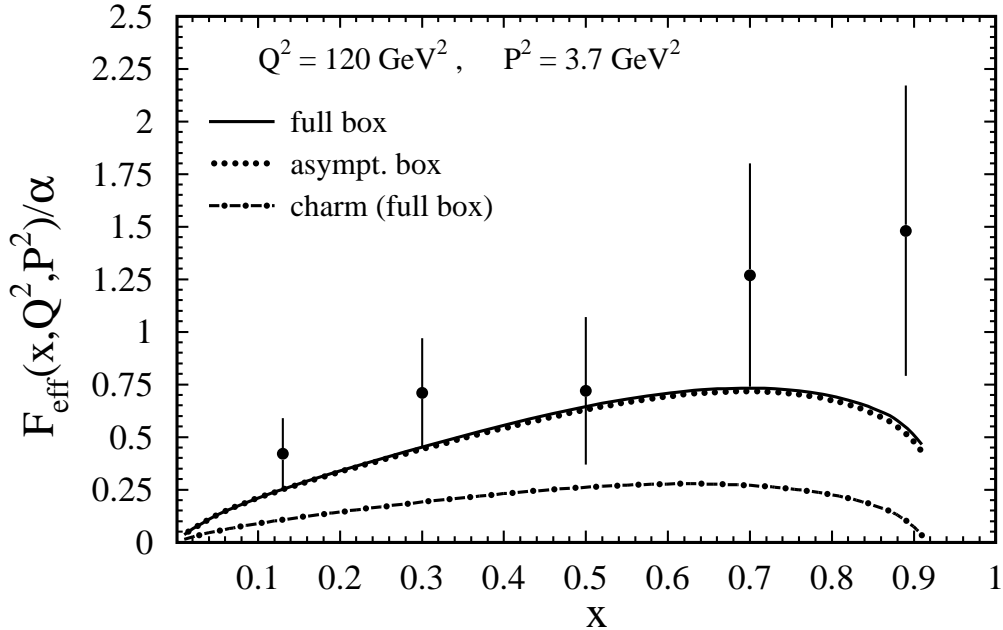,width=14.0cm}
\vspace*{-0.5cm}
\caption{\sf
      As in Fig.\ \ref{fig:feffbox_fig1}, but for $Q^2=120$ GeV$^2$ and $P^2=3.7$ GeV$^2$ 
      appropriate for the LEP--L3 data \protect\cite{Acciarri:2000rw}. 
}
\label{fig:feffbox_fig2}
\end{figure}

We now turn to a comparison of the effective structure function $\Feff$ obtained
in lowest order perturbation theory with
all presently available $e^+e^-$ data of PLUTO  
\cite{Berger:1984xu} and the recent one of LEP--L3 \cite{Acciarri:2000rw}.  
In addition to the full LO--box calculation of $\Feff$ (solid line) we show the
'asymptotic  box' results (dotted line) where the light quark contributions have been calculated
according to (\ref{eq:vgam2.6}). 
As can be seen the 'full' and the 'asymptotic' curves are very similar and
coincide in Fig.~\ref{fig:feffbox_fig2} over the entire $x$--range.
Furthermore, the charm contribution (dash--dotted line) is shown separately to 
demonstrate its relative importance 
due to the charge 
factor $e_q^4$ in Eq.~(\ref{eq:LO-box}) suppressing contributions from down--type quarks
by a factor of 16 relative to the up--type quarks.
Of course, in Fig.~\ref{fig:feffbox_fig1} the charm contribution is restricted by
the available phase space for producing a $c \bar{c}$ pair. 
Generally,
the LO--box predictions for $F_{\rm eff}$ in (\ref{eq:vgam2.3}) 
shown in Figs.\ \ref{fig:feffbox_fig1} and \ref{fig:feffbox_fig2} are in
agreement with the present low statistics data.\footnote{Note also, that the rightmost 
data point in Fig.~\protect\ref{fig:feffbox_fig2} is already close to the boundary
of the phase space $x_{\mathrm{max}} = (1+P^2/Q^2)^{-1}$.} 
This is not unexpected in the case of the L3 data shown in Fig.~\ref{fig:feffbox_fig2}
since the target photon is deeply virtual, $P^2 = 3.7\ \gevsq$, such that perturbation
theory is applicable.
On the other hand the PLUTO data (Fig.~\ref{fig:feffbox_fig1}) with $P^2 = 0.35\ \gevsq$
are just in the transition region from deeply virtual ($P^2 \gg \Lambda^2$) to
real ($P^2 = 0$) photons where non--perturbative effects are expected to become
increasingly important, especially in the small--x region.
Clearly, more precise data for $\Feff$ in this 'transition region' with intermediate 
target virtualities are highly desirable.


Two questions naturally arise:
\begin{enumerate}
\item[(i)]
Firstly, it would be interesting to see if NLO corrections to the doubly virtual box
\cite{Cacciari:2000cb,*Frixione:2000cd,nlobox} can further 
improve the description of the data.
\item[(ii)]
The other interesting question is whether for $Q^2\gg P^2\gg \Lambda^2$
it is necessary to resum 'large' collinear logarithms $\ln Q^2/P^2$
occurring in Eq.~(\ref{eq:vgam2.6})\footnote{It has, however, already been 
noted in \protect\cite{cit:Rossi-PhD} that 
the term containing $\ln Q^2/P^2$ dominates only if $(Q^2/P^2)$ is very large.}
to all orders in 
perturbation theory via renormalization group (RG) techniques.
This issue will be addressed in Chapter \ref{chap:vgam}
where we compare RG--resummed parton model expectations for $\Feff$ with
the fixed order box results presented here.
\end{enumerate}

It should be emphasized that, at $Q^2\gg P^2\gg \Lambda^2$, the $\ln Q^2/P^2$ 
terms in (\ref{eq:vgam2.6}) are {\em finite}
which therefore need not necessarily be resummed
but only if a resummation is phenomenologically relevant.
This is 
in contrast to the situation for the real photon ($P^2=0$) with its 
well known mass singularities $\ln\, Q^2/m_q^2$, see Eq.~(\ref{eq:f2boxreal}), 
which {\em{afford}} 
the introduction of scale dependent (RG--improved) parton distributions which 
are a priori unknown 
unless one resorts to some model assumptions about their 
shape at some low resolution scale (see, e.g., \cite{cit:GRSc99} and the recent 
reviews \cite{Erdmann-9701,Nisius:1999cv}). 


In the next chapter we present 
the necessary theoretical background for describing 
(virtual) photon structure functions
within the QCD--improved (RG--improved) parton model.

%% file: phd_gam_parton.tex
\chapter{The Partonic Structure of Real and Virtual Photons: Theoretical Framework}
\label{chap:partonmodel}
The following chapters are mainly devoted to an analysis of the 
structure of the (real and virtual) photon within the framework of the 
renormalization group (RG) improved parton model.
It is the main purpose here  
to provide the necessary theoretical background for these phenomenological
studies where special emphasis is laid on a unified treatment of 
both real ($P^2=0$) and virtual ($P^2\ne 0$) photons.

%
After motivating our unified approach in Sec.~\ref{sec:fact2}
we turn to a detailed description of the unpolarized photon
structure functions within the {QCD improved parton model}  
in Sec.~\ref{sec:SFSPM}. This section provides the complete 
technical framework for our analysis of the parton content of real
and virtual photons in Chapter \ref{chap:bc}.
In Sec.~\ref{sec:evolution} we deal with
the RG--based evolution equations for photonic parton distributions
and their solutions. For completeness, we discuss boundary conditions
for a deeply virtual target photon and
from the evolution equations we derive leading and next--to--leading order 
expressions for the momentum sum carried by the photon's quark and gluon densities.
Finally, in Sec.~\ref{sec:longitudinal} we briefly discuss how to include
contributions from longitudinal target photons.
\section{Introduction}\label{sec:fact2}
\begin{figure}[htb]
\begin{center}
\setlength{\unitlength}{1pt}
\SetScale{1.8}
\vspace*{1.5cm}
\begin{picture}(80,70)
\SetColor{Black}
   \Line(0,50)(25,50)
   \Line(25,50)(70,65)
   \Photon(25,50)(65,37){-1.3}{7}
   \rText(45,75)[][]{\normalsize$q$}
   \rText(-25,85)[b][]{\normalsize e$(p_1)$}
\SetColor{NavyBlue}
   \Line(0,15)(25,15)
   \Line(25,15)(70,10)
   \rText(-25,35)[t][]{\normalsize e$(p_2)$}
   \Photon(25,15)(35,20){1.3}{2}
\SetColor{Black}
   \Photon(35,20)(45,25){1.3}{2}
   \rText(45,40)[][]{\normalsize$p$}
   \Line(45,20)(52,20)
   \Line(46.5,22)(52,22)
   \Line(47,27)(65,33)
   \Line(67,38)(75,41)
   \Line(68,35)(75,35)
   \Line(67,32)(75,29)
    \IfColor{\COval(45,25)(5,2)(0){Black}{Red}
         }{\GOval(45,25)(5,2)(0){0.5}}
   \GOval(65,35)(5,3)(0){0}

\SetColor{Black}
\SetWidth{1.0}
\Line(34,24)(38,17)
\Line(32,23)(36,16)
\SetWidth{1.0}
\end{picture}
\end{center}
\caption{\sf Deep inelastic electron--photon scattering (cf.~Fig.~\ref{fig:fact1}) 
in the parton model where
the target photon is produced by the bremsstrahlung process. 
According to the factorization theorems
\protect\cite{Collins:1989gx,Collins:1987pm} the $e\gamma$ cross section factorizes into
'long distance' (non--perturbative) parton distributions (grey oval) describing the 
internal structure
of the photon and 'short distance' (perturbatively calculable) 
electron--parton subprocesses (black oval).}
\label{fig:fact2}
\end{figure}

We have seen in Chapter \ref{chap:twogam} that the cross section for the
process $\eetoeeX$ factorizes in the Bjorken limit into fluxes 
of transversely (longitudinally) polarized photons and the cross sections for 
deep inelastic scattering on these transverse (longitudinal) target photons.
The deep inelastic scattering cross sections can in turn be 
expressed in terms of two independent structure functions, 
e.g.~$\sfs{2}{\gam[T](P^2)}(x,Q^2)$ and
$\sfs{L}{\gam[T](P^2)}(x,Q^2)$
($\sfs{2}{\gam[L](P^2)}(x,Q^2)$, $\sfs{L}{\gam[L](P^2)}(x,Q^2)$).

Furthermore, in Chapter \ref{chap:lobox} we have obtained these structure functions
(or the effective structure function $\Feff$) according to a perturbative calculation
of the doubly virtual box $\vgvg$ in lowest order perturbation theory.
As has been discussed there in the deeply virtual case 
$P^2 \gg \Lambda^2$ (where $\Lambda$ is a hadronic scale)
perturbation theory is expected to be reliable 
and we have compared these 
fixed order 'box results'
$\Feff^{\mathrm{box}}$ with presently available $e^+ e^-$ data leaving
the open question if resummations of the collinear logarithms $\ln Q^2/P^2$
might further improve the situation.
On the other hand, for $P^2 \lesssim \Lambda^2$ 
the perturbatively calculated structure functions (cross sections) 
are invalidated by non--perturbative (long--distance)
mass singularities which have to be subtracted.
This problem can be consistently solved by turning to a parton model
description (see Fig.~\ref{fig:fact2}):
Since we are considering the Bjorken limit, 
the structure functions 
can be described in terms of photonic parton distributions, quite
similar to the hadronic case where factorization theorems 
\cite{Collins:1989gx,Collins:1987pm}
state that the structure functions are given by convolutions
of parton distributions with Wilson coefficient functions.
The parton distributions contain the long--distance part
of the cross section and therefore can (in general) {\em not} be calculated
perturbatively and have to be fixed by experimental information or model
assumptions.
On the other hand the coefficient functions involve only short--distance
physics and are calculable in perturbation theory.
The main value of this approach lies in the fact that the
parton distribution functions (PDFs) are {\em universal}, i.e.~do {\em not}
depend on the particular hard scattering subprocess.
Once fixed by experimental information from, e.g., deep inelastic electron--photon
scattering the PDF's can be used to make predictions for other processes, for
example photo--and electroproduction of jets or heavy quarks.

Due to QED gauge invariance
the contributions (structure functions, cross sections) from
longitudinal target photons
have to vanish in the real photon limit ($P^2 \to 0$).
More specifically,
they are suppressed by a factor $P^2$ divided by a
typical hadronic scale, say
$P^2/{m_\rho}^2$ \cite{cit:Bud-7501},
for not too large virtualities $P^2$ of the target photon.
Therefore, we will mainly focus on the structure functions of 
{\em transverse} target photons.
The contributions from longitudinal photon targets will be
considered at the end of this chapter.
In the following we can therefore drop the index $\mathrm{T}$ and use instead
$\gam(P^2)$ to denote transverse target photons.
Our main interest will lie on
$\sfs{2}{\gam(P^2)}(x,Q^2)$ since up to now the cross section
has been measured only for small values of $y$ where the contribution
of $\sfs{L}\, (\propto y^2)$ is negligible.

The following parton content can be assigned to a {\em transverse}
photon target: $\qvg(x,Q^2)$, $\gvg(x,Q^2)$, and $\gamvg(x,Q^2)$
where $q$ and $g$ 
are the photonic analogues of the usual quark and gluon 
parton distributions inside hadrons.
In addition the target photon also contains an elementary
photon--parton $\gamvg$ reflecting the fact that the photon is a 
genuinely elementary particle which can directly enter a partonic subprocess.
As usual in a massless parton model approach 
(justified by the factorization theorems \cite{Collins:1989gx,Collins:1987pm})
the partonic subprocesses are calculated with {\em on--shell} partons
partly due to the $P^2/Q^2$ power suppression of effects invoked by the
off--shellness of the partons (which is of the order $\Ord(P^2)$).
While this is quite familiar for the quarks and gluons the latter rule also
refers to the photon--parton $\fivg[\Gamma](x,Q^2)$ in the target photon, i.e.,
the 'direct' subprocesses, e.g., $\gamma^\star(Q^2)\fivg[\Gamma]\to q \bar{q}$
have to be calculated with $P^2=0$ (cf.~rule (ii) in \cite{cit:GRSc99}).
Consequently we can use the {\em same} (massless) short distance coefficient
functions in our calculations {\em irrespective} of $P^2$.
All effects due to a non--zero target virtuality are entirely taken care of
by the $P^2$--dependence of the parton distributions $\fivg$ 
(and by the Weizs{\"a}cker--Williams flux factors)
which of course have to be given in the same factorization scheme as the
coefficient functions in order to obtain a physically meaningful, i.e.~factorization 
scheme independent, result.
By construction a physically smooth behavior in $P^2$
(including a smooth transition to the real photon case $P^2=0$)
of the observable structure functions can be automatically
achieved by demanding a smooth behavior of the boundary conditions for the
parton distributions.

The heavy quark contributions deserve special care due to the additional scale
$m_h$ provided by the heavy quark mass.
In this case terms $P^2/m_h^2$ are possibly not negligible 
and have to be kept in the heavy quark coefficient functions. We will come 
back to this point in Sec.~\ref{sec:heavy}.

\section{Photon Structure Functions in the QCD--Improved Parton Model}
\label{sec:SFSPM}
We shall introduce the decomposition
\begin{equation}\label{eq:totf2gam}
\sfs{i}{\gam(P^2)}(x,Q^2) = \sfs{i,$\ell$}{\gam(P^2)}(x,Q^2) + 
\sfs{i,h}{\gam(P^2)}(x,Q^2) 
\end{equation}
with $\sfs{i,$\ell$ (h)}{\gam(P^2)}$ denoting contributions with light quarks
$q = u,d,s$ (heavy quarks $h = c,b,t$) in the final state.
In NLO($\msbar$) $\sfs{2,$\ell$}{\gam(P^2)}(x,Q^2)$ and
$\sfs{1,$\ell$}{\gam(P^2)}(x,Q^2)$ are given by the following expressions:
\begin{align}\label{eq:f2gam}
\frac{1}{x} \sfs{2,$\ell$}{\gam(P^2)}(x,Q^2)\ & = 
\sum_{q = u,d,s} e_q^2 \bigg\{ q^{\gam(P^2)}(x,Q^2)+\bar{q}\,^{\gam(P^2)}(x,Q^2)  
+ \, \frac{\alpha_s(Q^2)}{2\pi}
\nonumber\\*
& \phantom{={}}
\times
\left[ \cq[2] \otimes (q+\bar{q})^{\gam(P^2)} + 2\, \cg[2] \otimes g^{\gam(P^2)}\right] +
\, \frac{\alpha}{\pi}\, e_q^2 \cgam[2](x) \bigg\}
\\
\sfs{1,$\ell$}{\gam(P^2)}(x,Q^2)\ & = \frac{1}{2} 
\sum_{q = u,d,s} e_q^2 \bigg\{ q^{\gam(P^2)}(x,Q^2)+\bar{q}\,^{\gam(P^2)}(x,Q^2)  
+ \, \frac{\alpha_s(Q^2)}{2\pi}
\nonumber\\*
& \phantom{={}}
\times
\left[ \cq[1] \otimes (q+\bar{q})^{\gam(P^2)} + 2\, \cg[1] \otimes g^{\gam(P^2)}\right] +
\, \frac{\alpha}{\pi}\, e_q^2 \cgam[1](x) \bigg\}
\end{align}
where $\otimes$ denotes the usual convolution integral of two functions
$f$ and $g$ defined on the interval $[0,1]$:
\begin{equation}\label{eq:Mellin-convolution}
(f \otimes g)(x) = \int_0^1 dx_1 \int_0^1 dx_2 f(x_1)g(x_2)\delta(x- x_1 x_2)
= \int_x^1 \frac{dz}{z} f(z) g(x/z)\ .
\end{equation}
Here 
$\bar{q}\,^{\gam(P^2)}(x,Q^2)=q^{\gam(P^2)}(x,Q^2)$ and $g^{\gam(P^2)}(x,Q^2)$
provide the so--called `resolved' contributions of 
$\gam(P^2)$ to $\sfs{i}{\gam(P^2)}$
with the $\msbar$ coefficient functions
\cite{cit:BBDM-7801,cit:AEM-7901,cit:FP-8201} 
\begin{align}\label{eq:wilsoncoeff}
\cq[2](x) & =  \cq[1](x) + \frac{4}{3}\, 2 x
\nonumber\\*
&=\frac{4}{3}\, \left[ \frac{1+x^2}{1-x}\, \left( \ln
   \frac{1-x}{x}-\frac{3}{4}\right) + \frac{1}{4}\, (9+5x)\right]_+
\nonumber\\
\cg[2](x) & = \cg[1](x) + \frac{1}{2}\, 4x(1-x)
\nonumber\\*
& = \frac{1}{2} \left[ \left(x^2+(1-x)^2\right) \ln \frac{1-x}{x} + 8x(1-x)-1 \right],
\end{align}
while $\cgam[2,1]$ in (\ref{eq:f2gam}) 
provides the `direct'
contribution as calculated according to the `box' diagram  
$\gamma^*(Q^2)\gamma \to q\bar{q}$ \cite{cit:BB-7901}:
\begin{equation}\label{eq:cgamma}
\cgam[i](x)=\frac{3}{(1/2)}\, \cg[i](x)
\end{equation}
with $\mathrm{i} = 1,2$.
The convolution with the $[\, ]_+$ distribution can be 
evaluated using, for example, Eq.~(A.21) in Ref.~\cite{Gluck:1992ng}:
\begin{equation}\label{eq:plusconv}
f_+ \otimes g = \int_x^1 \frac{dy}{y} f\left(\frac{x}{y}\right)
\left[g(y) - \frac{x}{y} g(x) \right] - g(x) \int_0^x dy f(y)\ .
\end{equation}

The coefficient functions $\cq[L]$ and $\cg[L]$ for the 
longitudinal structure function 
$\sfs{L}{\gam(P^2)}=\sfs{2}{\gam(P^2)}-2 x \sfs{1}{\gam(P^2)}$
may be deduced from Eqs.~(\ref{eq:wilsoncoeff}) and (\ref{eq:cgamma}) and are given 
by \cite{cit:GRS95}:
\begin{equation}
\cq[L](x) = \frac{4}{3}\, 2 x\, , \qquad \cg[L](x) = \frac{1}{2}\, 4x(1-x)\, , \qquad
\cgam[L](x)=\frac{3}{(1/2)}\, \cg[L](x)\ .
\end{equation}
\subsection{Scheme Choice}
In order to avoid the usual instabilities encountered in 
NLO($\msbar$)
in the large--$x$ region due to the $\ln(1-x)$ term in $\cgam[2](x)$
in Eq.~(\ref{eq:cgamma}), we follow
Ref.~\cite{cit:GRV-9201,cit:GRV-9202} and absorb such terms into 
the photonic $\msbar$
quark distributions in Eq.~(\ref{eq:f2gam}): this results in the 
so--called DIS$_{\gamma}$
factorization scheme which originally has been introduced for real
photons by absorbing the entire 'direct' $\cgam[2]$ term appearing 
in Eq.~(\ref{eq:f2gam}) into the NLO($\msbar$) quark densities 
$\fivg[q](x,Q^2) = \fivg[\bar{q}](x,Q^2)$, i.e.~
\begin{equation}\label{eq:schemetrafo1}
\begin{aligned}
(q+\bar{q})_{\disg}^{\gam{(P^2)}} & =  
   (q+\bar{q})_{\msbar}^{\gam{(P^2)}} + e_q^2\, \frac{\alpha}{\pi}\, \cgam[2](x)
\\
g_{\disg}^{\gam{(P^2)}} & =  g_{\msbar}^{\gam{(P^2)}}
\end{aligned}
\end{equation}
with $\cgam[2](x)$ given by Eq.~(\ref{eq:cgamma}).  
How much of the `finite' terms
in Eq.~(\ref{eq:cgamma}) is absorbed into the 
$\msbar$ distributions in Eq.~(\ref{eq:schemetrafo1}),
is of course arbitrary and a matter of convention \cite{cit:AFG-9401}.  
Since such different conventions \cite{cit:AFG-9401,cit:Vogt97} 
differ by terms of higher order and
turn out to be of minor
importance for our quantitative results to be discussed in Chap.~\ref{chap:bc}, we prefer
to stick to the original $\disg$ scheme 
\cite{cit:GRV-9201,cit:GRV-9202} as defined
in Eq.~(\ref{eq:schemetrafo1}).  
Furthermore, the redefinition of the parton densities in
Eq.~(\ref{eq:schemetrafo1}) implies that the NLO($\msbar$) splitting
functions $k_{q,g}^{(1)}(x)$ of the photon into quarks and gluons,
appearing in the inhomogeneous NLO RG evolution equations 
for $\fivg(x,Q^2)$, have to be transformed according to 
\cite{cit:GRV-9201,cit:GRV-9202,Gluck:1993zx}
\begin{equation}\label{eq:schemetrafo2}
\begin{aligned}
k_q^{(1)}|_{\disg} & =  k_q^{(1)} -
                e_q^2\, P_{qq}^{(0)} \otimes \cgam[2]
\\
k_g^{(1)}|_{\disg} & =  k_g^{(1)} -
                2\, \sum_q  e_q^2\, P_{gq}^{(0)}\otimes \cgam[2]
\end{aligned}
\end{equation}
where
\begin{align} 
k_q^{(1)}(x) & = \frac{1}{2}\, 3 e_q^2\, \frac{4}{3} 
   \biggl\{ -(1-2x)\,\ln^2x -(1-4x)\,\ln\,x + 4\,\ln(1-x)-9x+4 
\nonumber\\* 
&  +  \left[ x^2+(1-x)^2\right] \, \biggl[2\, \ln^2 x+ 2\, \ln^2(1-x)+ 
       4\, \ln\, x -4\, \ln\, x\, \ln(1-x)
\nonumber\\* 
&   -4\, \ln(1-x) + 10 -\frac{2}{3}\,\pi^2\biggr] \biggr\}
\nonumber\\ 
k_g^{(1)}(x) & =  3\sum_q e_q^2\frac{4}{3} 
    \left\{ -2(1+x)\ \ln^2 x-(6+10x)\, \ln\, x +\frac{4}{3x} + 
       \frac{20}{3}x^2+8x -16\right\} 
\end{align} 
with $k_q^{(1)}$ referring to each single (anti)quark flavor.  The 
LO splitting functions are given by 
$P_{qq}^{(0)}=\frac{4}{3}\left(\frac{1+ x^2}{1-x}\right)_+$ and 
$P_{gq}^{(0)}=\frac{4}{3}\left[1+(1-x)^2\right]/x$. 

In NLO the expression for $\sfs{2}{\gam(P^2)}$
in the above $\disg$
factorization scheme is given by retaining the $\cq[2], \cg[2]$ terms while
dropping the destabilizing $\cgam[2]$ term in 
Eq.\ (\ref{eq:f2gam}), which has
already been absorbed into the quark densities 
according to Eq.\ (\ref{eq:schemetrafo1}). 
Similarly, $\cq[1]$ and $\cg[1]$ remain unchanged while $\cgam[1]$ has to be modified
according to $\cgam[1]^\disg = \cgam[1]^\msbar - \cgam[2]^\msbar=-12 x(1-x)$. 

The LO expression for $\sfs{i}{\gam(P^2)}$ ($\mi=1,2$) are obviously entailed in 
Eq.~(\ref{eq:f2gam})
by simply dropping all NLO terms proportional to $\cq[i], \cg[i]$ as well as
$\cgam[i]$.  
\subsection{Heavy Flavor Contributions}\label{sec:heavy}
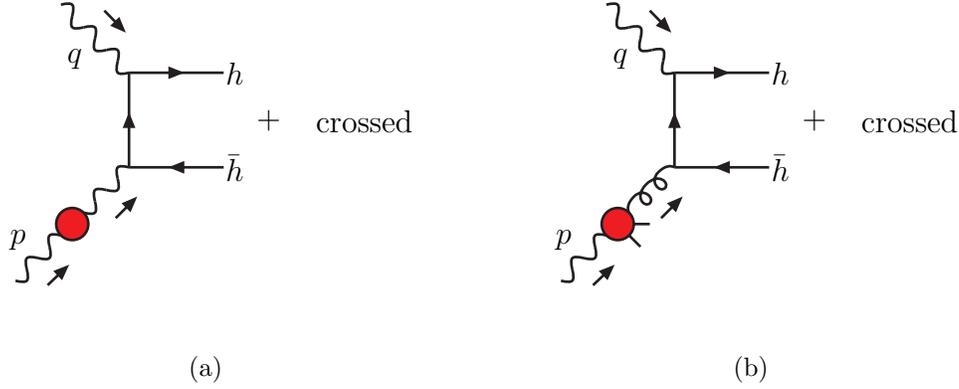
\begin{figure}
\begin{center}
\subfigure[]{
\begin{picture}(175,127.112)(0,0)
\SetWidth{1}
\SetOffset(-100.9415,-27.888)
\Photon(131.495,143.505)(157.165,117.835){3}{3}\Text(126.495,148.505)[]{}
\LongArrow(147.865,141.265)(154.935,134.015)
\Text(137.265,123.595)[]{$q$}
\Photon(114.383,39.383)(131.495,56.495){-3}{2}\Text(126.495,51.495)[]{}
\LongArrow(152.143,63.013)(159.213,70.083)
\Photon(140.053,65.053)(157.165,82.165){-3}{2}\Text(126.495,51.495)[]{}
\LongArrow(126.475,37.715)(133.545,44.415)
\Text(115.875,55.015)[]{$p$}
\IfColor{\COval(135.774,60.774)(6.05,6.05)(0){Black}{Red}
         }{\GOval(135.774,60.774)(6.05,6.05)(0){0.5}}
\ArrowLine(192.835,82.165)(157.165,82.165)\Text(197.835,82.165)[]{$\bar h$}
\ArrowLine(157.165,82.165)(157.165,117.835)
\ArrowLine(157.165,117.835)(192.835,117.835)\Text(197.835,117.835)[]{$h$}
\Text(210.67,100)[]{$+$}
\Text(228.505,100)[l]{crossed}
\end{picture}
\label{subfig:dircontr}}\qquad
\subfigure[]{
\begin{picture}(175,127.112)(0,0)
\SetWidth{1}
\SetOffset(-100.9415,-27.888)
\Photon(131.495,143.505)(157.165,117.835){3}{3}\Text(126.495,148.505)[]{}
\LongArrow(147.865,141.265)(154.935,134.015)
\Text(137.265,123.595)[]{$q$}
\Photon(114.383,39.383)(131.495,56.495){-3}{2}\Text(126.495,51.495)[]{}
\LongArrow(152.143,63.013)(159.213,70.083)
\Gluon(140.053,65.053)(157.165,82.165){3}{2}\Text(126.495,51.495)[]{}
\LongArrow(126.475,37.715)(133.545,44.415)
\Text(115.875,55.015)[]{$p$}
\IfColor{\COval(135.774,60.774)(6.05,6.05)(0){Black}{Red}
         }{\GOval(135.774,60.774)(6.05,6.05)(0){0.5}}
\Line(141.824,60.774)(147.874,60.774)
\Line(140.052,56.496)(144.33,52.218)
\ArrowLine(192.835,82.165)(157.165,82.165)\Text(197.835,82.165)[]{$\bar h$}
\ArrowLine(157.165,82.165)(157.165,117.835)
\ArrowLine(157.165,117.835)(192.835,117.835)\Text(197.835,117.835)[]{$h$}
\Text(210.67,100)[]{$+$}
\Text(228.505,100)[l]{crossed}
\end{picture}
\label{subfig:rescontr}}
\caption{Direct (a) and resolved (b) contribution of a heavy quark $h$ in deep inelastic
$\e\gam{(P^2)}$ scattering.}\label{fig:hqcontr}
\end{center}
\end{figure}

The photonic quark distributions discussed thus far and which appear
in Eq.\ (\ref{eq:f2gam}) are adequate for the $f=3$ light 
$u,\, d,\, s$ flavors.
Since heavy quarks $h=c,\,b,\,t$ will not be considered as `light'
partons in the photon (as in the case of the proton \cite{cit:GRV98} and the
pion \cite{cit:GRSc99pi,cit:GRS98pi}), their contributions to 
$\sfs{i}{\gam(P^2)}$ ($\mi=1,2,L$) 
have to be calculated in fixed order perturbation theory.
The heavy quark contribution consist of two parts, a 'direct' one and a 
'resolved' one as depicted in Fig.~\ref{fig:hqcontr}:
\begin{equation}\label{eq:dir+res}
\sfs{i,h}{\gam(P^2)}=\sfs{i,h}{\gam(P^2),\mathrm{dir}}+\sfs{i,h}{\gam(P^2),\mathrm{res}}\ .
\end{equation}
The `direct' contribution derives from the box--diagram 
$\gamma^*(Q^2)\,\gamma\to h\bar{h}$ expression, i.e.~the 
usual Bethe--Heitler cross section \cite{cit:Wit-7601,*cit:GR-7901}
\begin{align}\label{eq:f2gamh}
\frac{1}{x}\, \sfs{2,h}{\gam(P^2),\mathrm{dir}}(x,Q^2)
& =  3\, e_h^4\, \frac{\alpha}{\pi}\,\theta(\beta^2)\,
\left\{ \beta\,\left[ 8x(1-x)-1-x(1-x) \, \frac{4m_h^2}{Q^2}\right] \right. 
\nonumber\\*
& \left. + \left[ x^2+(1-x)^2 + x(1-3x)\, \frac{4m_h^2}{Q^2}-x^2\, 
        \frac{8m_h^4}{Q^4}\right] \, \ln \frac{1+\beta}{1-\beta}\, \right\}
\nonumber\\
2 \sfs{1,h}{\gam(P^2),\mathrm{dir}}(x,Q^2)
& =  3\, e_h^4\, \frac{\alpha}{\pi}\,\theta(\beta^2)\,
\left\{ \beta\,\left[ 4x(1-x)-1-x(1-x) \, \frac{4m_h^2}{Q^2}\right] \right. 
\nonumber\\*
& \left. + \left[ x^2+(1-x)^2 + x(1-x)\, \frac{4m_h^2}{Q^2}-x^2\, 
        \frac{8m_h^4}{Q^4}\right] \, \ln \frac{1+\beta}{1-\beta}\, \right\}
\end{align}
where $\beta^2\equiv1-4m_h^2/W^2=1-4m_h^2x/(1-x)Q^2$. 
A similar expression holds for the longitudinal structure function \cite{cit:GRS95} 
$F_L\equiv F_2-2xF_1$.  
The `resolved' heavy quark contribution 
to $\sfs{i}{\gam(P^2)}$
has to be calculated via \mbox{$\gamma^*(Q^2)g \to h\bar{h}$} and
is given by \cite{cit:GRS95} 
\begin{align}\label{eq:f2gamhres}
\sfs{2,h}{\gam(P^2),\mathrm{res}}(x,Q^2)
&= \int_{z_{min}}^1\frac{dz}{z}\, z \fivg[g](z,\mu_F^2)\, 
\hat{f}_2^{\gamma^*(Q^2)g\to h\bar{h}}\left(\frac{x}{z},\,Q^2 \right)
\nonumber\\
\sfs{1,h}{\gam(P^2),\mathrm{res}}(x,Q^2)
&= \int_{z_{min}}^1\frac{dz}{z}\, \fivg[g](z,\mu_F^2)\, 
\hat{f}_1^{\gamma^*(Q^2)g\to h\bar{h}}\left(\frac{x}{z},\,Q^2 \right)
\end{align}
where $\frac{1}{x}\, \hat{f}_2^{\gamma^*(Q^2)g\to h\bar{h}}(x,Q^2)$
and $2\hat{f}_1^{\gamma^*(Q^2)g\to h\bar{h}}(x,Q^2)$
are given by Eq.~(\ref{eq:f2gamh}) with $e_h^4\alpha\to e_h^2\alpha_s(\mu_F^2)/6$,
$z_{\min}=$ \mbox{$x(1+ 4m_h^2/Q^2)$} and $\mu_F^2\simeq 4m_h^2$ 
\cite{cit:GRS94}\footnote{An alternative choice would be $\mu_F^2 = Q^2 + 4m_h^2$
\protect\cite{Stratmann-PhD}
which satisfies the requirement $\mu_F^2 \gg P^2$ also for large $P^2$.}. 
To ease the calculations we shall keep these LO expressions 
in Eqs.~(\ref{eq:f2gamh}) and (\ref{eq:f2gamhres}) also in NLO, since the full NLO 
expressions for heavy quark 
production \cite{Laenen:1994ce,*Laenen:1995tz,*Laenen:1996ee} turn out to 
be a small correction to the 
already not too sizeable (at most about 20\%) 
contribution in LO.
Notice that such small corrections are not larger than ambiguities
due to different choices for $m_h$ and for the factorization scale
$\mu_F$.  For our purposes it is sufficient to include only the charm
contributions which will be calculated using $m_c=1.4$ GeV.

As already mentioned in the introduction
the heavy quark contributions deserve special care due to the additional scale
$m_h$ provided by the heavy quark mass.
For general $P^2 (\ll Q^2)$  terms $P^2/m_h^2$ are not always power--suppressed
and have to be kept in the heavy quark coefficient functions.
The corresponding 'direct' and 'resolved' heavy quark contributions can be found
in Eqs.~(18) and (19) of Ref.~\cite{cit:GRS95}.

\section{$Q^2$--Evolution}\label{sec:evolution}
\subsection{Evolution Equations}
As motivated in Sec.~\ref{sec:fact2} 
a {\em transverse} hadronic photon 
can be described by its partonic quark, gluon and photon content denoted by
$q_i^{\gamma(P^2)}(x,Q^2)\, (i=1,\ldots,f)$ with $q_1=u$, $q_2=d$, $q_3=s$ 
and so on\footnote{Note that due to charge conjugation invariance 
$\bar{q}_i^{\gamma(P^2)}(x,Q^2)=q_i^{\gamma(P^2)}(x,Q^2)$.}, 
$\gvg(x,Q^2)$, and $\gamvg(x,Q^2)$.
The general evolution equations for these parton densities are given
by (suppressing the $x$-- and $Q^2$--dependence)
\begin{align}\label{eq:genevol}
\frac{\der q_i^{\gamma(P^2)}}{\der \ln Q^2} & = 
\Pqigambar \otimes \Gamma^{\gamma(P^2)}
+ 2 \sum_{k=1}^f \Pqiqkbar \otimes q_k^{\gamma(P^2)}
       + \Pqigbar \otimes g^{\gamma(P^2)}
\nonumber\\
\frac{\der g^{\gamma(P^2)}}{\der \ln Q^2} & = 
\Pggambar \otimes \Gamma^{\gamma(P^2)}
+ 2 \sum_{k=1}^f \Pgqkbar \otimes q_k^{\gamma(P^2)}
       + \Pggbar \otimes g^{\gamma(P^2)}
\\
\frac{\der \Gamma^{\gamma(P^2)}}{\der \ln Q^2} & = 
\Pgamgambar \otimes \Gamma^{\gamma(P^2)}
+ 2 \sum_{k=1}^f \Pgamqkbar \otimes q_k^{\gamma(P^2)}
       + \Pgamgbar \otimes g^{\gamma(P^2)}
\nonumber
\end{align}
where $f$ denotes the number of active (massless) quark flavors.
The evolution kernels $\Pbar_{ij}$ are generalized splitting functions 
\begin{equation}
\Pbar_{i j}(x,\alpha,\alpha_s) = 
\sum_{l,m=0} \frac{\alpha^l \alpha_s^m}{(2 \pi)^{l+m}}
\Pbar_{i j}^{(l,m)}(x),
\label{eq:gensplit}
\end{equation}
with $\Pqiqkbar$ being the average of the quark--quark and antiquark--quark
splitting functions.
Note that terms 
due to photon radiation from the quark line 
are taken into account
in Eq.~(\ref{eq:genevol}) as it must be in order to guarantee momentum
conservation (see the erratum to \cite{cit:DeWitt-7901}).

Since the electromagnetic coupling constant $\alpha \ll 1$ 
we can safely neglect terms of order 
$\Ord(\alpha^2)$ in Eq.~(\ref{eq:genevol})\footnote{For this reason we can also
neglect the running of $\alpha(Q^2)$ and use $\alpha \simeq 1/137$.}.
Noticing that $\qvg$ and $\gvg$ are already 
of order $\Ord(\alpha)$ 
we can take $l=0$ in all generalized splitting functions in (\ref{eq:gensplit})
which are multiplied by $\qvg$ or $\gvg$
such that the functions $\Pbar_{ij}$ reduce in this case to the conventional QCD
splitting functions $P_{ij}(x,\alpha_s)$.
On the other hand in $\Pqigambar$, $\Pggambar$ and $\Pgamgambar$
the contributions with $l=0$ obviously vanish such that the only contribution
to order $\Ord(\alpha)$ comes from setting $l=1$ in Eq.~(\ref{eq:gensplit}) 
in this case.
Furthermore, the terms proportional to $\Pgamqk$ and $\Pgamg$ 
are of the order $\Ord(\alpha^2)$ and can be dropped
such that the
evolution equation for the photon distribution decouples and 
can be solved separately as will be discussed in Sec.~\ref{sec:MSR}.
Here it suffices to know
the photon density at order $\alpha^0$ only
in order to obtain the quark and gluon densities 
in Eq.~(\ref{eq:genevol})
and the
structure functions 
in Eq.~(\ref{eq:f2gam})
consistently to order $\Oalp$.
At $0$th order, however,
no splitting is possible and the photon carries its full momentum
with probability one: $\gamvg(x,Q^2) = \delta(1-x) +\Oalp$.

Therefore, Eq.~(\ref{eq:genevol}) can be reduced
to the following inhomogeneous evolution equations for the quark
and gluon content of the photon \cite{cit:DeWitt-7901}:
\begin{align}
\frac{\der \fivg[q_i]}{\der \ln Q^2} & = 
\Kqi
+ 2 \sum_{k=1}^f \Pqiqk \otimes \fivg[q_k] + \Pqig \otimes \gvg 
\nonumber\\
\frac{\der \gvg}{\der \ln Q^2} & = 
\Kg 
+ 2 \sum_{k=1}^f \Pgqk \otimes \fivg[q_k]+ \Pgg \otimes \gvg 
\label{eq:x-evol}
\end{align}
where we have utilized the standard notation 
$\Kqi \equiv \Pqigam$, $\Kg \equiv \Pggam$
for the photon--quark and photon--gluon splitting functions.
The splitting functions ${\mathrm P}_{ij}(x,\alpha_s)$ are presently known to
next--to--leading order in $\alpha_s$ and can be found in 
\cite{Curci:1980uw,*Furmanski:1980cm,*Floratos:1981hs,
Fontannaz:1992gj,cit:GRV-9201}.

The $Q^2$--evolution is most conveniently treated in the Mellin moment
space where the convolutions turn into ordinary products:
$(a \otimes b)(x) \rightarrow (a \otimes b)^n = a^n b^n$ where
the $n$--th Mellin moment of a function $f(x,Q^2)$ is defined by
\begin{equation}
f^n(Q^2) = \int_0^1 dx\ x^{n-1} f(x,Q^2)\ .
\label{eq:mellin}
\end{equation} 
Hence the evolution of $n$--moments is governed by
a set of ordinary coupled differential equations which can be
solved analytically \cite{cit:GRV-9201}.

In order to decouple the evolution equations as far as possible 
we introduce flavor  non--singlet (NS) quark combinations:
\begin{equation}
\qns = 2\left[\sum_{i=1}^j \fivg[q_i] -j \fivg[q_j] \right] 
\end{equation}
with $i, j = 1, \ldots, f$ and the group theoretical index $\ell = j^2 - 1$.
Thus, for three active flavors ($u,d,s$) we have two non--vanishing
NS combinations $q_{\mathrm{NS},3}= 2 (u - d)$ and
$q_{\mathrm{NS},8}= 2 (u+d-2s)$.
Furthermore, we define a singlet (S) vector which helps to write the
singlet evolution equations and their solutions in a compact matrix form:
\begin{equation}
\qs = \left( 
\begin{array}{c}
\fivg[\Sigma] 
\\
\gvg
\end{array}
\right)\ , \qquad \fivg[\Sigma] = 2 \sum_{i=1}^f \fivg[q_i] \ .
\end{equation}

We wish to make an expansion of the solutions of the evolution equations
and eventually of the structure functions in terms of $\alpha_s$.
For this reason the evolution equations will be rewritten as differential
equations in $\alpha_s$ with help of the
renormalization group equation for the strong coupling constant:
\begin{equation}\label{eq:running}
\frac{\der \alpha_s}{\der \ln Q^2} =
-\frac{\beta_0}{4 \pi} \alpha_s^2(Q^2) -\frac{\beta_1}{(4 \pi)^2} \alpha_s^3(Q^2) 
+ \Ord(\alpha_s^4)
\end{equation}
with $\beta_0 = 11 - (2/3) f$ and $\beta_1 = 102 - (38/3) f$.
Expanding the splitting functions in $\alpha_s$ in a form
which can be generically written as $P = \asb (P^{(0)} + \asb P^{(1)} + \ldots)$ and
$k = \aemb (k^{(0)} + \asb k^{(1)} + \ldots)$, respectively, and
neglecting terms of the order $\Ord(\alpha_s^2)$ the inhomogeneous evolution equations
can be brought into the following form: 
\begin{align}
\frac{\der}{\der \alpha_s} \qnsn &=
-\frac{\alpha}{\alpha_s^2}\frac{2}{\beta_0} k_\ell^{(0)n}
-\frac{\alpha}{\alpha_s}\frac{1}{\pi \beta_0} 
\left(k_\ell^{(1)n}-\frac{\beta_1}{2 \beta_0} k_\ell^{(0)n}\right)
\nonumber\\*
&
-\left[\frac{1}{\alpha_s}\frac{2}{\beta_0} P_{qq}^{(0)n}
+\frac{1}{\pi \beta_0} \left(P_{+}^{(1)n}-\frac{\beta_1}{2 \beta_0}P_{qq}^{(0)n}\right)
\right] \qnsn
\nonumber\\
\frac{\der}{\der \alpha_s} \qsn &=
-\frac{\alpha}{\alpha_s^2}\frac{2}{\beta_0} \vec{k}^{(0)n}
-\frac{\alpha}{\alpha_s}\frac{1}{\pi \beta_0} 
\left(\vec{k}^{(1)n}-\frac{\beta_1}{2 \beta_0} \vec{k}^{(0)n}\right)
\nonumber\\*
&
-\left[\frac{1}{\alpha_s}\frac{2}{\beta_0} \hat{P}^{(0)n}
+\frac{1}{\pi \beta_0} \left(\hat{P}^{(1)n}-\frac{\beta_1}{2 \beta_0}\hat{P}^{(0)n}\right)
\right] \qsn \ .
\label{eq:evol}
\end{align}
Here $\vec{k}^{(0)n}$ and $\vec{k}^{(1)n}$ denote the LO and NLO photon--to-parton
splitting functions in the singlet sector:
$\vec{k}^n = \left({k}^n_\Sigma, {k}^n_g \right)^T$ with 
$k^n_\Sigma = 2 \sum_{i=1}^f k^n_{q_i}$
while $k^{(0)n}_\ell$ and $k^{(1)n}_\ell$ are the LO and NLO parts 
of the non--singlet combination
$k^{n}_\ell = 2\left[\sum_{i=1}^j k^n_{q_i} -j k^n_{q_j} \right]$.
Furthermore, $\hat{P}^{(0)n}$ and $\hat{P}^{(1)n}$ refer to LO and NLO
$2 \times 2$ matrices of one-- and two--loop splitting functions whereas
$P_+^{(1)n}$ and $P_{qq}^{(0)n}$ occurring in the NS equation are 
scalar functions.
The complete set of LO and NLO (non--)singlet 
splitting functions can be found, e.g., in \cite{cit:GRV-9201,Vogt-PhD}.
\subsection{Analytic Solutions}
\label{sec:analyticsol}
The general solutions of the inhomogeneous evolution equations (\ref{eq:evol})
are formally identical to the ones of a real photon \cite{cit:GRV-9201} and
can be written, as usual, as a sum of a ('pointlike') particular solution $\qpl$ 
of the inhomogeneous problem
and a general solution $\qhad$ of the homogeneous ('hadronic') equations :
\begin{equation}
q = \qpl+\qhad \, , \qquad q = \qnsn, \qsn, \pdfn{}, \ldots \ .
\end{equation}
The particular solution is not unique and we fix it by the condition
$\qpl(Q_0^2) \equiv 0$ at the reference scale $Q_0^2$.

The NLO 'pointlike' (inhomogeneous) flavor--singlet solution is given 
by \cite{cit:GRV-9201,cit:GRS95}
\begin{align}
\qsnpl(Q^2) & = 
\frac{2 \pi}{\alpha_s(Q^2)}\left(1+\asbq \hat{U}^n\right)
\left[1-L^{1+\dn}\right]
\frac{1}{1+\dn}
\frac{\alpha}{\pi \beta_0} \vec{k}^{(0)n}
\nonumber\\*
& + \left[1-L^{\dn}\right]\frac{1}{\dn}\frac{\alpha}{\pi \beta_0}
\left(\vec{k}^{(1)n}-\frac{\beta_1}{2 \beta_0} \vec{k}^{(0)n}
-\hat{U}^n\vec{k}^{(0)n}\right)
\label{eq:inhomsol}
\end{align}
where $\dn \equiv -(2/\beta_0) \hat{P}^{(0)n}$, $L \equiv \XiL$ and the
$2 \times 2$ NLO evolution matrix $\hat{U}$ is fixed through the commutation relation
$[\hat{U}^n,\hat{P}^{(0)n}] = \tfrac{\beta_0}{2}\hat{U}^n+\hat{P}^{(1)n}
-\tfrac{\beta_1}{2 \beta_0}\hat{P}^{(0)n}$ \cite{cit:FP-8201,cit:GRV-9201,Vogt-PhD}.
The NLO 'hadronic' (homogeneous) solution
depending on input distributions $\qsnhad(Q_0^2)=\qsn(Q_0^2)$  
which will be specified in Chapter \ref{chap:bc}
is given by \cite{cit:GRV-9201,cit:GRS95}
\begin{align}
\qsnhad(Q^2) & = 
\left[L^{\dn} + \asbq \hat{U}^n L^{\dn} - \asbqz L^{\dn} \hat{U}^n\right]
\qsnhad(Q_0^2)\ .
\label{eq:homsol}
\end{align}
The LO results are obviously entailed in these expressions by simply
dropping all higher order terms ($\beta_1, \vec{k}^{(1)n},\hat{U}$).
The non--singlet solutions can be easily obtained from Eqs.~(\ref{eq:inhomsol}) and
(\ref{eq:homsol}) by the obvious replacements
$\hat{P}^{(0)} \rightarrow \Pqq^{(0)}$, $\hat{P}^{(1)} \rightarrow \mathrm{P}_{+}^{(1)}$,
$\vec{k} \rightarrow k_\ell$ together with
$\hat{U} \rightarrow U_{+,\ell} = - \tfrac{2}{\beta_0}\left(\mathrm{P}_{+}^{(1)} 
- \tfrac{\beta_1}{2 \beta_0} \Pqq^{(0)}\right)$ \cite{cit:GRV-9201,Vogt-PhD}.

Note that the solutions (\ref{eq:inhomsol}) and (\ref{eq:homsol}) 
are valid for a {\em fixed number} $f$ of active flavors.
Evoluting beyond an $\msbar$ 'threshold', set by the heavy quark
masses $m_c$, $m_b$ and $m_t$,
the number
of active flavors has to be increased by one, i.e., $f \rightarrow f+1$
in the calculation of $\alpha_s$.
The evolution of $\alpha_s^{(f)}(Q^2)$, corresponding to a 
number of $f$ active flavors, is obtained by exactly solving in
NLO($\msbar$)
\begin{equation}\label{eq:running2}
\frac{d\alpha_s^{(f)}(Q^2)}{d\ln Q^2} = -\frac{\beta_0^{(f)}}{4\pi}
 \left[\alpha_s^{(f)}(Q^2)\right]^2 - \frac{\beta_1^{(f)}}{16\pi^2}
  \left[\alpha_s^{(f)}(Q^2)\right]^3
\end{equation}
numerically \cite{cit:GRV98} using $\alpha_s^{(5)}(M_Z^2)=0.114$, rather
than using the more conventional approximate solution
\begin{equation}\label{eq:alphasnlo}
\frac{\alpha_s^{(f)}(Q^2)}{4\pi} \simeq \frac{1}
         {\beta_0^{(f)}\ln(Q^2/\Lambda^2)}
  -\frac{\beta_1^{(f)}}{(\beta_0^{(f)})^3} 
    \frac{\ln\ln(Q^2/\Lambda^2)}{[\ln (Q^2/\Lambda^2)]^2}
\end{equation}
which becomes sufficiently accurate only for 
$Q^2$ \raisebox{-0.1cm}{$\stackrel{>}{\sim}$} $m_c^2\simeq 2\ \gevsq$
with \cite{cit:GRV98} 
$\Lambda_{\msbar}^{(f=4,5,6)}=257, \, 173.4,\,  68.1\ \mev$, 
whereas in LO ($\beta_1\equiv0$) $\Lambda_{\rm LO}^{(4,5,6)} =
175, 132, 66.5\ \mev$.  
Furthermore, $\beta_0^{(f)}=11-2f/3$ and
$\beta_1^{(f)}=102-38f/3$.  For matching $\alpha_s$ at the
$\overline{\rm MS}$ `thresholds' $Q\equiv Q_f = m_f$, i.e.\
$\alpha_s^{(f+1)}(m_{f+1}^2)=\alpha_s^{(f)}(m_{f+1}^2)$, we have
used \cite{cit:GRV98} $m_c=1.4\ \gev$, $m_b=4.5\ \gev$ and $m_t=175\ \gev$.
On the other hand, we fix $f=3$ in the splitting functions 
$P_{ij}^{(0,1)}$ in Eqs.~(\ref{eq:inhomsol}) and (\ref{eq:homsol}) for consistency
since we treat the heavy quark sector ($c,b,\ldots$) by the 
perturbatively stable full production cross sections in fixed--order
perturbation theory, i.e. 
$\gamma^*(Q^2)\fivg[\Gamma]\to c\bar{c}$
and $\gamma^*(Q^2)g^{\gamma(P^2)} \to c\bar{c}$, etc., keeping
$m_c\neq 0$.

Thus the full solutions can be obtained iteratively by the following
procedure written out in detail in \cite{Gluck:2001rn}:
Between thresholds
we can use Eqs.~(\ref{eq:inhomsol}) and (\ref{eq:homsol}) for the
corresponding value of $f$.
Crossing a threshold $Q^2 = m_h^2$ we have to replace $f$ by $f+1$
in the following expressions occurring in
Eqs.~(\ref{eq:inhomsol}) and (\ref{eq:homsol}):
\begin{gather}
\alpha_s^{(f)}(Q^2 = m_h^2) \rightarrow \alpha_s^{(f+1)}(Q^2 = m_h^2) 
\nonumber\\
\dn^{(f)} =-(2/\beta_0^{(f)}) \hat{P}^{(0)n} \rightarrow \dn^{(f+1)}
\nonumber
\end{gather}
and the {\em full} solution at the threshold evolved with $f$ active flavors
so far
then serves as input at the new 'input' scale $Q_0^2 = m_h^2$ which 
subsequently has to be evolved 
using Eqs.~(\ref{eq:inhomsol}) and (\ref{eq:homsol}) 
with $f+1$ active flavors.
\subsection{Numerical Mellin--Inversion}
The solutions in $x$--space are then obtained 
by numerically inverting the Mellin transformation in Eq.~(\ref{eq:mellin}) according to
\begin{equation}\label{eq:inverse-mellin}
f(x,Q^2) = \frac{1}{2 \pi i} \int_c dn\ x^{-n} f^n(Q^2)
\end{equation}
where the integration contour $c$ 
has to enclose all poles in the complex $n$--plane. 
This can be realized, e.g., if the contour
ranges from negatively to positively infinite imaginary values 
and lies to the right of all singularities of $f^n$.
In the case of parton distributions and structure functions all singularities
are located on the real axis and hence we can parameterize the contour as
follows:
\begin{equation}
c(z) = c_0 + \e^{i \phi} z\ ,\qquad c_0 \in \mathbf{R}\ .
\end{equation}
For such contours the transformation (\ref{eq:inverse-mellin}) can be
written as \cite{cit:GRV-9201}
\begin{equation}\label{eq:inverse-mellin2}
f(x,Q^2) = \frac{1}{\pi} \int_0^\infty dz\ \imag 
\left\{\e^{i \phi} x^{-c(z)} f^{n=c(z)}(Q^2) \right\} \ .
\end{equation}
Employing $\phi > \tfrac{\pi}{2}$ the factor $x^{-z \exp(i\phi)}$ dampens
the integrand for increasing values of $z$ allowing for a smaller
upper limit $z_{\mathrm{max}}$ in the numerical calculation of
(\ref{eq:inverse-mellin2})
as compared to the standard integration contour with $\phi = \tfrac{\pi}{2}$.
We choose the following parameters in all practical applications 
of Eq.~(\ref{eq:inverse-mellin2}) \cite{cit:GRV-9201,Vogt-PhD}\footnote{The upper 
limit of integration $z_{\mathrm{max}}$ refers to the photon case only.
For inversions of pionic parton densities in Chap.~\protect\ref{pipdf} we have used
$z_{\mathrm{max}}(x \le 0.001) = 24$, $z_{\mathrm{max}}(0.001 < x\le 0.05) = 40$,
$z_{\mathrm{max}}(0.05 < x\le 0.2) = 56$, $z_{\mathrm{max}}(0.2 < x\le 0.4) = 72$, 
$z_{\mathrm{max}}(0.4 < x\le 0.7) = 88$ and $z_{\mathrm{max}}(0.7 < x < 1) = 136$.}:
\begin{equation*}
\phi = \frac{3}{4} \pi,\qquad 
z_{\mathrm{max}} = 5 + \frac{10}{\ln(1/x)},\qquad
c_0 = 
  \begin{cases}
  0.8 & \text{for non--singlet inversions}\\ 
  1.8 & \text{for singlet inversions} 
  \end{cases}   
\end{equation*}
where the difference between non--singlet and singlet inversions is due to
the different singularity structures of the corresponding splitting functions
the rightmost pole lying at $n=0$ and $n=1$, respectively.
%
\subsection{Boundary Conditions for a Deeply Virtual Target Photon}\label{sec:pert_bc}
For a deeply virtual target photon $\gam(P^2)$ with $\Lambda^2 \ll P^2 \ll Q^2$
the photon structure functions are perturbatively calculable
according to the doubly virtual box $\vgvg$ (and higher order corrections)
either within fixed order perturbation theory (FOPT)
or by resumming large collinear logarithms $\ln Q^2/P^2$ 
occurring in the fixed order calculation
with help
of renormalization group (RG) techniques to leading order \cite{Uematsu:1981qy}
and next--to--leading order \cite{Uematsu:1982je,Rossi:1984xz} accuracy.

More specifically, in Ref.~\cite{Uematsu:1982je} 
the framework of the RG--improved operator product expansion (OPE)
has been adopted 
in order to calculate predictions for the
unpolarized (twist--2) structure functions $\sfs{2,L}{<\gam(P^2)>}(x,Q^2)$
of a spin--averaged target photon.
These (OPE) results can be translated by a one--to--one correspondence 
into the framework of the RG--improved parton model (PM) by utilizing
'technical' boundary conditions given at the input scale $Q^2=P^2 \gg \Lambda^2$
such that the condition
\begin{equation}
\sfs{2}{\mathrm{OPE}}(x,P^2,Q^2) \overset{!}{=} \sfs{2}{<\gam(P^2)>,{\mathrm{PM}}}(x,Q^2)
\end{equation}
is fulfilled for $Q^2 \gg P^2$.
The appropriate boundary conditions are given by 
\cite{Uematsu:1982je,Rossi:1984xz,cit:GRS95}
\begin{equation}\label{eq:bclo}
\pdfb[f]{LO}{<\gam(P^2)>}(x,Q^2 = P^2) = 0
\end{equation}
in LO and by
\begin{align}\label{eq:bcdisg1}
\pdfb[q]{\disg}{<\gam(P^2)>}(x,Q^2 = P^2) & = \pdfb[\bar{q}]{\disg}{<\gam(P^2)>}(x,Q^2 = P^2) 
\nonumber\\*
&= N_c e_q^2\, \frac{\alpha}{2 \pi} \bigg\{
\left[ x^2+(1-x)^2\right] \ln \frac{1}{x^2} + 6x(1-x) - 2 \bigg\}
\nonumber\\*
\pdfb[g]{\disg}{<\gam(P^2)>}(x,Q^2 = P^2) & = 0 \, . 
\end{align}
in the NLO($\disg$) factorization scheme.
It should be noted, that these 'technical' boundary conditions have no
direct physical interpretation at $Q^2=P^2$
since the virtual target photon $\gam(P^2)$ is not resolved by the scale $Q^2 = P^2$.
Furthermore, at $Q^2=P^2$, the $x$--range is kinematically restrained 
to $x \le 1/2$ due to $0 \le x \le (1+P^2/Q^2)^{-1}$. However,
the boundary conditions (\ref{eq:bclo}) and (\ref{eq:bcdisg1}) generate the correct 
parton content at $Q^2 \gg P^2$ needed to
reproduce the purely perturbative results of 
\cite{Uematsu:1982je} in the Bjorken limit over the whole $x$--range 
$0 \le x \lesssim 1$.\footnote{To be precise, 
Eq.~(\ref{eq:f2gam}) supplemented with the boundary conditions in (\ref{eq:bcdisg1})
reproduces the OPE result of \protect\cite{Uematsu:1982je} up to spurious terms
of the order $\Ord(\alpha \alpha_s)$ which are of next--to--next--to--leading order (NNLO)
in a perturbative expansion of the photon structure function
$\sfs{2}{<\gam(P^2)>}$ in powers of $\alpha_s$.}

It is instructive to derive Eqs.~(\ref{eq:bclo}) and (\ref{eq:bcdisg1}) from
our fixed order calculation of the doubly virtual box $\vgvg$ in 
Chapter \ref{chap:lobox}.
The universal asymptotic ($Q^2 \gg P^2$) leading log (LL) part
of the box to the structure function
$\sfs{2,box}{<\gam(P^2)>}(x,Q^2)$
in Eq.~(\ref{eq:f2boxav}) (or Eq.~(\ref{eq:f2boxt})) may be used to define 
light (anti--)quark
distributions in the virtual photon target
\begin{align}\label{eq:def_bclobox}
\sfs{2,box}{<\gam(P^2)>}(x,Q^2)|_{\text{univ}} &= 
N_c \sum e_q^4\, \frac{\alpha}{\pi} x \left[ x^2+(1-x)^2\right] \, \ln \frac{Q^2}{P^2} 
\nonumber\\
& \equiv \sum_{q=u,d,s} x e_q^2 \left[\pdfb[q]{box}{<\gam(P^2)>}(x,Q^2)
+\pdfb[\bar{q}]{box}{<\gam(P^2)>}(x,Q^2)\right]
\end{align}
with
\begin{equation}\label{eq:bclobox}
\pdfb[q]{box}{<\gam(P^2)>}(x,Q^2) = \pdfb[\bar{q}]{box}{<\gam(P^2)>}(x,Q^2)
= N_c e_q^2 \frac{\alpha}{2\pi} \left[ x^2+(1-x)^2\right] \, \ln \frac{Q^2}{P^2} \ .
\end{equation}
Note that these fixed order box expressions do imply to this order 
a vanishing gluon component in the
virtual photon, $\pdfb[g]{box}{<\gam(P^2)>}(x,Q^2)=0$.

Higher powers of the finite but asymptotically large 
logarithm $\ln Q^2/P^2$ occurring in (\ref{eq:bclobox})
may be resummed to all orders in perturbation theory with help of the renormalization
group by imposing leading order boundary 
conditions\footnote{This is quite similar to the case of 
heavy quark production studied in Part I of this thesis where logarithms $\ln Q^2/m_h^2$ 
of the heavy quark mass $m_h$ may be resummed to all orders using 
perturbatively calculable boundary conditions
for the heavy quark distribution function in a hadron.}
\begin{equation}
\pdfb[f]{LO}{<\gam(P^2)>}(x,Q^2=P^2) = \pdfb[f]{box}{<\gam(P^2)>}(x,Q^2=P^2) = 0\, , \qquad
(f = q,g)\ .
\end{equation}

In order to derive the NLO input distributions in (\ref{eq:bcdisg1})
we have to take into account the full asymptotic box structure function
$\sfs{2,box}{<\gam(P^2)>}(x,Q^2)$ in Eq.~(\ref{eq:f2boxav}) 
following from Eq.~(\ref{eq:LO-box}) in the limit $P^2 \ll Q^2$ for
the light quarks ($m \equiv 0$) 
which can be used to define scheme--dependent
light (anti--)quark
distributions in the virtual photon target in NLO
\begin{align}
\sfs{2,box}{<\gam(P^2)>}(x,Q^2) 
&\equiv \sfs{2}{<\gam>,{\mathrm dir},\disg}(x,Q^2) 
+ 2 \sum e_q^2 x\ 
\pdfb[q]{box,\disg}{<\gam(P^2)>}(x,Q^2)  
\end{align}
where the 'direct' term $\sfs{2}{<\gam>,{\mathrm dir}}(x,Q^2)$ 
vanishes in the $\disg$ scheme implying
\begin{align}\label{eq:bcdisgbox}
\pdfb[q]{box,\disg}{<\gam(P^2)>}(x,Q^2)  = \pdfb[\bar{q}]{box,\disg}{<\gam(P^2)>}(x,Q^2)  
&= N_c e_q^2\, \frac{\alpha}{2\pi} \bigg\{\left[ x^2+(1-x)^2\right] \, \ln \frac{Q^2}{P^2} 
\nonumber\\
&+\left[ x^2+(1-x)^2\right] \, \ln \frac{1}{x^2} 
+ 6x(1-x)-2 
\bigg\}\, . 
\end{align}
The fixed order
box' distributions 
in the $\disg$ scheme
can again be resummed employing the NLO($\disg$) 
evolution equations with the input distributions
\begin{align}\label{eq:bcdisg}
\pdfb[q]{\disg}{<\gam(P^2)>}(x,Q^2=P^2) &= \pdfb[q]{box,\disg}{<\gam(P^2)>}(x,Q^2=P^2) 
\nonumber\\*
&= N_c e_q^2\, \frac{\alpha}{2\pi} \bigg\{
\left[ x^2+(1-x)^2\right] \, \ln \frac{1}{x^2} + 6x(1-x)-2 
\bigg\}
\nonumber\\
\pdfb[g]{\disg}{<\gam(P^2)>}(x,Q^2=P^2) &= \pdfb[g]{box,\disg}{<\gam(P^2)>}(x,Q^2=P^2) = 0 
\end{align}
which coincide with the results in Eq.~(\ref{eq:bcdisg1}).

So far, the discussion has been for a spin--averaged target photon.
However, the boundary conditions for a transverse target photon can be
obtained for free from 
$\sfs{2,box}{\gam[T](P^2)}(x,Q^2)$ given in Eq.~(\ref{eq:f2boxt})
by changing $6x(1-x)$ to $8x(1-x)$ in (\ref{eq:bcdisg}):
\begin{align}\label{eq:bcdisg2}
\pdfb[q]{\disg}{\gam[T](P^2)}(x,Q^2 = P^2) & = \pdfb[\bar{q}]{\disg}{\gam[T](P^2)}(x,Q^2 = P^2) 
\nonumber\\
&= N_c e_q^2\, \frac{\alpha}{2 \pi} \bigg\{
\left[ x^2+(1-x)^2\right] \ln \frac{1}{x^2} + 8x(1-x) - 2 \bigg\}
\nonumber\\
\pdfb[g]{\disg}{\gam[T](P^2)}(x,Q^2 = P^2) & = 0 \, . 
\end{align}
%
%

The corresponding results in the $\msbar$ scheme can either be obtained by the
factorization scheme transformation in (\ref{eq:schemetrafo1}) or by
repeating the above described steps in the $\msbar$ scheme with
the 'direct' term 
\begin{align}\label{eq:f2dirmsbar}
\sfs{2}{<\gam>,{\mathrm dir},\msbar}(x,Q^2)  &=
N_c \sum e_q^4\, \frac{\alpha}{\pi} x 
  \bigg\{ 
\left[ x^2+(1-x)^2\right] \ln \frac{1-x}{x} + 8x(1-x)-1 \bigg\}\, . 
\end{align}

Finally, let us stress that the {\em finite} collinear logarithm $\ln Q^2/P^2$
a priori needs {\em not} to be resummed. This issue has to be decided by
phenomenological relevance.
In addition, the predictive power of observables obtained from
perturbatively calculated boundary conditions generally appears to be weakened
by the freedom in choosing the input scale $Q_0^2$ \cite{Kretzer:1999zd}
(at least at low orders in perturbation theory).
Albeit $Q_0^2=P^2$ is a 'natural' choice for the input scale due to the
vanishing of the collinear logarithm $\ln Q^2/P^2$ it is by
no means compelling 
and a different input scale $Q_0^2 \ne P^2$ 
would result in equally well justified boundary conditions
\begin{align}
\pdfb[q]{LO}{\gam[T](P^2)}(x,Q_0^2) & = N_c e_q^2\, \frac{\alpha}{2 \pi} 
\left[ x^2+(1-x)^2\right] \ln \frac{Q_0^2}{P^2}
\nonumber\\
\pdfb[q]{\disg}{\gam[T](P^2)}(x,Q_0^2) & = 
N_c e_q^2\, \frac{\alpha}{2 \pi} \bigg\{\left[ x^2+(1-x)^2\right] \ln \frac{Q_0^2}{P^2} 
\nonumber\\
&\phantom{= N_c e_q^2\, \frac{\alpha}{2 \pi} \bigg\{}
+\left[ x^2+(1-x)^2\right] \ln \frac{1}{x^2} 
+ 8x(1-x) - 2 \bigg\}
\nonumber\\
\pdfb[g]{LO,\disg}{\gam[T](P^2)}(x,Q_0^2) & = 0
\end{align}
as long as $\ln Q_0^2/P^2$ is not large.
Writing $\ln Q^2/P^2 = \ln Q_0^2/P^2 + \ln Q^2/Q_0^2$ we see in this case
that a part of the collinear logarithm is kept in fixed order ($\ln Q_0^2/P^2$)
while the other part ($\ln Q^2/Q_0^2$) is resummed to all orders.

\subsection{Momentum Sum}\label{sec:MSR}
Due to energy--momentum conservation the momentum fractions carried by the 
individual partons must add up to one\footnote{This is certainly true 
for a transversely polarized target photon
considered here whereas the longitudinal degrees of freedom 
deserve possibly some attention.}:
\begin{equation}\label{eq:msr1}
\int_0^1 dx\ x \left[ \sum_{f=q,\bar{q},g} \fivg(x,Q^2) + \gamvg(x,Q^2)\right] = 1\ .
\end{equation}
However, as we have seen before $\gamvg(x,Q^2)$ drops out of the general evolution equations
in (\ref{eq:genevol})
leaving inhomogeneous evolution equations, see (\ref{eq:x-evol}) and (\ref{eq:evol}),
for the (resolved) partonic content of the photon and their
solutions require boundary conditions at some reference scale (input scale) $Q_0^2$.
A sum rule for the total momentum of the quarks and gluons in the photon 
\begin{equation}\label{eq:defM2}
\fivg[M_2](Q^2) \equiv \sum_{f=q,\bar{q},g} \int_0^1 dx\ x \fivg(x,Q^2) 
\end{equation}
would provide
useful information for constraining the input distributions. 
Eqs.~(\ref{eq:msr1}) and (\ref{eq:defM2}) can be rewritten in terms of Mellin moments 
(suppressing the obvious $Q^2$-- and $P^2$--dependence)
\begin{equation}\label{eq:msr2}
\mS + \mg + \mGam = 1, \qquad M_2 \equiv \mS + \mg = 1 - \mGam \ .
\end{equation}

To order $\alpha$ the evolution equation for $\gamvg$ reduces to
\begin{equation}\label{eq:gamevol}
\frac{\der \gamvg}{\der \ln Q^2} = \Pgamgam \otimes \gamvg 
\end{equation}
where the photon--photon splitting function $\Pgamgam$ can be constructed
from the second moments of $\Ksig\equiv 2 \sum \Kq$ and $\Kg$ as follows:
Using the evolution equations (\ref{eq:x-evol}) for the quark and gluon densities
and (\ref{eq:gamevol}) for the photon and exploiting in addition the conservation
of the hadronic energy--momentum tensor one can write
\begin{equation}
0 = \frac{\der}{\der \ln Q^2} (\mS+\mg+\mGam) = (\mkS+\mkg+\mPgamgam)\mGam \ .
\end{equation}
Furthermore, $\Pgamgam \propto \delta(1-x)$ to all orders in $\alpha_s$ 
because in order $\Ord(\alpha)$ only virtual diagrams contribute
to photon--photon splittings while
radiation of real photons starts at order $\Ord(\alpha^2)$ and we can write
\begin{equation}
\Pgamgam = \delta(1-x) \mPgamgam = -\delta(1-x) (\mkS+\mkg)\ .
\end{equation}
The $n$--moments of the photon--gluon and photon--quark splitting functions can be found, e.g.,
in \cite{cit:GRV-9201,Vogt-PhD}.
The second ($n=2$) moments then read
\begin{equation} 
\mkS[(0)n=2]+\mkg[(0)n=2]=\mkS[(0)n=2]= 2 \sum_q e_q^2,\qquad
\mkS[(1)n=2]+\mkg[(1)n=2]= 4 \sum_q e_q^2
\end{equation}
where the latter equation is valid both in the $\msbar$ and the $\disg$ scheme
as follows from Eq.~(\ref{eq:schemetrafo2}) with $\Pqq^{(0)n=2}+\Pgq^{(0)n=2}=0$.

Recalling the expansion 
$\mathrm{k}=\aemb (\mathrm{k}^{(0)} + \tfrac{\alpha_s}{2 \pi} \mathrm{k}^{(1)}+\ldots)$
we arrive at the following expression for the
photon--photon splitting function (to order $\alpha$)\footnote{Alternatively, $\Pgamgam$ 
can be
inferred from the Abelian ($C_A = 0$), diagonal (independent of $n$) part
of $\Pgg$ with the appropriate replacement
$T_R \equiv \tfrac{f}{2} \rightarrow N_c \sum_q e_q^2$.}:
\begin{equation}\label{eq:pgamgam}
\Pgamgam = - \delta(1-x)\ \frac{\alpha}{\pi} \sum_q e_q^2 \ 
\left[1 + \frac{\alpha_s}{\pi} + \Ord(\alpha_s^2)\right]\ .
\end{equation}

Eq.~(\ref{eq:gamevol}) can be solved analytically 
either in $n$--moment space or directly in
$x$--space since the convolution becomes trivial due to the $\delta(1-x)$ 
in (\ref{eq:pgamgam}) and one easily finds
\begin{align}\label{eq:Gamma}
\Gamma(x,Q^2) &= \Gamma(x,Q_0^2) \exp \left(-\frac{\alpha}{\pi} \sum_q e_q^2 
\left( \ln \frac{Q^2}{Q_0^2} + 4 I_1\right)\right)
\nonumber\\
& = \delta(1-x) \left[1 - \frac{\alpha}{\pi}\left(\sum_q e_q^2 \ln \frac{Q^2}{Q_0^2} 
+ c_1(Q_0^2,P^2)
+ 4 \sum_q e_q^2 I_1 
\right)\right]+\Ord(\alpha^2)
\end{align} 
with $c_1$ being a constant (depending on $Q_0^2$ and $P^2$) 
which automatically enters the description due to
the boundary condition $\Gamma(x,Q_0^2) = \delta(1-x)[1 + \Ord(\alpha)]$
and $I_1 = \int \tfrac{\alpha_s(Q^2)}{4 \pi} \der\ln Q^2$.
Employing the renormalization group equation for the strong coupling
in (\ref{eq:running}) $I_1$ can be trivially integrated and one
obtains
\begin{equation}\label{eq:I1}
I_1 = \frac{1}{\beta_0} \ln \frac{\alpha_s(Q_0^2)}{\alpha_s(Q^2)}
+ \frac{1}{\beta_0} 
\ln \frac{4 \pi \beta_0 + \beta_1 \alpha_s(Q^2)}{4 \pi \beta_0 + \beta_1 \alpha_s(Q_0^2)}\ .
\end{equation}
where the first logarithm in (\ref{eq:I1}) dominates over
the second one which is approximately 
$\tfrac{\beta_1}{4 \pi \beta_0} (\alpha_s(Q^2)-\alpha_s(Q_0^2))$ and hence 
parametrically of NNLO.
Note that contributions proportional to $\alpha_s(Q^2)-\alpha_s(Q_0^2)$ 
are generated also from terms of the order $\Ord(\alpha_s^2)$ in (\ref{eq:pgamgam}).

Eq.~(\ref{eq:msr1}) holds order by order in $\alpha$, so that (\ref{eq:Gamma})
implies for the total momentum carried by the (resolved) photonic partons
in LO--QCD \cite{cit:Vogt97}
\begin{equation}\label{eq:LO-msr}
\fivg[M_2](Q^2) = \frac{\alpha}{\pi}\left(\sum_q e_q^2 
\ln \frac{Q^2}{Q_0^2}
+ c_1(Q_0^2,P^2)\right)  
\end{equation}
and in NLO($\disg$)
\begin{align}\label{eq:NLO-msr}
\fivg[M_2](Q^2)& = \frac{\alpha}{\pi}\Bigg(\sum_q e_q^2 \ln \frac{Q^2}{Q_0^2}+ c_1(Q_0^2,P^2)
+ \frac{4}{\beta_0} \sum_q e_q^2 
\ln \frac{\alpha_s(Q_0^2)}{\alpha_s(Q^2)}
\nonumber\\*
&\phantom{=\frac{\alpha}{\pi}\Bigg({}}
+ \frac{4}{\beta_0} \sum_q e_q^2 
\ln \frac{4 \pi \beta_0 + \beta_1 \alpha_s(Q^2)}{4 \pi \beta_0 + \beta_1 \alpha_s(Q_0^2)}
\Bigg)  
\end{align}
where the sum $\sum_q e_q^2$ always extends over the 'light' $u$, $d$, and $s$ quarks
in the scheme adopted here (with $f=3$ fixed in the splitting functions)
whereas $\beta_0$, $\beta_1$ and $\alpha_s$ depend on the number of active flavors
(and therefore the heavy quark thresholds have to be taken into account by an iterative procedure
as described in Sec.~\ref{sec:analyticsol}).

The constant $c_1$ is tightly related to the momentum sum of the hadronic input
distributions at $Q^2=Q_0^2$: $M_2(Q^2=Q_0^2)/\alpha = c_1/\pi$.  
While $c_1$ is perturbatively calculable for large $P^2 \gg \Lambda^2$
from the boundary conditions in (\ref{eq:bclo}), (\ref{eq:bcdisg1}) or (\ref{eq:bcdisg2}) 
this is generally {\em not} the case for $P^2 \lesssim \Lambda^2$, especially
not for real $P^2=0$ photons and an important constraint on the parton densities
seems to be missing.

In the past few years some attempts were undertaken 
\cite{Frankfurt:1995bd,*Frankfurt:1996nz,*Frankfurt:1996gy,cit:SaS-9501,*Schuler:1996fc,
Gluck:1998fz}
to infer $c_1$ from elsewhere by relating Eq.~(\ref{eq:Gamma}) to the 
hadronic part of the photon vacuum polarization which in turn can be determined via
a dispersion relation from the well measured cross section 
$\sigma_h\equiv\sigma(e^+e^-\to \text{hadrons})$.
However, it turns out that the usefulness of such an approach is spoiled 
by higher--twist contributions
present in the experimental quantity \cite{Gluck:1998fz}. 
We will return to this issue when we discuss 
boundary conditions for the real photon in Chapter \ref{chap:bc}.

\section{Longitudinal Target Photons}
\label{sec:longitudinal}
Before turning to our analysis of the parton content of pions 
within a constituent quark model needed as basic ingredient for
our analysis of the parton content of
real and transversely polarized virtual photons within the framework of the radiative parton model 
in Chapter \ref{chap:bc}
let us discuss for completeness also the case of longitudinally polarized virtual
target photons. While these contributions vanish like, say, $P^2/m_{\rho}^2$
in the real photon limit due to gauge invariance they yield a {\em finite} contribution
if $P^2 \gg \Lambda^2$ and should be taken into account in
Eq.~(\ref{eq:fact2}).

As already mentioned in Sec.~\ref{sec:pert_bc}, the structure functions of
(transverse or longitudinal)
deeply virtual target photons $\gam(P^2)$ with $\Lambda^2 \ll P^2 \ll Q^2$
are reliably calculable in fixed order according to the doubly virtual box
$\vgvg$.
Nevertheless, it has been demonstrated recently 
\cite{Chyla:2000cu,Chyla:2000ue} that a partonic treatment
of $\gam[L](P^2)$ is phenomenologically useful and relevant.
The quark and gluon distribution functions 
$\pdfb[q]{}{\gam[L](P^2)}(x,Q^2)$ and
$\pdfb[g]{}{\gam[L](P^2)}(x,Q^2)$ of a {\em longitudinal}
target photon satisfy {\em homogeneous} evolution equations
\cite{Chyla:2000hp} in line with the expectation that
(to order $\Ord(\alpha)$) there is no (on--shell and therefore) transverse photon--parton
inside the longitudinal target photon, 
$\pdfb[\Gamma]{}{\gam[L](P^2)}(x,Q^2) = 0$.
In any case, 
the complete
theoretical framework for the longitudinal photon target $\gam[L]$
is obtainable from Sec.~\ref{sec:SFSPM} and \ref{sec:evolution}
by formally setting $\pdfb[\Gamma]{}{\gam[L]} = 0$, i.e
by dropping all 'direct' contributions 
($\Kq \otimes \pdfb[\Gamma]{}{\gam[L]}=0, 
\pdfb[\Gamma]{}{\gam[L]} \otimes C_{\gamma}=0$ etc)
and by using the perturbatively calculable boundary conditions
at $P^2 \gg \Lambda^2$ \cite{Chyla:2000hp}
\begin{align}\label{eq:bc_gamL}
\pdfb[q]{}{\gam[L](P^2)}(x,Q^2=P^2) &=
\pdfb[\bar{q}]{}{\gam[L](P^2)}(x,Q^2=P^2) =
N_c e_q^2\, \frac{\alpha}{2 \pi} 4 x (1-x)
\nonumber\\
\pdfb[g]{}{\gam[L](P^2)}(x,Q^2=P^2) &= 0\, ,
\end{align}
which follow directly from the asymptotic box expression in Eq.~(\ref{eq:f2boxl}).

%% file: phd_gam_pion.tex
\chapter{Pionic Parton Densities in a Constituent Quark Model}
\label{pipdf}
Before specifying our boundary conditions for the parton distributions of 
real and virtual photons we derive pionic parton distributions from
constituent quark model constraints. Being, of course, also interesting in 
their own rights, they will serve in the next chapter as input for the hadronic
component of the photon.
The analysis presented in this chapter is almost identical to Ref.~\cite{cit:GRSc99pi}.

%
The parton content of the pion is poorly known at present.  The main
experimental source about these distributions is mainly due to data of
Drell--Yan dilepton production in $\pi^-$--tungsten reactions 
\cite{cit:Freudenreich90,cit:NA10-85,cit:NA10-87,cit:E615-89,cit:SMRS92pi}, which 
determine the shape of the pionic valence
density $v^{\pi}(x,Q^2)$ rather well, and due to measurements of direct
photon production in $\pi^{\pm}p\to\gamma X$ 
\cite{cit:Freudenreich90,cit:NA24-87,cit:WA70-88,cit:E706-98} which
constrain the pionic gluon distribution $g^{\pi}(x,Q^2)$ only in the
large--$x$ region \cite{cit:GRV92pi}.  
In general, however, present data are not sufficient for fixing $g^{\pi}$
uniquely, in particular the pionic sea density $\bar{q}\,^{\pi}(x,Q^2)$
remains entirely unconstrained experimentally.  
Therefore in \cite{cit:GRS98pi} a constituent quark 
model \cite{cit:APR96} has been utilized
to relate $\bar{q}\,^{\pi}$ and $g^{\pi}$ to the much 
better known radiatively generated parton distributions $f^p(x,Q^2)$ of 
the proton \cite{cit:GRV94}. Therefore, previously \cite{cit:GRS98pi} a constituent quark 
model \cite{cit:APR96} has been utilized
to relate $\bar{q}\,^{\pi}$ and $g^{\pi}$ to the much 
better known radiatively generated parton distributions $f^p(x,Q^2)$ of 
the proton \cite{cit:GRV94}.
These relations arise as follows:  describing the constituent quark
structure of the proton $p=UUD$ and the pion, say $\pi^+=U\bar{D}$,
by the scale $(Q^2)$ independent distributions $U^{p,\pi^+}(x),\,\,
D^p(x)$ and $\bar{D}\,^{\pi^+}(x)$, and their universal (i.e.\ hadron 
independent) partonic content by $v_c(x,Q^2),\,\, g_c(x,Q^2)$ and 
$\bar{q}_c(x,Q^2)$, the usual parton content of the proton and the pion 
is then given by
\begin{eqnarray}
f^p(x,Q^2) & = & \int_x^1\frac{dy}{y}\left[ U^p(y)+D^p(y)\right] f_c
 \left( \frac{x}{y}, Q^2\right)
\label{pieqn1}
\\
f^{\pi}(x,Q^2) & = & \int_x^1\frac{dy}{y} \left[ U^{\pi^+}(y)+
  \bar{D}\,^{\pi^+}(y)\right] f_c \left( \frac{x}{y}, Q^2\right)
\label{pieqn2}
\end{eqnarray}
where $f=v,\,\bar{q},\,g$ with $v^p=u_v^p+d_v^p,\,\,\, 
\bar{q}\,^p=(\bar{u}\,^p+\bar{d}\,^p)/2,\,\,\, 
v^{\pi}=u_v^{\pi^+}+\bar{d}\,_v^{\pi^+},\,\,
\bar{q}\,^{\pi}=(\bar{u}\,^{\pi^+}+d^{\pi^+})/2$ and $\bar{u}\,^{\pi^+}
=d^{\pi^+}$ due to ignoring minor SU(2)$_{\rm{flavor}}$ breaking effects
in the pion `sea' distributions. Assuming these relations to apply at
the low resolution scale $Q^2=\mu^2$ ($\mu^2_{\rm{LO}} = 0.23$ GeV$^2,\,\,\,
\mu^2_{\rm{NLO}}=0.34$ GeV$^2$) of \cite{cit:GRV94} where the strange quark
content was considered to be negligible,  
\begin{equation}
 s^p(x,\mu^2) = \bar{s}\,^p(x,\mu^2) =s^{\pi}(x,\mu^2) =
  \bar{s}\,^{\pi}(x,\mu^2) = 0,
\label{pieqn3}
\end{equation}
one obtains from (\ref{pieqn1}) and (\ref{pieqn2}) the constituent quark independent
relations \cite{cit:GRS98pi}  
\begin{equation}
\frac{v^{\pi}(n,\mu^2)}{v^p(n,\mu^2)} =
\frac{\bar{q}\,^{\pi}(n,\mu^2)}{\bar{q}\,^p(n,\mu^2)} =
\frac{g^{\pi}(n,\mu^2)}{g^p(n,\mu^2)}
\label{pieqn4}
\end{equation}
where for convenience we have taken the Mellin $n$--moments of Eqs.\ (\ref{pieqn1}) 
and (\ref{pieqn2}), i.e.\ $f(n,Q^2)\equiv \int_0^1x^{n-1}f(x,Q^2)dx$.  Thus, as
soon as $v^{\pi}(x,\mu^2)$ is reasonably well \mbox{determined} from 
experiment,
our basic relations (\ref{pieqn4}) uniquely fix the gluon and sea densities of the
pion in terms of the rather well known parton distributions of the 
proton:
\begin{equation}
g^{\pi}(n,\mu^2) = \frac{v^{\pi}(n,\mu^2)}{v^p(n,\mu^2)}\, g^p(n,\mu^2),
\quad\quad
\bar{q}\,^{\pi}(n,\mu^2) = \frac{v^{\pi}(n,\mu^2)}{v^p(n,\mu^2)}\,
\bar{q}\,^p(n,\mu^2).
\label{pieqn5}
\end{equation}

Furthermore, the sum rules \cite{cit:GRS98pi}
\begin{eqnarray}
\int_0^1 v^{\pi}(x,Q^2) dx & = & 2
\label{pieqn6}\\
\int_0^1 xv^{\pi}(x,Q^2)dx & = & \int_0^1xv^p(x,Q^2)dx
\label{pieqn7}
\end{eqnarray}
impose strong constraints on $v^{\pi}(x,\mu^2)$ which are very useful
for its almost unambiguous determination from the $\pi N$ Drell--Yan
data.  Notice that Eq.\ (\ref{pieqn7}), together with (\ref{pieqn4}), implies the 
energy--momentum sum rule for $f^{\pi}$ to be manifestly satisfied.
In addition, Eq.\ (\ref{pieqn7}) implies that the valence quarks in the proton
and the pion carry similar total fractional momentum as suggested
by independent analyses within the framework of the radiative parton
model \cite{cit:GRV92pi,cit:GRV94}.

The relations in Eq.\ (\ref{pieqn5}) imply that any updating of $f^p(x,\mu^2)$
yields a corresponding updating of $f^{\pi}(x,\mu^2)$.  Recently an
updating of $f^p(x,\mu^2)$ within the framework of the radiative
(dynamical) parton model was undertaken \cite{cit:GRV98} utilizing additional
improved data on $F^p_2(x,Q^2)$ from HERA 
\cite{Aid:1996au,cit:H1-97,cit:H1-HEP97,Derrick:1996ef,cit:ZEUS-96b,cit:ZEUS-HEP97} 
and a somewhat
increased $\alpha_s(M_Z^2) = 0.114$ resulting in a slight increase in
$\mu^2\,\, (\mu^2_{\rm{LO}}=0.26$ GeV$^2,\,\,
\mu^2_{\rm{NLO}} = 0.40$ GeV$^2$).  An improved treatment of the
running $\alpha_s(Q^2)$ at low $Q^2$ was furthermore implemented by
solving in NLO($\msbar$)
Eq.~(\ref{eq:running2})
numerically \cite{cit:GRV98} rather than using the approximate NLO solution
in Eq.~(\ref{eq:alphasnlo})
as done in \cite{cit:GRV92pi,cit:GRS98pi,cit:GRV94}, which is sufficiently accurate only
for $Q^2$ \raisebox{-0.1cm}{$\stackrel{>}{\sim}$} $m_c^2\simeq2$ GeV$^2$ 
\cite{cit:GRV98}.  The LO and
NLO evolutions of $f^{\pi}(n,Q^2)$ to $Q^2>\mu^2$ are performed in
Mellin $n$--moment space, followed by a straightforward numerical
Mellin--inversion 
\cite{cit:GRV90} 
to Bjorken--x space as described
in Chapter \ref{chap:partonmodel}.  It should be noted
that the evolutions are always performed in the fixed (light) $f=3$
flavor factorization scheme \cite{cit:GRS94,cit:GRS98pi,cit:GRV94,cit:GRV98}, 
i.e.\ we refrain from generating radiatively massless `heavy' quark
densities $h^{\pi}(x,Q^2)$ where $h=c,\, b$, etc., in contrast to
\cite{cit:GRV92pi}.  Hence heavy quark contributions have to be calculated
in fixed order perturbation theory via, e.g., $g^{\pi}g^p\to h\bar{h},\,\,
\bar{u}\,^{\pi}u^p\to h\bar{h}$, etc.  (Nevertheless, rough estimates
of `heavy' quark effects, valid to within a factor of 2, say, can be
easier obtained with the help of the massless densities $c^{\pi}(x,Q^2)$
and $b^{\pi}(x,Q^2)$ given in \cite{cit:GRV92pi}.)

Using all these modified ingredients together with the new updated 
\cite{cit:GRV98} $f^p(x,\mu^2)$ in our basic predictions in Eq.\ (\ref{pieqn5}), the
present reanalysis of the available Drell--Yan data 
\cite{cit:NA10-85,cit:NA10-87,cit:E615-89},
closely following the procedure described in \cite{cit:GRS98pi}, yields
\begin{eqnarray}
v_{\rm{LO}}^{\pi}(x,\mu_{\rm{LO}}^2) & = & 1.129x^{-0.496}(1-x)^{0.349}
   (1+0.153\sqrt{x})
\label{pieqn10}\\
v_{\rm{NLO}}^{\pi}(x,\mu_{\rm{NLO}}^2) & = & 1.391x^{-0.447}(1-x)^{0.426}
\label{pieqn11}
\end{eqnarray}
where 
\cite{cit:GRV98} $\mu_{\rm{LO}}^2=0.26$ GeV$^2$ and $\mu_{\rm{NLO}}^2=0.40$
GeV$^2$.  These updated input valence densities correspond to total
momentum fractions
\begin{eqnarray}
\int_0^1x\, v_{\rm{LO}}^{\pi}(x,\mu_{\rm{LO}}^2)dx & = & 0.563
\label{pieqn12}\\
\int_0^1x\, v_{\rm{NLO}}^{\pi}(x,\mu_{\rm{NLO}}^2)dx & = & 0.559
\label{pieqn13}
\end{eqnarray}
as dictated by the valence densities of the proton \cite{cit:GRV98} via Eq.\ (\ref{pieqn7}).
\begin{figure}
\centering
\vspace*{-2.5cm}
\epsfig{figure=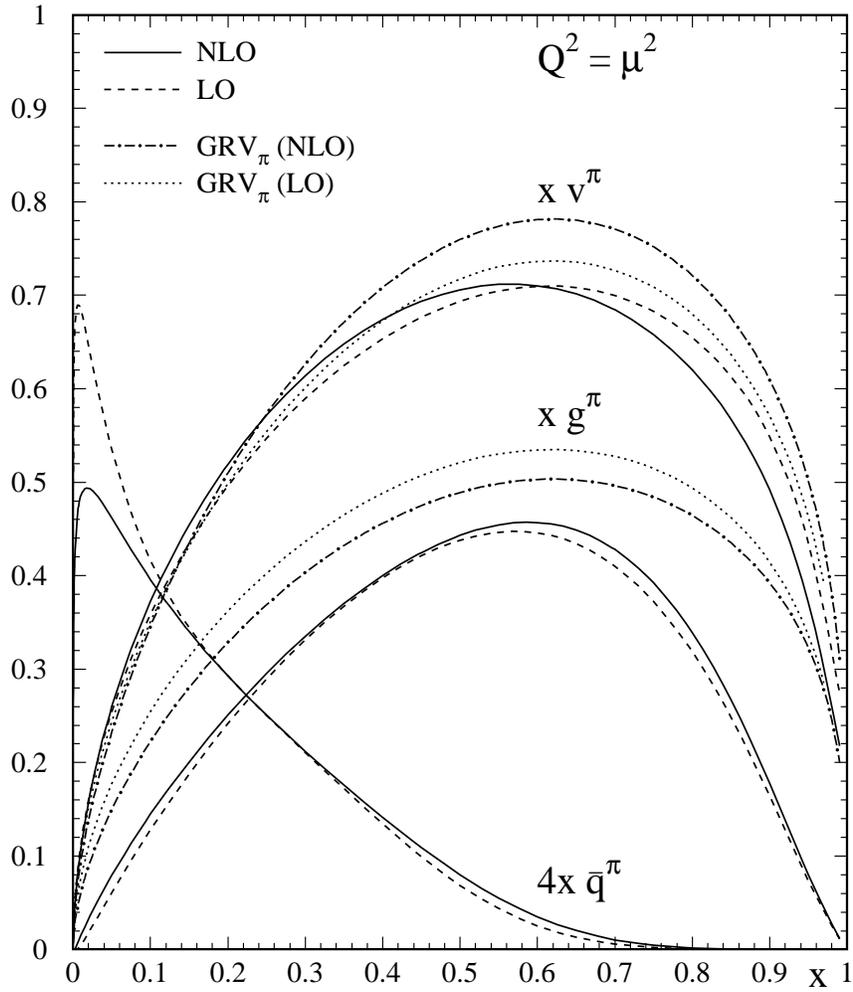,width=12cm}
%
\caption{\sf  The valence and valence--like input distributions
        $xf^{\pi}(x,Q^2=\mu^2)$ with $f=v,\, \bar{q},\,g$ as compared 
        to those of GRV$_{\pi}$ \protect\cite{cit:GRV92pi}.  
        Notice that GRV$_{\pi}$
        employs a vanishing SU(3)$_{\rm{flavor}}$ symmetric $\bar{q}\,^{\pi}$
        input at $\mu_{\rm{LO}}^2=0.25$ GeV$^2$ and 
        $\mu_{\rm{NLO}}^2=0.3$ GeV$^2$ \protect\cite{cit:GRV92pi}.  
        Our present 
        SU(3)$_{\rm{flavor}}$ broken sea densities refer to a vanishing 
        $s^{\pi}$ input in (\ref{pieqn3}), as for GRV$_{\pi}$ 
        \protect\cite{cit:GRV92pi}.}  
\label{pifig1}
\end{figure}

\begin{figure}
\centering
\vspace*{-1.5cm}
\epsfig{figure=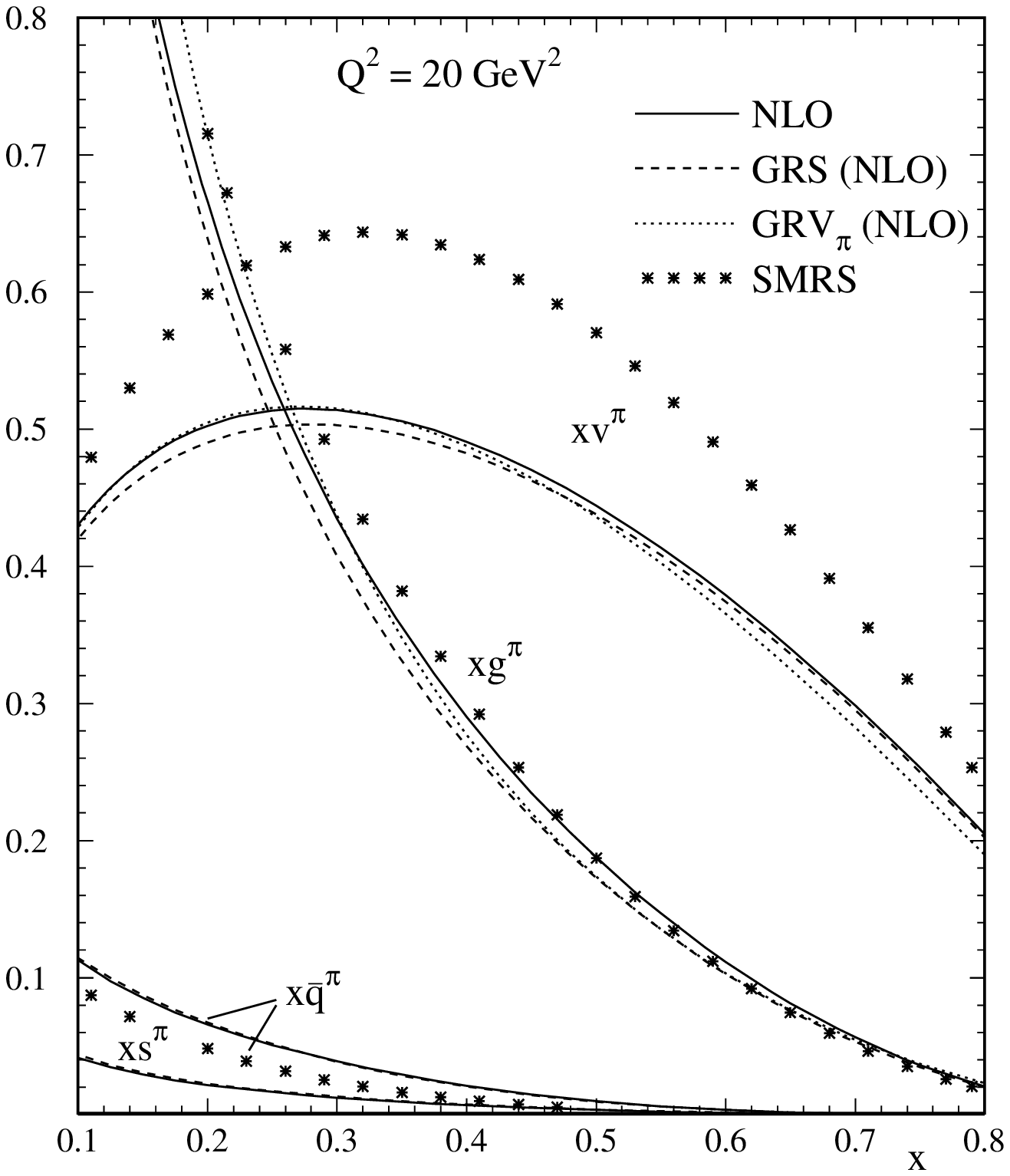,width=12cm}
%
\caption{\sf Comparison of our NLO valence distribution at 
        $Q^2=20$ GeV$^2$ with the one of GRV$_{\pi}$ 
        \protect\cite{cit:GRV92pi} and
        GRS \protect\cite{cit:GRS98pi}.  
        This density plays the dominant role for 
        describing presently available $\pi N$ Drell--Yan dimuon production
        data.  For illustration, the gluon and sea densities are shown as
        well.  The SU(3)$_{\rm{flavor}}$ symmetric GRV$_{\pi}$ sea
        $\bar{q}\,^{\pi}=s^{\pi}$ is not shown, since it is similar to
        $s^{\pi}$ of our present analysis and of GRS which are all
        generated from a vanishing input at 
        $Q^2=\mu^2$, cf.\ Eq.\ (\ref{pieqn3}).
        The SMRS \protect\cite{cit:SMRS92pi} results refer also to a 
        SU(3)$_{\rm{flavor}}$
        symmetric sea $\bar{q}\,^{\pi}\equiv \bar{u}\,^{\pi^+}=d^{\pi^+}
        =s^{\pi}=\bar{s}\,^{\pi}$.}
\label{pifig2}
\end{figure}
Our new updated input distributions in Eqs.\ (\ref{pieqn10}), (\ref{pieqn11}) and 
(\ref{pieqn5}) are rather
different than the original GRV$_{\pi}$ input \cite{cit:GRV92pi} in 
Fig.\ \ref{pifig1} which 
is mainly due to the vanishing sea input of GRV$_{\pi}$ in contrast to the
present one in Eq.\ (\ref{pieqn5}).  
On the other hand, our updated input in Fig.\ \ref{pifig1}
is, as expected, rather similar to the one of \cite{cit:GRS98pi}.  In both cases,
however, the valence and gluon distributions become practically 
indistinguishable from our present updated ones at scales relevant for
present Drell--Yan dimuon and direct--$\gamma$ production data, 
$Q^2\equiv M_{\mu^+\mu^-}^2\simeq20$ GeV$^2$, as illustrated in Fig.\ \ref{pifig2}.
Therefore our present updated pionic distributions give an equally good
description of all available $\pi N$ Drell--Yan data as the ones shown
in \cite{cit:GRS98pi}. Notice that the different gluon distributions presented 
in Fig.\ \ref{pifig2} can {\em not} be discriminated by present direct--photon
production data \cite{cit:NA24-87,cit:WA70-88,cit:E706-98} due to the 
uncertainty of the theoretically
calculated cross section arising from variations of the chosen 
factorization scale and from possible intrinsic $k_T$ contributions, cf.\
for example L.\ Apanasevich et al.\ \cite{cit:NA24-87,cit:WA70-88,cit:E706-98}.

For completeness let us mention that our basic predictions (\ref{pieqn5}) 
for the valence--like gluon and sea densities at $Q^2=\mu^2$, as shown 
in Fig.\ \ref{pifig1}, can be simply parametrized in Bjorken--$x$ space :
in LO at $Q^2=\mu_{\rm{LO}}^2=0.26$ GeV$^2$
\begin{eqnarray}
x\, g^{\pi}(x,\mu_{\rm{LO}}^2) & = & 7.326\,x^{1.433} (1-1.919\, 
       \sqrt{x}+1.524\,x)(1-x)^{1.326}\nonumber\\
x\, \bar{q}\,^{\pi}(x,\mu_{\rm{LO}}^2) & = &  0.522\,x^{0.160} 
   (1- 3.243\, \sqrt{x} + 5.206\, x) (1-x)^{5.20}\, ,
\label{pieqn14}
\end{eqnarray}
whereas in NLO at $Q^2=\mu_{\rm{NLO}}^2 = 0.40$ GeV$^2$ we get
\begin{eqnarray}
x\, g^{\pi}(x,\mu_{\rm{NLO}}^2) & = & 5.90\, x^{1.270}(1-2.074\,\sqrt{x}\,
     +1.824\, x)(1-x)^{1.290}\nonumber\\
x\, \bar{q}\,^{\pi}(x,\mu_{\rm{NLO}}^2) & = & 0.417\, x^{0.207}(1-2.466\,
     \sqrt{x}\, +3.855\, x)(1-x)^{4.454} .
\label{pieqn15}
\end{eqnarray}

\begin{figure}[t]
\centering
\epsfig{figure=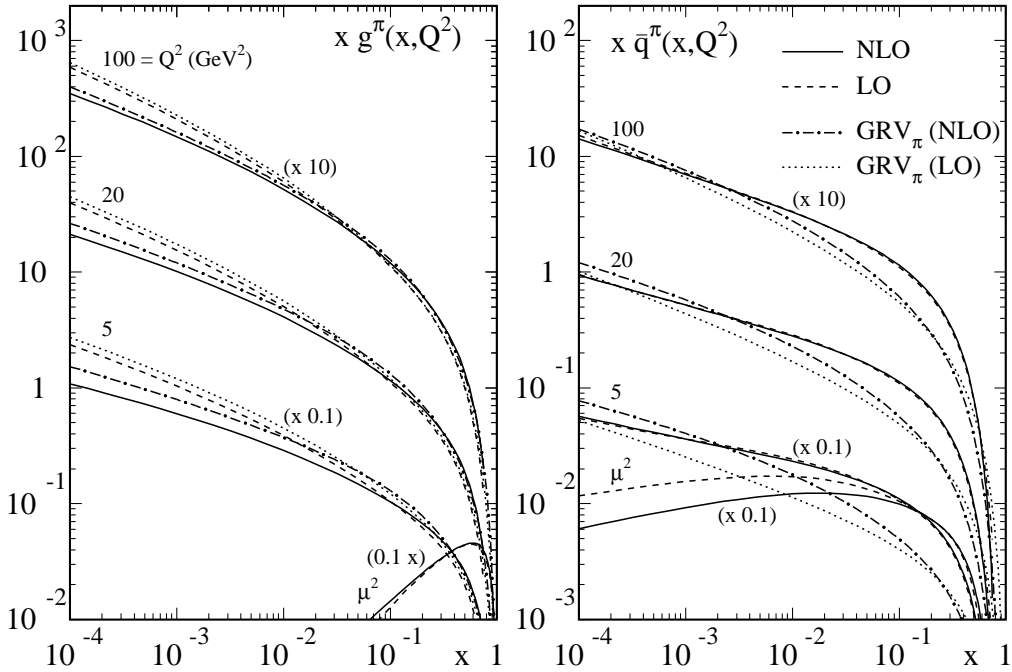,width=14cm}
\caption{\sf The small--$x$ predictions of our radiatively
        generated pionic gluon and sea--quark distributions in LO and NLO
        at various fixed values of $Q^2$ as compared to those of 
        GRV$_{\pi}$ \protect\cite{cit:GRV92pi}.   
        The valence--like inputs, according to
        Eq.\ (\ref{pieqn5}) as presented in Fig.\ \ref{pifig1}, are shown for 
        illustration by
        the lowest curves referring to $\mu^2$.  The predictions for
        the strange sea density $s^{\pi}=\bar{s}\,^{\pi}$ are similar to
        the GRV$_{\pi}$ results for $\bar{q}\,^{\pi}$.  The results are
        multiplied by the numbers indicated in brackets.}
\label{pifig3}
\end{figure}
Finally, Fig.\ \ref{pifig3} shows our resulting predictions for $x\, g^{\pi}(x,Q^2)$
and $x\,\bar{q}\,^{\pi}(x,Q^2)$ as compared to the former GRV$_{\pi}$ 
results \cite{cit:GRV92pi}.  The GRV$_{\pi}$ results for $x\, \bar{q}\,^{\pi}$
are significantly steeper and softer for $x$ \raisebox{-0.1cm}{$\stackrel{>}
{\sim}$} 0.01
due to the vanishing SU(3)$_{\rm{flavor}}$ symmetric (light) sea input
$x\,\bar{q}\,^{\pi}(x,\mu^2)=0$, in contrast to our present approach
\cite{cit:GRS98pi} based on a more realistic finite light sea 
input in Eq.\ (\ref{pieqn5}).
The valence--like gluon and sea inputs at $Q^2=\mu^2$, which become
(vanishingly) small at $x<10^{-2}$, are also shown in Fig.\ \ref{pifig3}.  This
illustrates again the purely dynamical origin of the small--$x$ structure
of gluon and sea quark densities at $Q^2>\mu^2$.  Our predictions for
$s^{\pi}=\bar{s}\,^{\pi}$, as evolved from the vanishing input in Eq.\ (\ref{pieqn3}),
are not shown in the figure since they practically coincide with
$\bar{q}\,^{\pi}(x,Q^2)$ of GRV$_{\pi}$ shown in Fig.\ \ref{pifig3} which also results
from a vanishing input \cite{cit:GRV92pi}. Simple analytic parametrizations
of our LO and NLO predictions for $f^{\pi}(x,Q^2)$ are given in 
Appendix \ref{app_para}.

To conclude let us recall that an improvement of $f^{\pi}(x,Q^2)$ is
particularly important in view of its central role in the construction
of the photon structure function and the photonic parton distributions
\cite{cit:GRV-9201,cit:GRV-9202,cit:GS-9201,*cit:GS-9701,
cit:AFG-9401,cit:SaS-9501,*Schuler:1996fc,cit:Vogt97}.  
Furthermore, recent (large rapidity
gap) measurements of leading proton and neutron production in deep inelastic
scattering at HERA \cite{cit:H1-99} allow, under certain (diffractive) model
assumptions, to constrain and test the pion structure functions for the
first time at far smaller vales of $x$ (down to about $10^{-3}$) than
those attained from fixed target $\pi N$ experiments.

%% file: phd_gam_photon.tex
\chapter{Radiatively Generated Parton Distributions of Real and Virtual Photons}
\label{chap:bc}
Having presented the general theoretical framework 
in Chapter \ref{chap:partonmodel}
we now turn to an analysis
of the partonic content of (virtual) photons within the 
phenomenologically successful framework
of the radiative parton model 
\cite{cit:GRV90,Gluck:1992ng,Gluck:1993im,cit:GRV94,cit:GRV98}.
This analysis follows Ref.~\cite{cit:GRSc99}.
The results in Sec.~\ref{sec:smallx} have been taken from \cite{cit:smallx}.
\section{Introduction}\label{sec:photonintro}
Modern theoretical QCD studies 
\cite{cit:GRV-9201,cit:GRV-9202,cit:GS-9201,*cit:GS-9701,Hagiwara:1995ag,
cit:AFG-9401} of the parton 
distributions of {\em{real}}, i.e.\ on--shell, photons 
$f^{\gamma}(x,Q^2),\,\, f=q,\, \bar{q},\, g$, agree surprisingly well 
with measurements of the (anti--)quark and gluon contents of the resolved 
real photons as obtained from $e^+e^-$ and $ep$ reactions at collider 
energies (for recent reviews, see \cite{Erdmann-9701,cit:Vogt97,gamref6}).  
For clarity
let us denote the resolved real target photon with virtuality 
$P^2\equiv -p^2\simeq0$ by $\gamma\equiv\gamma(P^2\simeq 0)$ which is
probed by the virtual probe photon $\gamma^*(Q^2),\, Q^2\equiv -q^2$,
via the subprocess $\gamma^*(Q^2)\gamma\to X$ as in $e^+e^-\to e^+e^-X$.
Here, $p$ denotes the four--momentum of the photon emitted from, say,
an electron in an $e^+e^-$ or $ep$ collider.  In the latter case it is
common to use $Q^2$ instead of $P^2$ for denoting the photon's virtuality,
but we prefer $P^2$ for the subprocess $\gamma(P^2)p\to X$ according
to the original notation used in $e^+e^-$ annihilations.  (Thus the
factorization scale in $f^{\gamma(P^2)}(x,Q^2)$ refers now to some
properly chosen scale of the produced hadronic system $X$, e.g.\ 
$Q\sim p_T^{\rm{jet}}$ in high--$p_T$ jet events, etc.).

In general one expects 
\cite{Uematsu:1981qy,Uematsu:1982je,Ibes:1990pj,
Rossi:1984xz,cit:Rossi-PhD,Borzumati:1993za,cit:GRS95,
cit:SaS-9501,*Schuler:1996fc,Drees:1994eu} 
also a {\em{virtual}}
photon $\gamma(P^2\neq 0)$ to possess a parton content $f^{\gamma(P^2)}
(x,Q^2)$.  It is a major problem to formulate a consistent set of
boundary conditions which allow for a calculation of $f^{\gamma(P^2)}
(x,Q^2)$ also in the next--to--leading order (NLO) of QCD
{\em{as well as}} for a smooth transition to $P^2=0$,\, i.e.\
to the parton distributions of a real photon (see Refs.\ {\cite{cit:GRS95,
cit:SaS-9501,*Schuler:1996fc,Stratmann:1998wx} for a detailed discussion).  
Indeed, experimental studies
of the transition of the deep inelastic (di--)jet cross section from
the real photon to the virtual photon region at HERA point to the
existence of a nonvanishing, though suppressed, parton content for
virtual photons \cite{ZEUS-IEC95,Adloff:1997nd,Adloff:1998st,H1-ICHEP98,Adloff:1998vc}.  
These measurements have triggered
various analyses of the dependence of the $ep$ jet production cross
section on the virtuality of the exchanged photon 
\cite{Gluck:1996ra,deFlorian:1996uk,*Chyla:1996ux}
and experimental tests of such predictions will elucidate the so far 
unanswered question as to when a deep inelastic scattering (DIS) $ep$ 
process is eventually dominated by the usual `direct' $\gamma^*\equiv
\gamma(P^2)$ induced cross sections, {\em{not}} contaminated
by the so far poorly known {\em{resolved}} virtual photon
contributions.  More recently, NLO calculations of the (di--)jet
rate in $ep$ (and $e\gamma$) scattering, which properly include the
contributions of resolved virtual photons, have become available
\cite{Klasen:1998jm,*Kramer:1998bc,*Potter:1998jt} and the 
resolved virtual photon contributions have
already been included in a Monte Carlo event generator \cite{Jung:1995gf}
as well.

It is the main purpose of the present chapter to formulate a consistent
set of boundary conditions, utilizing valence--like input parton
distributions at the universal target--mass independent 
\cite{cit:GRV-9201,cit:GRV-9202,Gluck:1992ng,Gluck:1993im,cit:GRV94,cit:GRV92pi} 
low resolution scale $Q_0^2=\mu^2\simeq 0.3$
GeV$^2$, which allow for a calculation of $f^{\gamma(P^2)}(x,Q^2)$ 
also in NLO--QCD as well as for a smooth transition to the parton
distributions of a real photon, $P^2=0$.  
We shall furthermore employ
the pionic parton distributions of Chapter \ref{pipdf},
$f^{\pi}(x,Q^2)$, 
which are required for describing, via vector meson
dominance (VMD), the hadronic components of the photon.  
It should be
recalled that the pionic gluon and sea densities, $g^{\pi}(x,Q^2)$ and
$\bar{q}\,^{\pi}(x,Q^2)$, have been uniquely derived 
from the
experimentally rather well known pionic valence density $v^{\pi}(x,Q^2)$
and the (recently updated \cite{cit:GRV98} dynamical) parton distributions
$f(x,Q^2)$ of the proton.  Thus we arrive at essentially 
parameter--{\em{free}} predictions for $f^{\gamma(P^2)}(x,Q^2)$
which are furthermore in good agreement with all present measurements
of the structure function of real photons, $F_2^{\gamma}(x,Q^2)$.

Sec.~\ref{sec:photoninput} is devoted to an analysis
of the parton distributions and structure functions of real photons and the
resulting predictions are compared with recent experiments. 
In Sec.~\ref{sec:smallx} we compare our
unique small--$x$ predictions --characteristic for the radiative parton model--
with very recent small--$x$ measurements of the photon structure
function $F_2^{\gamma}(x,Q^2)$ by the LEP--OPAL collaboration.  
Sec.~\ref{sec:vgaminput} contains the formulation of our model
for the parton distributions of virtual photons, together with some
quantitative predictions for structure functions as well as a comparison
with very recent data extracted from DIS dijet events.  Our conclusions  
are drawn in Sec.~\ref{sec:gamsum}.  
In App.~\ref{app_para} we present simple analytic
parametrizations of our LO and NLO predictions for the parton 
distributions of real and virtual photons.
\section{The Parton Content of Real Photons}
\label{sec:photoninput}
\subsection{Boundary Conditions}
The pionic parton distributions presented in the previous chapter
will now be utilized to construct LO and NLO input distributions
for the real photon via a vector meson dominance ansatz
at the low resolution scale $Q_0^2 = \mu^2 \simeq 0.3\ \gevsq$.

In NLO
the nonperturbative hadronic VMD input $f^{\gamma}(x,Q_0^2)$
refers to the partons in the DIS$_{\gamma}$ scheme which guarantees
the perturbative stability of the resulting $F_2^{\gamma}(x,Q^2)$ provided
this input is given by the NLO $f^{\pi}(x,Q_0^2)$ of Chap.~\ref{pipdf}, while
the corresponding input in LO is given via VMD by the LO  
$f^{\pi}(x,Q_0^2)$ of Chap.~\ref{pipdf}.
We shall assume that the input resolution scale 
$Q_0^2=\mu^2\simeq 0.3$ GeV$^2$ for the valence--like parton structure
is universal, i.e. independent of the mass of the considered targets
$p,\, \pi, \, \gamma$, etc. 
\cite{cit:GRV-9201,cit:GRV-9202,Gluck:1992ng,Gluck:1993im,cit:GRV94,
cit:GRV92pi,cit:GRS98pi}.  

The hadronic VMD ansatz for $\fv(x,\mu^2)$ is based on a 
coherent superposition of vector mesons \cite{cit:SaS-9501,*Schuler:1996fc}
\begin{equation}\label{eq:VMDansatz}
|\gamma\rangle_{\mu^2,\,\had}\simeq \frac{e}{f_\rho}\, 
  |\rho\rangle_{\mu^2} + \frac{e}f_\omega\, |\omega\rangle_{\mu^2}
\end{equation} 
where the $\phi$--meson contribution is considered to be strongly 
suppressed at $\mu^2\ll m_{\phi}^2$.  
As usual, we identify the parton distribution functions in the parton model 
with matrix elements of
local twist--2 operators within the operator product expansion (OPE) formalism
and we obtain using \eqref{eq:VMDansatz}
\begin{align}
\fv(\mu^2) & = \fv_{\had}(\mu^2)
            = \langle\gamma | \Op[f] | \gamma\rangle_{\mu^2,\had}
\nonumber\\*
&= \frac{e^2}{f_\rho^2} \langle\rho | \Op[f] | \rho\rangle_{\mu^2}
+\frac{e^2}{f_\rho f_\omega} ( \langle\rho | \Op[f] | \omega\rangle_{\mu^2}
                              +\langle\omega | \Op[f] | \rho\rangle_{\mu^2})
+\frac{e^2}{f_\omega^2} \langle\omega | \Op[f] | \omega\rangle_{\mu^2}
\nonumber
\end{align}
with $f = u,d,s,g$.
Assuming, within the OPE,
\begin{align}
\langle\rho|\Op[q]|\rho\rangle_{\mu^2} & 
=  \langle\omega|\Op[q]|\omega\rangle_{\mu^2} 
= \langle\pi^0|\Op[q]|\pi^0\rangle_{\mu^2}\ , \qquad
\langle\rho|\Op[q]|\omega\rangle_{\mu^2} 
=2\, I_{3q}\, e^{-i\theta}  \langle\pi^0|\Op[q]|\pi^0\rangle_{\mu^2}
\nonumber\\*
\langle\rho|\Op[g]|\rho\rangle_{\mu^2} & =  \langle\omega|\Op[g]|
   \omega\rangle_{\mu^2} = \langle\pi^0|\Op[g]|\pi^0\rangle_{\mu^2}\ , \qquad
\langle\rho|\Op[g]|\omega\rangle=0
\nonumber
\end{align}
where the last equation holds 
due to isospin conservation, 
one obtains
\begin{eqnarray}\label{eq:gam9}
(u+\bar{u})^{\gamma}(x,\mu^2) & = & \alpha(g_{\rho}^2 + g_{\omega}^2 +
  2\,g_{\rho}g_{\omega}\cos\theta)(u+\bar{u})^{\pi^0}(x,\mu^2)\nonumber\\
(d+\bar{d})^{\gamma}(x,\mu^2) & = & \alpha(g_{\rho}^2 + g_{\omega}^2 -
  2\,g_{\rho}g_{\omega}\cos\theta)(d+\bar{d})^{\pi^0}(x,\mu^2)\nonumber\\
(s+\bar{s})^{\gamma}(x,\mu^2) & = & \alpha(g_{\rho}^2 + g_{\omega}^2)\,
    (s+\bar{s})^{\pi^0}(x,\mu^2) = 0\nonumber\\
g^{\gamma}(x,\mu^2) & = & \alpha(g_{\rho}^2 + g_{\omega}^2)\, g^{\pi^0}
   (x,\mu^2) .
\end{eqnarray}
Here $\Op[q,g]$ refer to the leading twist--2 quark and gluon 
operators 
and $g_V^2\equiv 4\pi/f_V^2$ with
\begin{equation}
g_{\rho}^2 = 0.50\, , \quad\quad g_{\omega}^2 = 0.043\, ,
\end{equation}
i.e.\ $f_{\rho}^2/4\pi = 2.0$ and $f_{\omega}^2/4\pi =23.26$, as obtained 
from a zero--width calculation of the relevant leptonic widths
$\Gamma(V\to\ell^+\ell^-)=\alpha^2m_V g_V^2/3$ presented in \cite{Caso:1998tx}.
The omission of a finite--width correction for $g_{\rho}^2$ is due to
the central role \cite{Caso:1998tx} of the precise results in 
\cite{Antipov:1989nk} which
do not require such a correction in contrast to the situation for the
less precise resonance analysis at $e^+e^-$ colliders 
\cite{cit:Berger-Rep,Kolanoski:1987sn,Ioffe:1985ep}.

For the a priori unknown coherence factor (fit parameter) $\cos\theta$
in Eq.\ (\ref{eq:gam9}) we take $\cos\theta=1$, i.e.\ we favor a superposition of
$u$ and $d$ quarks  which maximally enhances the contributions of the
up--quarks to $F_2^{\gamma}$ in Eq.\ (\ref{eq:totf2gam}).  This favored value for 
$\cos\theta$ is also supported by fitting $\cos\theta$ in (\ref{eq:gam9}) to all
presently available data on $F_2^{\gamma}(x,Q^2)$, to be discussed 
below, which always resulted in $\cos\theta\simeq 1$ in LO as
well as NLO.  This is also in agreement with the LO results obtained
in Ref.~\cite{cit:SaS-9501,*Schuler:1996fc}.  
The LO and NLO input distributions $f^{\pi}(x,\mu^2)$ 
of the pion in (\ref{eq:gam9}) are taken from the analysis in the previous chapter
which correspond to 
\cite{cit:GRSc99pi,cit:GRV98} $\mu_{\rm{LO}}^2 = 0.26$ GeV$^2$ 
and $\mu_{\rm{NLO}}^2=0.40$ GeV$^2$ in LO and NLO, 
respectively.  Since by now all free input quantities have been fixed
in Eq.\ (\ref{eq:gam9}), we arrive at rather unique parameter--{\em{free}}
predictions for $f^{\gamma}(x,Q^2)$ and $F_2^{\gamma}(x,Q^2)$.

The calculation of $f^{\gamma}(x,Q^2)$ at $Q^2>\mu^2$ follows from the
well known inhomogeneous renormalization group (RG) evolution equations 
in LO and NLO 
which we solve, as usual, analytically
for the $n$--th Mellin moment of $f^{\gamma}(x,Q^2)$, followed by a 
straightforward Mellin--inversion to Bjorken--$x$ space
as described in detail in Chapter \ref{chap:partonmodel}.
The general structure of these solutions is 
\begin{equation}\label{eq:gam11}
f^{\gamma}(x,Q^2) = f_{\pl}^{\gamma}(x,Q^2) + f_{\had}^{\gamma} 
  (x,Q^2)\, .
\end{equation}
Here $f_{\pl}^{\gamma}$ denotes the perturbative `pointlike' solution
which vanishes at $Q^2=\mu^2$ and is driven by the pointlike photon
splitting functions $k_{q,g}^{(0,1)}(x)$ appearing in the inhomogeneous
evolution equations, while $f_{\had}^{\gamma}$ depends on the hadronic
input $f^{\gamma}(x,\mu^2)$ in Eq.\ (\ref{eq:gam9}) and evolves according to the
standard homogeneous evolution equations.  

The prescription for the VMD ansatz in Eq.\ (\ref{eq:gam9}) at the input scale
$\mu^2$, together with $\cos\theta=1$ as discussed above, yields a
simple expression for the general $Q^2$--dependence of $f^{\gamma}(x,Q^2)$:
\begin{equation}\label{eq:gam13}
f^{\gamma}(x,Q^2) = f_{\pl}^{\gamma}(x,Q^2)+\alpha\, \left[G_f^2\, f^{\pi}(x,Q^2)+
\delta_f\, \frac{1}{2}(G_u^2-G_d^2)\, s^{\pi}(x,Q^2)\right]
\end{equation}
with $\delta_u=-1$, $\delta_d=+1$ and $\delta_s=\delta_g=0$, and
where the index $\pi$ obviously refers to $\pi^0$ and 
\begin{eqnarray}\label{eq:gam14}
G_u^2 & = & (g_{\rho}+g_{\omega})^2 \simeq 0.836
\nonumber\\
G_d^2 & = & (g_{\rho}-g_{\omega})^2 \simeq 0.250
\nonumber\\
G_s^2 & = & G_g^2 = g_{\rho}^2 + g_{\omega}^2 = 0.543\, .
\end{eqnarray}
Simple analytic LO and NLO($\disg$) parametrizations for the
pointlike piece $f_{\pl}^{\gamma}(x,Q^2)$ are given in App.~\ref{sec:gampara},
whereas the ones for $f^{\pi}(x,Q^2)$ can be found in App.~\ref{sec:pipara}.  
\subsection{Quantitative Results}
\afterpage{\clearpage}
\begin{figure}[H]
\centering
\epsfig{figure=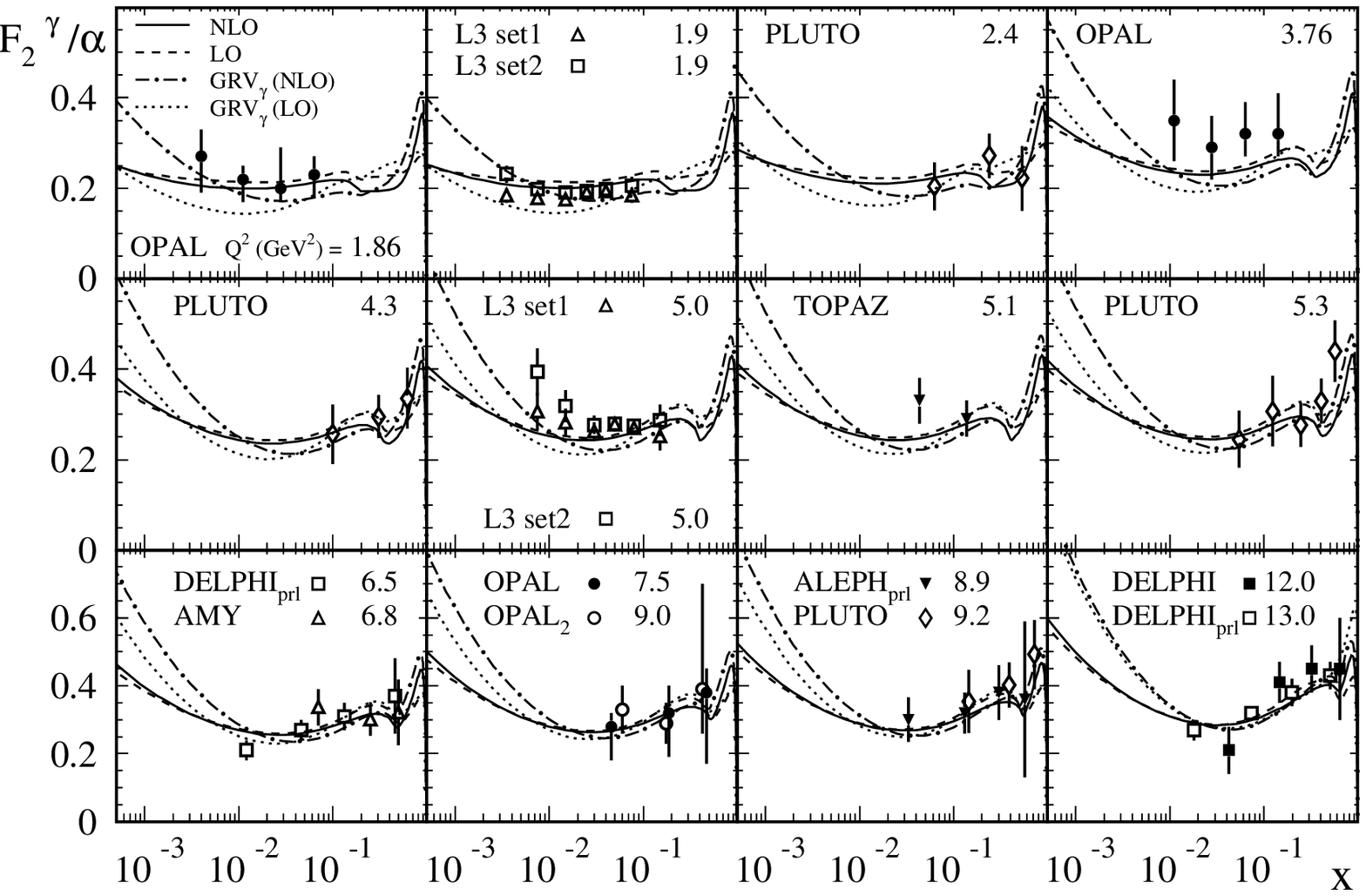,width=13.0cm}
\epsfig{figure=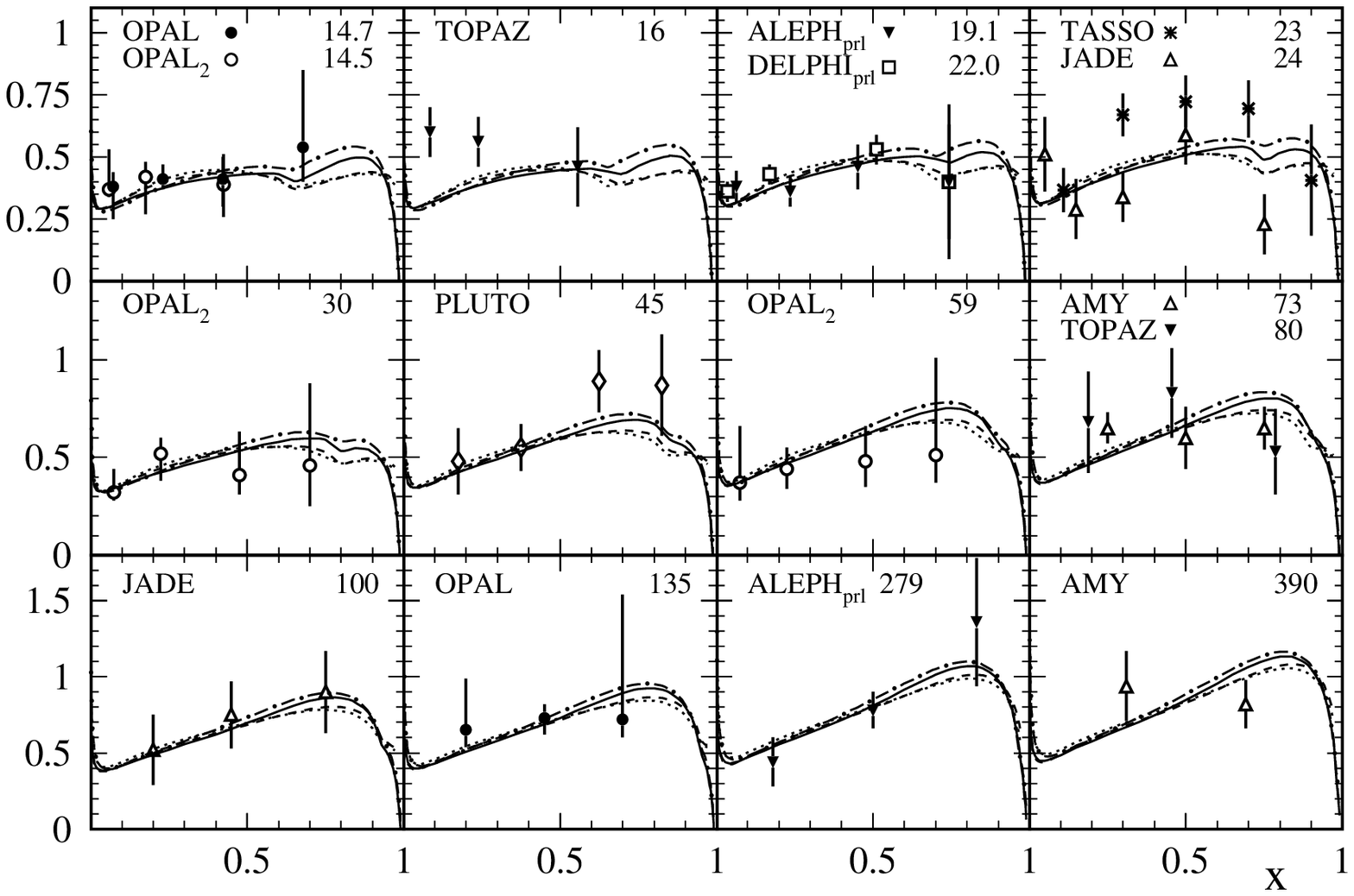,width=13.0cm}
\caption{\sf Comparison of our radiatively generated LO and 
  NLO($\disg$) predictions for $F_2^{\gamma}(x,Q^2)$, based 
  on the valence--like parameter--free VMD input in Eq.\ (\ref{eq:gam9}), with
  the data of Ref.~\protect\cite{Berger:1984xt,*Berger:1987ke,
Bartel:1984cg,Althoff:1986fi,Sahu:1995gj,
Kojima:1997wg,Muramatsu:1994rq,Ackerstaff:1996se,*Ackerstaff:1997ni,
*Ackerstaff:1997ng,Abreu:1996xt,*DELPHI:HEP97,ALEPH:HEP97,*ALEPH:Photon97,
*ALEPH:ICHEP98,Acciarri:1998ig}. 
  For our comparison the GRV$_{\gamma}$ \protect\cite{cit:GRV-9201,cit:GRV-9202} 
  results are shown as well. In both cases, the charm contribution has been 
  added, in the relevant kinematic region $W\geq 2m_c$, according to 
  Eqs.\ (\ref{eq:f2gamh}) and (\ref{eq:f2gamhres}).}  
\label{fig:gamfig1}
\end{figure}
Having outlined the theoretical basis for our photonic parton 
distributions, we now turn to the quantitative results.
First we apply our parameter--free predictions for $f^{\gamma}(x,Q^2)$
to the structure function of real photons which, according to
Eq.\ (\ref{eq:totf2gam}), is finally given by
\begin{eqnarray}\label{eq:gam17}
\frac{1}{x}\, F_2^{\gamma}(x,Q^2) & = & 2\sum_{q=u,d,s} e_q^2 \left\{ 
  q^{\gamma}(x,Q^2)+\,\frac{\alpha_s(Q^2)}{2\pi}\, \left[ \cq[2] \otimes 
    q^{\gamma} + \cg[2] \otimes g^{\gamma} \right] 
\right\}
\nonumber\\
& &  
+\, \frac{1}{x}\, F_{2,c}^{\gamma}(x,Q^2) + \, \frac{1}{x}\, 
     F_{2,c}^{g{^\gamma}}(x,Q^2) 
\end{eqnarray}
where $f^{\gamma}(x,Q^2)$ refers to the $\disg$ factorization
scheme defined in (\ref{eq:schemetrafo1}) and the charm contributions $F_{2,c}^{\gamma}$
and $F_{2,c}^{g^{\gamma}}$ are given by Eqs.\ (\ref{eq:f2gamh}) and 
(\ref{eq:f2gamhres}), respectively.

In Fig.\ \ref{fig:gamfig1} we compare our LO and NLO predictions 
with all available\footnote{January 1999}
relevant data 
\cite{Berger:1984xt,*Berger:1987ke,Bartel:1984cg,Althoff:1986fi,Sahu:1995gj,
Kojima:1997wg,Muramatsu:1994rq,Ackerstaff:1996se,*Ackerstaff:1997ni,
*Ackerstaff:1997ng,Abreu:1996xt,*DELPHI:HEP97,ALEPH:HEP97,*ALEPH:Photon97,
*ALEPH:ICHEP98,Acciarri:1998ig} 
for $F_2^{\gamma}$ of the real photon. 
Our present new NLO results are rather similar to the ones of AFG 
\cite{cit:AFG-9401},
but differ from the previous (GRV$_{\gamma}$) predictions \cite{cit:GRV-9201,cit:GRV-9202}
which are steeper in the small--$x$ region, as shown in Fig.\ \ref{fig:gamfig1}, because
the dominant hadronic (pionic) sea density $\bar{q}\,^{\pi}(x,Q^2)$ is
steeper since it has been generated purely dynamically from a vanishing
input at $Q^2=\mu^2$ \cite{cit:GRV-9201,cit:GRV-9202,cit:GRV92pi}.  
Similarly the SaS 1D \cite{cit:SaS-9501,*Schuler:1996fc}
expectations, for example, fall systematically below the data in the small
to medium $Q^2$ region around $Q^2\simeq 5$ GeV$^2$, partly due to a 
somewhat different treatment of the hadronic coherent VMD input as compared
to our results in Eqs.\ (\ref{eq:gam9}), (\ref{eq:gam13}), and (\ref{eq:gam14}).  
\begin{figure}[htb]
\centering
\epsfig{figure=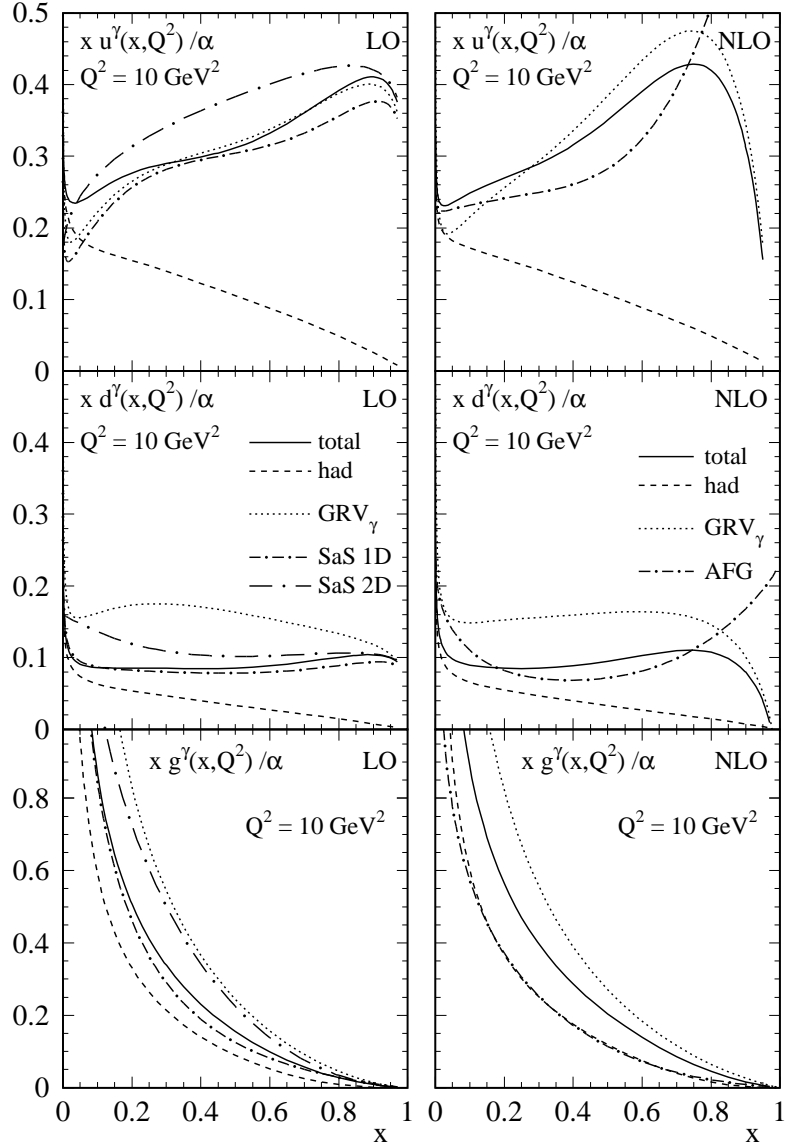,width=11.0cm}
\caption{\sf Comparison of our predicted LO and NLO($\disg$) distributions 
   $u^{\gamma}=\bar{u}\,^{\gamma},\,\, d^{\gamma}= \bar{d}\,^{\gamma}$ and 
   $g^{\gamma}$ at $Q^2=10$ GeV$^2$ with the LO/NLO GRV$_{\gamma}$ densities 
   \protect\cite{cit:GRV-9201,cit:GRV-9202}, the LO SaS 1D and 2D 
   \protect\cite{cit:SaS-9501,*Schuler:1996fc} and the NLO AFG 
   \protect\cite{cit:AFG-9401} distributions.
   The `hadronic' (pionic) components of our total LO and NLO
   results in Eq.\ (\ref{eq:gam11}) are displayed by the dashed curves.}
\label{fig:gamfig2}
\end{figure}

The relevant LO and NLO 
photonic parton densities are compared in Fig.\ \ref{fig:gamfig2} at $Q^2=10$ GeV$^2$.  For
illustration we also show the purely `hadronic' component (homogeneous
solution) in (\ref{eq:gam11}) of $f^{\gamma}$ which demonstrates the dominance of 
the `pointlike' component (inhomogeneous solution) in (\ref{eq:gam11}) for 
$u^{\gamma}$ and $d^{\gamma}$ in the large--$x$ region, $x>0.1\,$.  

\begin{figure}[thb]
\centering
\epsfig{figure=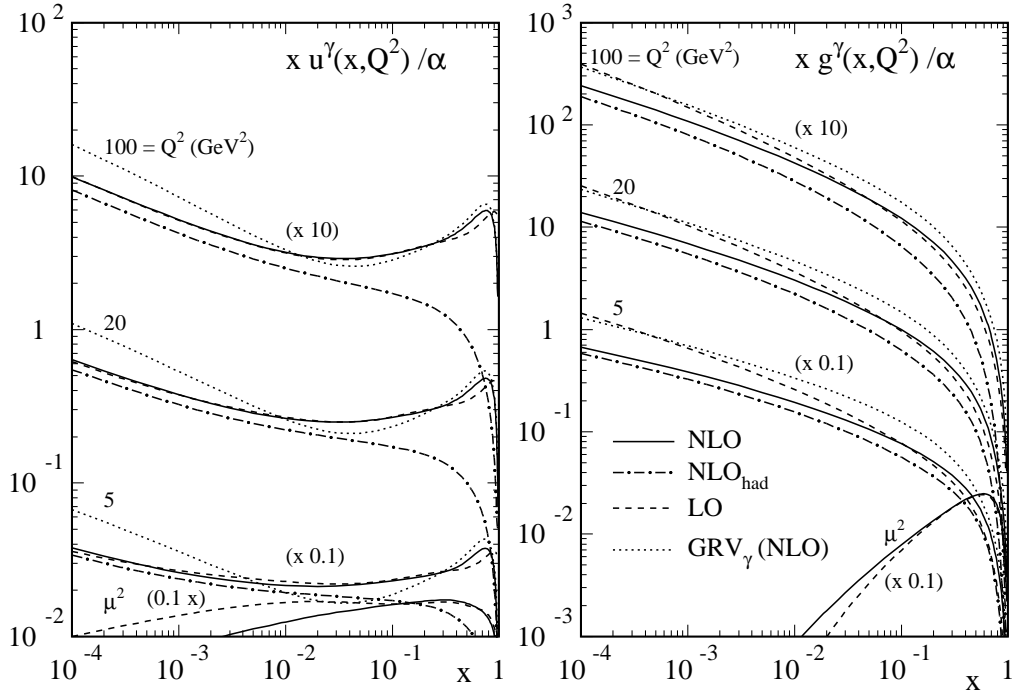,angle=0,width=14cm}
\caption{\sf Detailed small--$x$ (as well as large--$x$) 
   behavior and predictions of our radiatively generated 
   $u^{\gamma}=\bar{u}\,^{\gamma}$ and $g^{\gamma}$ distributions 
   in LO and NLO(DIS$_{\gamma}$) at fixed values of $Q^2$.  The 
   dashed--dotted curves show the hadronic NLO contribution 
   $f_{\had}^{\gamma}$ to $f^{\gamma}=f_{\pl}^{\gamma} +
   f_{\had}^{\gamma}$ in Eq.\ (\ref{eq:gam11}). The valence--like inputs
   at $Q^2=\mu_{\rm{LO,\,NLO}}^2$, according to Eq.\ (\ref{eq:gam9}), are
   shown by the lowest curves referring to $\mu^2$.  For comparison
   we show the steeper NLO GRV$_{\gamma}$ 
   \protect\cite{cit:GRV-9201,cit:GRV-9202} 
   expectations as well. The results have been multiplied by the number 
   indicated in brackets.}
\label{fig:gamfig3}
\end{figure}
\begin{figure}[htb]
\centering
\epsfig{figure=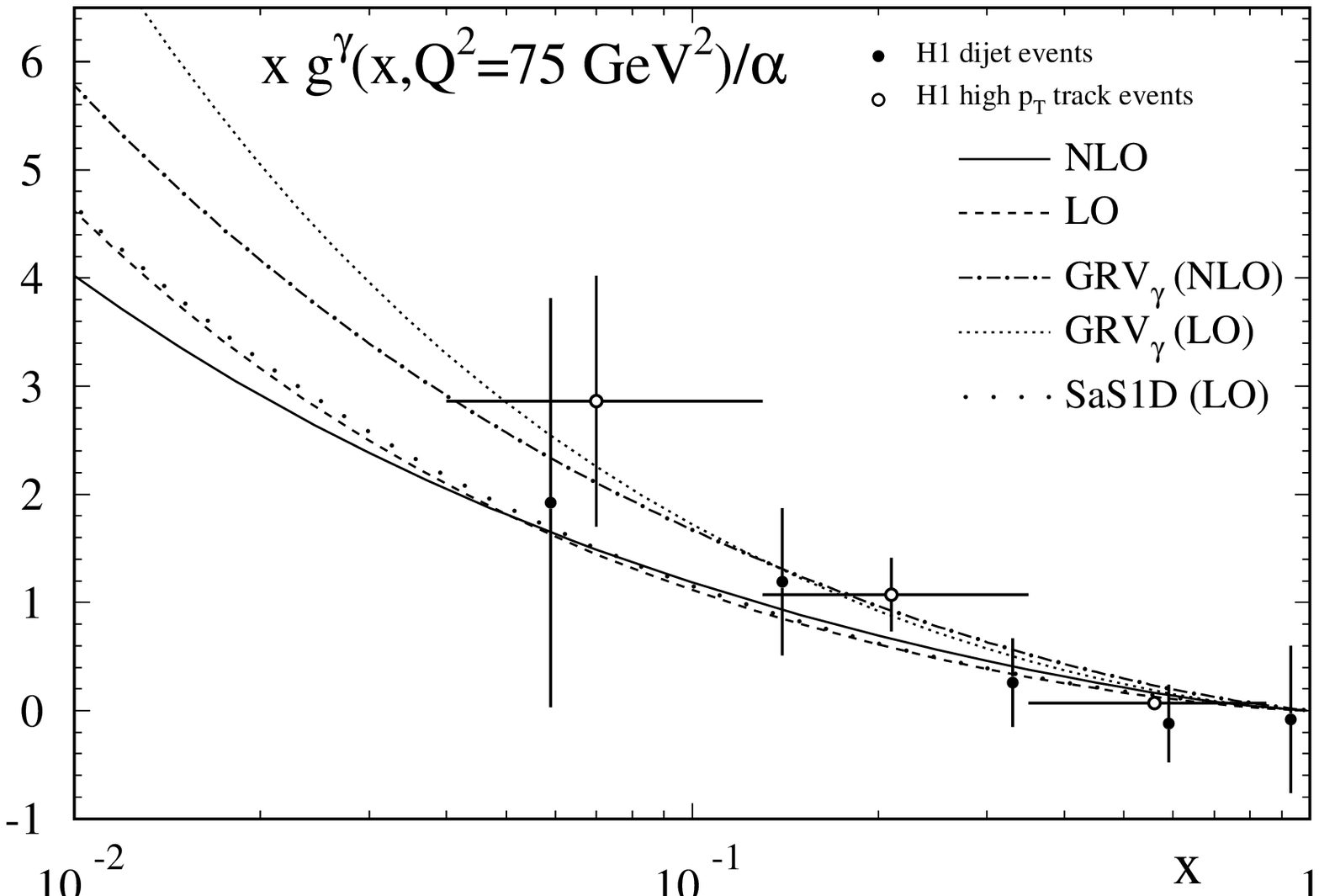,width=14cm}
\caption{\sf Comparison of our LO and NLO predictions for
   $xg^{\gamma}$ at $Q^2\equiv \langle (p_T^{\rm{jet}})^2 \rangle = 75$ GeV$^2$ 
   with HERA(H1) measurements \protect\cite{Ahmed:1995wv,Adloff:1998vt}.  
   The GRV$_{\gamma}$ and SaS expectations are taken from 
   Refs.\ \protect\cite{cit:GRV-9201,cit:GRV-9202} and 
   \protect\cite{cit:SaS-9501,*Schuler:1996fc}, respectively.}
\label{fig:gamfig4}
\end{figure}

In Fig.\ \ref{fig:gamfig3}  we show our predictions for $xu^{\gamma}(x,Q^2)$ and 
$xg^{\gamma}(x,Q^2)$.  The parton distributions of the photon behave,
in contrast to the ones of a hadron, very differently in the limits
of large and small $x$.  In the former case, the purely perturbative
pointlike part in (\ref{eq:gam11}) dominates for $x$ \raisebox{-0.1cm}
{$\stackrel{>}{\sim}$} 0.1, especially for the quark distributions.
On the other hand, this uniquely calculable contribution amounts at
most to about 20\% at very small $x$ where the hadronic VMD component
in (\ref{eq:gam11}) dominates, giving rise to a very similar increase for $x\to 0$
as observed in the proton case.  In Fig.\ \ref{fig:gamfig3} we also show our valence--like
inputs at $Q^2=\mu^2_{\rm{LO,\,NLO}}$ which become (vanishingly) small
at $x<10^{-2}$.  This illustrates the purely dynamical origin of the
small--$x$ increase at $Q^2>\mu^2$.  Also noteworthy is the perturbative
LO/NLO stability of $u^{\gamma}(x,Q^2)$ which is almost as good as the
one required for a physical quantity like $F_2^{\gamma}(x,Q^2)$ in Fig.\ \ref{fig:gamfig1}.
The situation is, as usual 
\cite{cit:GRV-9201,cit:GRV-9202,Gluck:1992ng,Gluck:1993im,cit:GRV94,cit:GRV98}, 
different for $g^{\gamma}(x,Q^2)$.  
Nevertheless, despite the sizable difference 
between the LO and NLO gluon distributions in Fig.\ \ref{fig:gamfig3} in the small--$x$
region, the directly measurable $F_2^{\gamma}$ and the gluon--dominated          
heavy quark contribution in Eq.\ (\ref{eq:f2gamhres}) shows a remarkable perturbative
stability \cite{cit:GRV98}.  

Next, we compare in Fig.\ \ref{fig:gamfig4} our predictions
for $xg^{\gamma}(x,Q^2)$ at $Q^2\equiv(p_T^{\rm{jet}})^2=75\ \gevsq$  
with recent HERA (H1) measurements \cite{Ahmed:1995wv,Adloff:1998vt}.  
Our somewhat flatter 
results for $xg^{\gamma}$ in the small--$x$ region, as compared to the
older GRV$_{\gamma}$ expectations \cite{cit:GRV-9201,cit:GRV-9202}, is caused by the recently
favored flatter gluon distribution in the proton \cite{cit:GRV98} which
determines $g^{\gamma}$ via $g^{\pi}$ \cite{cit:GRSc99pi}, \mbox{cf.\ Eq.\ (\ref{eq:gam9})},
at small $x$.

Finally it is interesting to consider the total momenta carried
by the photonic partons,
\begin{equation}
M_2^{\gamma}(Q^2)\equiv \sum_{f=q,\bar{q},g} \int_0^1x\, f^{\gamma}
  (x,Q^2)\, dx\, .
\end{equation}
Inspired by the ideas and suggestions put forward in Refs.\ 
\cite{Frankfurt:1995bd,*Frankfurt:1996nz,*Frankfurt:1996gy,cit:SaS-9501,*Schuler:1996fc}, 
it has been conjectured recently \cite{Gluck:1998fz} that
this leading twist--2 quantity $M_2^{\gamma}$ should satisfy, in 
LO-QCD,
\begin{equation}\label{eq:gam19}
M_2^{\gamma}(Q^2)\simeq \Pi_h(Q^2)
\end{equation}
where the well known dispersion relation relates the hadronic part of
the photon's vacuum polarization
\begin{equation}\label{eq:gam20}
\Pi_h(Q^2)=\frac{Q^2}{4\pi^2\alpha}\, \int_{4m_{\pi}^2}^{\infty}\,
   \frac{\sigma_h(s)}{s+Q^2}\, ds
\end{equation}
to $\sigma_h\equiv\sigma(e^+e^-\to$ hadrons).  It should be noted that
$\Pi_h(Q^2)$, being an experimental quantity, includes, besides the
usual twist--2 term, all possible nonperturbative higher--twist 
contributions.  The `consistency' relation (\ref{eq:gam19}) is, however, expected
to hold already at $Q^2$ \raisebox{-0.1cm}{$\stackrel{>}{\sim}$} 2 to
4 GeV$^2$ to within, say, 20 to 30\% 
where the twist--2 component
in $\Pi_h(Q^2)$ may become dominant, as possibly indicated by DIS $ep$
processes.  Indeed, our LO results imply 
$M_2^{\gamma}(2\,\,{\rm{GeV}}^2)/\alpha\simeq 0.976$ 
and $M_2^{\gamma}(4\,\,{\rm{GeV}}^2)/\alpha\simeq 1.123$
which compares favorably with \cite{Burkhardt:1995tt} $\Pi_h(2\,\,{\rm{GeV}}^2)/
\alpha=0.694\pm0.028$ and $\Pi_h(4\,\,{\rm{GeV}}^2)/\alpha=0.894\pm0.036$,
respectively.

%
%
\section{The Photon Structure Function at Small--$x$} 
\label{sec:smallx}
Recently the OPAL collaboration \cite{Abbiendi:2000cw} at the CERN--LEP 
collider has extended the measurements of the photon structure 
function $F_2^{\gamma}(x,Q^2)$ into the small--$x$ region down 
to $x\simeq 10^{-3}$, probing lower values of $x$ than ever 
before.  The observed rise of $F_2^{\gamma}$ towards low values 
of $x$, $x<0.1$, is in agreement with general QCD renormalization 
group improved expectations.  It has, however, been noted 
that the rising small--$x$ data at lower scales $Q^2\simeq 2-4$ 
GeV$^2$ lie above the original QCD expectations anticipated 
almost a decade ago \cite{cit:GRV-9202,cit:SaS-9501}. 
 
In the following we will demonstrate that our 
updated parameter--{\em{free}} QCD predictions 
for $F_2^{\gamma}(x,Q^2)$ \cite{cit:GRSc99}, discussed in Sec.~\ref{sec:photoninput},
are in general also consistent with 
the OPAL small--$x$ measurements at all presently accessible values 
of $Q^2$. 
 
Before presenting our results it is instructive to recapitulate 
briefly the main differences between the original GRV$_{\gamma}$ 
\cite{cit:GRV-9202} approach to the photonic parton distributions 
and 
our parameter--free predictions of Sec.~\ref{sec:photoninput}.
In the latter approach a coherent superposition of vector mesons 
has been employed, which maximally enhances the $u$--quark  
contributions to $F_2^{\gamma}$, for determining the hadronic 
parton input $f_{\rm had}^{\gamma}(x,Q_0^2)$ at a GRV--like 
\cite{cit:GRV98} input scale $Q_0^2\equiv \mu_{\rm LO}^2=0.26$ GeV$^2$ 
and $Q_0^2\equiv \mu_{\rm NLO}^2=0.40$ GeV$^2$ for calculating 
the (anti)quark and gluon distributions $f^{\gamma}(x,Q^2)$ of 
a real photon in leading order (LO) and next--to--LO (NLO) of 
QCD.  
Furthermore, in order to remove the ambiguity of the  
hadronic light quark sea and gluon input distributions of the 
photon (being related to the ones of the pion, $f^{\pi}(x,Q_0^2)$, 
via vector meson dominance), inherent to the older GRV$_{\gamma}$ 
\cite{cit:GRV-9202} and SaS \cite{cit:SaS-9501} parametrizations, 
predictions \cite{cit:GRS98pi,cit:GRSc99pi}
for $\bar{q}\,^{\pi}(x,Q^2)$ and $g^{\pi}(x,Q^2)$, cf.\ Chapter \ref{pipdf},
have been used in our present approach
which follow from constituent 
quark model constraints \cite{Altarelli:1974ff,Hwa:1980pn}.  These latter constraints 
allow to express $\bar{q}\,^{\pi}$ and $g^{\pi}$ entirely in terms 
of the experimentally known pionic valence density and the 
rather well known quark--sea and gluon distributions of the  
nucleon, using most recent updated valence--like 
input parton densities of the nucleon.  Since more recent  
DIS small--$x$ measurements at HERA imply somewhat less steep 
sea and gluon distributions of the proton \cite{cit:GRV98}, the  
structure functions of the photon will therefore also rise less 
steeply in $x$ than the previous GRV$_{\gamma}$ 
\cite{cit:GRV-9202} ones as was already discussed in the previous section and 
as will be seen in the figures shown below. 
In this way one arrives at truly parameter--free predictions 
for the structure functions and parton distributions of the  
photon. 

In Figs.\ \ref{fig:doth0104_fig1} and \ref{fig:doth0104_fig2} 
we compare our predictions of Section \ref{sec:photoninput}, 
denoted by GRSc \cite{cit:GRSc99}  
and the older GRV$_{\gamma}$ results \cite{cit:GRV-9202} 
with the recent small--$x$ OPAL measurements \cite{Abbiendi:2000cw} and, 
for completeness, some relevant L3 data \cite{Acciarri:1998ig} are shown 
as well.  Our parameter--free LO and NLO expectations  
are confirmed by the small--$x$ OPAL data\footnote{Note that the new OPAL data
\protect\cite{Abbiendi:2000cw}
at $Q^2 = 1.9$ and $3.7\ \gevsq$ supersede 
the older OPAL data shown in Fig.~\ref{fig:gamfig1}
at $Q^2 = 1.86$ and $3.76\ \gevsq$, see \protect\cite{Abbiendi:2000cw}.}
at {\em{all}} 
(small and large) experimentally accessible scales $Q^2$. 
This is in contrast to the GRV$_{\gamma}$ and SaS results  
which at LO are somewhat below the data at small $Q^2$ in Fig.\ \ref{fig:doth0104_fig1}  
and seem to increase too strongly at small $x$ in NLO, in  
particular at larger values of $Q^2$ as shown in Fig.\ \ref{fig:doth0104_fig2}.   
The main reason for this latter stronger and steeper $x$--dependence 
in LO and NLO derives from the assumed vanishing (pionic) quark--sea 
input at $Q_0^2=\mu_{\rm LO,NLO}^2$ for the anti(quark)  
distributions of the photon as well as from relating the 
hadronic gluon input of the photon directly to its (pionic) 
valence distribution \cite{cit:GRV-9202,cit:GRV92pi}.  This is in contrast 
to the more realistic (input) boundary conditions employed here. 
 
Clearly these small--$x$ measurements imply that the photon must 
contain \cite{Abbiendi:2000cw} a dominant hadron--like component at low 
$x$, since the simple direct `box' cross section (based on the 
subprocess $\gamma^*(Q^2)\gamma\to q\bar{q}\,$) yields  
$F_{2,{\rm box}}^{\gamma}\to 0$ as $x\to 0$, in contrast to the 
data for $x<0.1$ in Figs.\ \ref{fig:doth0104_fig1}  and \ref{fig:doth0104_fig2}.  
The QCD RG--improved parton 
distributions of the photon are thus essential for understanding 
the data on $F_2^{\gamma}(x,Q^2)$, with its dominant contributions 
deriving from $q^{\gamma}(x,Q^2)=\bar{q}\,^{\gamma}(x,Q^2)$. 
It would be also interesting and important to extend present 
measurements \cite{Ahmed:1995wv,Adloff:1998vt,Erdmann-9701,Nisius:1999cv} of the 
gluon distribution of 
the photon, $g^{\gamma}(x,Q^2)$, {\em{below}} the presently 
measured region  
0.1 \raisebox{-0.1cm}{$\stackrel{<}{\sim}$} $x<1$  
where similarly $g^{\gamma}(x<0.1,\,Q^2)$ is expected to be also 
somewhat flatter, see Fig.~\ref{fig:gamfig4}, in the small--$x$ region than 
previously anticipated \cite{cit:GRV-9202}. 

\begin{figure}
\centering
\epsfig{figure=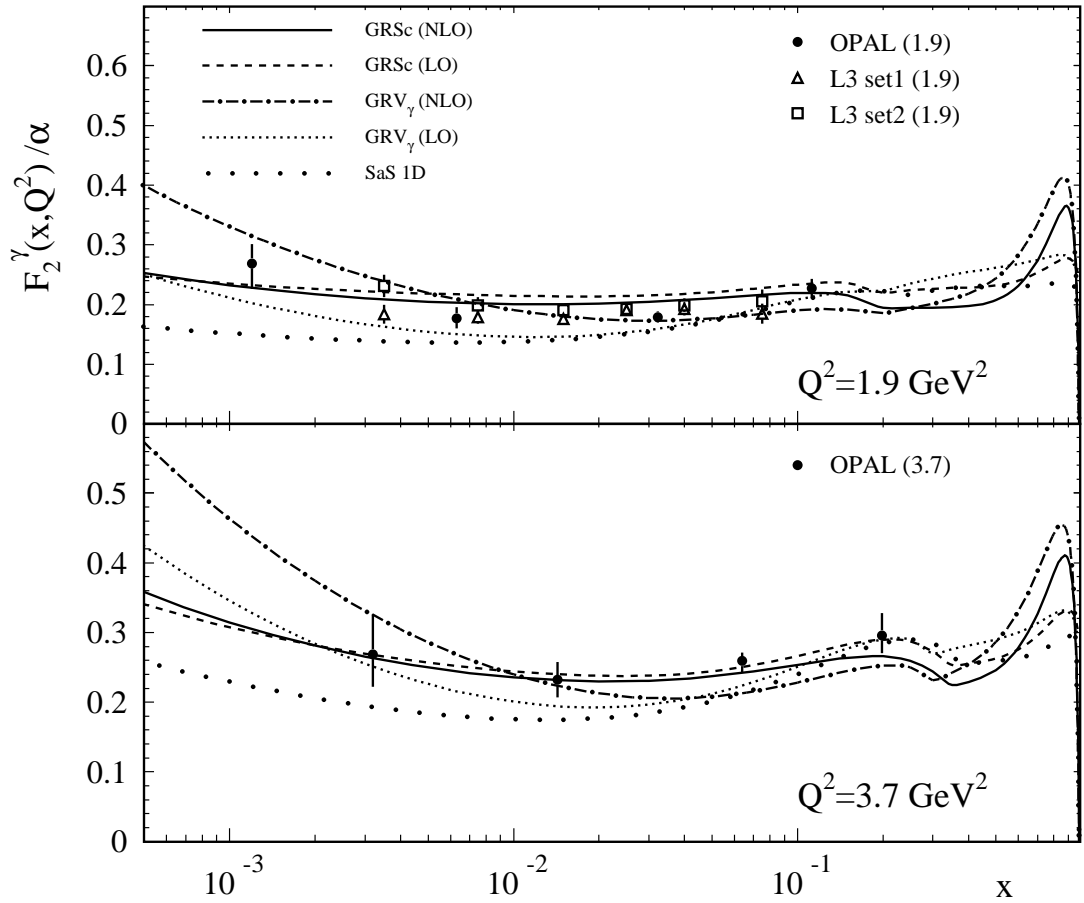,width=16cm}
\caption{\sf Comparison of our parameter--free predictions 
      from Sec.~\ref{sec:photoninput} denoted by GRSc
      \protect\cite{cit:GRSc99}, the previous GRV$_{\gamma}$ 
      \protect\cite{cit:GRV-9202} and SaS \protect\cite{cit:SaS-9501} results for 
      $F_2^{\gamma}(x,Q^2)$ with the recent OPAL ($1.9\ \gevsq$) 
      small--$x$ measurements 
      \protect\cite{Abbiendi:2000cw} at two fixed lower scales $Q^2$.  Some relevant 
      L3 data \protect\cite{Acciarri:1998ig} are displayed as well.}
\label{fig:doth0104_fig1}
\end{figure}

\begin{figure}
\centering
\epsfig{figure=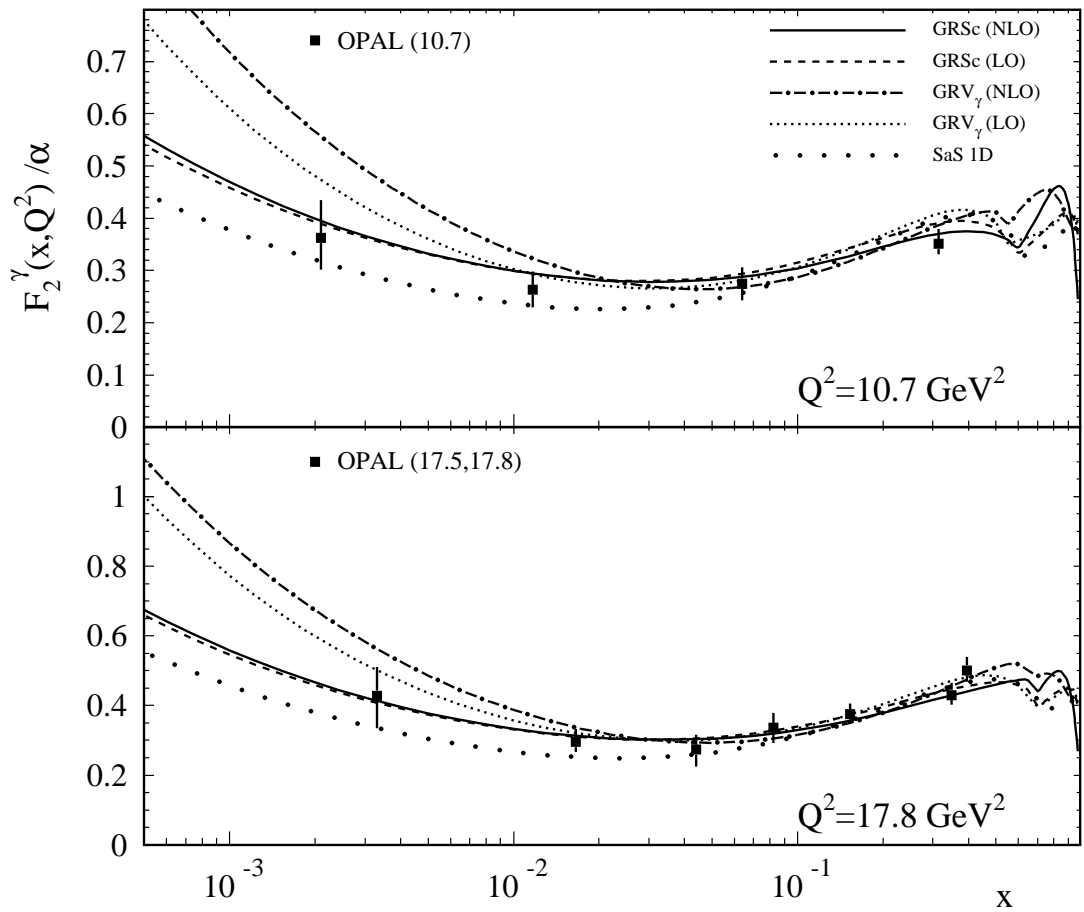,width=16cm}
\caption{\sf As in Fig.\ \protect\ref{fig:doth0104_fig1}  but at two fixed scales $Q^2$.  
         The recent OPAL small--$x$  data are taken from 
         Ref.\ \protect\cite{Abbiendi:2000cw}.} 
\label{fig:doth0104_fig2}
\end{figure}

\clearpage
\section{The Parton Content of Virtual Photons}
\label{sec:vgaminput}
Next we turn to the somewhat more speculative concept and models of
`resolved' virtual photons ($P^2\neq 0$). 
As motivated in Chap.~\ref{chap:partonmodel} the dependence on the
target virtuality $P^2$ will be entirely taken care of by the boundary
conditions whereas the partons will be treated as if they were on--shell.
Of course, this is quite familiar for the quarks and gluons, however, the latter rule also
refers to the photon--parton $\fivg[\Gamma](x,Q^2)$ in the target photon, i.e.,
the 'direct' subprocesses, e.g., $\gamma^\star(Q^2)\fivg[\Gamma]\to q \bar{q}$
have to be calculated with $P^2=0$ (cf.~rule (ii) in \cite{cit:GRSc99}).
Thus we
can implement the same $\disg$ factorization scheme as for real
photons in Eq.~(\ref{eq:schemetrafo1}), and the structure function 
$F_2^{\gamma(P^2)}(x,Q^2)$
of a transverse virtual target photon
becomes formally very similar to Eq.\ (\ref{eq:gam17}):
\begin{eqnarray}\label{eq:gam23}
\frac{1}{x}\, F_2^{\gamma(P^2)}(x,Q^2) & = & 2\sum_{q=u,d,s} e_q^2 \left\{ 
  q^{\gamma(P^2)}(x,Q^2)+\,\frac{\alpha_s(Q^2)}{2\pi}\, \left[ \cq[2] \otimes 
    q^{\gamma(P^2)} + \cg[2] \otimes g^{\gamma(P^2)} \right] 
\right\}
\nonumber\\
& &  
+\, \frac{1}{x}\, F_{2,c}^{\gamma}(x,Q^2) + \, \frac{1}{x}\, 
     F_{2,c}^{g{^{\gamma(P^2)}}}(x,Q^2) 
\end{eqnarray}
with $\cq[2](x), \cg[2](x)$ given in (\ref{eq:wilsoncoeff}).  
The `direct' heavy (charm) quark 
contribution is given by Eq.~(\ref{eq:f2gamh}) as for real photons and
the  `resolved' charm contribution
$F_{2,c}^{g^{\gamma(P^2)}}$ is as in Eq.\ (\ref{eq:f2gamhres}) with the gluon 
distribution $g^{\gamma}(z,\mu_F^2)\to g^{\gamma(P^2)}(z,\mu_F^2)$.

It is physically compelling to expect a smooth behavior 
of the observable structure function in Eq.~(\ref{eq:gam23})
with respect to $P^2$.
Especially, in the real photon limit Eq.~(\ref{eq:gam17}) has to 
be recovered imposing the condition
$\fvg(x,Q^2) \to \fv(x,Q^2)$ as $P^2 \to 0$ for 
the parton distributions.
Note that this requirement is automatically satisfied as soon as
it has been implemented at the input scale $Q_0^2$ since the PDFs
are governed by the {\em same} evolution equations which is a
consequence of our unified approach.
\subsection{Boundary Conditions}
The above smoothness requirements are fulfilled by the following
boundary conditions for $\fvg$, cf.\ Eq.\ (\ref{eq:gam13}),
\begin{equation}\label{eq:gam24}
f^{\gamma(P^2)}(x,Q^2=\tilde{P}^2) = 
   f_{\had}^{\gamma(P^2)}(x,\tilde{P}^2) =
     \eta(P^2) f_{\had}^{\gamma}(x,\tilde{P}^2)
\end{equation}
in LO {\em{as well as}} in NLO.  Here $\tilde{P}^2={\rm{max}}\,(P^2,\mu^2)$ 
as dictated by continuity in $P^2$ \cite{cit:GRS95} and 
$\eta(P^2)=(1+P^2/m_{\rho}^2)^{-2}$ is a dipole suppression factor with
$m_{\rho}^2=0.59$ GeV$^2$.  
The scale $\tilde{P}^2$
is dictated not only by the above mentioned continuity requirement, but 
also by the fact that the hadronic component of $\fvg(x,Q^2)$
is probed at the scale $Q^2=\tilde{P}^2$ 
\cite{Borzumati:1993za,cit:GRS95,cit:SaS-9501,*Schuler:1996fc} where
the pointlike component vanishes by definition.  
The boundary condition
in Eq.\ (\ref{eq:gam24}) guarantees, as should be evident, a far 
better perturbative
stability as compared to the situation in \cite{cit:GRS95} where the NLO
input differed drastically from its LO counterpart (cf.\ Eq.\ (8) in
Ref.~\cite{cit:GRS95}).  

The evolution to $Q^2>\tilde{P}^2$ is now analogous to the case of 
real photons in the previous section and the general solution for the
resulting parton distributions is similar to the one in 
Eq.\ (\ref{eq:gam11}) and Eq.\ (\ref{eq:gam13}),
\begin{eqnarray}\label{eq:gam25}
f^{\gamma(P^2)}(x,Q^2) & = & f_{\pl}^{\gamma(P^2)}(x,Q^2) + 
    f_{\had}^{\gamma(P^2)}(x,Q^2)\\
 & =  & f_{\pl}^{\gamma(P^2)}(x,Q^2)+\eta(P^2)\alpha \left[G_f^2\, f^{\pi}(x,Q^2)+
\delta_f\, \frac{1}{2}(G_u^2-G_d^2)\, s^{\pi}(x,Q^2)\right]\nonumber
\end{eqnarray}
with $\delta_f$ as in Eq.\ (\ref{eq:gam13}) and
where $f_{\had}^{\gamma(P^2)}$ refers again to the solution of the
homogeneous RG evolution equations, being driven by the hadronic input
in (\ref{eq:gam24}), which is explicitly given by 
Eq.\ (\ref{eq:homsol}).
Its parametrization is fixed by the available parametrization 
for $f^{\pi}(x,Q^2)$ in App.~\ref{sec:pipara}.  
The inhomogeneous `pointlike' solution in (\ref{eq:gam25}) is explicitly 
given by Eq.~(\ref{eq:inhomsol}) 
where $L=\alpha_s(Q^2)/\alpha_s(\tilde{P}^2)$.  A parametrization of 
$f_{\pl}^{\gamma(P^2)}(x,Q^2)$ in LO is thus easily obtained from the 
one for the real photon $f_{\pl}^{\gamma}(x,Q^2)$ in (\ref{eq:gam13}) 
in terms of 
$\ln \, L^{-1}=\ln\,\left[\alpha_s(\mu^2)/\alpha_s(Q^2)\right]$, where now 
$\alpha_s(\mu^2)$ has simply to be replaced by $\alpha_s(\tilde{P}^2)$ 
as described in detail in App.~\ref{sec:gampara}.
Furthermore, since our NLO predictions for $\fvg(x,Q^2)$
turn out to be rather similar to the LO ones, as will be shown below,
the simple analytic LO parametrizations for $\fvg(x,Q^2)$
can be used for NLO calculations as well.  This is certainly sufficiently
accurate and reliable in view of additional model ambiguities inherent
in the parton distributions of virtual photons.

It should be emphasized that the RG resummed results in (\ref{eq:gam25}) are 
relevant when-\- ever $P^2\ll Q^2$, typically \cite{cit:GRS95,Gluck:1996ra} 
$P^2\simeq\frac{1}{10}\,\,Q^2$, so as to suppress power--like (possibly 
higher twist) terms $(P^2/Q^2)^n$ which would spoil the dominance of 
the resummed logarithmic contributions. 
For $P^2$ approaching $Q^2$, the $e^+e^-\to e^+e^- X$
reaction, for example, should be simply described by the full fixed
order box $\vgvg$ keeping all $(P^2/Q^2)^n$
terms (cf.\ Chap.~\ref{chap:lobox}).  
A calculation of
the full perturbative $\Ord(\alpha_s)$ corrections
to this virtual box \cite{nlobox}
will offer the possibility to determine reliably at what values of $P^2$
(and possibly $x$) this $\Ord(\alpha_s)$ 
corrected virtual box becomes the more appropriate and correct description.
Similar remarks hold for a DIS process $ep\to eX$, i.e. 
$\gamma(P^2)\, p\to X$, where $\Ord(\alpha_s)$ corrections to pointlike
virtual $\gamma(P^2)$--parton subprocesses have to be analyzed in 
detail in order to decide at what $P^2$ these pointlike expressions
become the more appropriate description and the virtual photonic
parton distributions (i.e., resummations) become irrelevant.  

Our unified approach implies that the `direct'
photon contribution to any process whatsoever should always be calculated
as if this photon is real apart from the fact that its flux should be
evaluated according to Eq.\ (\ref{eq:photonfluxes}) with $P^2\neq 0$ 
\cite{Drees:1994eu,Gluck:1996ra,deFlorian:1996uk,*Chyla:1996ux}.
This differs from the somewhat inconsistent procedure adopted by the 
\mbox{HERA--H1} collaboration 
\cite{ZEUS-IEC95,Adloff:1997nd,Adloff:1998st,H1-ICHEP98,Adloff:1998vc} 
where exact $e\,q\to e\,q\,g$ 
and $e\,g\to e\,q\,\bar{q}$ matrix elements were used for the direct 
photon contribution to the dijet cross section.  As long as 
\mbox{$P^2$ \raisebox{-0.1cm}{$\stackrel {<}{\sim}$} $\frac{1}{10}\,Q^2$,} 
the exact treatment of matrix elements,
however, should not differ too much 
\cite{Klasen:1998jm,*Kramer:1998bc,*Potter:1998jt} from the more appropriate
treatment described above.  
To conclude let us stress that our unified approach, as illustrated 
by the foregoing examples,
is not a free option but a {\em{necessary consistency condition}}
for introducing the concept of the resolved parton content of the 
virtual photon as an {\em{alternative}} to a non--resummed fixed
order perturbative analysis at $P^2\neq 0$.  This consistency requirement
is related to the fact that all the resolved contributions due to
$\fvg(x,Q^2)$ are calculated (evoluted) as if these partons
are massless $[7-11]$ (i.e.\ employing photon splitting
functions for real photons, etc.) in spite of the fact that their actual
virtuality is given by $P^2\neq 0$.  Thus the direct photon contribution
should obviously be also treated accordingly.

\subsection{Quantitative Results}
\begin{figure}[htb]
\centering
\vspace*{-1.0cm}
\subfigure{\epsfig{figure=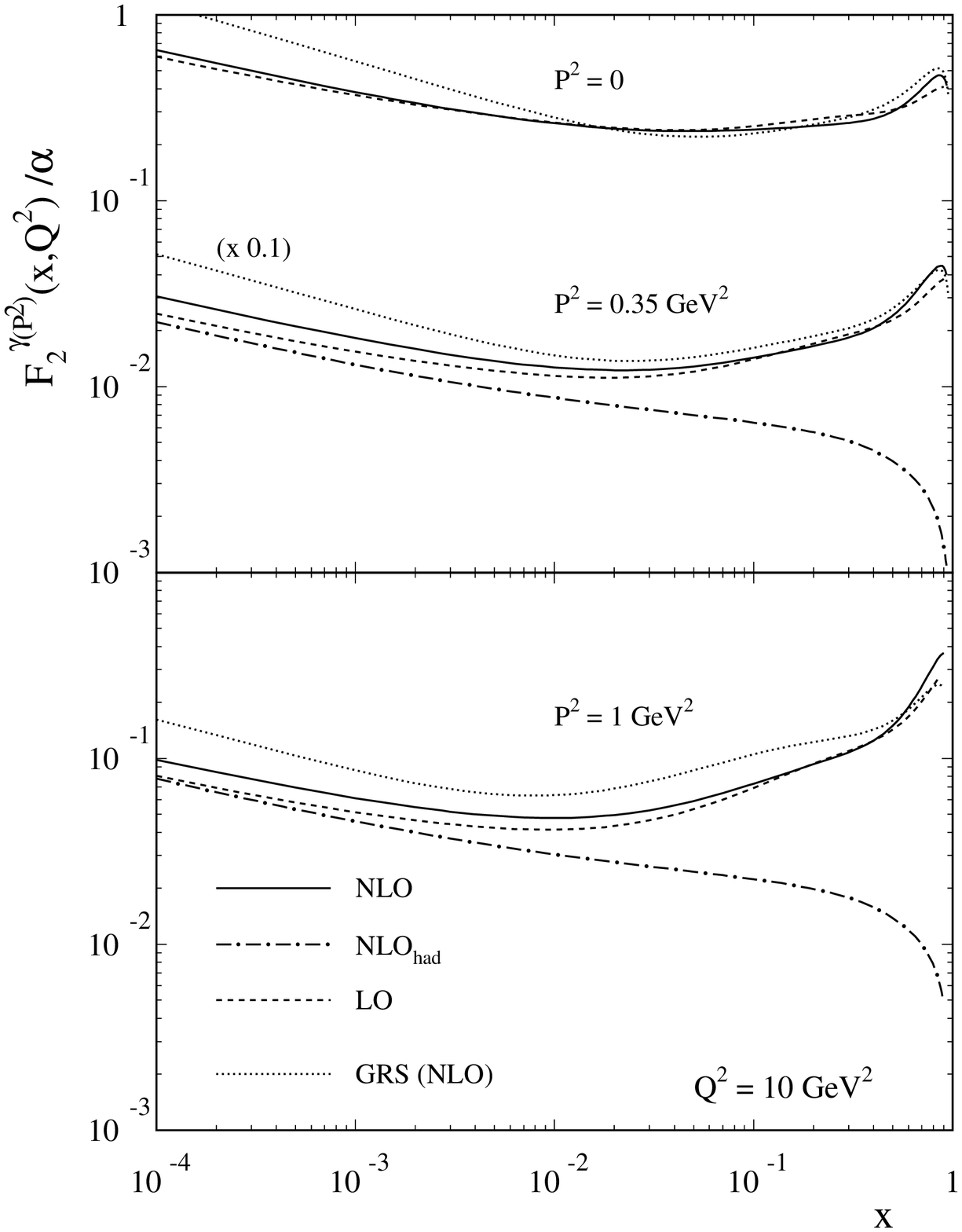,height=10cm}}
\subfigure{\epsfig{figure=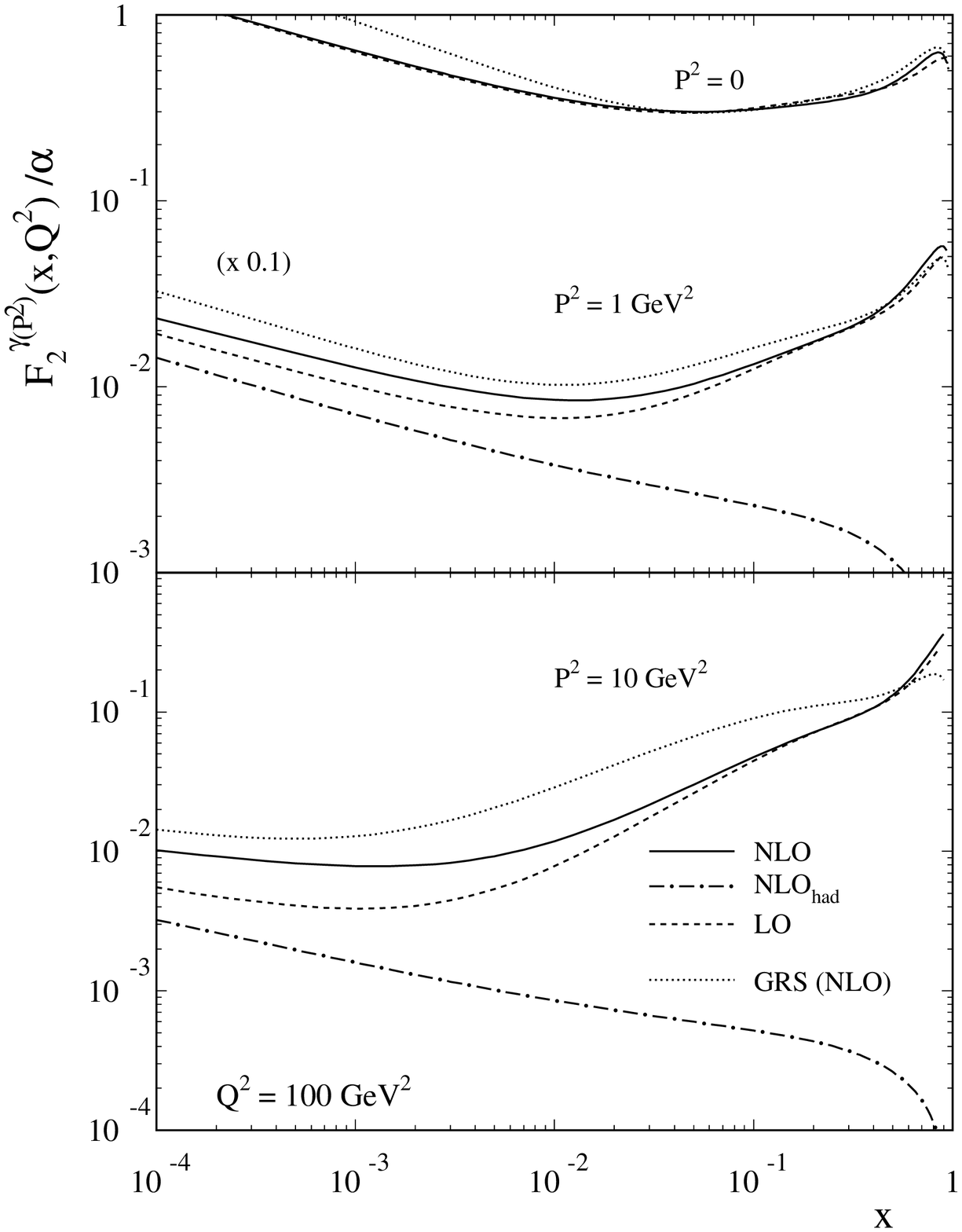,height=10cm}}
\vspace*{-0.5cm}
\caption{\sf LO and NLO predictions for the $x$ dependence of the
   virtual photon structure function $F_2^{\gamma(P^2)}$ at 
   $Q^2=10$ and $100\ \gevsq$ and various fixed values of $P^2\ll Q^2$,
   neglecting any heavy quark contribution.  For comparison we also
   show the NLO Gl{\"u}ck--Reya--Stratmann (GRS) \protect\cite{cit:GRS95} results 
   as well as the predictions
   for a real ($P^2=0$) photon.  Notice that the dotted curve for
   $P^2=0$ referred to as GRS obviously coincides with the 
   GRV$_{\gamma}$ result \protect\cite{cit:GRV-9201,cit:GRV-9202}.  
   The results have been multiplied by the number indicated in brackets.}
\label{fig:gamfig6a}
\end{figure}
Now we turn to our quantitative predictions and we first present
in Fig.\ \ref{fig:gamfig6a} 
detailed predictions for $F_2^{\gamma(P^2)}(x,Q^2)$ 
for various virtualities $P^2$ and scales $Q^2$.
Since the `pointlike' component in (\ref{eq:gam25}) is uniquely calculable 
perturbatively, a detailed measurement of the $x$ and $P^2$ dependence
at various fixed values of $Q^2$, as shown in 
Fig.\ \ref{fig:gamfig6a}, would
shed light on the theoretically more speculative and far less understood
nonperturbative `hadronic' contribution in Eq.\ (\ref{eq:gam25}) and eventually
establish the absolute perturbative predictions.  Our LO and NLO
predictions in Fig.\ \ref{fig:gamfig6a} show a 
remarkable perturbative stability
throughout the whole $x$--region shown, except perhaps for $P^2\gg 1$
GeV$^2$ where the perturbatively very stable \cite{cit:GRSc99pi,cit:GRS98pi} `hadronic'
component in (\ref{eq:gam25}) becomes strongly suppressed with respect to the 
`pointlike' solution which is less stable in the small $x$ region,
$x<10^{-2}$, as is evident from the right hand side of Fig.\ \ref{fig:gamfig6a}.

\begin{figure}[ht]
\centering
\epsfig{figure=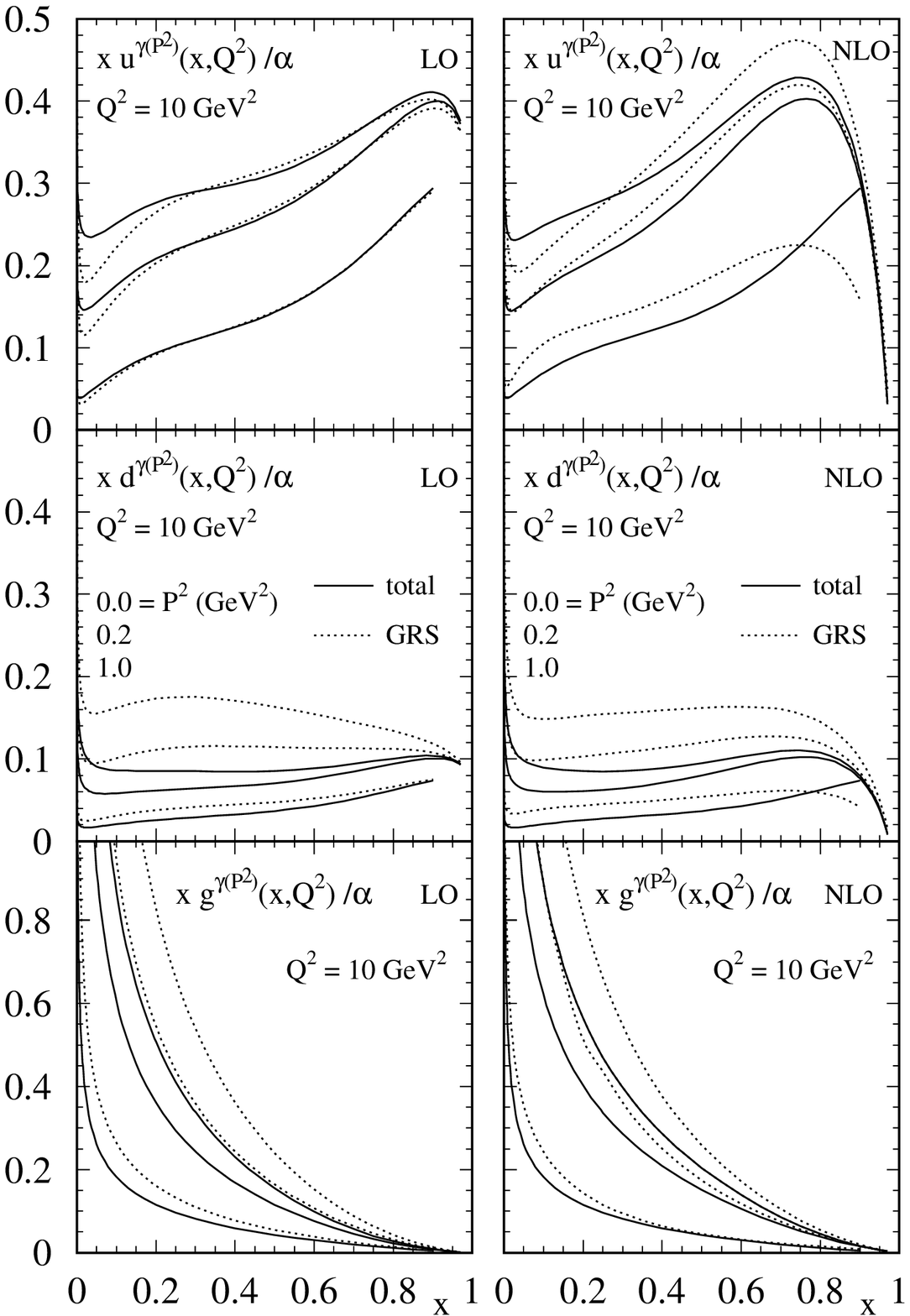,height=12cm}
\caption{\sf Comparison of our predicted LO and NLO(DIS$_{\gamma})$
   distributions of the virtual photon at $Q^2=10$ GeV$^2$ and various
   fixed values of $P^2\ll Q^2$ with 
   the GRS densities \protect\cite{cit:GRS95}.
   The curves refer from top to bottom to $P^2=0,\,\, 0.2$ and 1 GeV$^2$, 
   respectively.  
   The results for the real photon ($P^2=0$) are very similar to the ones 
   in Fig.\ \ref{fig:gamfig2} where the GRV$_{\gamma}$ curves practically coincide with GRS.}
\label{fig:gamfig7a}
\end{figure}
\begin{figure}[ht]
\centering
\epsfig{figure=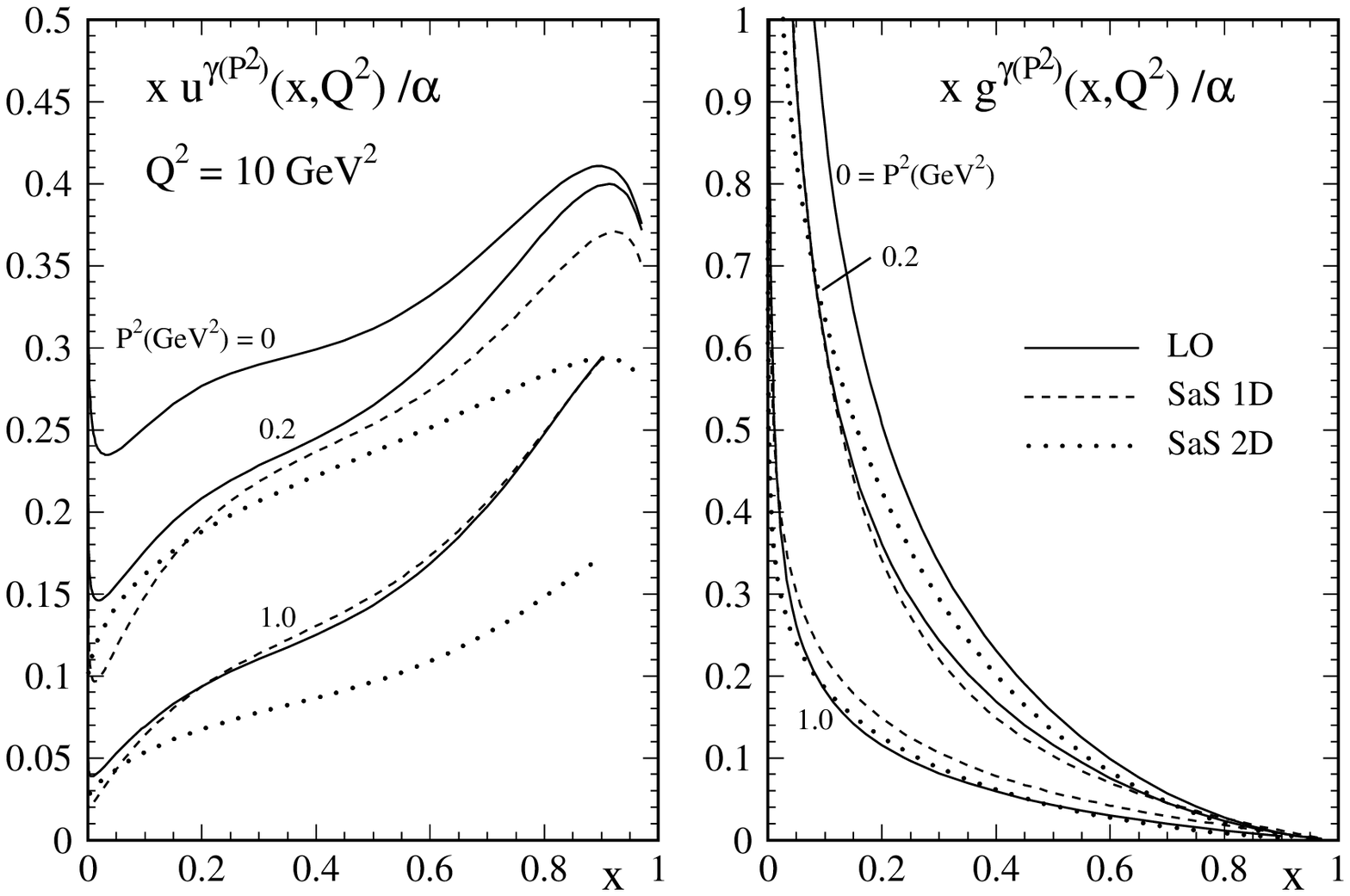,angle=0,width=12cm}
\caption{\sf Our LO distributions as in Fig.~\ref{fig:gamfig7a} compared with
        the ones of SaS \protect\cite{cit:SaS-9501,*Schuler:1996fc}.}
\label{fig:gamfig7b}
\end{figure}

The individual LO and NLO parton distributions of the virtual photon
at $Q^2=10\ \gevsq$ are shown in Fig.\ \ref{fig:gamfig7a} where they are compared with
the ones of 
Gl{\"u}ck, Reya and Stratmann (GRS) \cite{cit:GRS95}.  
The LO SaS expectations \cite{cit:SaS-9501,*Schuler:1996fc}
are compared with our LO predictions in Fig.\ \ref{fig:gamfig7b}.  

\begin{figure}[ht]
\centering
\subfigure{\epsfig{figure=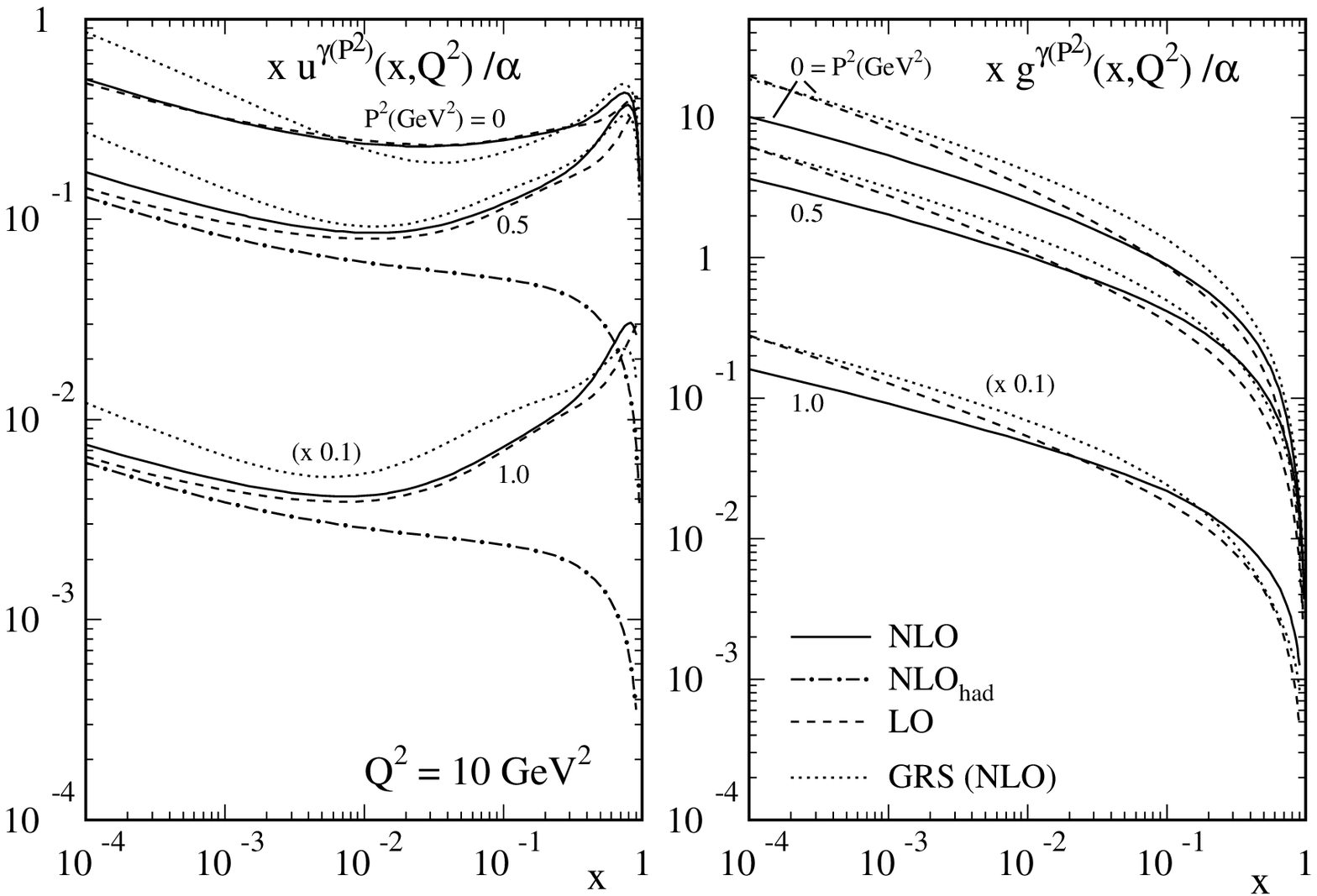,angle=0,width=12cm}}
\subfigure{\epsfig{figure=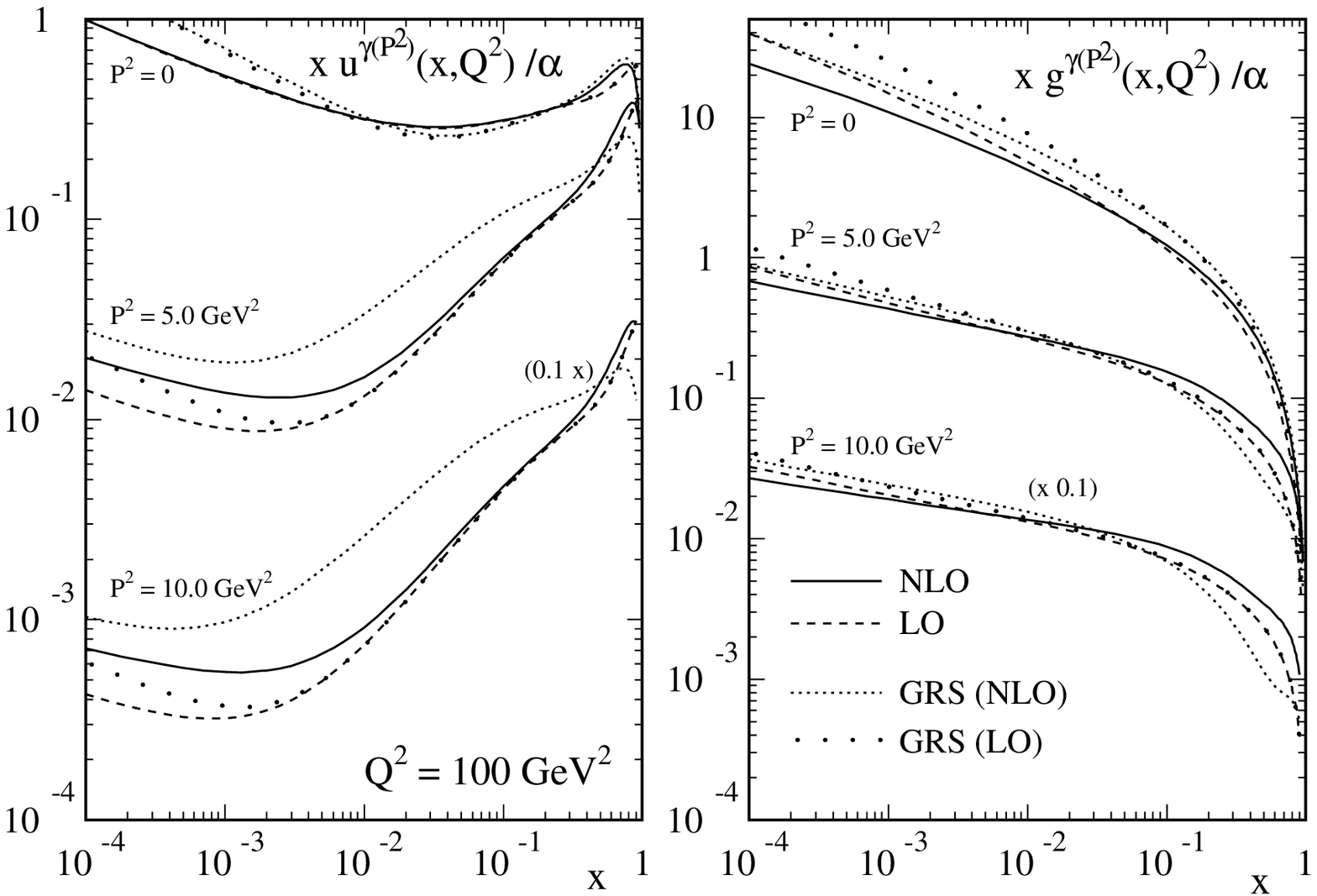,angle=0,width=12cm}}
\caption{\sf LO and NLO predictions for the up--quark and gluon distributions
   of a virtual photon $\gamma(P^2)$ at $Q^2=10$ and $100\ \gevsq$.  
   For comparison the results for the real photon ($P^2=0$) are shown as well.  
   In the upper half	   
   the NLO `hadronic' contribution in (\ref{eq:gam25}) is also shown separately.  
   The GRS expectations are taken from Ref.~\protect\cite{cit:GRS95}.
   The DIS$_{\gamma}$ results for $u^{\gamma(P^2)}$ can be easily converted to 
   the $\overline{\rm{MS}}$ scheme with the help of Eq.\ (\ref{eq:schemetrafo1}), whereas 
   $g^{\gamma(P^2)}$ remains the same in both schemes.  The results have 
   been multiplied by the numbers indicated in brackets.}
\label{fig:gamfig8}
\end{figure}

In Fig.\ \ref{fig:gamfig8} 
we show our predictions for $xu^{\gamma(P^2)}(x,Q^2)$ and
$xg^{\gamma(P^2)}(x,Q^2)$ with particular emphasis on the very small
$x$  region.  For comparison we also show the results for a real
$(P^2=0)$ photon.  Plotting the `hadronic' component in (\ref{eq:gam25}) separately
in the upper two panels 
of Fig.\ \ref{fig:gamfig8} demonstrates that the perturbative `pointlike' component in
(\ref{eq:gam25}) dominates for $x>10^{-2}$.
In the lower two panels of Fig.~\ref{fig:gamfig8}, at $P^2 = 5, 10\ \gevsq$, the
hadronic part is nearly vanishing even at small $x$ 
[$\eta(P^2(\gevsq) = 5, 10) = 0.01, 0.003$].
Furthermore the expected perturbative
stability of our present LO and NLO predictions is fulfilled.  This
is in contrast to the GRS results which are unstable \cite{cit:GRS95}
throughout the whole \mbox{$x$--region} for $P^2$ \raisebox{-0.1cm}
{$\stackrel{>}{\sim}$} 1 GeV$^2$, as illustrated in the lower half of 
Fig.\ \ref{fig:gamfig8} at $Q^2=100$ GeV$^2$,
due to the very different perturbative (box) input in LO and NLO 
\cite{cit:GRS95}.  In general, however, as soon as the perturbatively
very stable `hadronic' component in (\ref{eq:gam25}) becomes suppressed for 
$P^2\gg 1$ GeV$^2$, the remaining perturbatively less stable `pointlike'
component destabilizes the total results for $q^{\gamma(P^2)}(x,Q^2)$ 
in the 
very small $x$ region, $x$ \raisebox{-0.1cm}{$\stackrel{<}{\sim}$}
$10^{-3}$, as can be seen in Fig.\ \ref{fig:gamfig8} for $u^{\gamma(P^2)}$ at 
$Q^2=100$ GeV$^2$ (cf.\ Fig.\ \ref{fig:gamfig6a}).

\begin{figure}[ht]
\centering
\epsfig{figure=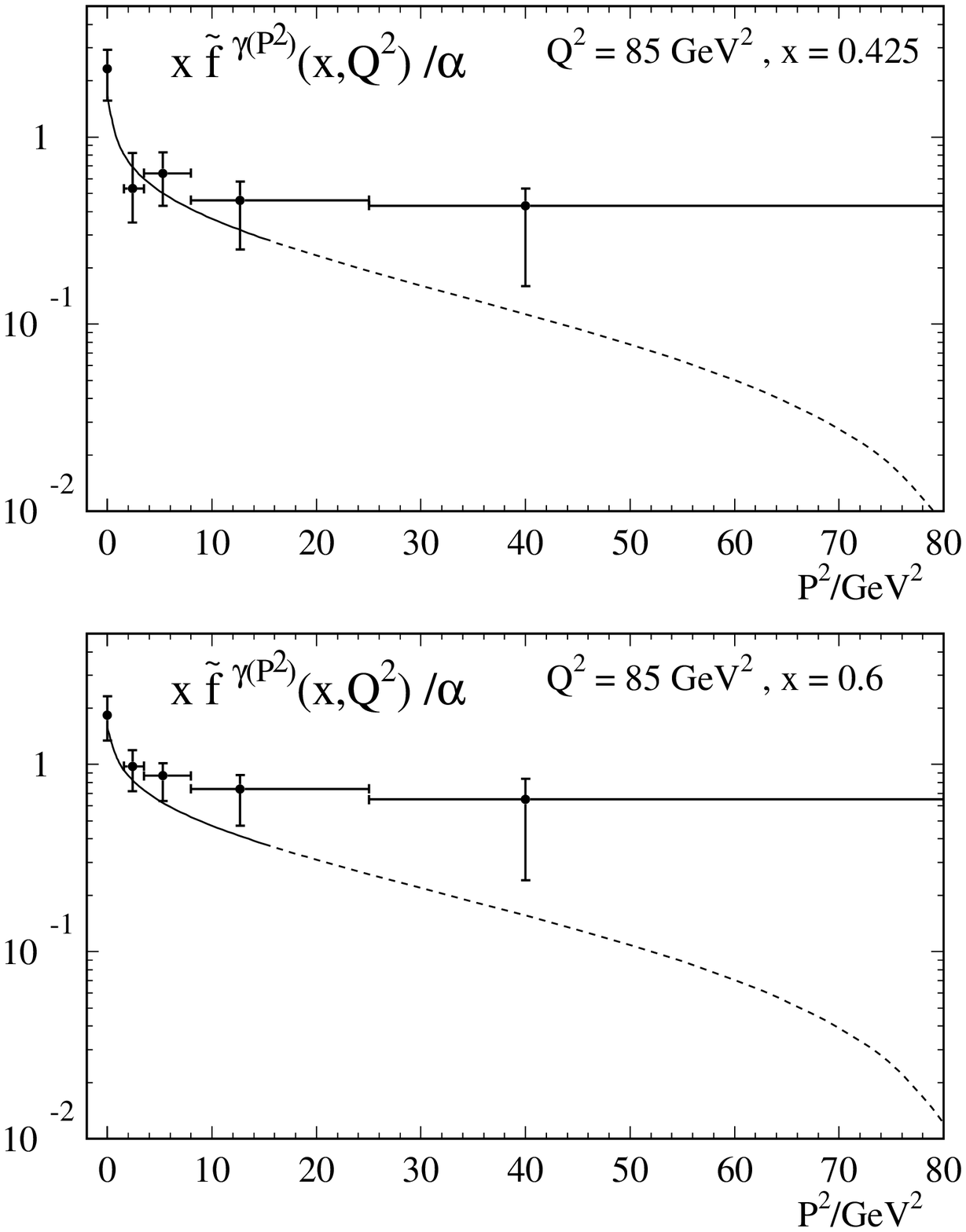,width=12cm}
\caption{\sf Predictions for the LO effective parton density 
        $x\tilde{f}\,^{\gamma(P^2)}(x,Q^2)$, defined in Eq.~(\ref{eq:gam29}), at
        the scale $Q^2\equiv\left(p_T^{\rm{jet}}\right)^2=85$ GeV$^2$
        and at two fixed values of $x$.  
        The H1 data 
        \protect\cite{ZEUS-IEC95,Adloff:1997nd,Adloff:1998st,H1-ICHEP98,Adloff:1998vc} 
        have
        been extracted from DIS $ep$ dijet production.  The solid curves
        refer to our predictions in the theoretically legitimate region
        $P^2\ll Q^2\equiv\left( p_T^{\rm{jet}}\right)^2$, whereas the
        dashed curves extend into the kinematic region of larger $P^2$
        approaching $Q^2$ where the concept of parton distributions of
        virtual photons is not valid anymore (see text).}
\label{fig:gamfig10}
\end{figure}

Finally in Fig.\ \ref{fig:gamfig10} we confront our LO predictions for 
$f^{\gamma(P^2)}(x,Q^2)$ with the effective parton density
\begin{equation}\label{eq:gam29}
\tilde{f}\,^{\gamma(P^2)}(x,Q^2) = \sum_{q=u,d,s}\left(q^{\gamma(P^2)}
   + \bar{q}\,^{\gamma(P^2)}\right) +\, \frac{9}{4}\,g^{\gamma(P^2)}
\end{equation}
extracted in LO from DIS dijet data by the HERA--H1 collaboration
\cite{ZEUS-IEC95,Adloff:1997nd,Adloff:1998st,H1-ICHEP98,Adloff:1998vc} very recently.  
The predicted dependence on the photon's
virtuality $P^2$ at the scale $Q^2\equiv\left( p_T^{\rm{jet}}\right)^2=85$
GeV$^2$ agrees reasonably well with the measurements in the relevant
kinematic region $P^2\ll Q^2$.  This is also the case at other scales
$Q^2\equiv\left( p_T^{\rm{jet}}\right)^2$ and fixed values of $x$ 
\cite{ZEUS-IEC95,Adloff:1997nd,Adloff:1998st,H1-ICHEP98,Adloff:1998vc} not shown in 
Fig.\ \ref{fig:gamfig10}.  As discussed above, it should be
kept in mind, however, that for larger values of $P^2$ approaching $Q^2$,
which refer to the dashed curves in Fig.\ \ref{fig:gamfig10}, the whole concept of RG
resummed parton distributions of virtual photons is {\em{not}}
appropriate anymore.  Since the resolved contributions of a virtual
photon with virtuality as large as $P^2=10-15$ GeV$^2$ are by a factor
of about 10 smaller than the ones of a real ($P^2=0$) photon, it is
reasonable to conclude from Fig.\ \ref{fig:gamfig10} that for $P^2$ \raisebox{-0.1cm}
{$\stackrel{>}{\sim}$} 10 GeV$^2$ the DIS $ep\to eX$ process considered
is dominated by the usual direct $\gamma^*\equiv\gamma(P^2)$ exchange
cross sections and {\em{not}} `contaminated' anymore by resolved
contributions.  
%
%

\clearpage
\section{Summary and Conclusions}
\label{sec:gamsum}
The main purpose of the present chapter was to formulate a consistent
set of boundary conditions which allow for a perturbatively stable
LO and NLO calculation of the photonic parton distributions
$f^{\gamma(P^2)}(x,Q^2)$ as well as for a smooth transition to the
parton densities of a real $(P^2=0)$ photon.  
Employing the 
pionic distributions $f^{\pi}(x,Q^2)$ of Chapter \ref{pipdf}, required
for describing, via VMD, the nonpointlike hadronic components of a 
photon, we arrive at essentially parameter--{\em{free}}
predictions for $f^{\gamma(P^2)}(x,Q^2)$ which are furthermore in
good agreement with all present measurements of the structure function
$F_2^{\gamma}(x,Q^2)$ of real photons $\gamma\equiv\gamma(P^2=0)$.
It should be noted that the experimentally almost unconstrained pionic
gluon and sea distributions, $g^{\pi}(x,Q^2)$ and $\bar{q}\,^{\pi}(x,Q^2)$,
have been uniquely derived 
from the experimentally 
rather well known pionic valence density $v^{\pi}(x,Q^2)$ and the 
(recently updated \cite{cit:GRV98} dynamical) parton distributions
of the proton.  We have furthermore implemented these hadronic 
components by using a VMD ansatz for a coherent superposition of 
vector mesons which maximally enhances the contributions of the 
up--quarks to $F_2^{\gamma}$ as favored by all present data.  Since
these hadronic contributions are generated from the valence--like 
input parton distributions at the universal target--mass independent
low resolution scale $Q_0^2=\mu^2\simeq 0.3$ GeV$^2$, we arrive, at
least for real ($P^2=0$) photons, at unique small--$x$ predictions 
for $x$ \raisebox{-0.1cm}{$\stackrel{<}{\sim}$} $10^{-2}$ at $Q^2>\mu^2$
which are of purely dynamical origin, as in the case of hadrons. 
As has been demonstrated in Sec.~\ref{sec:smallx} these small--$x$
predictions are in perfect agreement with recent small--$x$ measurements
of the photon structure function $F_2^\gamma(x,Q^2)$ \cite{Abbiendi:2000cw}
at all presently accessible values of $Q^2$.
Furthermore, since our universal input scale $\mu^2$ fixes also
uniquely the perturbative pointlike part of the photonic parton
distributions, which dominates for $x$ 
\raisebox{-0.1cm}{$\stackrel{>}{\sim}$} 0.1, the large--$x$ 
behavior of photonic structure functions
is unambiguously predicted as well.

Our expectations for the parton content of virtual ($P^2\neq0$) photons
are clearly more speculative, depending on how one models the hadronic
component (input) of a virtual photon.  The latter is usually assumed
to be similar to the VMD input for a real photon, times a dipole
suppression factor which derives from an effective vector--meson
$P^2$--propagator, cf.\ Eq.\ (\ref{eq:gam24}).  
Whenever a virtual photon is probed
at a scale $Q^2\gg P^2$, with its virtuality being entirely taken 
care of by the (transverse) equivalent photon flux factor
and the boundary conditions for the photonic parton distributions, 
all cross sections of partonic subprocesses (resolved {\em and} direct) 
should be calculated as if $P^2=0$.
In other words, the treatment and expressions for $f^{\gamma(P^2)}(x,Q^2)$
as {\em{on}}--shell transverse partons obeying the usual RG
$Q^2$--evolution equations (with the usual splitting functions of 
{\em{real}} photons, etc.) dictate an identification of the 
relevant 
sub--cross--sections $f^{\gamma(P^2)}X\to X'$ with
that of the {\em{real}} photon, $\hat{\sigma}(f^{\gamma(P^2)}
X\to X')=\hat{\sigma}(f^{\gamma}X\to X')$.  In particular, the calculation
of $F_2^{\gamma(P^2)}(x,Q^2)$ requires the {\em{same}} photonic
Wilson coefficient $C_{\gamma}(x)$ as for $P^2=0$.
%
This allows to
formulate similar boundary conditions in LO and NLO which give rise
to perturbatively stable parton distributions, cross sections (i.e.\
also structure functions) of virtual photons $\gamma(P^2)$ as long
as they are probed at scales $Q^2\gg P^2$ where 
$Q^2\equiv 4m_c^2,\,\,\left(p_T^{\rm{jet}}\right)^2$, etc., and typically 
\mbox{$P^2$ \raisebox {-0.1cm}{$\stackrel{<}{\sim}$} $\frac{1}{10}\,Q^2$.}  
It should be
emphasized that only in this latter kinematic range $P^2\ll Q^2$ is
the whole concept of RG resummed parton distributions of (resolved)
virtual photons appropriate and relevant.  Parton distributions of
virtual photons extracted recently from DIS $ep$ dijet data are in
good agreement with our (parameter--free) predictions.

Finally, we present simple analytic parametrizations of our
predicted LO and NLO(DIS$_{\gamma}$) parton distributions of real
photons.  From these LO parametrizations one can easily obtain also
the ones for a virtual photon which, whithin sufficient accuracy,
may also be used in NLO.  Our NLO($\disg$) parametrizations
of the parton densities of the real photon can be easily transformed
to the $\msbar$ scheme according to Eq.\ (\ref{eq:schemetrafo1}) which might
be relevant for future NLO analyses of resolved photon contributions
to hard processes where most NLO subprocesses have so far been
calculated in the $\msbar$ scheme.

A FORTRAN package containing our most recent parametrizations
can be obtained by electronic mail on request.

%% file: phd_gam_vgam.tex
\chapter{Has the QCD RG--Improved Parton Content of Virtual 
Photons been Observed?}
\label{chap:vgam}

\noindent It is demonstrated that present $e^+e^-$ and DIS $ep$ 
data on the structure of the virtual photon can be understood 
entirely in terms of the standard `naive' quark--parton model box 
approach.  Thus the QCD renormalization group (RG) improved parton 
distributions of virtual photons, in particular their gluonic 
component, have not yet been observed.  The appropriate 
kinematical regions for their future observation are pointed out 
as well as suitable measurements which may demonstrate their 
relevance. The results presented in this chapter are taken 
from \cite{Gluck:2000sn}.

\section{Introduction}\label{sec:vgam1}
 Recent measurements and experimental studies of dijet events in 
deep inelastic $ep$ \cite{Adloff:1998vc} and of double--tagged $e^+e^-$ 
\cite{Acciarri:2000rw} reactions have indicated a necessity for assigning a 
(QCD resummed) parton content of virtual photons $\gamma(P^2)$ as 
suggested and predicted theoretically 
\cite{Uematsu:1981qy,Uematsu:1982je,Ibes:1990pj,Rossi:1984xz,
cit:Rossi-PhD,Borzumati:1993za,cit:GRS95,cit:SaS-9501,*Schuler:1996fc,
Drees:1994eu,cit:GRSc99}.
In particular the DIS dijet production data \cite{Adloff:1998vc} appear to 
imply a sizeable gluon component $g^{\gamma(P^2)}(x,Q^2)$ in the 
derived effective parton density of the virtual photon, where 
$Q^2$ refers to the hadronic scale of the process, $Q\sim p_T^{\rm {jet}}$, 
or to the virtuality of the probe photon $\gamma^*(Q^2)$ 
which probes the virtual target photon $\gamma(P^2)$ in 
$e^+e^-\to e^+e^-X$.  It is the main purpose of this chapter to 
demonstrate that this is {\em{not}} the case and that all 
present data on virtual photons can be explained entirely in 
terms of the conventional QED doubly virtual box contribution 
$\vgvg$ in fixed order 
perturbation theory --sometimes also referred to as the 
quark--parton model (QPM). 
 
This is of course in contrast to the well known case of a real 
photon $\gamma\equiv\gamma(P^2\equiv 0)$ whose (anti--)quark and 
gluon content has been already experimentally established (for 
recent reviews see \cite{Erdmann-9701,Nisius:1999cv}) 
which result mainly from  
resummations (inhomogeneous evolutions) of the pointlike mass 
singularities proportional to $\ln\, Q^2/m_q^2$ occurring in the  
box diagram of $\vgg$ for the light 
$q=u,\, d,\, s$ quarks.  This is in contrast to a virtual photon 
target where $\vgvg$ does 
{\em not} give rise to collinear (mass) singularities but 
instead just to finite contributions proportional to $\ln\, 
Q^2/P^2$ which a priori need not be resummed to all orders in QCD. 

This chapter is organized as follows:\\
In Sec.~\ref{sec:vgam4} we compare fixed order box and QCD resummed expectations
for the effective structure function $\Feff$ with present $e^+e^-$ data,
while Sec.~\ref{sec:vgam5} contains a comparison 
of an effective leading order parton density for the virtual photon
with DIS $ep$ data.  
Suggestions of experimental signatures which can 
probe the QCD parton content, in particular the gluon content of 
virtual photons are presented in Sec.\ \ref{sec:vgam6} and our conclusions are 
finally drawn in Section \ref{sec:vgam7}. 

\section{RG--Improved Parton Model Expectations for the Effective 
Structure Function $\Feff$}  
\label{sec:vgam4}
\begin{figure}
\centering
\epsfig{figure=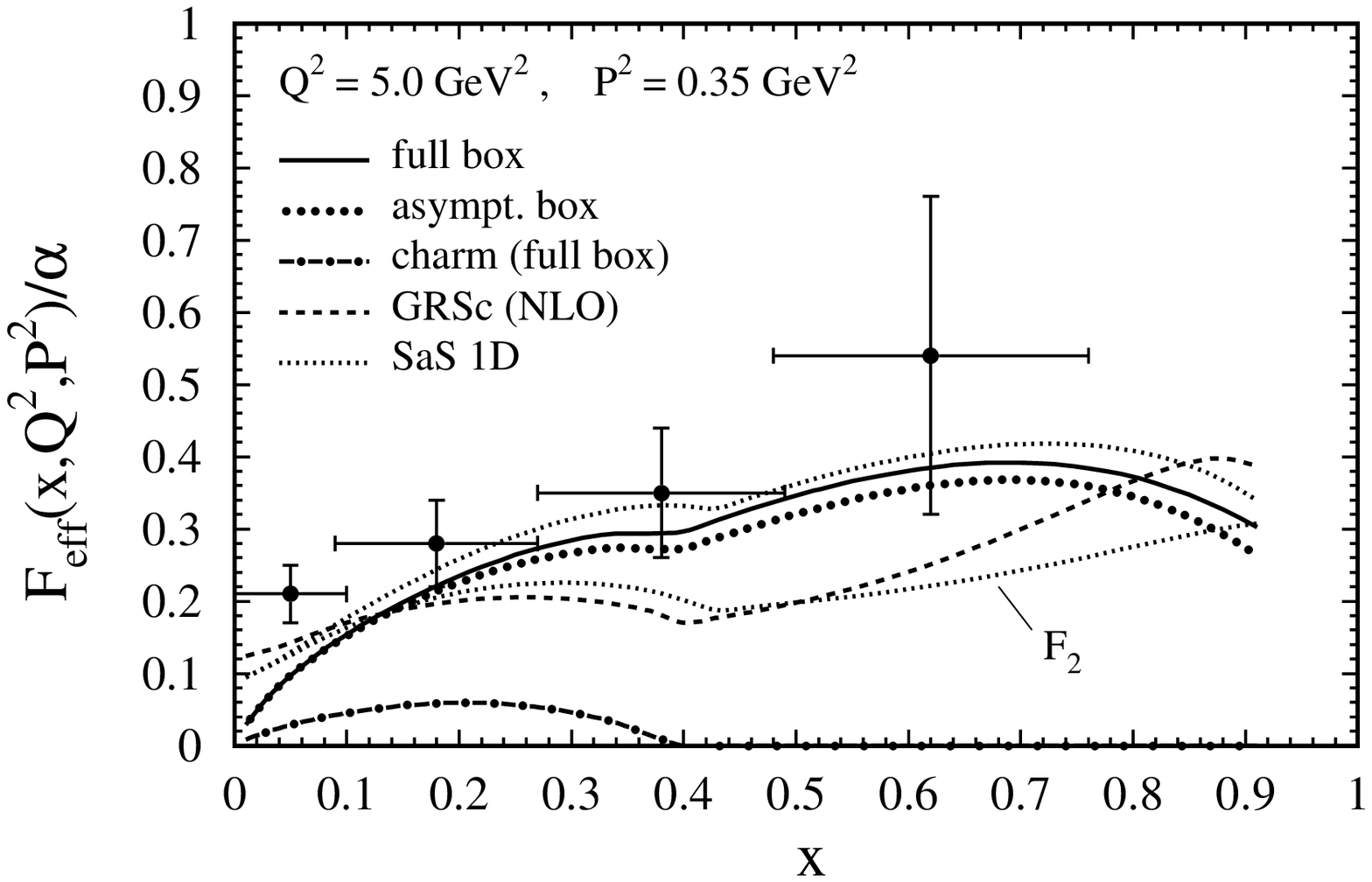,width=13cm}
\vspace*{-0.5cm}
\caption{\sf
      Predictions for $\Feff$ as defined in (\ref{eq:vgam2.3}). The box 
      results are as in Fig.~\ref{fig:feffbox_fig1}.
      The QCD resummed NLO expectations of 
      GRSc \protect\cite{cit:GRSc99} for $F_2$ in (\ref{eq:vgam3.4}) turn out to be 
      similar to the LO ones \protect\cite{cit:GRSc99}.  
      Also shown are the LO--resummed results of SaS 1D 
      \protect\cite{cit:SaS-9501,*Schuler:1996fc} for 
      $F_2$ and $\Feff = F_2 +\frac{3}{2}\, F_{\rm LT}$ (see text). The total 
      charm contribution to the latter two QCD results involves also a `resolved' 
      component $\sfs{2,h}{\gam(P^2),\mathrm{res}}(x,Q^2)$ 
      according to Eq.~(\ref{eq:f2gamhres}) which 
      turns out to be small as compared 
      to the box contribution shown which dominates in the kinematic region  
      considered. The PLUTO data are taken from \protect\cite{Berger:1984xu}. 
}
\label{fig:vgamfig1}
\end{figure}
\begin{figure}
\centering
\epsfig{figure=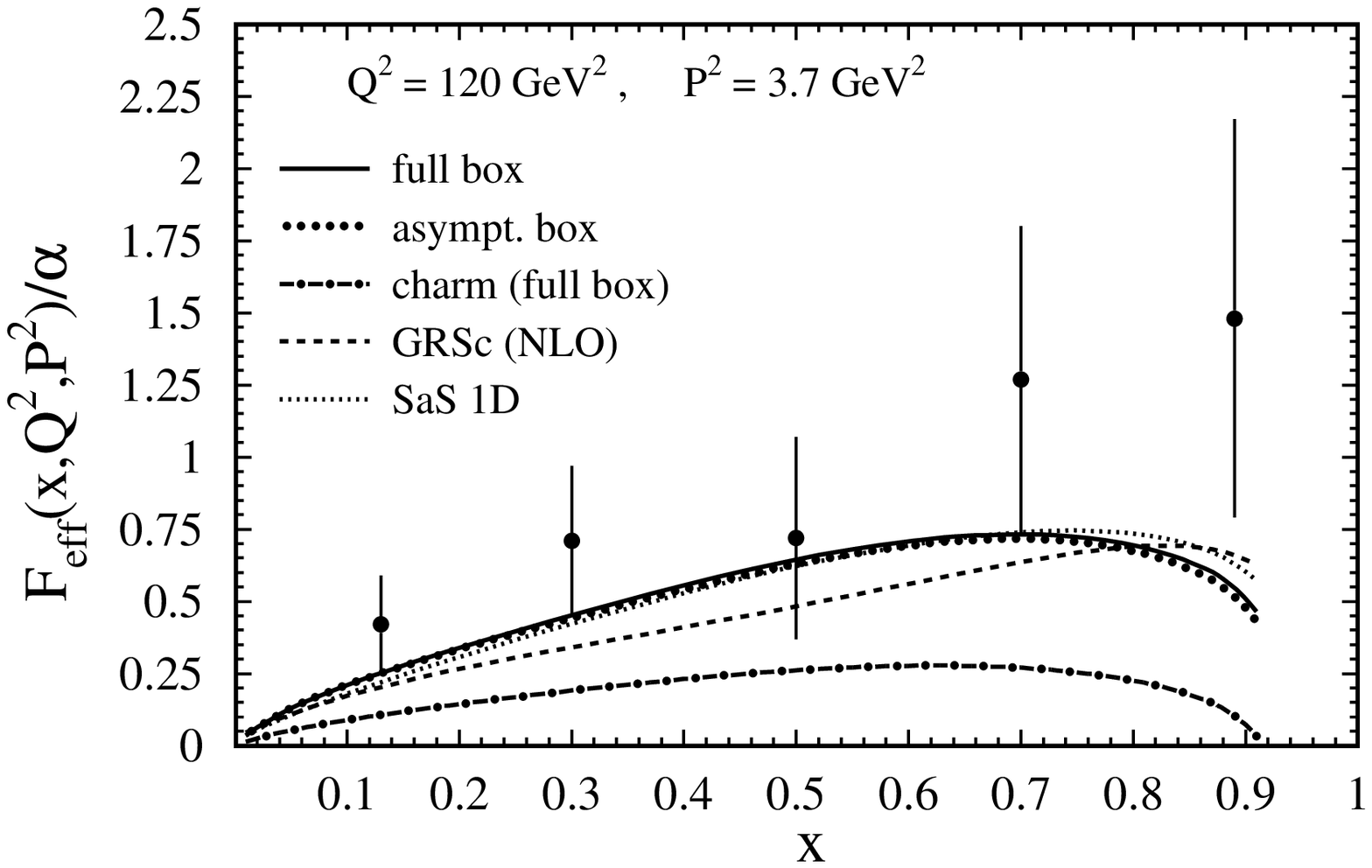,width=13cm}
\vspace*{-0.5cm}
\caption{\sf
      As in Fig.\ \ref{fig:vgamfig1}, but for $Q^2=120$ GeV$^2$ and $P^2=3.7$ GeV$^2$ 
      appropriate for the LEP--L3 data \protect\cite{Acciarri:2000rw}. 
}
\label{fig:vgamfig2}
\end{figure}

As promised in Chap.~\ref{chap:lobox}
we shall now turn to a quantitative study of the various QED--box and QCD  
$Q^2$--evoluted structure function expectations for a virtual photon target 
and confront them with all presently available $e^+e^-$ data of PLUTO  
\cite{Berger:1984xu} and the recent one of LEP--L3 \cite{Acciarri:2000rw}.  
The 'asymptotic' and 'full' box results for $\Feff$ have already been 
presented in Chap.~\ref{chap:lobox} and details of their calculation can be found there. 
The QCD resummed expectations for $\Feff$ 
according to our model in the previous chapter are given by
\begin{equation}\label{eq:vgam3.4}   
\Feff(x,Q^2,P^2) =  \sfs{2}{\gam(P^2)}(x,Q^2)
\end{equation} 
since we neglect any effects due to longitudinal target photons in our approach
\cite{cit:GRSc99}.

A different approach has been suggested by Schuler and Sj\"ostrand 
\cite{cit:SaS-9501,*Schuler:1996fc}. 
Apart from using somewhat different input scales $Q_0$ and parton densities,  
the perturbatively exactly calculable box expressions for $\Lambda^2\ll P^2\ll Q^2$ 
(cf.~App.~\ref{sec:bjorken2}) are, together with 
their LO--QCD $Q^2$--evolutions, extrapolated  
to the case of real photons $P^2=0$ by employing some dispersion--integral--like 
relations.  These link perturbative and non--perturbative contributions and  
allow a smooth limit $P^2\to 0$. (Note, however, that the LO $Q^2$--evolutions 
are performed by using again the splitting functions of real photons and  
on--shell partons.)  Since one works here explicitly with virtual ($P^2\neq 0)$ 
expressions, the longitudinal contributions of the virtual photon target  
should be also taken into account when calculating 
$\Feff = F_2 + \frac{3}{2} x F_{\rm LT}$ with 
$F_{ab} \equiv Q^2/(4 \pi^2 \alpha) (x \vtwo)^{-1} \sigma_{ab}$ ($a = (\mathrm{L,T})$,
$b = (\mathrm{L,T})$) and where $\sigma_{ab}$ is given in Eq.~(\ref{eq:LO-box}), as 
described for example in \cite{cit:GRS95}, which 
is in contrast to our approach in (\ref{eq:vgam3.4}).\footnote{ 
In an alternative approach \protect\cite{Chyla:2000hp,Chyla:2000cu} 
one may consider the longitudinal 
component of the virtual photon target $\gam[L](P^2)$ to possess,  
like the transverse component, a universal process independent hadronic content 
obtained radiatively via the standard homogeneous (Altarelli--Parisi)  
$Q^2$--evolution equations with the boundary conditions for the pointlike  
component at $Q^2=P^2$ given in Eq.~(\ref{eq:bc_gamL}).  
We have checked that the  
predictions for $\Feff(x,Q^2,P^2)$ obtained in this approach differ 
only slightly (typically about 10\% or less) 
 from those of the standard fixed order perturbative approach at 
presently relevant kinematical regions  
$(P^2$ \raisebox{-0.1cm}{$\stackrel{<}{\sim}\,$} $\frac{1}{10}Q^2$, 
$x$ \raisebox{-0.1cm}{$\stackrel{>}{\sim}$} 0.05) 
due to the smallness of $F_{\rm TL}^{\ell}$ relative to $F_{\rm TT}^{\ell}$ 
deriving from $\sigtl^{\ell}$ and $\sigtt^{\ell}$, respectively, in (\ref{eq:vgam2.6}).} 

Despite the limited 
statistics of present data the box predictions for $\Feff$ in (\ref{eq:vgam2.3}) 
shown in Figs.\ \ref{fig:vgamfig1} and \ref{fig:vgamfig2} appear to be in even better 
agreement with present 
measurements than the QCD resummed expectations of 
SaS \cite{cit:SaS-9501,*Schuler:1996fc} and  
GRSc \cite{cit:GRSc99}.  Typical QCD effects like the increase in the small--x  
region in Fig.\ \ref{fig:vgamfig1}, being partly caused by the presence of a finite gluon 
content $\gvg(x,Q^2)$, cannot be delineated with the present poor 
statistics data. 
 
These results clearly demonstrate that the naive QPM predictions derived 
from the doubly virtual box $\vgvg$ 
fully reproduce all $e^+e^-$ data on the structure of virtual 
photons $\gamma(P^2)$.  In other words, there is {\em{no}} sign of a 
QCD resummed parton content in virtual photons in present data, in particular 
of a finite gluon content $\gvg(x,Q^2)$ which is absent in the  
`naive' box (QPM) approach. 
 
Characteristic possible signatures for QCD effects which are caused by the 
presence of a finite and dominant gluon component $\gvg$ will be  
discussed in Sec.\ \ref{sec:vgam6}. 

\section[Comparison with DIS $ep$ Data and Effective Quark Distributions]{Comparison of Theoretical Expectations with DIS $ep$ Data and Effective 
Quark Distributions of Virtual Photons} 
\label{sec:vgam5}
\begin{figure}
\centering
\epsfig{figure=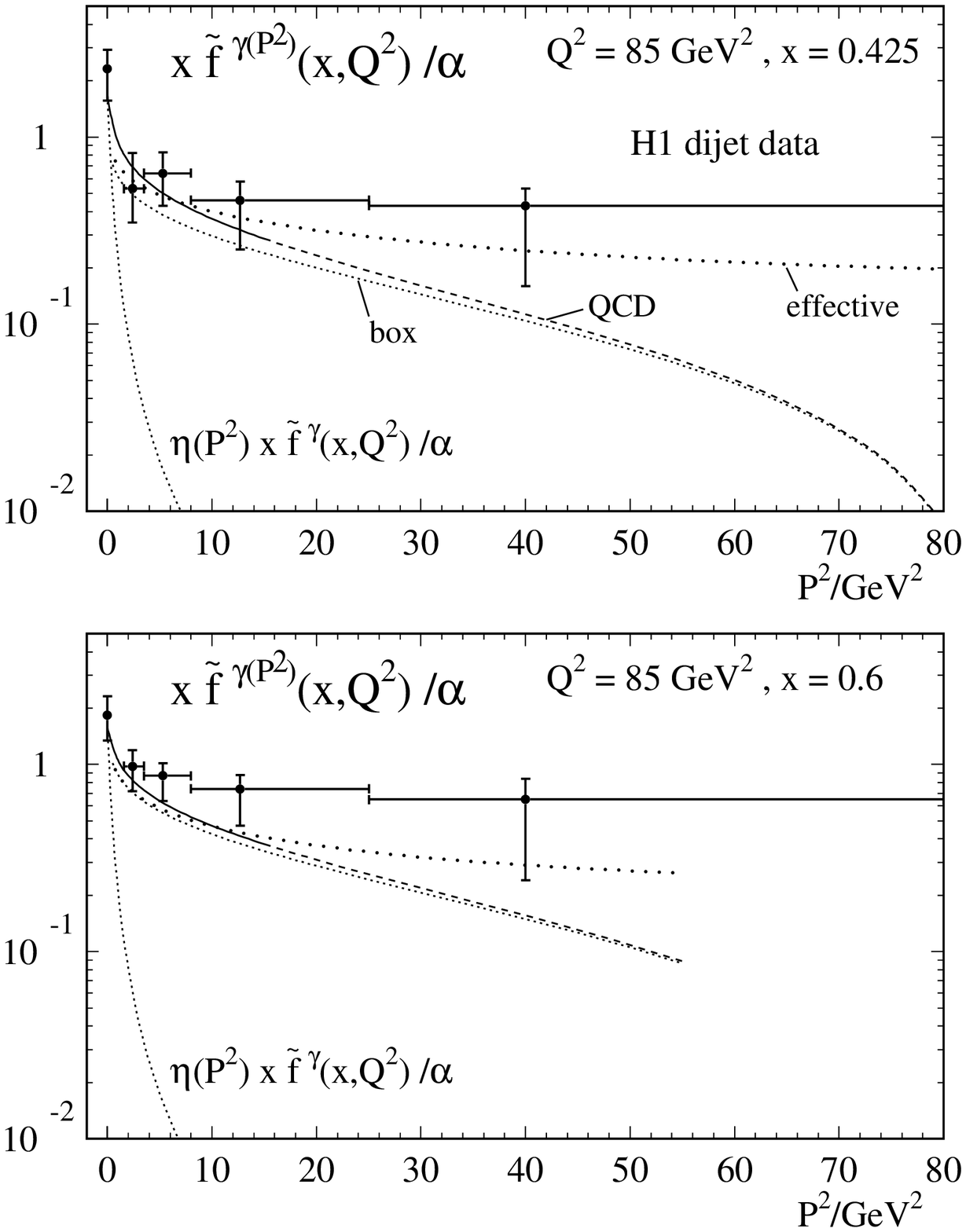,width=14cm}
\caption{\sf
      Predictions for the effective parton density defined in Eq.\ (\ref{eq:vgam5.1}). 
      The `box' results refer to the universal $q_{\rm box}^{\gamma(P^2)}$ in 
      (\ref{eq:vgam2.10}), 
      and the `effective' ones to $q_{\rm eff}^{\gamma(P^2)}$ 
      as defined in (\ref{eq:vgam5.2}) as  
      derived from the full box expressions (\ref{eq:LO-box}) 
      including all $\Ord(P^2/Q^2)$ contributions.  
      The LO--QCD predictions of GRSc \protect\cite{cit:GRSc99} are shown by the solid 
      curves which refer to the predictions in the theoretically legitimate region 
      $P^2\ll Q^2$, whereas the dashed curves extend into the kinematic region of larger 
      $P^2$ approaching $Q^2$ where the concept of QCD--resummed parton distributions 
      of virtual photons is not valid anymore. (Note that the results for $x=0.6$ 
      terminate at $P^2\simeq 54\ \gevsq$ 
      due to the kinematic constraint $W^2>0$, with $W^2$ being defined in 
      (\ref{eq:variables}), i.e.\ $x<(1+P^2/Q^2)^{-1}$.) For illustration we also show the 
      effective LO--QCD parton density $\tilde{f}^{\gamma}$ of a real photon 
      $\gamma\equiv \gamma(P^2=0)$ of GRSc \protect\cite{cit:GRSc99} multiplied by the 
      simple $\rho$--pole suppression factor $\eta(P^2)$ in (\ref{eq:gam24}) 
      which clearly underestimates the H1--data \protect\cite{Adloff:1998vc}. 
}
\label{fig:vgamfig3}
\end{figure}

In oder to extract the parton densities of virtual photons from DIS $ep$ dijet 
data, the H1 collaboration \cite{Adloff:1998vc} has adopted the `single effective 
subprocess approximation' \cite{Combridge:1984jn} which exploits the fact that the 
dominant contributions to the cross section in LO--QCD comes from the $2\to 2$ 
parton--parton hard scattering subprocesses that have similar shapes and thus 
differ mainly by their associated color factors.  Therefore the sum over the 
partonic subprocesses can be replaced by a single effective subprocess cross 
section and effective parton densities for the virtual photon given by 
\begin{equation}\label{eq:vgam5.1}    
 \tilde{f}\,^{\gamma(P^2)}(x,Q^2) = \sum_{{\rm q=u,\, d,\, s}} 
  \left[ q^{\gamma(P^2)}(x,Q^2)\, + \bar{q}\,^{\gamma(P^2)}(x,Q^2) \right] 
   + \frac{9}{4}\,g^{\gamma(P^2)}(x,Q^2)  
\end{equation} 
with a similar relation for the proton $\tilde{f}\,^p(x,Q^2)$ which is  
assumed to be known.  It should be emphasized that such an effective 
procedure does not hold in NLO where all additional (very different) $2\to 3$ 
subprocesses contribute 
\cite{Klasen:1998jm,*Kramer:1998bc,*Potter:1998jt,*Potter:1997ii}. 
This NLO analysis affords therefore a 
confrontation with more detailed data on the triple--differential dijet 
cross section as compared to presently available data \cite{Adloff:1998vc} which 
are not yet sufficient for examining the relative contributions of  
$\qvg(x,Q^2)$ and $\gvg(x,Q^2)$.  In Fig.\ \ref{fig:vgamfig3} we  
compare our LO RG--resummed predictions for $\tilde{f}\,^{\gamma(P^2)}(x,Q^2)$ 
with the naive non--resummed universal (process independent) box expressions 
[cf.~Sec.~\ref{sec:pert_bc}]
\begin{equation}\label{eq:vgam2.10}
\pdfb[q]{box}{\gam(P^2)}(x,Q^2) = \pdfb[\bar{q}]{box}{\gam(P^2)}(x,Q^2)
= 3 e_q^2 \frac{\alpha}{2\pi} \left[ x^2+(1-x)^2\right] \, \ln \frac{Q^2}{P^2} \ .
\end{equation}
Although the fully QCD--resummed results are sizeable and somewhat 
larger in the small $P^2$ region than the universal box expectations, present 
H1 data \cite{Adloff:1998vc} at $Q^2\equiv (p_{\rm T}^{\rm jet})^2 = 85$ GeV$^2$ 
cannot definitely distinguish between these predictions.  It should be furthermore 
noted that the QCD gluon contribution $\gvg(x,Q^2)$ is suppressed 
at the large values of $x$ shown in Fig.\ \ref{fig:vgamfig3}.  
Therefore present data \cite{Adloff:1998vc} 
cannot discriminate between the finite QCD resummed component 
$\gvg(x,Q^2)$ and the non--resummed $g_{\rm box}^{\gamma(P^2)}(x,Q^2) = 0$. 
 
It is obvious that these two results shown in Fig.\ \ref{fig:vgamfig3} are only appropriate  
for virtualities $P^2\ll Q^2$, typically $P^2=10$ to $20\ \gevsq$ at 
$Q^2 = 85\ \gevsq$, since $\Ord(P^2/Q^2)$ contributions are neglected in  
RG--resummations as well as in the definition (\ref{eq:vgam2.10}).  
In order to demonstrate 
the importance of $\Ord(P^2/Q^2)$ power corrections in the large $P^2$  
region let us define, generalizing the definition (\ref{eq:def_bclobox}), some effective 
(anti)quark distributions as common via 
\begin{equation}\label{eq:vgam5.2}     
\frac{1}{x}\, 
\sfs{2,box}{\ell}(x,Q^2) \equiv 
  \sum_{\rm q=u,\, d,\, s} e_q^2 \left[ q_{\rm eff}^{\gamma(P^2)}(x,Q^2) 
      + \bar{q}_{\rm eff}^{\gamma(P^2)}(x,Q^2)\right] 
\end{equation} 
where, of course, $q_{\rm eff}^{\gamma(P^2)}=\bar{q}_{\rm eff}^{\gamma(P^2)}$ 
and the full box expression for 
$\sfs{2,box}{\ell}(x,Q^2)$ 
can be obtained from the general definition in Eq.~(\ref{eq:sfs_av}) and 
the full box results in Eq.~(\ref{eq:LO-box})
utilizing $m\equiv m_q=0$, i.e.\ $\lambda =0$ and summing over the light quarks.
The full box expressions imply again $g_{\rm eff}^{\gamma(P^2)}(x,Q^2)=0$ 
in contrast to the QCD resummed gluon distribution.  The $q_{\rm eff}^{\gamma(P^2)}$ 
introduced in (\ref{eq:vgam5.2}) is, in contrast to (\ref{eq:vgam2.10}), of course 
non--universal.  The `effective' results shown in 
Fig.\ \ref{fig:vgamfig3} clearly demonstrate the importance of 
the $\Ord(P^2/Q^2)$ terms at larger values of  
$P^2$ \raisebox{-0.1cm}{$\stackrel{<}{\sim}\,$} $Q^2$  
which are not taken into account by the QCD resummations and by the universal 
box expressions in (\ref{eq:vgam2.10}) also shown in Fig.\ \ref{fig:vgamfig3}.  
It is interesting that the 
non--universal $q_{\rm eff}^{\gamma(P^2)}$ defined via 
$F_2$ in (\ref{eq:vgam5.2}) describes 
the H1--data at large values of $P^2$ in 
Fig.\ \ref{fig:vgamfig3} remarkably well.  This may be 
accidental and it remains to be seen whether future LO analyses will indicate 
the general relevance of $q_{\rm eff}^{\gamma(P^2)}(x,Q^2)$ in the large $P^2$ 
region. 
 
As we have seen, present DIS dijet data cannot discriminate between the universal 
naive box and QCD--resummed expectations in the theoretically relevant region 
$P^2\ll Q^2$, mainly because these data are insensitive to the gluon content in 
$\gamma(P^2)$ generated by QCD--evolutions which is absent within the naive box 
approach.  Therefore we finally turn to a brief discussion where such typical 
QCD effects may be observed and delineated by future experiments. 
 
\section{Possible Signatures for the QCD Parton Content of Virtual Photons} 
\label{sec:vgam6}
\begin{figure}
\centering
\epsfig{figure=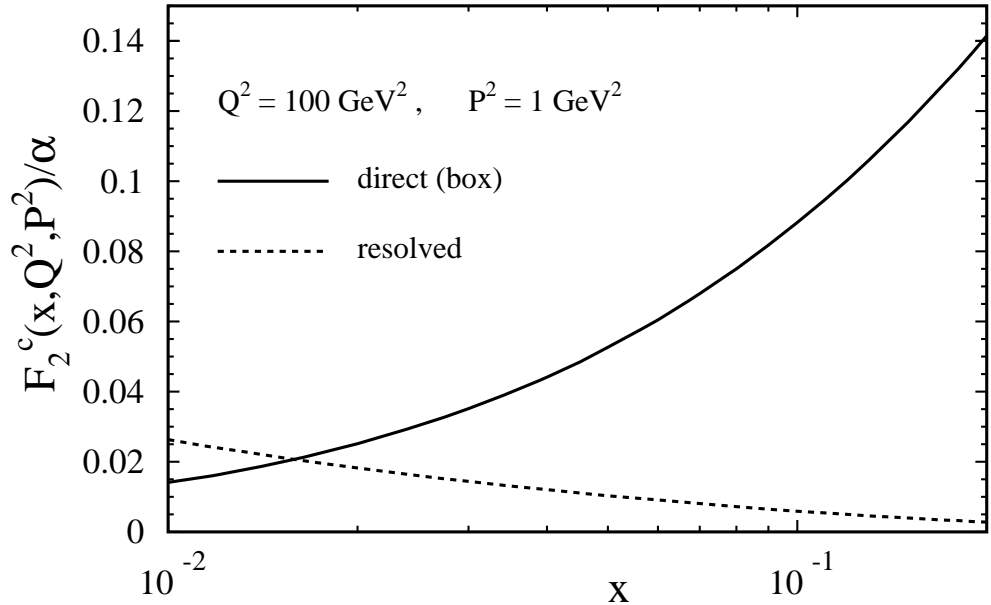,width=14cm}
\caption{\sf
      Expected charm contributions to $F_2$.  The naive `direct (box)' 
      result refers to (\ref{eq:f2gamh}) [or (\ref{eq:vgamA.8})]
      and the LO--QCD  `resolved' 
      prediction is given  
      in (\ref{eq:f2gamhres}) with $g^{\gamma(P^2)}(x,\, 4m_c^2)$ taken 
      from GRSc \protect\cite{cit:GRSc99}.  
      This 
      latter `resolved' contribution is absent in the naive box (QPM) approach. 
}
\label{fig:vgamfig4}
\end{figure}

\begin{figure}
\centering
\epsfig{figure=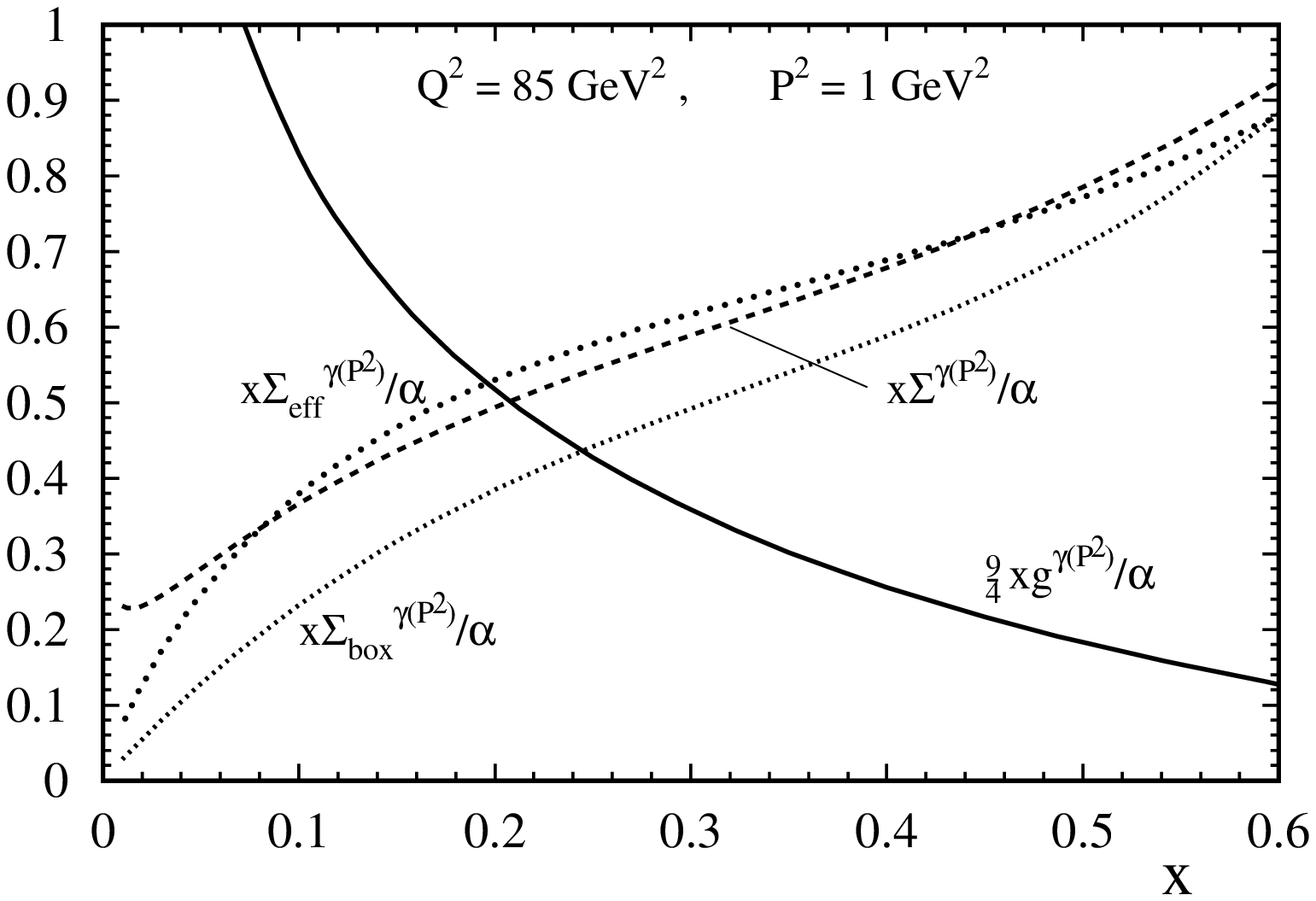,width=14cm}
\epsfig{figure=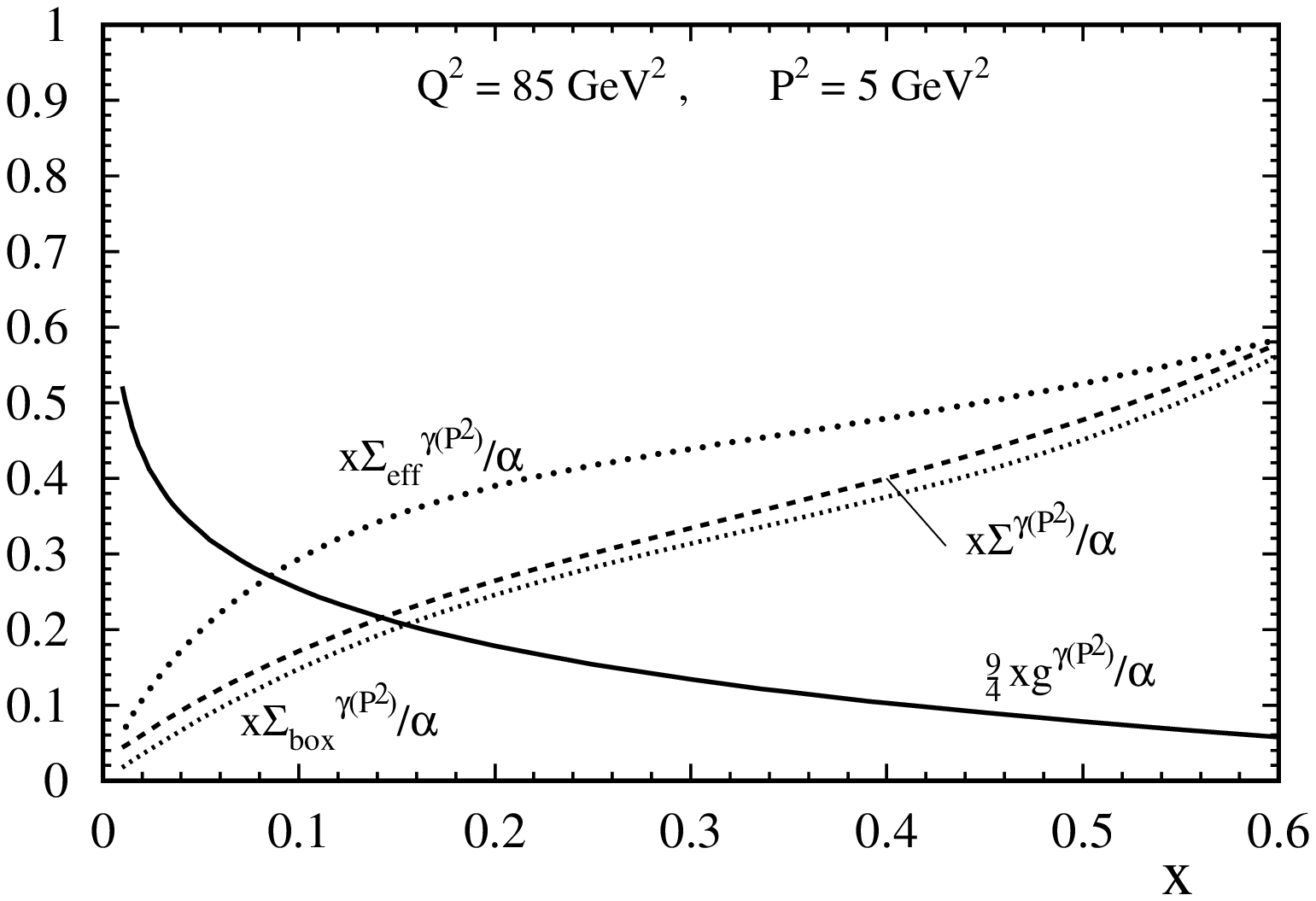,width=14cm}
\caption{\sf
      Predictions for the total light quark 
      $\Sigma^{\gamma(P^2)}\equiv 2\Sigma_{\rm{q = u,\, d,\, s}}\,q^{\gamma(P^2)}$ 
      and gluon contributions to the effective parton density in (\ref{eq:vgam5.1}) 
      at a fixed scale 
      $Q^2=85\ \gevsq$ and two fixed virtualities $P^2=1$ and $5\ \gevsq$.  The naive box 
      results refer to the universal $q_{\rm box}^{\gamma(P^2)}$ defined 
      in (\ref{eq:vgam2.10}) and 
      to $q_{\rm eff}^{\gamma(P^2)}$ defined in (\ref{eq:vgam5.2}).  
      The LO--QCD RG--resummed predictions 
      are denoted by $\Sigma^{\gamma(P^2)}$ and $g^{\gamma(P^2)}$ according 
      to GRSc \protect\cite{cit:GRSc99}. The latter gluon contribution is 
      absent in the naive box (QPM) approach. 
}
\label{fig:vgamfig5}
\end{figure}

Since $e^+e^-$ and DIS $ep$ dijet data cannot, at present, delineate the  
QCD--resummed parton content of a virtual photon, in particular not its gluon 
content, we shall now propose and discuss a few cases where such typical QCD 
effects may be observed and possibly confirmed by future experiments. 
 
Charm production in $e^+e^-\to e^+e^-\, c\bar{c}X$ would be a classical possibility 
to delineate such effects due to a nonvanishing $\gvg(x,Q^2)$.  In  
Fig.\ \ref{fig:vgamfig4} we compare the usual (fixed order) `direct' box 
contribution to $F_2^c$ 
with the  `resolved' gluon--initiated one as 
given by (\ref{eq:f2gamhres}).  The 
`direct' box contribution entirely dominates in the large--$x$ region,   
$x$ \raisebox{-0.1cm}{$\stackrel{>}{\sim}$} 0.05, accessible by present experiments 
(cf.\ Figs.\ \ref{fig:vgamfig1} and \ref{fig:vgamfig2}), whereas 
the typical QCD--resummed `resolved' contribution 
becomes comparable to the  `direct' one and eventually dominates in the small--$x$ 
region, $x<0.05$.  Thus a careful measurement of the charm contribution to $F_2$ 
at  
$x$ \raisebox{-0.1cm}{$\stackrel{<}{\sim}$} 0.05 would shed some light on the QCD 
parton (gluon) content of virtual photons, since such a `resolved' contribution 
in Fig.\ \ref{fig:vgamfig4} would be absent within the naive box approach. 
 
The effective parton distribution 
$\tilde{f}^{\gamma(P^2)}(x,Q^2)$ in (\ref{eq:vgam5.1}) at 
not too large values of $x$ and $P^2$, as may be extracted in LO from DIS $ep$ dijet 
data, would be another possibility to observe QCD--resummation effects due to a 
nonvanishing gluon component $\gvg(x,Q^2)$.  
In Fig.\ \ref{fig:vgamfig5} we show the quark 
and gluon contributions to $\tilde{f}^{\gamma(P^2)}$ in (\ref{eq:vgam5.1}) separately.  
The box 
(anti--)quark contributions, which are similar to the QCD--resummed ones, entirely 
dominate over the QCD--resummed gluon contribution in the large--$x$ region,  
$x$ \raisebox{-0.1cm}{$\stackrel{>}{\sim}$} 0.4,  
accessible to present experiments (cf.\ Fig.\ \ref{fig:vgamfig3}).  
Only {\em below} $x\simeq 0.3$ 
does the QCD gluon contribution become comparable to the (anti--)quark components  
and dominates, as usual, for  
$x$ \raisebox{-0.1cm}{$\stackrel{<}{\sim}$} 0.1. 
It should be remembered that $g_{\rm box}^{\gamma(P^2)}(x,Q^2)=0$.  Furthermore, 
the increase of the RG--resummed   
$\qvg(x,Q^2)$ at small $x$ in Fig.\ \ref{fig:vgamfig5} 
is induced by the vector--meson--dominance--like input for the $Q^2$--evolution 
of the `hadronic' component of the photon's parton distribution 
\cite{cit:SaS-9501,*Schuler:1996fc,cit:GRSc99} and 
is disregarded in our naive box (QPM) analysis. 
 
Thus a measurement of dijets produced in DIS $ep$ reactions in the not too large 
$x$ region,  
$x$ \raisebox{-0.1cm}{$\stackrel{<}{\sim}$} 0.3, 
would probe the QCD parton content of virtual photons, in particular their gluon 
content which is absent in the naive QPM box approach.  In this region, and at not 
too large photon virtualities  
$P^2$ \raisebox{-0.1cm}{$\stackrel{<}{\sim}$} 5 GeV$^2$ 
shown in Fig.\ \ref{fig:vgamfig5}, the  `resolved' gluon--dominated contribution 
of the virtual photon to high $E_{\rm T}$ jet production at scales 
$Q\equiv E_{\rm T} \simeq 5 - 10$ GeV exceeds by far the `direct' box--like 
contribution of a pointlike virtual photon \cite{Gluck:1996ra}.      
The production of prompt photons at HERA via a tagged DIS process
$e p \to e \gamma X$ may offer additional probes of the gluonic
content of virtual photons \cite{Krawczyk:1999eq}.
\section{Summary and Conclusions} 
\label{sec:vgam7}
Virtual photons $\gamma(P^2)$, probed at a large scale $Q^2\gg P^2$, may be 
described either by fixed order perturbation theory, which in lowest order of  
QCD yield the quarks and anti--quarks generated by the universal part of the `box' 
diagram, or alternatively by their renormalization group (RG) improved counterparts 
including particularly the gluon distribution $\gvg(x,Q^2)$. 
 
The results in Sections \ref{sec:vgam4} and \ref{sec:vgam5} demonstrate 
that all presently available $e^+e^-$ 
and DIS $ep$ dijet data can be fully accounted for by the standard doubly virtual 
QED box diagram and are not yet sensitive to RG resummation effects which are manifest 
only in the presently unexplored low--$x$ region of the parton distributions in 
$\gamma(P^2)$. In fact, as shown in Section \ref{sec:vgam6}, these 
resummation effects start to 
dominate only at $x<0.3$ and may be observed by future measurements at 
$P^2=\Ord(1\ \gevsq)$
of $\sigma(ep\to e\, jjX)$ or $\sigma(e^+e^- \to e^+e^-\, c\bar{c}X)$  
at high energy 
collisions.  These measurements could finally discriminate between the fixed order 
and RG--improved parton distributions of the virtual photon.

%% file: phd_gam_summary.tex
\chapter{Summary}
\label{chap:phd_gam_summary}
It was the main theme of Part II of this thesis to analyze the
parton content of pions and real and virtual photons 
in leading order (LO) and next--to--leading order (NLO) of QCD
within
the framework of the recently updated \cite{cit:GRV98}
radiative parton model 
\cite{cit:GRV90,Gluck:1992ng,Gluck:1993im,cit:GRV94,cit:GRV98}.

We started with a detailed consideration of the $\eetoeeX$ 
scattering process being presently the main source of information
on the structure of real and virtual photons.
We defined structure functions for virtual photons in a 
model independent way and related them to the photon--photon
scattering cross sections.
These structure functions can be measured in
deep inelastic electron--photon scattering ($\DISeg$) 
and we demonstrated the factorization of the cross section
for the process $\eetoeeX$ into a flux of photons (produced
by bremsstrahlung off the incident lepton) times the $\DISeg$
cross section for the general $P^2 \ne 0$ case in the Bjorken
limit $P^2 \ll Q^2$.
However, we found that the neglected terms are of the
order $\Ord(\sqrt{x P^2/Q^2})$ only and not of the order 
$\Ord(x P^2/Q^2)$ as might have been naively expected.
Clearly, a numerical comparison 
of the factorization formula with the exact $\eetoeeX$ cross section 
would be interesting in order to clarify when $x P^2/Q^2$ is small
enough such that the factorization formula is a good approximation.

From these general kinematical considerations we turned to a 
calculation of the photon--photon cross sections $\sigma_{ab}$ 
according to the doubly virtual box $\vgvg$ in lowest order
perturbation theory.
We utilized these results in order to compute the effective
structure function $\Feff(x,Q^2,P^2)$ in fixed order perturbation
theory (FOPT) which we compared with all present $e^+ e^-$ virtual
photon data \cite{Berger:1984xu,Acciarri:2000rw}. 
It came out that
the LO--box predictions for $\Feff$ are in agreement with the
present low statistics data 
even in the case of the PLUTO data
\cite{Berger:1984xu} with $P^2 = 0.35\ \gevsq$ which is just 
in the transition
region from deeply virtual ($P^2 \gg \Lambda^2$) to real ($P^2 \simeq 0$)
photons where non--perturbative effects are expected to become
increasingly important, especially in the small--$x$ region.
More precise data for $\Feff$ in this 'transition region' with intermediate 
target virtualities would be highly desirable in order to observe
the onset of such non--perturbative effects.
On the other hand, NLO corrections to the doubly virtual box
\cite{nlobox} offer the possibility to investigate 
the perturbative stability of
the fixed order results and to see if the description of the
data further improves in NLO accuracy.

%
%
After presenting the complete theoretical framework
we started our phenomenological studies 
of radiatively generated parton distributions
with an analysis of the parton content of the pion in LO and NLO QCD.
Since only the pionic valence density $v^{\pi}(x,Q^2)$ is experimentally rather
well known at present, we utilized a constituent quark model
\cite{Altarelli:1974ff,Hwa:1980pn}  
in order to unambiguously relate the pionic light sea 
$\bar{q}\,^{\pi}(x,Q^2)$
and gluon $g^{\pi}(x,Q^2)$ 
to the 
much better known (recently updated) parton distributions of the 
proton \cite{cit:GRV98}
and the pionic valence density $v^{\pi}(x,Q^2)$.

Next we formulated a consistent
set of boundary conditions which allowed for a perturbatively stable
LO and NLO calculation of the photonic parton distributions
$f^{\gamma(P^2)}(x,Q^2)$ as well as for a smooth transition to the
parton densities of a real $(P^2=0)$ photon.  
Employing the above summarized 
pionic distributions $f^{\pi}(x,Q^2)$, required
for describing --via vector meson dominance (VMD)-- the 
nonpointlike hadronic components of a 
photon, we arrived at essentially parameter--{\em{free}}
predictions for $f^{\gamma(P^2)}(x,Q^2)$ which turned out to be in
good agreement with all present measurements of the structure function
$F_2^{\gamma}(x,Q^2)$ of real photons $\gamma\equiv\gamma(P^2=0)$.
More specifically, we implemented these hadronic 
components by using a VMD ansatz for a coherent superposition of 
vector mesons which maximally enhances the contributions of the 
up--quarks to $F_2^{\gamma}(x,Q^2)$ as favored by all present data.  Since
these hadronic contributions are generated from the valence--like 
input parton distributions at the universal target--mass independent
low resolution scale $Q_0^2=\mu^2\simeq 0.3$ GeV$^2$, we arrived, at
least for real ($P^2=0$) photons, at unique small--$x$ predictions 
for $x$ \raisebox{-0.1cm}{$\stackrel{<}{\sim}$} $10^{-2}$ at $Q^2>\mu^2$
which are of purely dynamical origin, as in the case of hadrons. 
A comparison with most recent small--$x$ measurements
of the photon structure function $F_2^\gamma(x,Q^2)$ \cite{Abbiendi:2000cw}
showed that these small--$x$ predictions are in perfect agreement 
with the data at all presently accessible values of $Q^2$.
The agreement with the data on $F_2^\gamma(x,Q^2)$ confirms 
our radiatively generated quark distributions 
$q^{\gamma}(x,Q^2)=\bar{q}^{\gamma}(x,Q^2)$ while the gluon is probed
only indirectly at small--$x$ due to the evolution.
It would be also interesting and important to extend present 
measurements \cite{Ahmed:1995wv,Adloff:1998vt,Erdmann-9701,Nisius:1999cv} of the 
gluon distribution of 
the photon, $g^{\gamma}(x,Q^2)$, below the presently 
measured region  
0.1 \raisebox{-0.1cm}{$\stackrel{<}{\sim}$} $x<1$  
where $g^{\gamma}(x<0.1,\,Q^2)$ is expected to be 
somewhat flatter in the small--$x$ region than 
previously anticipated \cite{cit:GRV-9202}. 
Our expectations for the parton content of virtual ($P^2\neq0$) photons
are clearly more speculative, depending on how one models the hadronic
component (input) of a virtual photon.  We assumed the latter
to be similar to the VMD input for a real photon times a dipole
suppression factor which derives from an effective vector--meson
$P^2$--propagator representing somehow the simpliest choice.  
We formulated similar boundary conditions in LO and NLO which gave rise
to perturbatively stable parton distributions and cross sections
(i.e.\ also structure functions of virtual photons) as long as they
are probed at scales $Q^2 \gg P^2$ where $Q^2$ is a hard scale of the
process under consideration.
We showed that
effective parton distributions of virtual photons extracted recently
from DIS $ep$ dijet data are in good agreement with our (parameter--free) 
predictions.

Finally, we turned to more detailed tests of our model for the parton content
of virtual photons and we compared these QCD--resummed expectations with
the above summarized fixed order box expressions.
Our results demonstrated 
that all presently available $e^+e^-$ 
and DIS $ep$ dijet data can be fully accounted for by the standard doubly virtual 
QED box diagram and are not yet sensitive to RG resummation effects which are manifest 
only in the presently unexplored low--$x$ region of the parton distributions in 
$\gamma(P^2)$. 
It turned out that these
resummation effects start to 
dominate only at $x \lesssim 0.3$ (the larger $P^2$ the smaller $x$)
and may be observed by future measurements at 
$P^2=\Ord(1\ \gevsq)$
of $\sigma(ep\to e\, jjX)$ or $\sigma(e^+e^- \to e^+e^-\, c\bar{c}X)$  
at high energy 
collisions.  These measurements could finally discriminate between the fixed order 
and RG--improved parton distributions of the virtual photon.  

FORTRAN packages containing simple analytic parametrizations
of our most recent parton distributions of the pion and the (real and virtual) 
photon are available upon request by electronic mail from 
schien@zylon.physik.uni-dortmund.de.

%% file: phd_app_hqini2.tex
\chapter{Deep Inelastic Scattering on Massive Quarks at $\alpsi$} 
\label{hqini}

\section{Real Gluon Emission}
\label{rge}

We define a partonic tensor
\begin{equation}
\label{ptensor}
{\hat{\omega}}^{\mu\nu} \equiv 
{\overline{\sum_{{\rm{color}}}}} \sum_{\rm{spin}}
\langle Q_2(p_2),g(k)  \left| 
{\overline{Q}}_2 \gamma^\mu (V-A \gamma_5) Q_1
\right| Q_1(p_1) \rangle\ \times \langle \mu \rightarrow \nu \rangle^\ast
\end{equation} 
which can be decomposed into its different tensor components as usual
\begin{equation}
\label{comp}
{\hat{\omega}}^{\mu\nu}=
{-\hat{\omega}_1^Q}\ g^{\mu\nu}+{\hat{\omega}_2^Q}
\ p_1^{\mu} p_1^{\nu}+i{\hat{\omega}_3^Q}
\ \varepsilon_{\alpha\beta}^{\ \ \mu\nu}
p_1^{\alpha} q^{\beta}+{\hat{\omega}_4^Q}
\ q^{\mu} q^{\nu} + {\hat{\omega}_5^Q}
\ (q^{\mu} p_1^{\nu}+q^{\nu} p_1^{\mu})
\ \ \ .
\end{equation}
${\hat{\omega}}_{\mu\nu}$ can be easily calculated from the general Feynman
rules for invariant matrix elements which are customarily expressed 
as functions
of the Mandelstam variables ${\hat{s}}\equiv (p_1+q)^2$ and 
${\hat{t}}\equiv (p_1-k)^2$
to which we will refer in the following.  
Projection onto the individual ${\hat{\omega}_{i=1,2,3}^Q}$ in 
Eq.\ (\ref{comp}) is performed for 
nonzero
masses and in $n=4+2\varepsilon$ dimensions with the following operators
\begin{eqnarray} \nonumber
P_1^{\mu\nu} &=& \frac{-1}{2(1+\varepsilon)}\ \Big\{\ g^{\mu\nu}+\big[
\ m_1^2\ q^{\mu} q^{\nu}-Q^2\ p_1^{\mu}\ p_1^{\nu}-
(p_1\cdot q)(q^{\mu}p_1^{\nu}
+p_1^{\mu}q^{\nu})\big] \\ \nonumber
&\times& 4 \Delta^{-2}[m_1^2,{\hat{s}},-Q^2]\ \Big\} \\ \nonumber
P_2^{\mu\nu} &=& 2\ \Big[ -g^{\mu\nu} Q^2+4\ q^{\mu}q^{\nu}
\ \frac{2(1+\varepsilon)(p_1\cdot q)^2
- m_1^2Q^2}{\Delta^2[m_1^2,{\hat{s}},-Q^2]} \\ \nonumber
&+&4(3+2\varepsilon)Q^2\ \frac{
Q^2\ p_1^{\mu}p_1^{\nu}+(p_1\cdot q)(q^{\mu}p_1^{\nu}+p_1^{\mu}q^{\nu})}
{\Delta^2[m_1^2,{\hat{s}},-Q^2]}\Big]
\ \Big\{ (1+\varepsilon)\Delta^2[m_1^2,{\hat{s}},-Q^2]  
\Big\}^{-1}\\ 
P_3^{\mu\nu} &=&\frac{-2 i}{\Delta^2[m_1^2,{\hat{s}},-Q^2]}
\ \varepsilon^{\mu\nu}_{\ \ \lambda\kappa}
\ p_1^{\lambda}\ q^{\kappa}
\label{projectors}
\end{eqnarray}
such that $P_i \cdot {\hat{\omega}} = {\hat{\omega}}_i^Q$. 
The normalization in 
Eqs.\ (\ref{ptensor}), (\ref{comp}) is 
such that real gluon emission contributes $F_i^{R}$ to the hadronic structure 
functions via
\begin{eqnarray} \nonumber
F_1^{R} &=& \frac{1}{8\pi}\ \int_\chi^1 \frac{d \xi}{\xi}\ Q_1(\xi) \int 
d{\rm{\widehat{PS}}}\ {\hat{\omega}}_1^Q \\ \nonumber
F_2^{R} &=& \frac{2 x}{16 \pi} 
\ \int_\chi^1 \frac{d \xi}{\xi}\ \frac{\Delta^2[m_1^2,{\hat{s}},-Q^2]}{2 Q^2}
\ Q_1(\xi) \int 
d{\rm{\widehat{PS}}}\ {\hat{\omega}}_2^Q \\ 
F_3^{R} &=& \frac{1}{8\pi}\ \int_\chi^1 \frac{d \xi}{\xi}
\ \Delta[m_1^2,{\hat{s}},-Q^2]
\ Q_1(\xi) \int 
d{\rm{\widehat{PS}}}\  {\hat{\omega}}_3^Q
\label{hadronic}
\end{eqnarray}
where \cite{Gottschalk:1981rv}
\begin{equation}
\label{ps}
\int\ d{\rm{\widehat{PS}}} = \frac{1}{8\pi}\ \frac{{\hat{s}}-m_2^2}{{\hat{s}}}
\ \frac{1}{\Gamma(1+\varepsilon)}
\ \left[ \frac{({\hat{s}}-m_2^2)^2}{4\pi {\hat{s}}}
\right]^\varepsilon\ \int_0^1\ \left[y(1-y)\right]^\varepsilon\ dy
\end{equation}
is the partonic phase space.
In Eq.\ (\ref{ps}) $y$ is related to the partonic center of 
mass scattering angle
$\theta^\ast$ and the partonic Mandelstam variable ${\hat{t}}$ via
\begin{eqnarray} \nonumber
y &\equiv& \frac{1}{2}\ (1+\cos \theta^\ast) \\
  &=& \frac{1}{2\Delta[m_1^2,{\hat{s}},-Q^2]}\ \left[Q^2+m_1^2+{\hat{s}}+
\Delta[m_1^2,{\hat{s}},-Q^2]+
\frac{2{\hat{s}}({\hat{t}}-m_1^2)}{{\hat{s}}-m_2^2}\right]\ \ \ .
\end{eqnarray}
We have chosen dimensional regularization for
the soft gluon poles stemming from ${\hat{s}}\rightarrow m_2^2$ 
which arise from propagators 
in the ${\hat{\omega}}_i$ times phase space factors in $d{\rm{\widehat{PS}}}$.
In Eq.\ (\ref{hadronic}) we use \cite{cit:AEM-7901}
\begin{equation}   
(\hats-m_2^2)^{2\varepsilon-1} \sim
\left(1-\frac{\chi}{\xi}\right)^{2\varepsilon-1}=\frac{1}{2\varepsilon}
\ \delta (1-\chi/\xi)+
\frac{1}{(1-\chi/\xi)_+} + {\cal{O}}(\varepsilon)
\label{aem}
\end{equation}
which separates hard gluon emission ($\sim {\hat{f}}_i^Q$) from soft gluon
($S_i$) contributions in Eq.\ (\ref{coefficient}).  
Note that in Eqs.\ (\ref{QS1}), (\ref{coefficient}) the integration variable 
$\xi$, which is implicitly defined
in Eq.\ (\ref{ansatz}), has been changed to $\xip\equiv \chi/\xi$ for an easier
handling of the distributions. For the relation between ${\hat{s}}$ and
$\xip$ see Eq.\ (\ref{s1}) below.

Since all quark masses are kept nonzero, no poles in $y$ 
(collinear singularities)
are contained in the integration volume.
The ${\hat{f}}_i^Q$ which occur in Eq.\ (\ref{coefficient}) and which 
are given below in Eq.\ (\ref{fis}) are therefore 
straightforward integrals of the ${\hat{\omega}}_i^Q$
\begin{equation}
{\hat{f}}_i^Q = \left(g_s^2\ C_F\right)^{-1}
\ \int_0^1\ dy\ {\hat{\omega}}_i^Q
\end{equation} 
and the $S_i$ in Eqs.\ (\ref{coefficient}), (\ref{soft}) pick up the pole 
in Eq.\ (\ref{aem})
\begin{equation}
S_i \sim \frac{1}{\varepsilon}
\ \int_0^1 dy\ \left[y(1-y)\right]^{\varepsilon}
\ \left.\left[{\hat{\omega}}_i^Q (\hats-m_2^2)^2\right]
\right|_{\xi=\chi}\ \ \ ,
\end{equation}
where the proportionality is given by kinematical and phase space factors
which must be kept up to ${\cal{O}}(\varepsilon)$.

The normalization of our hadronic structure functions in Eq.\ (\ref{hadronic}) 
can be clearly inferred from the corresponding LO results in 
Eq.\ (\ref{LO}). Nevertheless, for definiteness we also
give the hadronic differential cross section to which it corresponds
\begin{eqnarray} \nonumber
\frac{d^2 \sigma^{l,{\bar{l}}}}{dx dy} &=& \frac{1}{n_l}
\ \frac{(G_l^{B,B^\prime})^2\ (G_q^{B,B^\prime})^2\ 2 M_N E_{l}}{2 \pi}
\\ &\times& \left[S_{l,+}\ (1-y)F_2 + S_{l,+}\ y^2xF_1 
\pm R_{l,+}\ 2 y(1-\frac{y}{2}) xF_3 \right]\ \ ,
\end{eqnarray} 
where 
$(G_{l,q}^{B,B^\prime})^2 = [ {g_{l,q}^B}^2/(Q^2+M_B^2)\ {g_{l,q}^{B^\prime}}^2/(Q^2+M_{B^\prime}^2)]^{1/2}$ 
is the effective squared gauge coupling --including the gauge boson propagator--
of the $\gamma_\mu (V-A\gamma_5)$ lepton and quark current, 
respectively, and $n_l$ counts the spin
degrees of freedom of the lepton, e.g.\ $n_l=1,2$ for $l=\nu,e^-$.
The leptonic couplings $S_{l,+},R_{l,+}$ are defined analogous to 
the quark couplings in Eq.\ (\ref{couplings}). As noted below 
Eq.\ (\ref{couplings}),
$B=B^\prime$ for non--interference (pure $B$ scattering). 


\section{Vertex Correction}
\label{vcorr}
\subsection{Results}
We have calculated the vertex correction in $n=4+2\varepsilon$ dimensions
at ${\cal{O}}(\alpha_s^1 )$ 
for general masses and couplings
using the Feynman gauge.
The unrenormalized vertex $\Lambda^\mu_0$ [Fig.~\ref{qsfeyn}(c.1)] has the structure
\begin{eqnarray}\label{eq:vertexstruc}
\Lambda^\mu_0 &=& C_F \as \Gamma(1-\varepsilon) \left(\frac{Q^2}{4 \pi \mu^2}
\right)^{\varepsilon}
\Bigg\{C_{0,-}\ \gamma^\mu L_5 + C_{+}\ \gamma^\mu R_5 
\nonumber\\
&+& C_{1,-}\ m_2\ p_1^\mu\ L_5 
+ C_{1,+}\ m_1\ p_1^\mu\ R_5 
+\ C_{q,-}\ m_2 q^\mu\ L_5 
+ C_{q,+}\ m_1 q^\mu\ R_5 
\Bigg\} 
\end{eqnarray}
with
$L_5=(V-A\ \gamma_5)$, $R_5=(V+A\ \gamma_5)$.
The coefficients read
\begin{eqnarray}\label{eq:vertexformfactors}
C_{0,-}&=&\frac{1}{\varepsilon}(-1-\spp I_1)
+\Bigg[\frac{\Delta^2}{2 Q^2}
+\spp \left(1+\ln\left(\frac{Q^2}{\Delta}\right)\right)\Bigg]I_1
\nonumber\\
&+&\frac{1}{2} \ln \left(\frac{Q^2}{m_1^2}\right)
+\frac{1}{2} \ln \left(\frac{Q^2}{m_2^2}\right)
+\frac{m_2^2-m_1^2}{2 Q^2}\ln \left(\frac{m_1^2}{m_2^2}\right)
+\frac{\spp}{\Delta} \\ \nonumber &\times&
\Bigg\{
\frac{1}{2} \ln^2 \left|\frac{\Delta-\spm}{2 Q^2}\right|
+\frac{1}{2} \ln^2 \left|\frac{\Delta-\smp}{2 Q^2}\right|
-\frac{1}{2} \ln^2 \left|\frac{\Delta+\spm}{2 Q^2}\right|
-\frac{1}{2} \ln^2 \left|\frac{\Delta+\smp}{2 Q^2}\right|
\nonumber\\
&-&\ {\rm Li}_2 \left(\frac{\Delta-\spm}{2 \Delta}\right)
-{\rm Li}_2 \left(\frac{\Delta-\smp}{2 \Delta}\right)
+{\rm Li}_2 \left(\frac{\Delta+\spm}{2 \Delta}\right)
+{\rm Li}_2 \left(\frac{\Delta+\smp}{2 \Delta}\right)
\Bigg\} \nonumber
\\ \nonumber
C_{+}&=& 2 m_1 m_2 I_1
\\ \nonumber
C_{1,-}&=&\frac{-1}{Q^2}\left[\spm I_1+\ln \left(\frac{m_1^2}{m_2^2}\right)\right]
\\ \nonumber
C_{1,+}&=&\frac{-1}{Q^2}\left[\smp I_1-\ln \left(\frac{m_1^2}{m_2^2}\right)\right]
\\ \nonumber
C_{q,-}&=&\frac{1}{Q^4}\left[\left(\Delta^2-2 m_1^2 Q^2\right) I_1-2 Q^2 +\spm
\ln \left(\frac{m_1^2}{m_2^2}\right)\right]
\\ 
C_{q,+}&=&\frac{1}{Q^4}\left[\left(-\Delta^2+2 m_2^2 Q^2
-\smp Q^2\right) I_1+2 Q^2 +(\smp+Q^2)
\ln \left(\frac{m_1^2}{m_2^2}\right)\right]
\end{eqnarray}
with
\begin{eqnarray}\label{eq:I1b}
I_1=\frac{1}{\Delta}\ln\left[\frac{\spp+\Delta}{\spp-\Delta}\right]
\end{eqnarray}
\begin{eqnarray}\label{eq:I2b}
I_2&=&I_1 \ln \Delta-\frac{1}{\Delta} \\ \nonumber &\times&
\Bigg\{
\frac{1}{2} \ln^2 \left|\frac{\Delta-\spm}{2 Q^2}\right|
+\frac{1}{2} \ln^2 \left|\frac{\Delta-\smp}{2 Q^2}\right|
-\frac{1}{2} \ln^2 \left|\frac{\Delta+\spm}{2 Q^2}\right|
-\frac{1}{2} \ln^2 \left|\frac{\Delta+\smp}{2 Q^2}\right|
\nonumber\\ 
&-&\ {\rm Li}_2 \left(\frac{\Delta-\spm}{2 \Delta}\right)
-{\rm Li}_2 \left(\frac{\Delta-\smp}{2 \Delta}\right)
+{\rm Li}_2 \left(\frac{\Delta+\spm}{2 \Delta}\right)
+{\rm Li}_2 \left(\frac{\Delta+\smp}{2 \Delta}\right)
\Bigg\}. \nonumber
\end{eqnarray}

The renormalized vertex [Fig.\ \ref{qsfeyn} (c.1)--(c.3)]
is obtained by wave function renormalization:
\begin{equation}
\Lambda^\mu_R=\gamma^\mu L_5 (Z_1 - 1)+\Lambda^\mu_0+{\cal{O}}(\alpha_s^2)
\end{equation}
where 
$Z_1=\sqrt{Z_2(p_1)Z_2(p_2)}$.
The fermion wave function renormalization constants are defined on mass shell:
\begin{eqnarray}
Z_2(m_i)=1+C_F\as \Gamma(1-\varepsilon)
\left(\frac{m_i^2}{4 \pi \mu^2}\right)^{\varepsilon}
\frac{1}{\varepsilon}[3 - 4 \varepsilon + {\cal{O}}(\varepsilon^2)]
\end{eqnarray}
such that 
\begin{eqnarray}
Z_1= 1+C_F\as \Gamma(1-\varepsilon)
\left(\frac{Q^2}{4 \pi \mu^2}\right)^{\varepsilon}
\left[\frac{3}{\varepsilon}-\frac{3}{2} \ln \left(\frac{Q^2}{m_1^2}\right)
-\frac{3}{2} \ln \left(\frac{Q^2}{m_2^2}\right)-4 \right]\ \ \ .     
\end{eqnarray}

The final result for the renormalized vertex $\Lambda^\mu_R$ reads
\begin{eqnarray}
\Lambda^\mu_R &=& C_F \as \Gamma(1-\varepsilon) 
\left(\frac{Q^2}{4 \pi \mu^2}\right)^{\varepsilon}
\Bigg\{C_{R,-} \gamma^\mu L_5 + C_{+} \gamma^\mu R_5 
\nonumber\\
&+& C_{1,-}\ m_2\ p_1^\mu\ L_5 
+ C_{1,+}\ m_1\ p_1^\mu\ R_5 
+\ C_{q,-}\ m_2\ q^\mu\ L_5 
+ C_{q,+}\ m_1\ q^\mu\ R_5 
\Bigg\} 
\label{ren}
\end{eqnarray}
with $C_{+}$, $C_{1,\pm}$, $C_{q,\pm}$ as given above and
\begin{eqnarray}
C_{R,-}&=&
\frac{1}{\varepsilon}(2-\spp I_1)
+\Bigg[\frac{\Delta^2}{2 Q^2}
+\spp \left(1+\ln\left(\frac{Q^2}{\Delta}\right)\right)\Bigg]I_1 \\ \nonumber
&+&\frac{m_2^2-m_1^2}{2 Q^2}\ln \left(\frac{m_1^2}{m_2^2}\right)
- \ln \left(\frac{Q^2}{m_1^2}\right)
- \ln \left(\frac{Q^2}{m_2^2}\right)
- 4
+\ \frac{\spp}{\Delta} \\ \nonumber &\times&
\Bigg\{
\frac{1}{2} \ln^2 \left|\frac{\Delta-\spm}{2 Q^2}\right|
+\frac{1}{2} \ln^2 \left|\frac{\Delta-\smp}{2 Q^2}\right|
-\frac{1}{2} \ln^2 \left|\frac{\Delta+\spm}{2 Q^2}\right|
-\frac{1}{2} \ln^2 \left|\frac{\Delta+\smp}{2 Q^2}\right|
\nonumber\\
&-&\ {\rm Li}_2 \left(\frac{\Delta-\spm}{2 \Delta}\right)
-{\rm Li}_2 \left(\frac{\Delta-\smp}{2 \Delta}\right)
+{\rm Li}_2 \left(\frac{\Delta+\spm}{2 \Delta}\right)
+{\rm Li}_2 \left(\frac{\Delta+\smp}{2 \Delta}\right)
\Bigg\}. \nonumber
\end{eqnarray}

%
\subsection{Calculation}
Basic ingredient in the calculation of one--loop virtual corrections
are one--loop integrals classified according
to the number $N$ of propagator factors in the denominator and the number
$P$ of integration momenta in the numerator.
Integrals with $N=1,2,3,4$ are usually called one--, two--, three-- and
four--point functions ($\mathrm{1PF} \ldots \mathrm{4PF}$).
For $P + 4 -2 N \ge 0$ these integrals are UV--divergent.
In dimensional regularization these divergences are regulated by evaluating the integrals in
general dimensions $n \ne 4$ and the divergences become manifest as poles 
in the limit $n \to 4$.

We define the following one--loop tensor integrals\footnote{Note that this definition is 
related to Eq.~(4.1) in \protect\cite{Denner:1993kt} as follows:\\
$T^N_{\mu_1\cdots\mu_P}(p_1,\ldots,p_{N-1},m_0,\ldots,m_{N-1})=
T^{N,\text{Denner}}_{\mu_1\cdots\mu_P}(\bp[1],\ldots,\bp[N-1],m_0,\ldots,m_{N-1})$.} 
(see for example \cite{Beenakker-PhD,Denner:1993kt}):
\begin{align}\label{eq:tensor_tensorint}
T^N_{\mu_1\cdots\mu_P}(p_1,\ldots,p_{N-1},m_0,\ldots,m_{N-1})
\equiv \NT \int \dps[n]{l}\ \frac{l_{\mu_1}\cdots l_{\mu_P}}{D_0 D_1 \cdots D_{N-1}}
\end{align}
where all the external momenta $p_i$ are defined to be incoming
and the denominators stemming from the propagators in the Feynman diagram are given by
\begin{align}\label{eq:tensor_Di}
D_i &= (l+ \bp[i])^2 - m_i^2 + i \vep\ , \qquad \text{with} \qquad
\bp[i] \equiv \sum_{j=1}^i p_j \ , \qquad (i=0,\ldots, N-1)\ .
\end{align}
To achieve a cyclic symmetry we furthermore identify
the 0--th with the $N$--th propagator: $D_N \equiv D_0$, $m_N \equiv m_0$ 
and $\bp[N] = \bp[0] = 0$.
Furthermore it is convenient to define
\begin{equation}\label{eq:tensor_pij}
p_{i0} = \bp[i]\, , \qquad p_{ij} = \bp[i]-\bp[j]\ .
\end{equation}

The $+i \vep$ part in the denominators with an infinitesimal $\vep > 0$
is needed to regulate singularities of the integrand and its 
specific choice ensures causality.
After integration it determines the correct imaginary parts of the logarithms and 
dilogarithms.
The arbitrary mass
scale $\mu$ has been introduced such that the integrals have
an integer mass dimension.
Conventionally $T^N$ is denoted by the $N$--th character of the alphabet, i.e.
$T^1 \equiv A$, $T^2 \equiv B$, $\ldots$, and the scalar integrals carry an index 0.

The tensor integrals in (\ref{eq:tensor_tensorint}) can be related to the scalar
integrals $A_0$, $B_0$, $C_0$ and $D_0$ by a Passarino--Veltman decomposition
\cite{Passarino:1979jh} which will be described in the following.
Due to Lorentz covariance the tensor integrals (\ref{eq:tensor_tensorint})
can be decomposed into tensors constructed from the external 
momenta $p_i$ and the metric tensor $g_{\mu\nu}$.
The choice of the tensor basis is not unique and we will 
stick to the following form \cite{Beenakker-PhD,Bojak-PhD}:
\begin{equation}
\label{eq:pvdec}
\begin{aligned}
B^\mu=~&p_1^\mu B_1 
\\
B^{\mu\nu}=~&p_1^\mu p_1^\nu B_{21}+g^{\mu\nu} B_{22} 
\\
&\\
C^\mu=~&p_1^\mu C_{11}+p_2^\mu C_{12} 
\\
C^{\mu\nu}=~&p_1^\mu p_1^\nu C_{21}+p_2^\mu p_2^\nu C_{22}+
  \{p_1 p_2\}^{\mu\nu} C_{23}+g^{\mu\nu} C_{24} 
\\
C^{\mu\nu\rho}=~&p_1^\mu p_1^\nu p_1^\rho C_{31}+
  p_2^\mu p_2^\nu p_2^\rho C_{32}+\{p_1 p_1 p_2\}^{\mu\nu\rho} C_{33}
  +\{p_1 p_2 p_2\}^{\mu\nu\rho} C_{34}
\\
  &{}+\{p_1 g\}^{\mu\nu\rho} C_{35}+
  \{p_2 g\}^{\mu\nu\rho} C_{36} 
\\
&\\
D^\mu=~&p_1^\mu D_{11}+p_2^\mu D_{12}+p_3^\mu D_{13} 
\\
D^{\mu\nu}=~&p_1^\mu p_1^\nu D_{21}+p_2^\mu p_2^\nu D_{22}+
  p_3^\mu p_3^\nu D_{23}+\{p_1 p_2\}^{\mu\nu} D_{24}
  +\{p_1 p_3\}^{\mu\nu} D_{25}
\\
  &{}+\{p_2 p_3\}^{\mu\nu} D_{26}+g^{\mu\nu} D_{27} 
\\
D^{\mu\nu\rho}=~&p_1^\mu p_1^\nu p_1^\rho D_{31}+
  p_2^\mu p_2^\nu p_2^\rho D_{32}+p_3^\mu p_3^\nu p_3^\rho D_{33}+
  \{p_1 p_1 p_2\}^{\mu\nu\rho} D_{34}+\{p_1 p_1 p_3\}^{\mu\nu\rho}D_{35}
\\
  &{}+\{p_1 p_2 p_2\}^{\mu\nu\rho}D_{36}+\{p_1 p_3 p_3\}^{\mu\nu\rho} D_{37}
  +\{p_2 p_2 p_3\}^{\mu\nu\rho}D_{38}+\{p_2 p_3 p_3\}^{\mu\nu\rho}D_{39}
\\
  &{}+\{p_1 p_2p_3\}^{\mu\nu\rho} D_{310}+\{p_1 g\}^{\mu\nu\rho} D_{311}
  +\{p_2 g\}^{\mu\nu\rho} D_{312}+\{p_3 g\}^{\mu\nu\rho} D_{313}
\end{aligned}
\end{equation}
where we have used the shorthand notations
\begin{equation*}
\{p_i p_j p_k\}_{\mu\nu\rho} \equiv \sum_{\sigma(i,j,k)} p_{\sigma(i)\mu}
p_{\sigma(j)\nu}p_{\sigma(k)\rho}
\end{equation*} 
with $\sigma(i,j,k)$ denoting all {\em different} permutations of $(i,j,k)$ 
and
\begin{equation*}
\{p_i g\}_{\mu\nu\rho} \equiv 
p_{i\mu} g_{\nu \rho}+p_{i\nu} g_{\mu \rho}+p_{i\rho} g_{\mu \nu}\ .
\end{equation*} 
For example
$\{p_1 p_1 p_2\}^{\mu\nu\rho}=p_1^\mu p_1^\nu p_2^\rho+
p_1^\nu p_1^\rho p_2^\mu+p_1^\rho p_1^\mu p_2^\nu$
and 
$\{p_1 g\}^{\mu\nu\rho}=p_1^\mu g^{\nu\rho}+p_1^\nu g^{\rho\mu}+
p_1^\rho g^{\mu\nu}$.
The scalar coefficients $B_{j(k)}$, $C_{jk}$ and $D_{jk(l)}$ can depend 
on all possible invariants (built from the leg momenta) and the masses 
$m_i$.

Using the Lorentz decomposition (\ref{eq:pvdec}) of the tensor integrals 
all scalar coefficients can be iteratively reduced to scalar integrals
(with equal or less propagator factors) \cite{Beenakker-PhD,Bojak-PhD}, see 
\cite{Denner:1993kt} for a general treatment. 
With other words, all one--loop tensor integrals can be expressed in
terms of scalar integrals (with equal or less propagator factors).
In simple cases the reduction can be done by hand, however, in general
the reduction algorithm has to be automized on a computer.\footnote{We are grateful 
to Dr.~I.~Bojak for providing us 
his \Mathematica\ package for the reduction of the tensor integrals.
This package is described in \protect\cite{Bojak-PhD}.}

We now present some details of the derivation of 
the unrenormalized vertex in Eq.~(\ref{eq:vertexstruc}).
In Feynman gauge the
vertex correction diagram in [Fig.~\ref{qsfeyn}(c.1)] is given 
in $n$ dimensions by
\begin{equation}\label{eq:vertex1}
\begin{aligned}
\Lambda^\mu_0 & = 
\CF \as \frac{(2 \pi \mu)^{4-n}}{i \pi^2}
\int \dps[n]{k}\
\gamma^\rho \frac{\ksl+\ptwosl+m_2}{(k+p_2)^2-m_2^2+}
\frac {\gamma^\mu L_5}{k^2} 
\frac{\ksl+\ponesl+m_1}{(k+p_1)^2-m_1^2} \gamma_\rho
\end{aligned}
\end{equation}
where $L_5 = (V - A \gamma_5)$, $\CF= 4/3$ and the scale $\mu$ has been 
introduced in order to maintain the coupling $g_s$ dimensionless (see above).

In the following we consider on--shell fermions, i.e.
$p_1^2 = m_1^2$ and $p_2^2 = m_2^2$, and a spacelike gauge boson with 4--momentum
$q = p_2-p_1$, $q^2 < 0$.
Furthermore, since $\Lambda^\mu_0$ will be eventually sandwiched between external
spinors $\bar{u}(p_2)\Lambda^\mu_0 u(p_1)$ we apply the Dirac equations
$\bar{u}(p_2) \ptwosl = m_2 \bar{u}(p_2)$ and $\ponesl u(p_1) = m_1 u(p_1)$
whenever possible which will be indicated by a '$\hat{=}$' instead of a '$=$'.
Utilizing an anti--commuting $\gamma_5$, $\{\gamma_\mu,\gamma_5\}=0$ 
\cite{Jegerlehner:2000dz}, 
the numerator (under the integral) in Eq.(\ref{eq:vertex1}) can be written as
\begin{equation}
\begin{aligned}
\mathrm{Num}\ &\hat{=}\ \Big\{\gamma^\mu \left[(n-6)k^2 + 2[(k+p_1)^2-m_1^2]
+2[(k+p_2)^2-m_2^2] + 4 p_1\cdot p_2\right]
\\*
&\phantom{\hat{=}\ \Big\{}+ 2 m_2 \gamma^\mu \ksl - 2 \ksl\ [2 q^\mu +
(n-2) k^\mu + 4 p_1^\mu]\Big\}L_5 
- 2 m_1 \gamma^\mu \ksl R_5 + 4 m_1 k^\mu R_5
\end{aligned}
\end{equation} 
with $R_5 = (V + A \gamma_5)$.

Inserting the numerator into Eq.~(\ref{eq:vertex1}) and
utilizing the Passarino--Veltman--decomposition (\ref{eq:pvdec}) of the 
encountered tensor integrals (and again exploiting the Dirac equation
in order to eliminate $\ponesl$ and $\ptwosl$)
there remain
only Dirac structures as given in Eq.~(\ref{eq:vertexstruc}).
Comparing the corresponding coefficients 
we obtain 
\begin{eqnarray}
C_{0,-}&\hat{=}& \left[\Gamma(\tfrac{6-n}{2}) 
\left(\tfrac{Q^2}{4 \pi \mu^2}\right)^{(n-4)/2}\right]^{-1}
\Big[(n-6)B_0(q,m_1,m_2)+ 2 B_0(p_1,0,m_1)
\nonumber\\ 
& &
+ 2 B_0(p_1+q,0,m_2)+ 4(m_1^2+p_1 \cdot q) C_0 
\nonumber\\ 
& &
- 2 m_1^2 C_{11} + 2(m_1^2-m_2^2)C_{12}-2(n-2)C_{24}\Big]
\nonumber\\
C_{+\phantom{0,}} &\hat{=}& 2 m_1 m_2 C_{11}
\nonumber\\
C_{1,-}&\hat{=}& -4 C_{12} - 2(n-2) C_{23}
\nonumber\\ 
C_{1,+}&\hat{=}& -4 C_{11} + 4 C_{12} - 2(n-2)(C_{21}-C_{23})
\nonumber\\ 
C_{q,-}&\hat{=}& -2(n-2) C_{22}
\nonumber\\ 
C_{q,+}&\hat{=}& -4 C_{11} + 4 C_{12} + 2(n-2)(C_{22}-C_{23})
\end{eqnarray}
with $C_{ij,0}=C_{ij,0}(p_1,q,0,m_1,m_2)$. 

As described above the coefficients of the Passarino--Veltman--decomposition can 
be expressed in terms of the scalar integrals
\begin{eqnarray*}
D\ C_{11} 
&=& \frac{q \cdot p_2}{2} [B_0(p_2,0,m_2)-B_0(q,m_1,m_2)]
-\frac{q \cdot p_1}{2} [B_0(p_1,0,m_1)-B_0(q,m_1,m_2)]
\\
D\ C_{12} 
&=& -\frac{p_1 \cdot p_2}{2} [B_0(p_2,0,m_2)-B_0(q,m_1,m_2)]
+\frac{m_1^2}{2} [B_0(p_1,0,m_1)-B_0(q,m_1,m_2)]
\\
D\ C_{21} 
&=& A_0(m_1) \frac{p_1 \cdot q}{4 m_1^2}-A_0(m_2) \frac{p_1 \cdot q+q^2}{4 m_2^2}
+ B_0(q,m_1,m_2) q^2 \frac{n-3}{2(n-2)}
\\
D\ C_{22} 
&=& (A_0(m_1)-A_0(m_2)) \frac{p_1 \cdot q}{4 q^2}+A_0(m_2) \frac{p_1 \cdot q+m_1^2}{4 m_2^2}
\\
&&
+ B_0(q,m_1,m_2) \left(\frac{m_1^2}{2(n-2)}-\frac{p_1 \cdot q}{4}+
\frac{(m_2^2-m_1^2)p_1\cdot q}{4 q^2}\right)
\\
D\ C_{23} 
&=& A_0(m_1) \frac{-1}{4}+A_0(m_2) \frac{p_1 \cdot q+m_1^2}{4 m_2^2}
- B_0(q,m_1,m_2) p_1\cdot q \frac{n-3}{2(n-2)}
\\
C_{24} &=& B_0(q,m_1,m_2) \frac{1}{2(n-2)}
\end{eqnarray*}
with $D = m_1^2 q^2 - (p_1 \cdot q)^2$.
It is easy to convince oneself that the coefficients
$C_{11},\ldots,C_{23}$ are finite, whereas $C_{24}$ is UV--divergent. 
Furthermore, $C_0(p_1,q,0,m_1,m_2)$ is IR--divergent.

At this place the problem has reduced to the calculation
of the scalar integrals $A_0$, $B_0$ and $C_0$.
Introducing Feynman parameters and performing a Wick rotation the 
$N$--point scalar
integral can be brought into the following form:
\begin{align}
T_0^N &\equiv T_0^N(p_1,\ldots,p_{N-1},m_0,\ldots,m_{N-1})  
= \NT \int \dps[n]{l}\ \prod_{i=1}^N\ \frac{1}{D_i} 
\nonumber\\
&=(-1)^N\ \Gamma(N-n/2)\ (4 \pi \mu^2)^{\frac{4-n}{2}}\
\int_0^1 [d\alpha]_N\ \left(M^2 \right)^{n/2 - N}
\end{align}
with
\begin{align}
[d\alpha]_N &= d\alpha_1 \cdot \ldots \cdot d\alpha_N\ \delta\left(1-\sum_{i=1}^N \alpha_i\right)
\nonumber\\
M^2 & = - \sum_{N\ge i > j \ge 1} \alpha_i \alpha_j p_{ij}^2 
+ \sum_{i=1}^N \alpha_i m_i^2 - i \vep 
\end{align}
where $p_{ij}$ has been introduced in Eq.~(\ref{eq:tensor_pij}).

The calculation of the scalar one--point function $A_0(m)$ is trivial:
\begin{equation}\label{eq:1PF}
A_0(m)= - m^2 \left(\frac{m^2}{4 \pi \mu^2} \right)^{\frac{n-4}{2}} \Gamma\left(1- \frac{n}{2}\right)
=m^2 \left(\DUV - \ln \frac{m^2}{\mu ^2} + 1 \right) + {\cal O}(n-4)
\end{equation}
with the UV--divergence contained in
\begin{equation}
\begin{aligned}
\DUV &= \frac{-2}{n-4} - \eulergam + \ln 4 \pi
\end{aligned}
\end{equation}
and \eulergam\ is Euler's constant.
Note that Eq.~(\ref{eq:1PF}) implies $A_0(0) = 0$.
The terms of order $\Ord(n-4)$ are only relevant for 
two-- or higher--loop calculations.

The evaluation of the needed two--point functions is still easy
\begin{equation}\label{eq:2PFb}
\begin{aligned}
B_0(q,m_1,m_2)|_{q^2 <0} &=\DUV + 2 - \ln \frac{m_1 m_2}{\mu^2}
+ \frac{m_1^2 - m_2^2}{q^2} \ln \frac{m_2}{m_1} 
\\
& \phantom{=}+ \beta \left(1 - \frac{(m_1 - m_2)^2}{q^2}\right)
\ln \left(- \frac{1-\beta}{1 + \beta}\right)
\end{aligned}
\end{equation}
with 
\begin{displaymath}
\beta^2 = \frac{(m_1+m_2)^2 - q^2}{(m_1-m_2)^2 - q^2} 
\end{displaymath}
and
\begin{align}
B_0(p,0,m)|_{p^2=m^2} & = \DUV - \ln \frac{m^2}{\mu^2} + 2\ .
\end{align}

Finally the required three--point function 
is given in $n = 4 + 2 \vep$ dimensions by
($M^2 = m_1^2 x + m_2^2 (1-x) - q^2 x(1-x) - i \vep$)
\begin{align}
C_0(p_1,q,0,m_1,m_2)|_{p_1^2=m_1^2,(p_1+q)^2=m_2^2,q^2<0} =
(4 \pi \mu^2)^{-\vep}\ \Gamma(1-\vep)\ \frac{1}{2 \vep}\ I
\end{align}
with
\begin{align}
I &= \int_0^1 dx\ [m_1^2 x + m_2^2 (1-x) - q^2 x(1-x)]^{-1+\vep}\ .
\end{align}
The Feynman parameter integral $I$ is finite at $\vep = 0$ 
since $\XX \equiv M^2 \equiv m_1^2 x + m_2^2 (1-x) - q^2 x(1-x) > 0$ 
for $x \in [0,1]$ as long as $m_1^2 > 0$ and $m_2^2 > 0$.
Therefore we can expand I in $\vep$ up to negligible terms
of the order $\Ord(\vep^2)$
\begin{align}
I &= \int_0^1 dx\ \XX^{-1+\vep} 
= \int_0^1 dx\ \frac{1 + \vep \ln \XX}{\XX} +\Ord(\vep^2)=I_1+\vep I_2
+\Ord(\vep^2)\ .
\end{align}
The integral $I_1 = \int_0^1 dx\ \frac{1}{\XX}$ is elementary and
can be found in Eq.~(\ref{eq:I1b}).
$I_2 = \int_0^1 dx\ \frac{\ln \XX}{\XX}$ 
can be calculated by writing $\XX$ in linear factors and by
partial fractioning the denominator.
After some algebra one finds the result given in Eq.~(\ref{eq:I2b}).
\section{Real and Virtual Contributions to Structure Functions}
\label{rvstruct}

The soft real contributions $S_i$ to the coefficient functions in Eq.\
(\ref{coefficient}) are given by
\begin{eqnarray} \nonumber
S_1 &=& \frac{1}{\varepsilon}(-2+\spp I_1)+
2+\frac{\spp}{\Delta} \Big[ \Delta\ I_1 
+ {\rm{Li}}_2\left(\frac{2\Delta}{\Delta-\spp}\right)
- {\rm{Li}}_2\left(\frac{2\Delta}{\Delta+\spp}\right)
\Big] \\ \nonumber
&+&\ln\frac{\Delta^2}{m_2^2 Q^2}\ (-2+\spp I_1) 
\\  S_{2,3}&=&S_1 \label{soft}
\end{eqnarray}
with $I_1$ given in Appendix \ref{vcorr} and
where $\chi$ is given in Eq.\ (\ref{chi}).
The virtual contributions are derived from the 
renormalized vertex in Eq.\ (\ref{ren})
by using the projectors in Eq.\ (\ref{projectors}): 
\begin{eqnarray} \nonumber
V_1 &=& C_{R,-} + \frac{S_- \spp-2S_+ m_1 m_2}{S_+ \spp-2S_- m_1 m_2}
\ C_+ \\ \nonumber
V_2 &=&  C_{R,-} + \frac{1}{2} \left( m_1^2\ C_{1,+} + m_2^2
\ C_{1,-}\right)+ \frac{S_-}{S_+}\left[C_+ +\frac{m_1
m_2}{2}\left(C_{1,+}+C_{1,-}\right)\right] \\ 
V_3 &=& C_{R,-} + \frac{R_-}{R_+}\ C_+
\label{virt} 
\end{eqnarray}
where the $C$s are given in Appendix \ref{vcorr}. 
Note that the soft poles
($1/\varepsilon$) of $S_i$, $V_i$ cancel in the sum $S_i+V_i$ in 
Eq.\ (\ref{coefficient}) as must be. 

\enlargethispage{2\baselineskip}
The massive matrix elements $\hat{f}_i^Q(\xip)$ are most conveniently given 
as functions of the Mandelstam variable
$\hsi(\xip)\equiv (p_1+q)^2-m_2^2$, i.e. $\hat{f}_i^Q(\xip)
\equiv \hat{f}_i^Q[\hsi(\xip)]$ with
\begin{equation}
\label{s1}
\hsi(\xip)\equiv {\hat{s}}-m_2^2 =
\frac{1-\xip}{2 \xip}[(\Delta-\spm)\xip+\Delta+\spm]\ \ \ .
\end{equation}
From the real graphs of Fig.\ \ref{qsfeyn} (b) one obtains
\begin{eqnarray} \nonumber
\hat{f}_1^Q(\hsi)&=&
\frac{8}{{\Delta^\prime}^2}\Bigg\{ 
 -\Delta^2 (\Sp \spp -2 m_1 m_2 \Sm) I_{\xip}
+ 2 m_1 m_2 \Sm \Bigg( 
\frac{1}{\hsi} [{\Delta^\prime}^2 + 4 m_2^2 \spm] \\ \nonumber
&+& 2\spm - \smp
+\frac{\spp+\hsi}{2}
+\frac{\hsi+m_2^2}{{\Delta^\prime} \hsi}
\left[{\Delta^\prime}^2 + 2 \spm \spp+ \left(m_2^2+Q^2\right)  
\hsi \right]\ L_{\xip} \Bigg)  
\\ \nonumber
&+& \Sp \Bigg(
\frac{-m_2^2 \spp}{(\hsi+m_2^2)\hsi}(\Delta^2+4 m_2^2 \spm)
-\frac{1}{4 (\hsi+m_2^2)}\Big[3 \spp^2 \smp+4 m_2^2(10 \spp \spm
\\ \nonumber
&-&\spm \smp
- m_1^2 \spp)
+\hsi[-7 \spp \smp+18 \Delta^2-4 m_1^2(7 Q^2-4 m_2^2+ 7 m_1^2)]
\\ \nonumber
&+& 3 \hsi^2[\spm - 2 m_1^2]-\hsi^3\Big]
+\frac{\hsi+m_2^2}{2 {\Delta^\prime}}
\left[\frac{-2}{\hsi} \spp \left(\Delta^2 +2 \spm \spp \right) \right.
\\ \nonumber
&+&\left.\left(4 m_1^2 m_2^2 - 7 \spm \spp \right)
-4\spm\hsi -\hsi^2\right]\ L_{\xip} \Bigg) 
\Bigg\}
\\ \nonumber
\hat{f}_2^Q(\hsi)&=&
\frac{16}{{\Delta^\prime}^4}\Bigg\{ 
 -2 \Delta^4 \Sp I_{\xip}
+ 2 m_1 m_2 \Sm \Bigg(\frac{\hsi+m_2^2}{\Delta^\prime}
\left({\Delta^\prime}^2 - 6 m_1^2 Q^2 \right)
\ L_{\xip} \\ \nonumber
&-&\frac{{\Delta^\prime}^2 (\hsi+\spp)}{2 (\hsi+m_2^2)}
+\left(2{\Delta^\prime}^2 
-3 Q^2 \left(\hsi+\spp \right)\right)\Bigg)  
+ \Sp \Bigg(-2(\Delta^2-6 m_1^2 Q^2)
(\hsi+m_2^2) \\ \nonumber
&-&2\left(m_1^2 + m_2^2\right)\hsi^2
- 9 m_2^2 \spm^2 
+\Delta^2 \left(2\spp-m_2^2 \right)  
+2\hsi \left(2 \Delta^2+\left(m_1^2-5 m_2^2 \right)\spm \right)  
\\ \nonumber
&+& \frac{\left({{\Delta^\prime}}^2 -6 Q^2\left(m_2^2+\hsi \right)\right)\spp 
\left(\hsi+\spp \right)}{2(\hsi+m_2^2)}
-\frac{2 \Delta^2}{\hsi} \left(\Delta^2+2(2 m_2^2+\hsi) \spm \right)
\\ \nonumber
&+& \frac{(\hsi+m_2^2)}
{{\Delta^\prime}}[\frac{-2}{\hsi}\Delta^2(\Delta^2+2\spm \spp)
-2\hsi(\Delta^2-6 m_1^2 Q^2)
\\ \nonumber
&-& ({\Delta^\prime}^2 -18 m_1^2 Q^2) \spp 
- 2 {\Delta^2} \left(\spp+2 \spm \right)]\ L_{\xip}
\Bigg)  
\Bigg\}
\\ \nonumber
\hat{f}_3^Q(\hsi)&=&
\frac{16}{{\Delta^\prime}^2}\Bigg\{ 
- 2 \Delta^2 \Rp I_{\xip}
+2 m_1 m_2 \Rm \left(1-\frac{\smp}{\hsi}+\frac{(\hsi+m_2^2) 
\left(\hsi+\spm \right)}{{\Delta^\prime} \hsi}\ L_{\xip} \right)
\\ \nonumber
&+& {\Rp} \Bigg(\smp-3\spm  - \frac{2}{\hsi} 
\left(\Delta^2 +2 m_2^2 \spm\right) 
- \frac{(\hsi-\smp)(\hsi+\spp) }{2 (\hsi+m_2^2)} 
\\ 
&+& \frac{\hsi+m_2^2}{{\Delta^\prime} \hsi}  
\left[ -\hsi^2 + 4 \left(m_1^2 \smp-\Delta^2\right)  
- 3 \hsi \spm \right]\ L_{\xip}
\Bigg)
\Bigg\}
\label{fis}
\end{eqnarray}
with 
\begin{displaymath}
\displaystyle L_{\xip} \equiv \LN
\end{displaymath}
and
\begin{displaymath}
I_{\xip}=\left(\frac{\hsi+2 m_2^2}{\hsi^2} + 
\frac{\hsi+m_2^2}{\Delta^\prime \hsi^2}\spp\ L_{\xip}\right)\ .  
\end{displaymath}
$\Delta$ is given below Eq.\ (\ref{pmpm}) and
$\Delta^\prime\equiv\Delta[m_1^2,{\hat{s}},-Q^2]$.

Finally, the normalization factors in Eq.\ (\ref{coefficient}) are
\begin{eqnarray}
N_1&=&\frac{\Sp \spp- 2m_1 m_2 \Sm}{2 \Delta},
\qquad N_2=\frac{2 \Sp \Delta}{(\Delta^\prime)^2},\qquad
N_3=\frac{2 \Rp}{\Delta^\prime}\ .
\end{eqnarray}

\section[Comparison with E.\ Hoffmann and R.\ Moore, \mbox{Z.\ Phys.\ {\bf{C20}}, 71 (1983)}]
{Comparison with E.\ Hoffmann and R.\ Moore, \mbox{Z.\ Phys.\ {\bf{C20}}, 71 (1983):}\\ 
Detailed List}
\label{homocomp}
\begin{flushleft}

HM: Refers to E.\ Hoffmann and R.\  Moore, Z.\ Phys.\ {\bf{C20}}, 71 (1983) \\
KS: Refers to S.\ Kretzer and I.\ Schienbein, Phys.\ Rev.\ {\bf D58}, 094035 (1998),
as documented in this thesis in Chapter \ref{hqcontrib} and the above Appendices 
\ref{rge}--\ref{rvstruct}.

Replacements:\\ 
$m_{1,2}{\rm{(KS)}} \rightarrow m$ \\
$V, A {\rm{(KS)}} \rightarrow 1, 0$ 

As in HM: $\lambda\equiv m^2/Q^2$.

The bracketed numbers refer to the equations in the Hoffmann \& Moore article
and in this Thesis, respectively.

\end{flushleft}

\subsection*{Vertex Correction}

\begin{eqnarray*}
\left.
\begin{array}{ccc}
\Lambda_R^{\mu} &=& \Gamma^\mu\\
{\rm{(\ref{ren})_{{}\atop{\tiny{KS}}}}} &=& {\rm{(31)_{{}\atop{\tiny{HM}}}}}  
\end{array}
\right\} {\mbox{up to a Gordon decomposition}}
\end{eqnarray*}

\subsection*{Virtual Contributions}

\begin{eqnarray*}
\left.
\begin{array}{ccc}
\frac{\alpha_s}{2\pi}\ C_F\ \Gamma (1-\varepsilon)\ \left(\frac{Q^2}
{4\pi\mu^2}\right)^\varepsilon V_2\ \delta (1-\xip) &=& \sqrt{1+4 \lambda}
\ \sigma_{1V}^{(2)} \\
\frac{\alpha_s}{2\pi}\ C_F\ \Gamma (1-\varepsilon)\ \left(\frac{Q^2}
{4\pi\mu^2}\right)^\varepsilon 
{\rm{(\ref{virt})_{{}\atop{\tiny{KS}}}}}\ \delta (1-\xip) 
&=& \sqrt{1+4 \lambda}\ {\rm{(38)_{{}\atop{\tiny{HM}}}}} \\ && \\
\frac{\alpha_s}{2\pi}\ C_F
\ (V_2-V_1)\ \delta (1-\xip) &=& -2\ \sqrt{1+4 \lambda}\ \sigma_{1V}^{(L)} \\
\frac{\alpha_s}{2\pi}\ C_F\ \left[ 
\Delta{\rm{(\ref{virt})}} \right]_{\rm{KS}} 
\ \delta (1-\xip) &=& -2\ \sqrt{1+4 \lambda}\ {\rm{(39)_{{}\atop{\tiny{HM}}}}}
\end{array}
\right\}
\begin{array}{ccc}
&\delta (1-\xip)& \\ &=& \\ &\sqrt{1+4 \lambda}\ \delta (1-z)& 
\end{array}
\end{eqnarray*}

\subsection*{Soft Real Contributions}

\begin{eqnarray*}
\frac{\alpha_s}{2\pi}\ C_F\ \Gamma (1-\varepsilon )\ \left(\frac{Q^2}
{4\pi\mu^2}\right)^\varepsilon\ \left. S_2 \right.
\ \delta (1-\xip) 
&=& \sqrt{1+4 \lambda}\ \left[
\left. \sigma_{1 B}^{(2)}\right|_{{\mbox{{\tiny{}}}}\atop{\sim\delta
(1-z)}}+f(\lambda )\right]\ \delta (1-z)\\
\frac{\alpha_s}{2\pi}\ C_F\ \Gamma (1-\varepsilon )\ \left(\frac{Q^2}
{4\pi\mu^2}\right)^\varepsilon\ \left. 
{\rm{(\ref{soft})_{{}\atop{\tiny{KS}}}}}
\right.
\ \delta (1-\xip)
&=& \sqrt{1+4 \lambda}\ \left[
\left. {\rm{(48)_{{}\atop{\tiny{HM}}}}}
\right|_{{\mbox{{\tiny{}}}}\atop{\sim\delta
(1-z)}}+f(\lambda)
\right]\ \delta (1-z)
\end{eqnarray*}
where the discrepancy $f(\lambda)$ 
\begin{eqnarray*}
f(\lambda)=\frac{\alpha_s}{2\pi}\ C_F\ 2\ \ln(1+4\lambda)
\left(-1-\frac{1+2\lambda}{\sqrt{1+4\lambda}}\ \ln\frac{\sqrt{1+4\lambda}-1}
{\sqrt{1+4\lambda}+1}\right)\rightarrow 0;\ \ \lambda\rightarrow 0
\end{eqnarray*}
arises from the different choice of the convolution variables 
and is exactly cancelled if the identity
\begin{eqnarray*}
\frac{1}{(1-z)_+}=\frac{1}{(1-\xip)_+}\ \left(\frac{1-\xip}{1-z}\right)
+\ln \sqrt{1+4 \lambda}\ \delta (1-z)
\end{eqnarray*} 
is used in Eq.\ ${\rm{(48)_{{}\atop{\tiny{HM}}}}}$ or 
in Eq.\ ${\rm{(\ref{coefficient})_{{}\atop{\tiny{KS}}}}}$. 

\subsection*{Hard Real Contributions}

\begin{eqnarray*}
\frac{\alpha_s}{2\pi}\ \left.{\hat{H}}_2^q\right|_{\xi^\prime\neq 1} &=&
\frac{1+4\lambda z^2}{\sqrt{1+4\lambda}}\ \left.\sigma_1^{(2)}\right|_{z\neq 1}
\\
\frac{\alpha_s}{2\pi}\ \left.
{\rm{(\ref{coefficient})_{{}\atop{\tiny{KS}}}}}\right|_{\xi^\prime\neq 1} &=&
\frac{1+4\lambda z^2}{\sqrt{1+4\lambda}}\ \left.{\rm{(51)_{{}\atop{\tiny{HM}}}}}
\right|_{z\neq 1}
\end{eqnarray*}
\begin{eqnarray*}
\frac{\alpha_s}{2\pi}\ \left.{\hat{H}}_1^q\right|_{\xi^\prime\neq 1} &=&
\sqrt{1+4\lambda }\ \left.(-2\ \sigma_1^{(L)}+\sigma_1^{(2)})\right|_{z\neq 1}
\\
\frac{\alpha_s}{2\pi}\ \left.
{\rm{(\ref{coefficient})_{{}\atop{\tiny{KS}}}}}\right|_{\xi^\prime\neq 1} &=&
{\sqrt{1+4\lambda}}\ \left.
\left[-2\ {\rm{(52)_{{}\atop{\tiny{HM}}}}}+{\rm{(51)_{{}\atop{\tiny{HM}}}}}\right]
\right|_{z\neq 1}\ \ \ ,
\end{eqnarray*}

where ${\rm{(52)_{{}\atop{\tiny{HM}}}}}$ should be multiplied by an obvious factor $z$.

\subsection*{Summary}

We (KS) agree with HM up to: 
\begin{itemize}
\item
The sign of the virtual contribution to the 
longitudinal structure function (which is irrelevant for $F_2^c$)
\item
The overall normalization of $F_1^c$. According to the above equations our
(KS) normalization differs from HM by a factor $\sqrt{1+4 \lambda}$. This does
not really mean a discrepancy but stems from the choice of normalizing
the $\alpsi$ coefficient function relative to a Born term contribution of
$\delta (1-z)$ (HM) or $\delta (1-\xip)$ (KS), corresponding to 
${\rm{(37)_{{}\atop{\tiny{HM}}}}}$ or 
${\rm{(\ref{def})_{{}\atop{\tiny{KS}}}}}$.
However, we regard our normalization as preferable since it corresponds
according to ${\rm{(\ref{ansatz})_{{}\atop{\tiny{KS}}}}}$ directly to the notion of
$c(\xi)$ carrying a light cone fraction $\xi$ of the nucleon's momentum
and is therefore directly related to
\begin{itemize}
\item  the original derivation of the conventionally
used intrinsic charm component in Phys.\ Rev.\ {\bf{D23}}, 2745 (1981)
({\it{Intrinsic heavy--quark states}})
\item the all--order proof of {\it Hard--scattering factorization with heavy
quarks} in Phys.\ Rev.\ {\bf D58}, 094002 (1998)
\end{itemize}

\item
The normalization of $F_2^c$, which is --besides the $\sqrt{1+4 \lambda}$ 
normalization ambiguity-- wrong by a factor $(1+4 \lambda z^2)/(1+4 \lambda)$
in HM. The HM expression corresponds to the partonic structure function 
multiplying the $\sim p_c^\mu p_c^\nu 2z^2/Q^2$ partonic tensor. This 
normalization is only correct in the massless case and neglects finite mass terms
which arise when contracting with the leptonic tensor and extracting the $F_2^c$
structure function as the properly normalized coefficient of the $\sim (1-y)$ term.
\end{itemize}

%% file: phd_app_hqfrag.tex
\chapter{Matrix Elements for Real Gluon Emission off Massive Quarks}
\label{rgediff}

The projections $\hat{f}_i^Q$ of the partonic Matrix Element onto the structure functions
are most conveniently given in the Mandelstam variables $\hsi$ and $\hti$
which are defined below Eq.\ (\ref{Hiq}):
\begin{eqnarray} \nonumber
\hat{f}_1^Q(\hsi,\hti)&=&
\frac{8}{{\Delta^\prime}^2}\Bigg\{
 -\Delta^2 (\Sp \spp -2 m_1 m_2 \Sm)  
\left(\frac{m_2^2}{\hsi^2}+\frac{m_1^2}{\hti^2}+\frac{\spp}{\hsi \hti}\right)  
\\ \nonumber
&+&  2 m_1 m_2 \Sm \Bigg(
\frac{m_1^2 \hsi (\hsi +2\spm)}{\hti^2}
+ \frac{{\Delta^\prime}^2 +(m_2^2+Q^2)\hsi + 
2 \spm \spp}{\hti}
\\ \nonumber
&+& \frac{{\Delta^\prime}^2 -\hsi (m_1^2+Q^2 +\hsi)+2 m_2^2 \spm}{\hsi} 
- \hti \frac{(m_2^2+\hsi)}{\hsi} \Bigg)
\\ \nonumber
&+& \Sp \Bigg(
- \frac{{m_1}^2 \hsi \spm (\hsi+2 \spp)}{\hti^2}
+ \frac{-\hsi^3 -4\hsi^2 \spm + \hsi (4 m_1^2 m_2^2- 7\spm \spp)}{2 \hti}
\\ \nonumber
&+& \frac{2 \spp (-\Delta^2 - 2 \spm \spp)}{2 \hti}     
+\Bigg[4 m_1^4+ 2 m_1^2 \hsi - \spm (m_2^2 + \spm) 
\\ \nonumber
&-& 
\frac{(\Delta^2 + 2 m_2^2 \spm)  \spp}{\hsi}\Bigg]- \hti \frac{{\Delta^\prime}^2 
- 2 (m_2^2+\hsi) \spp}{2 \hsi}
\Bigg) \Bigg\}
\\ \nonumber
\hat{f}_2^Q(\hsi,\hti)&=&
\frac{16}{{\Delta^\prime}^4}\Bigg\{
 -2 \Delta^4 \Sp 
\left(\frac{m_2^2}{\hsi^2}+\frac{m_1^2}{\hti^2}+\frac{\spp}{\hsi \hti}\right)  
+ 2 m_1 m_2 \Sm 
\Bigg(\frac{(\delp^2-6 m_1^2 Q^2)\hsi}{\hti}
\\ \nonumber
&+& \left[2(\delp^2-3 Q^2(\hsi+\spp))\right]
+ \hti \frac{\delp^2-6 Q^2(m_2^2+\hsi)}{\hsi} \Bigg)
\\ \nonumber
&+& \Sp \Bigg(\frac{-2 m_1^2 \hsi [(\Delta^2-6 m_1^2 Q^2)\hsi
+2 \Delta^2 \spm]}{\hti^2}
+\frac{-2 \Delta^2(\Delta^2+2\spm\spp)}{\hti}
\\ \nonumber
&+& \frac{- \hsi[2(\Delta^2-6 m_1^2 Q^2)\hsi
+(\delp^2-18 m_1^2 Q^2)\spp+2 \Delta^2(3\spp-4 m_1^2)]}{\hti}
\\ \nonumber
&+& \Big[-2(m_1^2+m_2^2)\hsi^2-9 m_2^2\spm^2-\frac{2 \Delta^2(\Delta^2+2 m_2^2\spm)}{\hsi}
+ 2\hsi[2\Delta^2
\\ \nonumber
&+& (m_1^2- 5 m_2^2)\spm] + \Delta^2(2\spp-m_2^2)\Big]
-\hti \frac{[\delp^2-6 Q^2(m_2^2+\hsi)]\spp}{\hsi} 
\Bigg) \Bigg\}
\\ \nonumber
\hat{f}_3^Q(\hsi,\hti)&=&
\frac{16}{{\Delta^\prime}^2}\Bigg\{
 -2 \Delta^2 \Rp 
\left(\frac{m_2^2}{\hsi^2}+\frac{m_1^2}{\hti^2}+\frac{\spp}{\hsi \hti}\right)  
+ 2 m_1 m_2 \Rm 
\Bigg(\frac{\hsi+\spm}{\hti}+\frac{\hsi-\smp}{\hsi}\Bigg)
\\ \nonumber
&+& \Rp \Bigg(
\frac{- 2 m_1^2 \hsi \spm}{\hti^2}
+\frac{-\hsi^2-4 (\Delta^2- m_1^2 \smp) -3 \hsi \spm}{\hti}
\\ 
&+& \left[2(m_1^2-m_2^2)-\frac{2(\Delta^2+m_2^2 \spm)}{\hsi} \right] 
+\hti \frac{\hsi-\smp}{\hsi}
\Bigg) \Bigg\}
\end{eqnarray}
where we conveniently use
the shorthands $\Delta \equiv \Delta[m_1^2,m_2^2,-Q^2]$ 
and $\Delta^\prime \equiv \Delta[m_1^2,\hat{s},-Q^2]$. 
In order to obtain the inclusive results $\hat{f}_i^Q(\hsi)$ in
Appendix \ref{hqini} the 
$\hat{f}_i^Q(\hsi,\hti)$ have to be integrated over $0\le y \le 1$, i.e.
\begin{equation}
\hat{f}_i^Q(\hsi)= \int_0^1 dy\ \hat{f}_i^Q(\hsi,\hti)\ \ \ ,
\end{equation}
where $y$ is defined via the partonic center of 
mass scattering angle $\theta^\ast$ and related to $\hti$ through
\begin{eqnarray} \nonumber
y &\equiv& \frac{1}{2}\ (1+\cos \theta^\ast) \\
\hti&=&
\frac{\hsi}{\hsi+m_2^2}\ \delp\ (y-y_0)\ \ \ ,
\end{eqnarray}
with $y_0=[1+(\spp+\hsi)/\delp]/2$ being the would--be collinear pole of the
${\hat{t}}$--channel propagator.

%% file: phd_app_box.tex
\chapter{Limits of the Doubly Virtual Box}\label{app:limits}
Due to the particular choice of variables, the results in Eq.~(\ref{eq:LO-box}) 
are especially well suited for deriving various important limits 
wide--spread over
the literature which are relevant for some of the chapters in this thesis:
\begin{itemize}
\item Most important for our purposes is the Bjorken limit 
($Q^2 \to \infty, \nu \to \infty, x={\rm const}$) 
in which we can study structure functions of the real and virtual photon
as has already been discussed in Chapter \ref{chap:twogam}. 
\item Also needed is the real photon $P^2=0$ ($\xp = 0$) case 
in which the general virtual box results in (\ref{eq:LO-box}) reduce to the
standard box--diagram $\vgg$ expressions
for a {\em real} photon $\gamma \equiv \gamma(P^2 = 0)$. 
Keeping the full mass dependence in (\ref{eq:LO-box}) we obtain expressions relevant
for the heavy quark contribution to the photon structure functions. 
\item Finally, the general light quark mass limit (for arbitrary $P^2$, $Q^2$) can be
easily obtained from (\ref{eq:LO-box}) by setting $m=0$ ($\lambda = 0$, $\vone = 1$) and
needs no separate discussion.
\end{itemize}

In the following, our main concern will lie on the Bjorken limit.
Practically this limit means that $Q^2$ is much larger than the other scales $m^2$ and $P^2$.
Beside the general case $m^2,P^2 \ll Q^2$ which will be studied in Section \ref{sec:bjorken1}
additional orderings 
$m^2=0, P^2 \ll Q^2$ (Section \ref{sec:bjorken2}) and $P^2=0, m^2 \ll Q^2$ (Section \ref{sec:bjorken3})
are of interest and the expressions further simplify 
under these circumstances.
One must be careful in handling these limits because for some of the
box cross sections the result depends on which of the two limits $m \to 0$ and
$P^2 \to 0$ is taken first.
For example, below we will find the following asymptotic expressions for $\sigtl$ derived
from (\ref{eq:LO-box}): 
\begin{equation*}
0 = \lim_{m^2 \to 0} \lim_{P^2 \to 0} \sigtl \neq \lim_{P^2 \to 0} \lim_{m^2 \to 0} \sigtl 
= N_c e_q^4 \frac{4 \pi \alpha^2}{Q^2} 4 x^2 (1-x)  
\end{equation*}
Mathematically, the origin of such a behavior is easily identified.
Terms like $\displaystyle \frac{4 x \xp}{4 x \xp + \lambda \vtwo^2}$ 
occurring in (\ref{eq:LO-box})
(not only as the argument of the logarithm) require a careful treatment.  
In general they are {\em not} negligible even for small $m^2$ and $P^2$ 
and, viewed as a function of $m^2$ and $P^2$, they are discontinuous at
$(m^2,P^2)=(0,0)$.
Finally, in Section \ref{sec:bjorken3} we also derive expressions
for the heavy quark contributions to the photon structure
functions in the real photon limit by setting $P^2 = 0$ ($\xp=0$) but keeping the full mass 
dependence in (\ref{eq:LO-box}). 

All results for the photon--photon cross sections will be given for a single quark 
with charge $e_q$ and mass $m$.
The photon structure functions $\sfs{i}{},\, (\mathrm{i}=1,2,\mathrm{L})$ can be obtained from
these expressions with help of the relations in Section \ref{sec:photonsfs} which
simplify in the Bjorken limit ($P^2 \ll Q^2 \Rightarrow \vtwo \simeq 1$):
\begin{alignat}{2}\label{eq:sfs1}
\sfs{2}{\gam[T]} &= \Normc \sigiit\, , & \qquad \sfs{L}{\gam[T]} &= \Normc \siglt 
\nonumber\\
\sfs{2}{\gam[L]} &= \Normc \sigiil\, , & \qquad \sfs{L}{\gam[L]} &= \Normc \sigll 
\end{alignat} 
with $\sigiit = \sigtt + \siglt$, $\sigiil = \sigtl + \sigll$.
The commonly utilized expressions for a spin--averaged target photon are given by
\begin{equation}\label{eq:sfs2}
\sfs{i}{<\gam>}=\sfs{i}{\gam[T]} - \tfrac{1}{2}\sfs{i}{\gam[L]}, \, (\mathrm{i}=1,2,\mathrm{L})\ .
\end{equation}
Finally, the structure function $\sfs{1}{}$ can be deduced from 
$\sfs{L}{} = \sfs{2}{}- 2 x \sfs{1}{}$.
Since the structure functions are (apart from the normalization factor $Q^2/4 \pi^2 \alpha$)
simple linear combinations of the photon--photon cross sections they will only 
be written out in those cases where they are needed explicitly in the thesis. 
%
\section{General Bjorken Limit: $m^2,P^2 \ll Q^2$}\label{sec:bjorken1}
In the general Bjorken limit the normalization factor $N$ given in Eq.~(\ref{eq:sigab}) and
the logarithm $L$ from Eq.~(\ref{eq:log}) are given by
\begin{equation}
\begin{gathered}
N = 4\pi N_c e_q^4 \frac{4 \pi \alpha^2}{Q^2} x\ (1+\Ord(\xp))
\\
L = \ln \frac{4}{4x\xp+\lambda}+\Ord(\xp,\lambda)\ .
\end{gathered}
\end{equation}
Keeping this in mind and using
$\vone = 1 + \Ord(\lambda)$, $\vtwo = 1 + \Ord(\xp)$
the following results can be easily deduced from
Eq.~(\ref{eq:LO-box}): 
\begin{align}\label{eq:Bj1}
\sigtt & = 
\Norm
\left\{\Big[x^2+(1-x)^2\Big]L+4x(1-x)-1
-\frac{4 x \xp}{4x\xp+\lambda}+\Ord(\xp,\lambda)\right\}
\nonumber\\
\siglt & =\Norm \left\{4x(1-x)+\Ord(\xp,\lambda)\right\}
\nonumber\\
\sigtl & = \Norm
\left\{4 x(1-x) \frac{4x\xp}{4x\xp+\lambda}+\Ord(\xp,\lambda)\right\}
\nonumber\\
\tautt & = \Norm
\left\{-2 x^2+\Ord(\xp,\lambda)\right\}
\nonumber\\
\tauatt & = \Norm
\left\{(2x-1)L+3-4x-\frac{4x\xp}{4x\xp+\lambda}
+\Ord(\xp,\lambda)\right\}
\ .
\end{align}
The remaining expressions are suppressed by powers of $\xp = x P^2/Q^2$ 
(see Eq.~(\ref{eq:LO-box})). 

At the price of a slightly worse approximation (at larger $x$) one can further use
\begin{equation}
\begin{gathered}
\frac{4 x \xp}{4 x \xp + \lambda} = \frac{P^2 x (1-x)}{P^2 x(1-x) + m^2} 
+\Ord(\tfrac{P^2}{W^2},\lambda)\, ,
\\
L = \ln \frac{Q^2 (1-x)}{x [P^2 x (1-x) + m^2]} 
+\Ord(\tfrac{P^2}{W^2},\lambda)\, .
\end{gathered}
\end{equation}
For example we can write: 
\begin{align}\label{eq:Bj1b}
\sigtt & = \Norm
\bigg\{\Big[1-2x(1-x)\Big]\ln \frac{Q^2 (1-x)}{x [P^2 x (1-x) + m^2]} 
+4 x (1-x)-1 
\nonumber\\
&-\frac{P^2 x (1-x)}{P^2 x(1-x) + m^2} 
+\Ord(\tfrac{P^2}{W^2},\lambda)\bigg\}\ .
\end{align}
%
\section{$m^2=0, P^2 \ll Q^2$}\label{sec:bjorken2}
In this section we consider the
asymptotic virtual ($P^2 \ne 0$) box expressions for light quarks in the Bjorken limit
for which the expressions in Eq.~(\ref{eq:Bj1}) further reduce.
Noticing that the logarithm $L$ is given by $L = \ln \frac{Q^2}{P^2 x^2}$ we can
write in this case (neglecting terms of the order $\Ord(\xp)$): 
\begin{align}
\sigtt & \simeq  \Norm \left\{[x^2+(1-x)^2]\ln \frac{Q^2}{P^2 x^2} +4x(1-x)-2\right\}
\nonumber\\
\sigtl & \simeq \siglt \simeq \Norm [4 x (1-x)]
\nonumber\\
\tautt & \simeq \Norm [-2 x^2]
\nonumber\\
\tauatt & \simeq  \Norm \left\{(2 x-1)\ln \frac{Q^2}{P^2 x^2}+2-4x\right\}\ .
\end{align}

Summing over $q=u,d,s$ and utilizing Eqs.~(\ref{eq:sfs1}) and (\ref{eq:sfs2})
we recover the well known asymptotic results for the virtual ($P^2 \ne 0$) box
structure functions for the light $q=u,d,s$ quarks in the Bjorken limit $P^2/Q^2 \ll 1$:
\begin{align}
\sfs{2,box}{\gam[T],\ell}(x,Q^2,P^2) &\simeq 
N_c \sum e_q^4\, \frac{\alpha}{\pi} x 
  \left\{ \left[ x^2+(1-x)^2\right] \, \ln \frac{Q^2}{P^2x^2} + 8x(1-x)-2 \right\}
\label{eq:f2boxt}\\
\sfs{2,box}{\gam[L],\ell}(x,Q^2,P^2) &\simeq 
N_c \sum e_q^4\, \frac{\alpha}{\pi} x \left\{ 4 x (1-x) \right\} 
\label{eq:f2boxl}\\
\sfs{2,box}{<\gam>,\ell}(x,Q^2,P^2) &\simeq 
N_c \sum e_q^4\, \frac{\alpha}{\pi} x 
  \left\{ \left[ x^2+(1-x)^2\right] \, \ln \frac{Q^2}{P^2x^2} + 6x(1-x)-2 \right\}\, . 
\label{eq:f2boxav}
\end{align}

%
\section{Real Photon Limit: $P^2=0$}\label{sec:bjorken3}
For $P^2=0$ the virtual box results in (\ref{eq:LO-box}) reduce to
the standard box--diagram $\vgg$ expressions
for a {\em real} photon $\gam \equiv \gam(P^2=0)$.

The heavy quark contribution becomes,
utilizing $\xp = 0$, $\vtwo = 1$ and $\displaystyle \lambda = \frac{4 m_h^2 x}{Q^2 (1-x)}$,
\begin{align}\label{eq:really-heavy}
\sigtt & = \Normh
\Theta(\vone^2)\ \bigg\{\left[x^2+(1-x)^2+x(1-x) \frac{4 m_h^2}{Q^2} -x^2\frac{8 m_h^4}{Q^4}\right]
\ln \frac{1+\vone}{1-\vone}
\nonumber\\
&+ \vone\left[4x(1-x)-1-x(1-x)\frac{4 m_h^2}{Q^2}\right]
\bigg\}
\nonumber\\
%
\siglt & =\Normh
\Theta(\vone^2)\ 
\bigg\{-x^2 \frac{8 m_h^2}{Q^2} \ln \frac{1+\vone}{1-\vone}
+\vone\ 4x(1-x)
\bigg\}
\nonumber\\
%
%
%
%
\tautt & =\Normh
\Theta(\vone^2)\ \bigg\{
\left[-x^2 \frac{8 m_h^2}{Q^2}-x^2 \frac{8 m_h^4}{Q^4}\right]\ln \frac{1+\vone}{1-\vone}
-\vone \left[2 x^2+x(1-x) \frac{4 m_h^2}{Q^2}\right]
\bigg\}
\nonumber\\
%
%
\tauatt & = \Normh
\Theta(\vone^2)\ \bigg\{(2x-1)\ln \frac{1+\vone}{1-\vone}
+ \vone(3-4x)
\bigg\}\ .
%
\end{align}
i.e., according to (\ref{eq:sfs2}) (or (\ref{eq:sfs1}))
\begin{align}\label{eq:vgamA.8}  
\sfs{2,box}{\gam,h}(x,Q^2)  & =  3\, e_h^4\, \frac{\alpha}{\pi}x\, \Theta(\vone^2) 
     \left\{ \left[ x^2+(1-x)^2+x(1-3x)\, \frac{4m_h^2}{Q^2}
-x^2\, \frac{8m_h^4}{Q^4}\right] \, \ln\, \frac{1+\vone}{1-\vone} \,\right.
\nonumber\\ 
& \left. 
+ \vone\left[ 8x(1-x)-1-x(1-x)\,  \frac{4m_h^2}{Q^2}\right] \right\}
\nonumber\\ 
\sfs{L,box}{\gam,h}(x,Q^2)  & =  3\, e_h^4\, \frac{\alpha}{\pi}x\, \Theta(\vone^2) 
\left\{-x^2 \frac{8 m_h^2}{Q^2} \ln \frac{1+\vone}{1-\vone}
+\vone\ 4x(1-x)
\right\}
\end{align} 
which are the 
familiar massive Bethe--Heitler expressions 
\cite{cit:Wit-7601,*cit:GR-7901} relevant for the  
heavy quark contributions to the structure functions of real photons 
(cf.\ \cite{cit:GRSc99}, for example).  


In the light quark sector where $\lambda \ll 1$, i.e. $m^2\equiv m_q^2 \ll Q^2$, 
the logarithm can be written as
\begin{displaymath}
L = \ln \frac{1+\vone}{1-\vone} = \ln \frac{Q^2 (1-x)}{m_q^2 x} +\Ord(\lambda)\ .
\end{displaymath}
and  we obtain from (\ref{eq:Bj1}) or (\ref{eq:really-heavy}) the following 
results (neglecting terms of the order $\Ord(\lambda)$):
\begin{align}
\sigtt & \simeq \Norm
 \bigg\{[x^2+(1-x)^2]\ln \frac{Q^2 (1-x)}{m_q^2 x} + 4 x (1-x) - 1\bigg\}
\nonumber\\
\siglt & \simeq \Norm [4 x (1-x)]
\nonumber\\
%
%
\tautt & \simeq \Norm [-2 x^2]
\nonumber\\
%
%
\tauatt & \simeq \Norm \bigg\{(2x-1)\ln \frac{Q^2 (1-x)}{m_q^2 x}+3-4x\bigg\}\ ,
%
\end{align}
i.e., according to (\ref{eq:sfs2}) (or (\ref{eq:sfs1}))
\begin{equation}\label{eq:f2boxreal}
\sfs{2,box}{\gam,\ell}(x,Q^2)  \simeq N_c \sum e_q^4\, \frac{\alpha}{\pi} x  
 \bigg\{[x^2+(1-x)^2]\ln \frac{Q^2 (1-x)}{m_q^2 x} + 8 x (1-x) - 1\bigg\}\ .
\end{equation}
Note also that $\sigtl$ vanishes like $\sigtl \propto P^2/m_q^2$. 

%% file: phd_app_para.tex
\chapter{Parametrizations}\label{app_para}
\section{Pion Distributions}
\label{sec:pipara}
\subsection{Parametrization of LO Parton Distributions}
Defining \cite{cit:GRV98}

\begin{equation}
s\equiv \ln \frac{\ln [Q^2/(0.204\, {\rm{GeV}})^2]}
    {\ln [\mu_{\rm{LO}}^2/(0.204\, {\rm{GeV}})^2]}
\label{pipara1}
\end{equation}

\noindent to be evaluated for $\mu_{\rm{LO}}^2=0.26$ GeV$^2$, all our 
resulting
pionic parton distributions can be expressed by the following simple
parametrizations,
valid for $0.5$ \raisebox{-0.1cm}{$\stackrel{<}{\sim}$} $Q^2$ 
\raisebox{-0.1cm} {$\stackrel{<}{\sim}$} $10^5$ GeV$^2$ 
\mbox{(i.e. $0.31 \leq s$ \raisebox{-0.1cm}{$\stackrel{<}{\sim}$} 2.2)} 
and $10^{-5}$ \raisebox{-0.1cm}{$\stackrel{<}{\sim}$} $x\,<\, 1$.  
For the valence distribution we take
\begin{equation}
x\, v^{\pi}(x,Q^2) = N\, x^a(1+A\sqrt{x}+Bx)(1-x)^D
\label{pipara2}
\end{equation}
with
\begin{eqnarray}
N & = & 1.212 + 0.498\,s + 0.009\, s^2 \nonumber\\
a & = & 0.517 - 0.020\,s \nonumber\\
A & = & -0.037 - 0.578\,s\nonumber\\
B & = & 0.241 + 0.251\,s\nonumber\\
D & = & 0.383 + 0.624\,s\, .
\label{pipara3}
\end{eqnarray}
The gluon and light sea--quark distributions are parametrized as
\begin{equation}
x\, w^{\pi}(x,Q^2)=\left[ x^a \left( A+B\sqrt{x} +Cx \right) 
   \left(\ln \frac{1}{x} \right)^b + s^{\alpha} \,\,{\rm{exp}}
     \left( -E+\sqrt{E's^{\beta}\ln \frac{1}{x}} \right)\right]
       (1-x)^D.
\label{pipara4}
\end{equation}
%
For $w=g$
\begin{equation}
\begin{array}{lcllcl}
\alpha & = & 0.504, & \beta & = & 0.226,\\[1mm]
a & = & 2.251 - 1.339\,\sqrt{s}, &  b & = & 0,\\[1mm]
A & = & 2.668 - 1.265\, s + 0.156\, s^2, & B  & = & -1.839 + 0.386
  \,s,\\[1mm]
C & = & -1.014 + 0.920\,s - 0.101\, s^2, &  D & = & -0.077 + 1.466
  \,s,\\[1mm]
E & = & 1.245 + 1.833\,s, & E' &  = & 0.510 + 3.844\,s\,,
\label{pipara5}
\end{array}
\end{equation}
and for the light sea $w=\bar{q}$
\begin{equation}
\begin{array}{lcllcl}
\alpha & = & 1.147,  & \beta & = & 1.241,\\[1mm]
a & = & 0.309 - 0.134\,\sqrt{s}, &  b & = & 0.893-0.264\, \sqrt{s},\\[1mm]
A & = & 0.219 - 0.054\, s, &  B & = & -0.593 + 0.240\,s,\\[1mm]
C & = & 1.100 - 0.452\,s, & D & = & 3.526 + 0.491\,s,\\[1mm]
E & = & 4.521 + 1.583\, s, &  E' & = & 3.102\, .
\label{pipara6}
\end{array}
\end{equation}
The strange sea distribution $s^{\pi}=\bar{s}\,^{\pi}$ is parametrized 
as
\begin{equation}
x\bar{s}\,^{\pi}(x,Q^2) = \frac{s^{\alpha}}{(\ln \frac{1}{x})^a}\,
\left( 1 + A\sqrt{x}+Bx\right)(1-x)^D\, {\rm{exp}}\, 
\left( -E+\sqrt{E's^{\beta}\ln \frac{1}{x}}\, \right)
\label{pipara7}
\end{equation}
with
\begin{equation}
\begin{array}{lcllcl}
\alpha & = & 0.823,  & \beta & = & 0.650,\\[1mm]
a & = & 1.036 - 0.709\,s, &  A & = & -1.245+0.713\,s,\\[1mm]
B & = & 5.580 - 1.281\,s, &  D & = & 2.746 - 0.191\,s,\\[1mm]
E & = & 5.101 + 1.294\,s, &  E' & = & 4.854-0.437\, s\, .
\label{pipara8}
\end{array}
\end{equation}
\vspace{0.5cm}
%


\subsection{Parametrization of NLO($\overline{{\rm{MS}}}$)
Parton Distributions}
Defining \cite{cit:GRV98}
\begin{equation}
s\equiv \ln \frac{\ln [Q^2/(0.299\, {\rm{GeV}})^2]}
      {\ln [\mu_{\rm{NLO}}^2/(0.299\, {\rm{GeV}})^2]}
\label{pipara9}
\end{equation}

\noindent to be evaluated for $\mu_{\rm{NLO}}^2=0.40$ GeV$^2$, our NLO 
predictions
can be parametrized as the LO ones and are similarly valid for
0.5 \raisebox{-0.1cm}{$\stackrel{<}{\sim}$} $Q^2$ 
\raisebox{-0.1cm}{$\stackrel{<}{\sim}$} $10^5$ GeV$^2$ 
(i.e. 0.14 \raisebox{-0.1cm}{$\stackrel{<}{\sim}$} s 
\raisebox{-0.1cm}{$\stackrel{<}{\sim}$} 2.38) 
and \mbox{$10^{-5}$ \raisebox{-0.1cm}{$\stackrel{<}{\sim}$} $x\,<\, 1$}.
The valence distribution is given by (\ref{pipara2}) with
\begin{eqnarray}
N & = & 1.500 + 0.525\,s - 0.050\, s^2 \nonumber\\
a & = & 0.560 - 0.034\,s \nonumber\\
A & = & -0.357 - 0.458\,s\nonumber\\
B & = & 0.427 + 0.220\,s\nonumber\\
D & = & 0.475 + 0.550\,s\, .
\label{pipara10}
\end{eqnarray}
The gluon and light sea distributions are parametrized as in (\ref{pipara4}) 
where for $w=g$
\begin{equation}
\begin{array}{lcllcl}
\alpha & = & 0.793, & \beta & = & 1.722,\\[1mm]
a & = & 1.418 - 0.215\sqrt{s}, &  b & = & 0,\\[1mm]
A & = & 5.392 + 0.553\,s - 0.385\, s^2, & B  & = & -11.928 + 1.844\,s,\\[1mm]
C & = & 11.548 - 4.316\,s + 0.382\, s^2, &  D & = & 1.347 + 1.135\,s,\\[1mm]
E & = & 0.104 + 1.980\,s, & E' &  = &2.375 - 0.188\,s,
\label{pipara11}
\end{array}
\end{equation}
and for the light sea $w=\bar{q}$
\begin{equation}
\begin{array}{lcllcl}
\alpha & = & 1.118,  & \beta & = & 0.457,\\[1mm]
a & = & 0.111 - 0.326\,\sqrt{s}, &  b & = & -0.978-0.488\,\sqrt{s},\\[1mm]
A & = & 1.035 - 0.295\,s, & B & = & -3.008 + 1.165 \,s,\\[1mm]
C & = & 4.111 - 1.575\,s, &  D & = & 6.192 + 0.705\,s,\\[1mm]
E & = & 5.035 + 0.997\,s, &  E' & = & 1.486 + 1.288\,s\, .
\label{pipara12}
\end{array}
\end{equation}
The strange sea distribution is parametrized as in (\ref{pipara7}) with
\begin{equation}
\begin{array}{lcllcl}
\alpha & = & 0.908,  & \beta & = & 0.812,\\[1mm]
a & = & -0.567 - 0.466\,s, &  A & = & -2.348 + 1.433\,s,\\[1mm]
B & = & 4.403, & D & = & 2.061,\\[1mm]
E & = & 3.796 + 1.618\,s, &  E' & = & 0.309 + 0.355\,s\, .
\label{pipara13}
\end{array}
\end{equation}

Let us recall that in the light quark sector $u_v^{\pi^+}=\bar{d}\,_v^{\pi^+}
= \bar{u}\,_v^{\pi^-}=d_v^{\pi^-},\,\,\,
\bar{u}\,^{\pi^+}=d^{\pi^+} = u^{\pi^-}=\bar{d}\,^{\pi^-}$ and
$f^{\pi^0}=(f^{\pi^+}+f^{\pi^-})/2$.

\section{Photon Distributions}
\label{sec:gampara}
Simple analytic parametrizations in LO and NLO of the `
hadronic' piece of the real and virtual photonic parton distributions, 
being proportional to $f^{\pi}(x,Q^2)$ in 
Eqs.\ (\ref{eq:gam13}) and (\ref{eq:gam25}), respectively,
are already known according to the 
parametrizations for $f^{\pi}(x,Q^2)$ in Appendix \ref{sec:pipara}.  
Therefore we only need to
parametrize the remaining `pointlike' components in Eqs.\ (\ref{eq:gam13}) and
(\ref{eq:gam25}).

\subsection{Parametrization of LO `pointlike' Photonic
Parton Distributions}

In LO the $Q^2$ dependence of the `pointlike' 
$f^{\gamma}_{\pl}(x,Q^2)$
term in (\ref{eq:gam13}) enters, apart from an overall $1/\alpha_s(Q^2)$ factor,
merely via the combination 
$L\equiv\alpha_s(Q^2)/\alpha_s(\mu_{\rm{LO}}^2)$ 
as is evident, for example, from Eq.\ (2.12) 
in Ref.\ \cite{cit:GRV-9201}. Therefore, we prefer to parametrize the
quantity $f_{\pl}^{\gamma}(x,Q^2)$ in terms of
\begin{equation}
s\equiv \ln\, \frac{\ln\,[Q^2/(0.204\, {\rm{GeV}})^2]}
    {\ln\,[\mu_{\rm{LO}}^2/(0.204\, {\rm{GeV}})^2]}
\label{gampara1}
\end{equation}

\noindent where \cite{cit:GRV98} $\mu_{\rm{LO}}^2=0.26$ GeV$^2$, which
will later provide us a parametrization also for the virtual `pointlike'
component in (\ref{eq:gam25}).  Our resulting `pointlike' distributions 
in Eq.\ (\ref{eq:gam13})
can be expressed by the following simple parametrizations, valid for
$0.5$ \raisebox{-0.1cm}{$\stackrel{<}{\sim}$} $Q^2$ 
\raisebox{-0.1cm} {$\stackrel{<}{\sim}$} $10^5$ GeV$^2$ 
(i.e.\  0.31 \raisebox{-0.1cm}{$\stackrel{<}{\sim}$} $s$  
\raisebox{-0.1cm}{$\stackrel{<}{\sim}$} $2.2$) and 
$10^{-5}$ \raisebox{-0.1cm}{$\stackrel{<}{\sim}$} $x < 1$ :   
\begin{eqnarray}
\frac{1}{\alpha}\, x\, f_{\pl}^{\gamma}(x,Q^2) & = & \frac{9}{4\pi}\, \ln\,
  \frac{Q^2}{(0.204\,\,{\rm{GeV}})^2}\, \Bigg[ s^{\alpha}x^a(A+B\sqrt{x} 
     + C\,x^b)  \nonumber\\ 
 & &    + s^{\alpha'}{\rm{exp}}\Bigg( -E+\sqrt{E's^{\beta}\ln\,
      \frac{1}{x}}\, \Bigg)\, \Bigg] \, (1-x)^D 
\label{gampara2}
\end{eqnarray}
where for $f_{\pl}^{\gamma}=u_{\pl}^{\gamma}=\bar{u}_{\pl}^{\gamma}$ 
\begin{equation}
\begin{array}{lcllcl}
\alpha & = & 0.897, & \alpha' & = & 2.626,\\[1mm]
\beta & = & 0.413, & & &\\[1mm]
a & = & 2.137 - 0.310\,\sqrt{s}, &  b & = & -1.049 + 0.113\, s,\\[1mm]
A & = & -0.785 + 0.270\,\sqrt{s}, &  B & = &  0.650 - 0.146\, s,\\[1mm]
C & = & 0.252 - 0.065\,\sqrt{s}, &  D & = & -0.116 + 0.403\, s -
   0.117\,s^2,\\[1mm]
E & = & 6.749 + 2.452\,s - 0.226\, s^2, & E' &  = & 1.994\,s - 0.216\,s^2\,,
\label{gampara3}
\end{array}
\end{equation}
for $f_{\pl}^{\gamma} = d_{\pl}^{\gamma} = \bar{d}\,^{\gamma}_{\pl}
  = s_{\pl}^{\gamma} = \bar{s}\,^{\gamma}_{\pl}$
\begin{equation}
\begin{array}{lcllcl}
\alpha & = & 1.084, & \alpha' & = & 2.811,\\[1mm]
\beta & = & 0.960, & & &\\[1mm]
a & = & 0.914,  &  b & = & 3.723 - 0.968\, s,\\[1mm]
A & = & 0.081 - 0.028\,\sqrt{s}, &  B & = &  -0.048,\\[1mm]
C & = & 0.094 - 0.043\,\sqrt{s}, &  D & = &  0.059 + 0.263\, s 
          - 0.085\,s^2,\\[1mm]
E & = & 6.808 + 2.239\,s - 0.108\, s^2, & E' &  = & 1.225 + 0.594\,s
          - 0.073\,s^2\,,
\label{gampara4}
\end{array}
\end{equation}
and for $f_{\pl}^{\gamma}=g_{\pl}^{\gamma}$
\begin{equation}
\begin{array}{lcllcl}
\alpha & = & 1.262, & \alpha' & = & 2.024,\\[1mm]
\beta & = & 0.770, & & &\\[1mm]
a & = & 0.081,  &  b & = & 0.848,\\[1mm]
A & = & 0.012 + 0.039\,\sqrt{s}, &  B & = &  -0.056 - 0.044\,s,\\[1mm]
C & = & 0.043 + 0.031\,s, &  D & = &  0.925 + 0.316\,s,\\[1mm]
E & = & 3.129 + 2.434\,s - 0.115\, s^2, & E' &  = & 1.364 + 1.227\,s
          - 0.128\,s^2\,.
\label{gampara5}
\end{array}
\end{equation}

With these parametrizations at hand in terms of $s$ in (\ref{gampara1}), the
appropriate ones for the `pointlike' distributions 
$f_{\pl}^{\gamma(P^2)}(x,Q^2)$ of a {\underline{virtual}} photon appearing 
in Eq.\ (\ref{eq:gam25}) are 
simply given by the same expressions above where in 
(\ref{gampara1}) $\mu_{\rm{LO}}^2$
has to be replaced by $\tilde{P}^2={\rm{max}}(P^2,\mu_{\rm{LO}}^2)$.
As discussed in Sec.~\ref{sec:vgaminput}, these parametrizations can, within sufficient
accuracy, also be used for the parton distribution of {\underline{virtual}}
photons in NLO.

\subsection{Parametrization of NLO `pointlike' Photonic
Parton Distributions}

In NLO the $Q^2$ dependence of the `pointlike' distributions
of real photons in (\ref{eq:gam13}), $f_{\pl}^{\gamma}(x,Q^2)$, can be easily
described in terms of the following `effective' logarithmic ratio
\begin{equation}
s\equiv \ln\, \frac{\ln\,[Q^2/(0.299\, {\rm{GeV}})^2]}
    {\ln\,[\mu_{\rm{NLO}}^2/(0.299\, {\rm{GeV}})^2]}
\label{gampara6}
\end{equation}
to be evaluated for $\mu_{\rm{NLO}}^2=0.40$ GeV$^2$.  Our 
NLO(DIS$_{\gamma}$) predictions can now be parametrized as the LO ones
and are similarly valid for 
$0.5$ \raisebox{-0.1cm}{$\stackrel{<}{\sim}$} $Q^2$ \raisebox{-0.1cm}
{$\stackrel{<}{\sim}$} $10^5$ GeV$^2$ (i.e.\ 0.14 \raisebox{-0.1cm}
{$\stackrel{<}{\sim}$} $s$ \raisebox{-0.1cm}{$\stackrel{<}{\sim}$}
$2.38$) and $10^{-5}$ \raisebox{-0.1cm}{$\stackrel{<}{\sim}$} $x < 1$.
For convenience we include now the NLO $\alpha_s$ in the r.h.s.\ of
(\ref{gampara2}), i.e.
\begin{eqnarray}
\frac{1}{\alpha}\, x\, f_{\pl}^{\gamma}(x,Q^2) & = & \Bigg[ s^{\alpha} x^a
  (A+B\sqrt{x}+Cx^b) \nonumber\\ 
& &  + s^{\alpha'}{\rm{exp}} \Bigg( -E + \sqrt{E's^{\beta}
   \,\ln\,\frac{1}{x}}\, \Bigg) \Bigg]\, (1-x)^D
\label{gampara7}
\end{eqnarray}
where for $f_{\pl}^{\gamma}=u_{\pl}^{\gamma}=\bar{u}_{\pl}^{\gamma}$ 
\begin{equation}
\begin{array}{lcllcl}
\alpha & = & 1.051,  & \alpha' & = & 2.107,\\[1mm]
\beta & = & 0.970, & & &\\[1mm]
a & = & 0.412 - 0.115\,\sqrt{s}, &  b & = & 4.544 - 0.563\, s,\\[1mm]
A & = & -0.028\,\sqrt{s} + 0.019\,s^2, &  B & = &  0.263 + 0.137\, s,\\[1mm]
C & = & 6.726 - 3.264\,\sqrt{s} - 0.166\,s^2, & D & = & 1.145 - 0.131\, 
   s^2,\quad\quad\quad\,\,\\ [1mm]
E & = & 4.122 + 3.170\,s - 0.598\, s^2, & E' &  = & 1.615\,s - 0.321\,s^2\,,
\label{gampara8}
\end{array}
\end{equation}
for $f_{\pl}^{\gamma} = d_{\pl}^{\gamma} = \bar{d}\,^{\gamma}_{\pl}
  = s_{\pl}^{\gamma} = \bar{s}\,^{\gamma}_{\pl}$
\begin{equation}
\begin{array}{lcllcl}
\alpha & = & 1.043, & \alpha' & = & 1.812,\\[1mm]
\beta & = & 0.457, & & &\\[1mm]
a & = & 0.416 - 0.173\,\sqrt{s}, &  b & = & 4.489 - 0.827\, s,\\[1mm]
A & = & -0.010\,\sqrt{s} + 0.006\,s^2, &  B & = &  0.064 + 0.020\,s,\\[1mm]
C & = & 1.577 - 0.916\,\sqrt{s}, &  D & = &  1.122 - 0.093\,s - 0.100
         \,s^2,\\[1mm]
E & = & 5.240 + 1.666\,s - 0.234\,s^2, & E' &  = & 1.284\,s,
\label{gampara9}
\end{array}
\end{equation}
and for $f_{\pl}^{\gamma}=g_{\pl}^{\gamma}$
\begin{equation}
\begin{array}{lcllcl}
\alpha & = & 0.901, & \alpha' & = & 1.773,\\[1mm]
\beta & = & 1.666, & & &\\[1mm]
a & = & 0.844 - 0.820\,\sqrt{s},  &  b & = & 2.302 - 0.474\,s,\\[1mm]
A & = & 0.194, &  B & = &  -0.324 + 0.143\,s,\\[1mm]
C & = & 0.330 - 0.177\,s, &  D & = &  0.778 + 0.502\,s -
         0.154\,s^2,\\[1mm]
E & = & 2.895 + 1.823\,s - 0.441\, s^2, & E' &  = & 2.344 - 0.584\,s\, .
\label{gampara10}
\end{array}
\end{equation}

%% file: phd_acknowledgements.tex
\chapter*{Acknowledgements}

It is a pleasure to thank 
Prof.\ Dr.\ M.\ Gl\"{u}ck and Prof.\ Dr.\ E.\ Reya for their instructive 
scientific guidance and for a fruitful collaboration. 
I have benefitted from their
continuous advice on all topics covered in this thesis.

I am by no means less grateful to Dr.\ Stefan Kretzer for a collaboration 
on the ACOT calculations and for being a pleasant 
office--mate over the 3 1/2 years in which we shared the same office. 
Furthermore, I wish to thank Christoph Sieg for an enjoyable collaboration on 
issues about the photon tensor and the QED--factorization.
I am thankful to the last two persons and Dr.\ Werner Vogelsang
for critically reading parts of the manuscript.

I would like to express my gratitude 
to Jens Noritzsch for generously taking over
part of my teaching load in this semester and
for help concerning computer and \LaTeX\ questions.

Thanks to 
Dr.\ Ingo Bojak for numerous enjoying (physics) discussions
and for providing his \Mathematica\ implementation of the 
Passarino--Veltman--decomposition.

Finally, my thanks go to  
all other members of our working group TIV
and to TIII
and to our secretary Susanne Laurent
for a pleasant working atmosphere.